\documentclass[11pt,letterpaper,titlepage,twoside,openright
,english
,headexclude,footexclude,automark,komastyle,bibtotoc,hyperindex
,headsepline,footsepline,plainfootsepline,bookmarks,BCOR15mm,DIV12,pagesize,tbtags
,abstracton]{scrreprt}

\usepackage{amsmath}
\numberwithin{equation}{section}
\usepackage{babel}
\usepackage{graphicx}
\usepackage{epstopdf}
\usepackage{mathpazo}
\usepackage{amsfonts}
\usepackage{amssymb}
\usepackage{eucal}
\usepackage[format=plain,labelfont=bf,font=normalsize]{caption}
\usepackage{setspace}
\usepackage{units}
\usepackage{scrpage2}
	\clearscrheadfoot
	
	\automark[section]{chapter}
	\setheadsepline{.5pt}
	\setfootsepline{.5pt}
	\ohead[]{\headmark}
	\ofoot[\pagemark]{\pagemark}
	\setfootwidth[0pt]{text}
	\setheadwidth[0pt]{text}
	\typearea[current]{current}
\usepackage{dcolumn}
	\newcolumntype{e}[1]{D{.}{.}{#1}}
\usepackage{hyperref}
	\hypersetup{
		pdftitle={Classical and Quantum Features of the Mixmaster Singularity },
		pdfsubject={},
		pdfauthor={Giovanni Montani, Marco Valerio Battisti, Riccardo Benini, Gianpaolo Imponente},
		pdfkeywords={}
		}

\newcommand\be{\begin{equation}}
\newcommand\ee{\end{equation}}
\newcommand\bea{\begin{eqnarray}}
\newcommand\eea{\end{eqnarray}}
\newcommand\bseq{\begin{subequations}} 
\newcommand\eseq{\end{subequations}}
\newcommand\bcas{\begin{cases}}
\newcommand\ecas{\end{cases}}
\newcommand{\p}{\partial}
\newcommand{\f}{\frac}
\newcommand\beq{\begin{equation}}
\newcommand\eeq{\end{equation}}
\newcommand\tr{\textrm{tr}}
\newcommand\nR{\mathbb{R}}
\newcommand{\reffcite}[1]{\cite{#1}}
\newcommand{\tbl}[1]{\caption{#1}}
\newcommand{\toprule}{\\\hline}
\newcommand{ \botrule}{\hline}
\newcommand{\colrule}{\null}

\newcommand{\ADDRESS}[1]{{\footnotesize #1}}

\newcommand{\sectionric}[1]{\chapter{#1}}
\newcommand{\subsectionric}[1]{\section{#1}}
\newcommand{\subsubsectionric}[1]{\subsection{#1}}

\newenvironment{romanlist}{\begin{itemize}}{\end{itemize}}

\newcommand{\cqg}{Classical and Quantum Gravity}
\newcommand{\prl}{Physical Review Letters}
\newcommand{\pra}{Physical Review A}

\newcommand{\prd}{Physical Review D}
\newcommand{\pre}{Physical Review E}
\newcommand{\apj}{Astrophysical Journal}
\newcommand{\apss}{Astrophysics and Space Science}
\newcommand{\physrep}{Physics Reports}
\newcommand{\mnras}{Monthly Notices of the Royal Academy Society}
\newcommand{\azh}{Astronomicheskii Zhurnal}
\newcommand{\aap}{Astronomy and Astrophysics}
\newcommand{\nat}{Nature Physical Science}
\newcommand{\ptp}{Progress of Theoretical Physics}
\newcommand{\ijtp}{International Journal of Theoretical Physics}
\newcommand{\grg}{General Relativity and Gravitation}
\newcommand{\pla}{Physics Letters A}
\newcommand{\plb}{Physics Letters B}

\newcommand{\ijmpa}{International Journal of Modern Physics A}

\newcommand{\ijmpd}{International Journal of Modern Physics D}

\newcommand{\mpla}{Modern Physics Letters A}

\newcommand{\advphys}{Advances in Physics}
\newcommand{\sovJETP}{Soviet Physics JETP}
\newcommand{\ncb}{Nuclear Physics B}
\newcommand{\JETPl}{JETP Letters}
\newcommand{\zetf}{Zhurnal Eksperimental noi i Teoreticheskoi Fiziki}
\newcommand{\pzetf}{Pis ma Zhurnal Eksperimental noi i Teoreticheskoi Fiziki}
\newcommand{\sjetp}{ Soviet Journal of Experimental and Theoretical Physics}

\begin{document}

\markboth{Montani et al.}
{Classical and Quantum Features of the 
	Mixmaster Singularity}

	\title{Classical and Quantum Features of the 
	Mixmaster Singularity }

		\author{\null\\Giovanni Montani, \protect\\
		\ADDRESS{ ICRA, International Center for Relativistic 
	Astrophysics}\\
	\ADDRESS{ ENEA C.R. Frascati, Dipartimento F.P.N., 
	Frascati, Roma, Italy}\\
	 \ADDRESS{ Dipartimento di Fisica 
		Universit\'a ``Sapienza'', P.le a. Moro 5, 00185 Roma, Italy}\\
	\ADDRESS{ ICRANet, International Relativistic Astrophysics Network, Pescara, Italy}\\
	\ADDRESS{\textit{montani@icra.it}}\\\null\\
		Marco Valerio Battisti, \protect\\
		\ADDRESS{ ICRA, International Center for Relativistic 
	Astrophysics}\\
 	\ADDRESS{ Dipartimento di Fisica 
	 Universit\'a ``Sapienza'', P.le a. Moro 5, 00185 Roma, Italy}\\
\ADDRESS{\textit{battisti@icra.it}}\\\null\\
		Riccardo Benini \protect\\
		\ADDRESS{ ICRA, International Center for Relativistic 
	Astrophysics}\\
 	\ADDRESS{ Dipartimento di Fisica 
	Universit\'a ``Sapienza'', P.le a. Moro 5, 00185 Roma, Italy}\\
	\ADDRESS{ \textit{riccardo.benini@icra.it}}\\\null\\
		Giovanni Imponente\\
		\ADDRESS{ Museo Storico della Fisica e Centro Studi e Ricerche ``E. Fermi'', Roma, Italy}\\
	\ADDRESS{ Dipartimento di Fisica 
	 Universit\'a ``Sapienza'', P.le a. Moro 5, 00185 Roma, Italy}\\
\ADDRESS{\textit{giovanni.imponente@cern.ch}}}
		\date{}
	\maketitle


\begin{abstract}

This review article is devoted to analyze the main properties
characterizing the cosmological singularity associated to the
homogeneous and inhomogeneous Mixmaster model. 
After the introduction
of the main tools required to treat the cosmological
issue,  we review in details  the main results got
along the last forty years on the Mixmaster topic. 
We firstly assess  the classical picture of the homogeneous chaotic 
cosmologies and, after a presentation of the canonical method for 
the quantization, we develop the quantum Mixmaster behavior. 
Finally, we  extend both the classical and quantum features to the
fully inhomogeneous case.
Our survey analyzes the fundamental framework
of the Mixmaster picture and completes it by accounting
for recent and peculiar outstanding results.

\vspace{10mm}

\begin{center}
\end{center}
\end{abstract}

\tableofcontents
\chapter{Preface}

The formulation by Belinski, Khalatnikov and
Lifshitz (BKL) at the end of the Sixties about the
general character of the cosmological singularity
represents one of  the most important contributions
of the Landau school to the development of
Theoretical Cosmology, after the  work by Lifshitz
on the isotropic Universe stability. 
The relevance
of the BKL work relies on two main points:
on one hand, this study provided a piecewise
analytical solution of the Einstein equations, 
improving and implementing
the topological issues of the  Hawking-Penrose
theorems; on the other hand, the dynamical features of
the corresponding  generic cosmological solution exhibit 
a chaotic profile.\\
Indeed the idea that the behavior of the actual
Universe is some how related to an 
inhomogeneous chaotic cosmology appears very far from
what physical insight could suggest. 
In fact,
the Standard Cosmological Model is based  on the
highly symmetric Friedman-Robertson-Walker (FRW)
model, and observations agree with these assumptions,
from the nucleosynthesis of light elements up to the
age when structures formation takes place in the
non-linear regime.

Furthermore a reliable (but to some extent, 
model-dependent) correspondence between theory and
data exists even for an inflationary scenario and its
implications. 
Several questions
about the interpretation of surveys from which inferring the
present status of the matter distribution and dynamics
of the Universe remain open.
However, in the light of an inflationary scenario, 
a puzzling plan appears unless the observation
of early cosmological gravitational waves become viable, 
since such  dynamics cancels out
much of the pre-existent information thus 
realizing local homogeneity and isotropy.\\
The validity of the BKL regime must be settled down just
in this pre-inflationary Universe
evolution, although unaccessible to present observations.
These considerations partly  determined the persistent
idea that this field of investigation could
have a prevalent academical nature, according to the 
spirit of the original derivation of the BKL
oscillatory regime.

In this paper, we review in details and under an updated  point
of view the main features of the Mixmaster model. 
Our aim is to provide a precise picture
of this research line and how it developed over the last
four decades in the framework of General Relativity
and canonical quantization of the gravitational field.

We start with a pedagogical formulation of the fundamental
tools required to understand the treated topics  and
then we pursue a consistent description which,
passing through the classical achievements, touches
the review of very recent understandings in this field.
The main aim of this work is to attract interest,
especially from researchers working on fundamental physics,
to the fascinating features of the classical and quantum inhomogeneous
Mixmaster model. 
Our goal would be to convince
the reader that this dynamical regime is physically well 
motivated and is not only mathematical cosmology, 
effectively concerning
the very early stages of the Universe evolution,
which matches the Planckian era with the inflationary
behavior.

The main reason underlying the relevance of
the Mixmaster dynamics can be identified as
follows: the pre-inflationary evolution fixes the
initial conditions for all the subsequent dynamical stages as they come out
from the quantum era of the Universe, and hence
the oscillatory regime
concerns the mechanism of transition to a classical
cosmology providing information on the origin
of the actual notion of space-time.

Moreover, we emphasize three points strongly supporting
that the very early Universe dynamics was
described by more general paradigms than the isotropic
model, indeed concerning the generic cosmological solution.
\vspace{-2mm}
\begin{itemize}
\item[-]The FRW model is backward unstable with
respect to tensor perturbations, which increase as the
inverse of the scale factor. 
Recent studies have shown how 
such instability holds for scalar perturbations too,
as far as the presence of bulk viscosity is taken into
account.

\item[-]The inflationary scenario offers an efficient
isotropization mechanism, able to reconcile the
primordial inhomogeneous Mixmaster with the local
high isotropy of the sky sphere at the recombination age.

\item[-]In the Planckian era, the quantum fluctuations
can be correlated at most on the causal scale, thus we
should regard global symmetries as approximated
toy models. In this respect, the assumption that the
Universe was born in the homogeneous and isotropic
configuration does not appear well grounded. On a more
realistic point of view, the quantum dynamics has
to be described in absence of any special symmetry, i.e.
by the inhomogeneous Mixmaster Universe.
\end{itemize}
\vspace{5mm}

This article  is organized as follows:\\
in Section 1 we review the fundamental tools for geometrodynamics, 
with particular relevance to the Hamiltonian formulation 
of General Relativity and to Singularity Theorems.\\
In Section 2 we widely discuss the classifications of the 
homogeneous spaces and in particular the dynamics of the types 
I, II and VII of the Bianchi classification.\\
In Section 3 the classical features of the Mixmaster model are 
analyzed in details, both in the field equations formalism and 
in the Hamiltonian one. 
Particular attention is paid to the chaotic properties and to 
the cosmological implications of the homogeneous dynamics, 
also pointing out the effects  on chaos of matter fields and 
of the number of dimensions.\\
In Section 4 we face the quantization of the Mixmaster model 
in different framework: after reviewing the standard Wheeler-DeWitt 
approach, we derive the full energy spectrum in the Dirac quantization 
scheme, and finally we discuss the more recent Loop Quantum Gravity 
approach and the Generalized Uncertainty Principle.\\
In Section 5 we  generalize the previous results to the inhomogeneous case, 
both on the classical and on the quantum level. 
We focus the attention also on several  important topics of the 
Generic Cosmological Solution of the Einstein equations and its 
identification with the Mixmaster model.

\sectionric{Fundamental Tools}

In this Section we will analyze and discuss some of the fundamental 
features of the classical theory of gravity, i.e. the Einstein theory 
of General Relativity (GR). 
In particular, after presenting the Einstein field equations, 
the way how the macroscopic matters fields are treated in such theory 
and a discussion about the tetradic formalism, we will analyze in 
some details the Hamiltonian formulation of the dynamics, 
which will be fundamental in the following. 
Then we will discuss the synchronous reference 
and 
finally 
review the fundamental singularity theorems although 
without entering in the ri\-go\-rous proofs.

Let us fix our convention, following Landau and Lifschitz \cite{LF}: 
the space-time 
indices are given by the Latin letters in the middle of the alphabet, 
i.e. $i,j,k...$ and run from $0$ to $3$ 
(the Lorentzian indices are labeled as $a,b,c$). 
The spatial indices are the Greek letters $\alpha,\beta,\gamma...$ and 
run as $1,2,3$. 
We adopt the signature $(+,-,-,-)$, unless differently specified. 


\subsectionric{Einstein Equations\label{sec:ee}}

The main issue of the Einstein theory of gravity is the dynamical character 
of the space-time metric, described within a fully covariant scheme. 
Assigned a four-dimensional manifold $M$, endowed with space-time coordinates 
$x^i$ and a metric tensor $g_{ij}(x^l)$, its line element reads as
\be
ds^2=g_{ij}dx^idx^j.
\label{1:line}
\ee
This quantity fixes the Lorentzian notion of distances. 
The motion of a free test particle on $M$ corresponds to 
the geodesic equation
\be
\f{du^i}{ds}+\Gamma^i_{jl}u^ju^l=0 \, ,
\label{1:geo}
\ee  
where $u^i\equiv dx^i/ds$ is its four-velocity, 
defined as the vector tangent  to the curve,
 and $\Gamma^i_{jl}=g^{im}\Gamma_{jlm}$ 
are the Christoffel symbols given  by 
\be
\Gamma_{j~lm}=\Gamma_{l~jm}=\f12\left(\p_jg_{lm}+\p_lg_{jm}-\p_mg_{lj}\right) \, .  
\ee
The geodesic character of a curve requires to deal with a 
parallelly transported tangent vector $u^i$. 
However, for a Riemannian manifold this curve extremizes the  
distance functional, i.e. it is provided by the variational principle
\be
\delta\int_M ds=\delta\int_M ds\sqrt{g_{ij}u^iu^j}=0.
\label{1:varpr}
\ee
If a test particle has zero rest mass, its motion is given by $ds=0$ 
and therefore an affine parameter must be introduced to describe the 
corresponding trajectory.\\
The equivalence principle is here recognized as the possibility to 
have vanishing Christoffel symbols at a given point of $M$ 
(or along a whole geodesic curve). 
The space-time curvature is ensured by a non-vanishing Riemann tensor
\be
R^i\,_{jkl}=\p_k\Gamma^i_{jl}-\p_j\Gamma^i_{kl}
+\Gamma^m_{jl}\Gamma^i_{mk}-\Gamma^m_{jk}\Gamma^i_{ml},
\label{riem}
\ee
which has the physical meaning of tidal forces acting on two 
free-falling observers and their effect is expressed by
the geodesic deviation equation
\be
u^l\nabla_l(u^k\nabla_ks^i)=R^i\,_{jlm}u^ju^ls^m,
\label{geodev}
\ee
where $s^i$ is the connecting vector between two nearby geodesics.\\
The Riemann tensor obeys the algebraic cyclic sum
\be
R_{ijkl}+R_{iljk}+R_{iklj}=0
\label{bianchiid1}
\ee
and the first order equations
\be
\nabla_mR_{ijkl}+\nabla_lR_{ijmk}+\nabla_kR_{ijlm}=0.
\label{bianchiid2}
\ee
The constraint in Eq. (\ref{bianchiid2}) is called the Bianchi 
identity, being identically satisfied by the metric tensor, meanwhile 
it is an equation for the Riemann tensor 
where covariant derivatives are expressed in terms of the metric. 
Contracting the Bianchi identity with $g^{ik}g^{jl}$, we get the equation
\be
\nabla_jG^j_i=0 \, ,
\ee
where the Einstein tensor $G_{ij}$  reads as
\be
G_{ij}\equiv R_{ij}-\f12Rg_{ij} \, ,
\label{ein_t}
\ee
in terms of the Ricci tensor  $R_{ij}\equiv R^l\,_{ilj}$ 
and of the scalar of curvature $R\equiv g^{ij}R_{ij}$.\\
In order to get the Einstein equations in the presence of 
matter fields described by a Lagrangian density $\Lambda_m$, 
we must fix a proper action for the gravitational field, 
i.e. for  the metric tensor of the manifold $M$. 
A gravity-matter action, satisfying  the fundamental requirements 
of a covariant geometrical physical theory of the space-time, 
takes the Einstein-Hilbert matter form
\be
S=S_{E-H}+S_m=-\f1{2\kappa}\int_Md^4x\sqrt{-g}\left(R-2\kappa\Lambda_m\right) \, ,
\label{eh_act}
\ee
where $g$ is the determinant of the metric tensor $g_{ij}$ 
and $\kappa$ is the Einstein constant. 
The variation of action (\ref{eh_act}) with respect to $g^{ij}$ can 
be expressed in terms of the vector $\delta W^j$  as \cite{LF}
\begin{equation}
\delta W^j = g^{ik} \delta \Gamma^j_{~ik} - g^{ij} \delta \Gamma^k_{~ik} \, ,
\label{eq:w}
\end{equation}
for which
the relation 
\begin{equation}
\sqrt{-g}g^{ij}\delta R_{ij}=\nabla_j\delta W^j
\end{equation} 
holds (for detailed calculation see \reffcite{wald,Ruff,LF,MAC79}), 
providing  the  field equations in presence of matter 
\be
\label{Eineq}
G_{ij}=\kappa T_{ij}.
\ee
Here $T_{ij}$ denotes the energy-momentum tensor of the 
matter field and reads as
\be
\label{enten}
T_{ij}=\f2{\sqrt{-g}}\left(\f{\delta(\sqrt{-g}\Lambda_m)}{\delta g^{ij}}-\p_l\f{\delta(\sqrt{-g}\Lambda_m)}{\delta(\p_lg^{ij})}\right) \,.
\ee
As a consequence of the Bianchi identity we find the conservation 
law $\nabla_j T^j_i=0$, which describes the motion of  matter 
and arises from the Einstein equations. The gravitational equations
thus imply the equations of motion for the matter itself.\\
Finally, by comparing the static weak field limit of the Einstein 
equations ({\ref{Eineq}}) with the Poisson equation of the Newton 
theory of gravity, we easily get the form of the Einstein constant 
in terms of the Newton constant $G$ as $\kappa=8\pi G$.

The whole analysis discussed so far regards the Einstein geometrodynamics
 formulation of gravity. 
In the gravity-matter action, the cosmological constant 
term is considered as vanishing, although it would be allowed 
by the paradigm of General Relativity. This choice is based  
on the idea that such term should come out from the matter 
contribution itself, eventually  on a quantum level.

\subsectionric{Matter Fields\label{sec:matter}}

In General Relativity, continuous matter fields are described 
by the energy-momentum tensor $T_{ij}$ \cite{wald,LF}. 
We will  focus our attention only on the tensor fields and will 
not discuss   more general entities as the spinor fields 
for which we recommend the exhaustive discussion in \reffcite{wald} and for 
the related bi-spinor calculus  \reffcite{LL4}. 
In particular, we briefly review the cases of the perfect fluid, 
the scalar and the electromagnetic fields. 

The definition of the symmetric tensor $T_{ij}$ as in Eq. (\ref{enten}) 
differs from the energy-momentum ten\-sor ob\-tained from the 
Eu\-le\-ro\--La\-gran\-ge equa\-tions. 
In fact, in the Minkowski space-time ($\nR^4,\eta_{ij}$) we have an 
alternative way to construct the energy-momentum tensor 
for a generic  field Lagrangian. 
However, such tensor in general has not the symmetry requirement, 
thus we adopt the definition given as in Eq. (\ref{enten}). 
Let us stress that, from another point of view,  
the equations of motion  of the matter field 
\be
\nabla^i T_{ij}=0
\label{derT}
\ee
are obtained requiring  the  action for the matter $S_m$ introduced 
in Eq. (\ref{eh_act}) to be invariant under diffeomorphisms. 
For the Einstein equations, the Bianchi identities 
$\nabla^i G_{ij}=0$ 
may be viewed as a consequence of the invariance of the Hilbert 
action under diffeomorphisms. 
Let us now list the principal aspects of the three fields under 
consideration.

\begin{romanlist}
\item 

The energy-momentum tensor of a perfect fluid is given by
\be 
T_{ij}=(p+\rho)u_iu_j-p\, g_{ij},
\ee   
where $u_i$ is a unit time-like vector field representing the 
four-velocity of the fluid. 
The scalar functions $p$ and $\rho$ are the energy density and 
the pressure, respectively, as measured by an observer 
in a locally inertial frame moving with the fluid, and are related 
 by an equation of state $p= p(\rho)$.
Since no terms describing heat conduction or viscosity are 
introduced the fluid is considered as perfect. 
For the early Universe we have 
$p= (\gamma -1) \rho$, where $\gamma$ is 
the 
polytropic index.
In the particular case where $p=0$ we deal with dust, whose 
fluid elements follow geodesic trajectories. 

\item

The Lagrangian density for the linear, relativistic, scalar 
field theory reads as 
\be
\Lambda_m=\f12\left(\partial^k\phi\partial_k\phi-m^2\phi^2\right)
\ee
and in a curved space-time it is obtained by the minimal substitution rule 
$\eta_{ij}\rightarrow g_{ij}$ and $\p_i\rightarrow\nabla_i$. 
The corresponding  energy-momentum tensor is expressed  as 
\be
T_{ij}=\nabla_i\phi\nabla_j\phi-\f1 2g_{ij}\left(\nabla^k\phi\nabla_k\phi-m^2\phi\right) \, ,
\ee  
and the Klein-Gordon equation becomes
\be
\left(\nabla^k\nabla_k+m^2\right)\phi=0.
\ee

\item 

The Maxwell field is described by the Lagrangian density
\be
\Lambda_m=-\f1{16\pi}F_{ij}F^{ij}
\ee
and  the electromagnetic energy-momentum tensor reads as
\be\label{elten}
T_{ij}=\f1 {4\pi}\left(-F_{ik}F^k_{~j}+\f14 g_{ij}F_{kl}F^{kl}\right) \, ,
\ee
where $F=dA$ is the curvature 2-form associated to the connection 
1-form $A=A_i dx^i$, i.e. $F_{ij}=\nabla_iA_j-\nabla_jA_i$. 
We note that the trace of the energy-momentum tensor defined as in 
Eq. (\ref{elten}) is identically vanishing. 
Thus, the Maxwell equations in a curved space-time become
\bseq
\label{eppa}
\begin{align}
\nabla^lF_{lk}    &=   -4\pi j_k\\
\nabla_{[i}F_{jk]}&=    0 \, ,
\end{align}
\eseq
where $j_k$ is the current density four-vector of 
electric charge and $[~]$ denote the antisymmetric sum,
or in the forms language equations (\ref{eppa}) 
are written as 
\bseq
\begin{align}
d\star F &=4\pi\star j  \\
    dF   &=0
\end{align}
\eseq
where $d$ is the exterior derivative and $\star$ is 
the Hodge star operator \cite{wald,MTW}. 
\end{romanlist}

\subsectionric{Tetradic Formalism\label{sec:tetrad}}

Let $M$ be a four-dimensional manifold and $e$ a one-to-one correspondence 
on it, $\hat{e}:M\rightarrow TM$,  mapping tensor fields on $M$ 
to  tensor fields in the Minkowski tangent space $TM$. 
The four linearly independent fields $e^a_i$ (tetrads or \textit{vierbein}) 
($a$ is the Lorentzian index) are an orthonormal basis for 
the local Minkowskian space-time and  satisfy the  condition
\begin{equation}
\label{vior}
e^i_a e_{bi}=\eta_{ab} \,
\end{equation}
only.
In Eq. (\ref{vior}) $\eta_{ab}$ is a symmetric, constant matrix 
with signature $(+, -, -, -)$ and  $\eta^{ab}$ is  the inverse 
matrix of $\eta_{ab}$. 
Let us define the reciprocal (dual) vectors $e^{ai}$, such 
that $e^a_i e_b^i=\delta^{a}_{b}$. By definition of $e^{ai}$ 
and by (\ref{vior}), the condition   $e^a_i e_a^j=\delta^j_i$ is also verified.\\
In this way, the Lorentzian index is lowered and raised with 
the matrix $\eta_{ab}$, and the metric tensor $g_{ij}$ assumes the form
\begin{equation}
\label{r tensore metrico nelle tetradi}
g_{ij} =\eta_{ab} e^a_i e^b_j \,.
\end{equation}
We denote  the projections of the vector $A^i$ along the four $e_a^i$ 
as ``vierbein components'',  $A_a =e^i_a A_i$ and $A^a=e_i^aA^i=\eta^{ab}A_b$. 
In particular, for the partial differential operator, we have 
$\partial_a=e^i_a \partial_i$, below denoted as a comma 
$(~)_{,a} \equiv \partial_a$.
The generalization to a tensor of any number of 
covariant  or 
contravariant indices is straightforward.

Let us  introduce a couple of quantities which will 
directly appear in the Einstein equations, the Ricci coefficients 
$\gamma_{abc}$, 
and their linear combinations 
$\lambda_{abc}$ by the following equations
\begin{equation}
\label{r gamma}
\gamma_{abc}= 
		\nabla_ke_{ai}e^i_b e^k_c,
\end{equation}
\begin{equation}
\label{r coefficienti di ricci}
\lambda_{abc}=
		\gamma_{abc} -\gamma_{acb}.
\end{equation}
With some algebra, the Riemann and the Ricci tensors can 
be  expressed in terms of  $\gamma_{abc}$ and of   $\lambda_{abc}$ as
\begin{equation}
\label{r riemann tetradico}
R_{abcd} =\gamma_{abc,d}-\gamma_{abd,c} + \gamma_{abf} 
\left(\gamma^{f}_{\phantom{f }cd}- \gamma^{f}_{\phantom{f }dc}\right) 
+ \gamma_{afc} \gamma^{f}_{\phantom{f }bd} -\gamma_{afd} \gamma^{f}_{\phantom{f }bc}\;,
\end{equation}
\begin{align}
\label{r ricci tetradico}
R_{ab} = -\displaystyle\frac{1}{2} 
	&\left(\lambda_{ab,\phantom{c }c}^{\phantom{ab,}c} 
	+ \lambda_{ba,\phantom{c\;}c}^{\phantom{ab,}c} 
	+ \lambda^{c}_{\phantom{c }ca, b} 
	+\lambda^{c}_{\phantom{c }cb, a} +\phantom{\frac{1}{1}}\right.\nonumber\\
	&\quad +\left. \lambda^{cd}_{\phantom{cd }b} \lambda_{cda} 
	+ \lambda^{cd}_{\phantom{cd }b} \lambda_{cda} 
	- \frac{1}{2} \lambda_{b}^{\phantom{b }cd} \lambda_{acd} 
	+\lambda^{c}_{\phantom{c }cd} \lambda_{ab}^{\phantom{ab }d}
	+ \lambda^{c}_{\phantom{c }cd} \lambda_{ba}^{\phantom{ba }d}\right)\;.
\end{align}
With the use of the tetradic fields $e^a_i$, 
we are able to rewrite the Lagrangian formulation of the 
Einstein theory in a more elegant and compact form. 
Let us introduce the connection 1-forms $\omega^a_b$ and  
the curvature 2-forms associated to it  by
\be
R^a_b=d\omega^a_b+\omega^a_c\wedge\omega^c_b \, ,
\ee
which is the so-called I Cartan structure equation. 
In this formalism, the action for GR in absence 
of matter field  reads as
\be
\label{actte}
S(e,\omega)= - \f1{4\kappa}\int_M\epsilon_{abcd} e^a\wedge e^b\wedge R^{cd},
\ee
where $\epsilon_{abcd}$ is the totally antisymmetric tensor on $TM$ such that 
$\epsilon_{ijkl}= \epsilon_{abcd}e^a_ie^b_je^c_ke^d_l$. 
We stress that here we are working {\it $\acute a$ la} Palatini, 
i.e. in a first order formalism where the tetrads $e^a_i$ and the spin 
connections $\omega^a_{~b}$ can be considered as independent variables. 
The variation of the action (\ref{actte}) with respect to the 
connections leads to the II Cartan structure equation, 
i.e. to the equations of motion for $\omega^a_b$
\be\label{IIC}
de^a+\omega^a_b\wedge e^b=0 \quad \Rightarrow \quad \omega=\omega(e)
\ee
and to the identity $e^b\wedge R^a_b=0$, while  
the variation  with respect to the tetrad $e^a$ leads 
to the Einstein vacuum equations
\be
\epsilon_{abcd}e^b\wedge R^{cd}=0 \, .
\ee
The solution to the II Cartan structure equation (\ref{IIC}) 
can be written as $\omega_i^{ab}=e^{aj}\nabla_i e^b_j$, 
which corresponds to the realtion 
$\omega^{ab}_i= \gamma^{ab}_c e^c_i$
for the Ricci coefficients. 
The two different notations are adopted  accordingly as $\gamma$ 
in \reffcite{LF}, and $\omega$ in \reffcite{wald}.

\subsectionric{Hamiltonian Formulation of the Dynamics\label{sec:hf}}

The aim of the first part of this paragraph is the Hamiltonian formulation 
of GR in the metric formalism \cite{wald,ishamtime,Thi,KT90}, 
and of the corresponding 
Hamilton-Jacoby theory in the following \cite{Horava,Kijo}. 
For the theory of constrained Hamiltonian systems we refer to the 
standard textbooks \cite{Dirac64,HT}.

\subsubsectionric{Canonical General Relativity\label{sec:cangr}}

\begin{figure}[ht]
	\begin{center}
		\includegraphics[width=6.6cm]{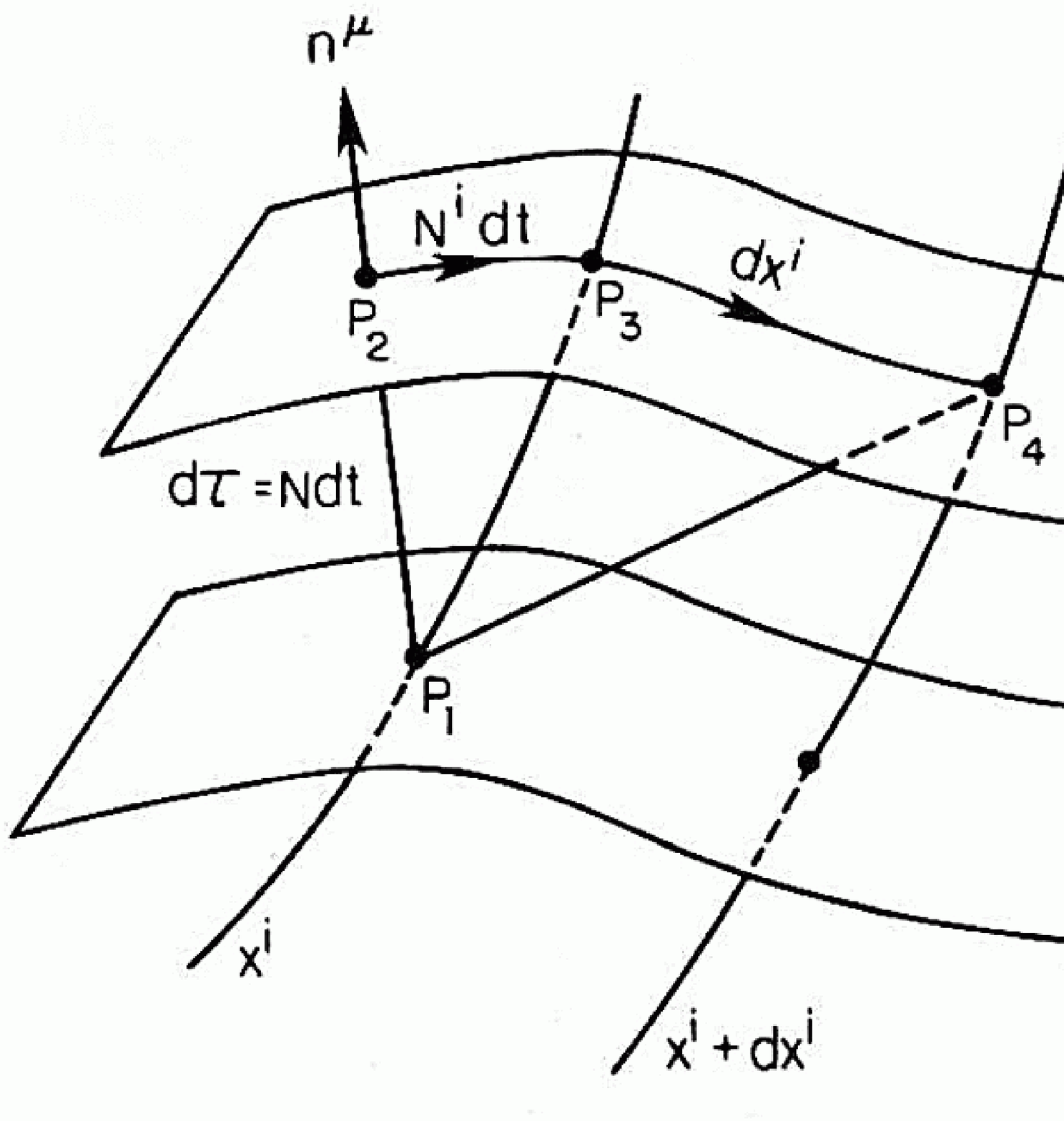}
		\caption[Construction of the lapse function and of the 
		shift vector]{The geometric interpretation of the lapse 
		function and of the shift vector (from \reffcite{KT90}). 
		In the figure latin indices denote spatial components, 
		while greek ones space-time components. \label{costruzione}}
	\end{center}
\end{figure}

The generally covariant system \textit{par excellence} 
is the gravitational field,
i.e. it is invariant under arbitrary changes of the space-time 
coordinates (four-dimensional diffeomorphisms). 
To perform  a canonical formulation of General Relativity one has 
to assume that the topology of $M$ (the physical space-time) is 
$M=\nR\times\Sigma$, where $\Sigma$ is a compact three-dimensional 
manifold (the three-space). 
As follows from standard theorems, all physically realistic 
space-times (globally hyperbolic) posses such topology; 
thus, $M$ can be foliated by a one-parameter family of 
embeddings $X_t:\Sigma\rightarrow M$, $t\in\nR$, of $\Sigma$ in $M$. 
As a consequence, the mapping 
$X:\nR\times\Sigma\rightarrow M$, defined by $(x,t)\rightarrow X_t$, 
is a diffeomorphism of $\nR\times\Sigma$ to $M$.
A useful parametrization of the embedding is given by the 
deformation vector field
\be\label{dv}
T^i(X)\equiv\f{\p X^i(x,t)}{\p t}=N(X)n^i(X)+N^i(X)\, ,
\ee
where  $N^i(X)\equiv N^\alpha X^i_\alpha$. 
The vector field $T^i$, satisfying 
$T^i\nabla_i T=1$, can be interpreted as the ``flow of time'' 
throughout the space-time. 
In Eq. (\ref{dv}), $n^i$ is the unit  vector field normal to 
$\Sigma_t$, i.e. the relations
\be
\bcas
g_{ij}n^in^j=1\\
g_{ij}n^i\p_\alpha X^j=0
\ecas
\ee
hold. 
The quantities $N$ and $N^\alpha$ are the lapse function 
and the shift vector, respectively. 
This way the space-time metric $g_{ij}$ induces a spatial metric, 
i.e. a three-dimensional Riemannian metric 
$h_{\alpha\beta}$ on each $\Sigma_t$, 
by $-h_{\alpha\beta}=g_{ij}\p_\alpha X^i\p_\beta X^j$. 
The space-time line-element adapted to this foliation 
can be written as
\be
\label{linevn}
ds^2=N^2dt^2-h_{\alpha\beta}(dx^\alpha+N^\alpha dt)(dx^\beta+N^\beta dt) \, .
\ee
This formallism was introduced by Arnowitt, Deser and Misner 
in 1962 \cite{ADM1962} and is known as the ADM procedure.\\
The geometrical meaning of $N$ and $N^\alpha$ is now clear: 
the lapse function specifies the proper time separation between 
the hypersurfaces $X_t(\Sigma)$ and $X_{t+\delta t}(\Sigma)$ 
measured in the direction normal to the first hypersurface. 
On the other hand, the shift vector  measures the displacement of 
the point $X_{t+\delta t}(x)$ from the intersection of the hypersurface 
$X_{t+\delta t}(\Sigma)$ with the normal geodesic drawn from the 
point $X_t(x)$ (see Fig.\ref{costruzione}). 
In order to have a future directed foliation of the space-time, we stress the
requirement for the lapse function $N$ to be positive everywhere in the domain 
of definition.

In the canonical analysis of General Relativity, the Riemannian metric 
$h_{\alpha\beta}$ on $\Sigma_t$ plays the role of the fundamental configuration 
variable. 
The rate of change of $h_{\alpha\beta}$ with respect to the 
time label  $t$ is related to the extrinsic curvature 
$K_{\alpha\beta}=-\nicefrac{1}{2}\mathcal L_nh_{\alpha\beta}$ of 
the hypersurface $\Sigma_t$ as
\be\label{ec}
K_{\alpha\beta}(x,t)= -\f 1{2N}\left(\p_th_{\alpha\beta}
	-(\mathcal L_{N^\alpha}h)_{\alpha\beta} \right) \, ,  
\ee
where $\mathcal L_a$ denotes the Lie derivative along the vector field $a$. \\
Let us now pull-back the Einstein Lagrangian density by the adopted foliation  
$X:\nR\times\Sigma\rightarrow M$ and express the result $X^\ast$ in terms of the 
extrinsic curvature, the three-metric $h_{\alpha \beta}$, $N$ and $N^\alpha$. 
This gives
\begin{multline}
\label{cod}
X^\ast \left(\sqrt{-g} {~}^{(4)}R \right) = 
	N\sqrt h\left(K_{\alpha\beta}K^{\alpha\beta}
	-(K_\alpha^\alpha)^2 +   {~}^{(3)}R\right)+\\ 
	-2\f d{dt}\left(\sqrt h K^\alpha_\alpha \right)-
	\p_\beta \left(K^\alpha_\alpha N^\beta
	-h^{\alpha\beta}\p_\alpha N \right) \, ,
\end{multline}
where ${~}^{(n)}R$ is the $n$-dimensional curvature scalar 
and $\sqrt{-g}=N\sqrt h$, 
being $h$ the determinant of the three-dimensional metric. 
We are able to re-cast the original Hilbert action into a $3+1$ form
by simply dropping the total differential in (\ref{cod}), i.e. the last two 
terms on the r.h.s. of  (\ref{cod}) 
\be
\label{act}
S(h,N,N^\alpha)=-\f 1 {2\kappa}\int dt\int_\Sigma d^3x 
	N\sqrt h\left(K_{\beta\gamma}K^{\beta\gamma}
	-(K_{~\beta}^\beta)^2+ {~}^{(3)}R\right) \, .
\ee
Performing a Legendre transformation of the Lagrangian density appearing in 
equation (\ref{act}) we obtain the corresponding Hamiltonian density. 
Let us note that the action (\ref{act}) does not depend on the time derivatives 
of $N$ and $N^\alpha$, and therefore 
using the definition (\ref{ec}) and the fact that ${~}^{(3)}R$ does not contain time 
derivatives,
we obtain for the conjugate momenta  
\begin{subequations}
\label{conjm}
\begin{align}
\label{conjm1}
\f 1 \kappa \Pi^{\alpha\beta}(x,t)& 
\equiv \f{\delta S}{\delta\dot h_{\alpha\beta}(x,t)}
	=\f{\sqrt h}{\kappa}
	\left(K^{\alpha\beta}-h^{\alpha\beta}K^\gamma_{~\gamma} \right)\\
\label{conjm2}
\Pi(x,t) &\equiv \f{\delta S}{\delta\dot N(x,t)}=0 \\
\label{conjm3}
\Pi_\alpha(x,t)&\equiv \f{\delta S}{\delta\dot N^\alpha(x,t)}=0 \,.
\end{align}
\end{subequations}
From equations (\ref{conjm}) follows that 
not all conjugate momenta are independent, i.e. one cannot 
solve for all velocities as functions of coordinates and momenta:
one can express $\dot h_{\alpha\beta}$ in terms of 
$h_{\alpha\beta}$, $N$, $N^\alpha$ and $\Pi^{\alpha\beta}$, but the same is not 
possible for $\dot N$ and $\dot N^\alpha$. 
By other words, we have the so-called {\it primary constraints} 
\be\label{pricon}
C(x,t)\equiv\Pi(x,t)=0, \qquad C^\alpha(x,t)\equiv\Pi^\alpha(x,t)=0 \, , 
\ee  
where ``primary''  emphasizes that the equations of motion 
are not used to obtain  relations (\ref{pricon}).

According to the theory of constrained Hamiltonian systems, let us 
introduce the new fields $\lambda(x,t)$ and $\lambda^\alpha(x,t)$ 
as the Lagrange multipliers for the primary constraints, making the Legendre 
transformations invertible and the corresponding 
action reading as
\be\label{azione}
S= \int dt\int_\Sigma d^3x\left[ \dot h_{\alpha\beta}\Pi^{\alpha\beta}
	+\dot N\Pi+\dot N^\alpha \Pi_\alpha
	-\left(\lambda C+\lambda^\alpha C_\alpha
	+N^\alpha H_\alpha+N H \right)\right],
\ee  
where
\begin{subequations}

\begin{align}
H_\alpha &\equiv-2h_{\alpha\gamma}\nabla_\beta \Pi^{\gamma\beta}\\
H        &\equiv\mathcal G_{\alpha\beta\gamma\delta}
		\Pi^{\alpha\beta}\Pi^{\gamma\delta}-\sqrt h {~}^{(3)}R 
\end{align}
in which
\be
\label{supermetrica} 
\mathcal G_{\alpha\beta\gamma\delta}\equiv \f{1}{2\sqrt h}
	\left(h_{\alpha\gamma}h_{\beta\delta}+
	h_{\beta\gamma}h_{\alpha\delta} -h_{\alpha\beta}h_{\gamma\delta}\right)
\ee
\end{subequations}
is the so-called {\it super-metric} on the space of the three-metrics, 
and the functionals $H_\alpha$ and  $H$ are the {\it super-momentum} and 
{\it super-Hamiltonian}, respectively.

The classical canonical algebra of the system is expressed in 
terms of the standard Poisson brackets as
\begin{subequations}

\begin{align}
\label{alg1}
\{h_{\alpha\beta}(x,t),h_{\gamma\delta}(x',t)\}&=0 \\
\label{alg2}
\{\Pi^{\alpha\beta}(x,t), \Pi^{\gamma\delta}(x',t)\}&=0  \\
\label{alg3}
\{h_{\gamma\delta}(x,t), \Pi^{\alpha\beta}(x',t)\}&=
	\kappa\delta^\alpha_{(\gamma}\delta^\beta_{\delta)}\delta^3(x-x') \, ,
\end{align}
\end{subequations}
where the parentheses $(~)$ denote symmetrized indices. 
We can analyze the meaning of the Hamiltonian of the system,  
bracketed in equation (\ref{azione})
\be\label{hamilgener}
\mathcal H\equiv\int_\Sigma d^3x \left(\lambda C+\lambda^\alpha C_\alpha
+N^\alpha H_\alpha+N H \right)
\equiv C(\lambda)+\vec C \left(\vec\lambda\right)+\vec H \left(\vec N \right)+H(N).
\ee
The variation of the action (\ref{azione}) with respect to 
the Lagrange multipliers $\lambda$ and $\lambda^\alpha$ 
reproduces the primary constraints (\ref{pricon}). 
The consistency of the dynamics is ensured by preserving them during the evolution 
of the system, i.e. requiring 
\be\label{eqmoto}
\dot C(x,t)\equiv\{C(x,t),\mathcal H\}=0, \qquad 
\dot C^\alpha(x,t)\equiv\{C^\alpha(x,t),\mathcal H\}=0 \, .
\ee
The Poisson brackets 
in equation (\ref{eqmoto})
do not vanish, but equal to $H(x,t)$ and $H^\alpha(x,t)$, 
respectively, and 
therefore the consistency of the motion leads to the 
{\it se\-con\-dary constraints}
\be\label{seccon} 
H(x,t)=0, \qquad H^\alpha(x,t)=0 \, ,
\ee
by means of the equations of motion.
Let us observe that the Hamiltonian in General Relativity is 
constrained as $\mathcal H\approx0$ being weakly zero, 
i.e. vanishing
on the constraint surface. 
This is not surprising  since we are dealing with a generally covariant system. 

A problem that can arise is that the constraint surface, i.e. 
the surface where the constraints hold, could not be preserved under the 
motion generated by the constraints themselves, but 
 this is not the case. 
In fact, the Poisson algebra of the 
super-momentum and super-Hamiltonian, computed using 
(\ref{alg1})-(\ref{alg3}), is closed. 
In other words, the set of constraints is of  {\it first class}, 
i.e. the Poisson brackets between the Hamiltonian $\mathcal H$ and any 
constraint weakly vanish, as arising from the relations
\be\label{rel1}
\{\mathcal H,\vec H(\vec f)\}=
	\vec H(\mathcal L_{\vec N}\vec f)- H(\mathcal L_{\vec f}N)
\ee
\be\label{rel2}
\{\mathcal H,H(f)\}=H(\mathcal L_{\vec N}f)+\vec H(\vec N(N,f,h)) \, ,
\ee
where $f$ is a smooth function and 
$\vec N^\alpha(N,f,h)=h^{\alpha\beta}(N\p_\beta f-f\p_\beta N)$. 
These equations are equivalent to the Dirac algebra \cite{Dirac64}.\\
Moreover, varying the action (\ref{azione}) with respect to the two 
conjugate momenta $\Pi$ and $\Pi^\alpha$, we obtain
\be
\dot N(x,t)=\lambda(x,t), \qquad \dot N^\alpha(x,t)=\lambda^\alpha(x,t)
\ee
 ensuring that the trajectories of the lapse function and of the shift 
vector in the phase space are completely arbitrary.

The Hamiltonian of the theory is not a true one but a 
linear combination of constraints. 
From  relations 
(\ref{rel1}) and (\ref{rel2}) it is possible to show that
rather than ge\-ne\-ra\-ting time translations,  
the Hamiltonian generates 
space-time diffeomorphisms, 
whose 
parameters are the completely arbitrary functions $N$ and $N^\alpha$, 
and  the corresponding motions on the phase space have to be regarded 
as gauge transformations. 
An observable is defined as a function on the constraint 
surface that is gauge invariant, 
and more precisely, in a system with  first class constraints
it can be described as a phase space function that has weakly 
vanishing Poisson brackets with the constraints. 
In our case, $A$ is an observable if and only if
\be
\{A,\mathcal H(\lambda,\lambda^\alpha,N^\alpha,N)\}\approx 0 \, ,
\ee
for generic $\lambda$, $\lambda^\alpha$, $N^\alpha$ and $N$. 
By this definition, one treats on the same footing ordinary gauge 
invariant quantities and constants of motion with respect to evolution 
along the foliation associated to $N$ and $N^\alpha$. 
The basic variables of the theory, $h_{\alpha\beta}$ 
and $\Pi^{\alpha\beta}$, are not observables as they are not gauge invariant. 
In fact, no observables for General Relativity are known, except for 
the particular situation with asymptotically flat boundary conditions.  

Let us remark that the equations of motion
\be
\dot h_{\alpha\beta}(x,t)=\f {\delta\mathcal H}{\delta P^{\alpha\beta}(x,t)}, \qquad \dot \Pi^{\alpha\beta}(x,t)=-\f {\delta\mathcal H}{\delta h_{\alpha\beta}(x,t)}
\ee 
together with the eight constraints (\ref{pricon}) and (\ref{seccon}) 
are completely equivalent to the vacuum Einstein equations, 
$R_{ij}=0$ \cite{Fis-Mar}.

\subsubsectionric{Hamilton-Jacobi Equations for Gravitational Field\label{sec:hjgrav}}

The formulation of the Hamilton-Jacobi theory for a covariant system is 
simpler than the conventional non-relativistic version. 
In fact, in such case the Hamilton-Jacoby equations are expressed as 
\be
H\left(q_a,\f{\p S}{\p q_a}\right)=0 \, ,
\ee 
where  $H$ and $S(q_a)$  denote the Hamiltonian and the Hamilton functions, 
respectively. 
For GR,  the Hamilton-Jacobi equations arising from 
the  super-Hamiltonian and  super-momentum read as
\begin{align}
\label{SHJE}
\widehat{HJ} ~S & \equiv\mathcal G_{\alpha\beta\gamma\delta} 
		\f{\delta S}{\delta h_{\alpha\beta}} \f{\delta S}{\delta h_{\gamma\delta}}
		-\sqrt h {~}^{(3)}R=0  \\
\widehat{HJ_\alpha} ~S & \equiv-2h_{\alpha\gamma}\nabla_\beta 
	\f{\delta S}{\delta h_{\gamma\beta}}=0 \,.
\end{align}
These four equations, together with the primary constraints (\ref{pricon}),
completely define the classical dynamics of the theory.

Let us point out  how through a change of variable we can define 
a time coordinate. 
In fact, writing $h_{\alpha\beta}\equiv\eta^{4/3}u_{\alpha\beta}$, 
with $\eta\equiv h^{1/4}$ and $\det u_{\alpha\beta}=1$, 
the  super-Hamiltonian Hamilton-Jacobi equation (\ref{SHJE}) 
can be written as
\be
\widehat{HJ} ~S=
-\f3{16}\kappa\left(\f{\delta S}{\delta\eta}\right)^2
+\f{2\kappa}{\eta^2}u_{\alpha\gamma}u_{\beta\delta}
\f{\delta S}{\delta u_{\alpha\beta}}\f{\delta S}{\delta u_{\gamma\delta}}-\f1{2\kappa}\eta^{2/3}V(u_{\alpha\beta},\nabla\eta,\nabla u_{\alpha\beta}),
\label{supesupe}
\ee
where the potential term $V$ comes out from the spatial Ricci scalar 
and $\nabla$ refers to 
spatial gradients only. As we can see from Eq. (\ref{supesupe}), $\eta$ has the 
correct signature for an internal  time variable candidate. 
We will show later how this variable is nothing but a power of the 
isotropic volume of the Universe.

\subsectionric{Synchronous Reference\label{sec:syref}}

In this paragraph we will focus our attention on one of the most 
interesting re\-fe\-ren\-ce system, i.e. the synchronous one. 
For a detailed discussion see \reffcite{LF}.
It is defined by the following choice for the metric tensor $g_{ij}$ 
\be\label{gp:sync}
g_{00}=1, \qquad g_{0\alpha}=0, 
\ee
thus in the $3+1$ framework we have to require $N=1$ and 
$N^\alpha=0$ in (\ref{linevn}). 
The first condition in (\ref{gp:sync}) is allowed from the 
freedom to rescale the variable $t$ with the transformation 
$\sqrt{g_{00}} dt$, in order to reduce $g_{00}$ to unity, and setting 
the time coordinate $x^0=t$ as the proper time at each point of space. 
The second one is allowed by the non-vanishing of $|g_{\alpha \beta}|$ 
and allows the synchronization of clocks at different points of space. 
The elementary line  interval is given by the expression
\begin{equation}
\label{gp:b0}
{ds}^2 ={dt}^2 -dl^2\, ,
\end{equation}
where
\begin{equation}
\label{gp:b1}
dl^2= h_{\alpha \beta}\left(t, x^{\gamma}\right)dx^{\alpha}dx^{\beta} \, ,
\end{equation}
in which the three-dimensional tensor $h_{\alpha \beta}$ defines 
the space metric.\\
In such a reference system, lines of equal times are geodesics 
in the four-space, as implied by the splitting definition. 
Indeed the four-vector $u^i=dx^i/ds$ which is tangent to the world 
line $(x^{\gamma}=\textrm{const.})$, has components $u^0 =1$, 
$u^{\alpha} =0$ and automatically satisfies the geodesic equations
\begin{equation}
\frac{du^i}{ds} + \Gamma^i_{kl}u^k u^l =\Gamma^i_{00}=0.
\label{c0}
\end{equation}
The choice of such a reference is always possible and moreover it 
is not unique. 
In fact, considering a generic infinitesimal displacement
\be
t'=t+\xi(t,x^\rho), \qquad x'^\alpha=x^\alpha+\xi^\alpha(t,x^\rho)
\ee
and the associated four-metric change 
$g'_{ij}=g_{ij}-2\nabla_{(i}\xi_{j)}$, 
the conditions of preserving the synchronous reference 
can be written as
\begin{subequations}
\begin{align}
\p_t\xi &=0 \Rightarrow t'=t+\xi(x^\rho)  \\
 \p_\alpha\xi&=0 \Rightarrow x'^\alpha= 
	x^\alpha+\p_\beta\xi\int h^{\alpha\beta}dt+\phi^\alpha(x^\rho) \, , 
\end{align}
\end{subequations}
$\phi^\alpha$ being generic space functions. 

In the reference defined by the metric  (\ref{gp:b0}), 
the Einstein equations are written in mixed components as
\begin{subequations}
\begin{align}
\label{gp:ea}
      R^0_0 &=-\frac{1}{2} \frac{\partial}{\partial t}
\kappa^{\alpha}_{\alpha}  -\frac{1}{4} \kappa^{\beta}_{\alpha} 
\kappa^{\alpha}_{\beta} = 8 \pi G\left(T^0_0 -\frac{1}{2}T\right) \\
\label{gp:eb}
    R^0_{\alpha} &= \frac{1}{2}\left(\kappa^{\beta}_{\alpha ; \beta} 
-\kappa^{\beta}_{\beta ; \alpha}\right) =8 \pi G T^0_{\alpha} \\
\label{gp:ec}
  R^{\beta}_{\alpha} &=  -P^{\beta}_{\alpha} 
-\frac{1}{2 \sqrt{h}}\frac{\partial}{\partial t}  
\left(\sqrt{h}  \kappa^{\beta}_{\alpha} \right) 
= 8 \pi G  \left(T^{\beta}_{\alpha} -\frac{1}{2} \delta^{\alpha}_{\beta}T\right)\, , 
\end{align}
\label{gp:ecce}
\end{subequations}
where 
\begin{equation}
\kappa_{\alpha \beta} =\frac{\partial h_{\alpha \beta}}{\partial t} ,
 \,\quad h \equiv \mid h_{\alpha \beta} \mid  \, ,
\label{gp:f0}
\end{equation}
$P_{\alpha \beta}$ is the three-dimensional Ricci tensor 
obtained through the metric $h_{\alpha\beta}$ which is used to 
raise and lower indices within the spatial sections. 

The metric $h_{\alpha\beta}$ allows to construct the three-dimensional Ricci tensor $P_{\alpha}^{\beta}=h^{\beta\gamma}P_{\alpha\gamma}$ as
\begin{equation}
P_{\alpha \beta} = 
\partial _{\gamma }{\lambda }^{\gamma }_{\alpha \beta } 
- \partial _{\alpha }{\lambda }^{\gamma }_{\beta \gamma } + 
{\lambda }^{\gamma }_{\alpha \beta }{\lambda }^{\delta }_{\gamma \delta } 
- {\lambda }^{\gamma }_{\alpha \delta }{\lambda }^{\delta }_{\beta \gamma } 
\label{gp:gr}
\end{equation}
in which appear the pure spatial Christoffel symbols 
\begin{equation}
\lambda^\gamma_{\alpha\beta}\equiv\frac{1}{2}h^{\gamma 
\delta }(\partial _{\alpha} h_{\delta \beta } + 
\partial _{\beta }h_{\alpha \delta } 
-\partial _{\delta }h_{\alpha \beta }) \, . 
\label{gp:hr} 
\end{equation} 
From (\ref{gp:ea}) it is straightforward to derive, even in the 
isotropic case, the Landau-Raychaudhuri theorem, stating that 
the metric determinant $h$ must  monotonically vanish in a 
finite instant of time. However, we want to stress that the 
singularity in this reference system is not physical and of course can 
be removed passing to another one.

\subsectionric{Singularity Theorems\label{sec:singth}}

In this Section we investigate space-time singularities. 
We will show how singularities are true, generic features 
of the Einstein theory of gravity and how they arise under 
certain, quite general, assumptions. We only enter in some 
details and we will  not give the  rigorous proofs, 
for which we refer to the standard literature \reffcite{HE,wald}. \\
In particular, in this Section only we follow the signature of \reffcite{wald}
$(-,+,+,+)$. 	\\
We will treat here the classical aspects, referring to 
\reffcite{Boj07,Got-Dem} and references therein 
for a discussion on quantum cosmological singularities. \\
After defining  what a space-time singularity means we will 
present some basic technology, and finally we will discuss the theorems.

\subsubsectionric{Definition of a Space-time Singularity\label{sec:singdef}}

Let us clarify the meaning of a singularity of the space-time. 
In analogy with field theory, we can represent such a singularity 
as a ``place'' of the  space-time where the curvature diverges, or 
where some similar pathological behavior of the geometric invariants 
 takes place. 
Therefore, a first problem   arises when characterizing  the singularity 
as  a ``place''. 
In fact, in General Relativity the  space-time consists of a manifold 
$M$ and a metric $g_{ij}$ defined everywhere on $M$: 
a singularity (as the Big-Bang singularity of the isotropic cosmological 
solution or the $r=0$ singularity in the Schwarzschild space-time) can 
not be considered as a part of the  manifold itself. 
We can speak of a physical event only when a manifold and a metric 
structure are defined around it. 
\textit{A priori}, it is possible to add points to the manifold 
in order to describe the singularity  as a real place 
(as a boundary of the manifold). 
But apart from very peculiar cases, no general notion or definition 
of a singularity boundary exists\cite{wald,GerochPenrose,GerochWald,Johnson}. 
Another problem is that singularities are not always 
accompanied by unbounded curvature as in the best known cases. 
There are several examples \cite{HE} 
of a singularity without diverging curvature. 
In fact, as we will see, this feature is not the basic mechanism 
behind singularity theorems.

The best way to clarify what a  singularity means is the 
geodesic incompleteness, i.e. the existence of  geodesics which are 
inextensible  at least in one direction, but have  only a finite range 
for the affine parameter. 
We can thus define a singular space-time as the one possessing  one 
incomplete geodesic curve.

\subsubsectionric{Preliminary notions\label{sec:singpn}}

We initially define the notions of expansion and contraction of the congruence 
of time-like geodesics. 
The treatment of null geodesic congruences  is conceptually 
similar and not reviewed here. \\
Let $O$ be an open set of a manifold $M$. 
A congruence in $O$ is defined as a family of curves such that only one 
curve of this family passes through each point $p\in O$. 
Let $\xi^i$ be the vector field tangent to the geodesics such that 
$\xi^i\xi_i=-1$, and define the tensor $B_{ij}=\nabla_j\xi_i$ 
to introduce the {\it expansion} $\theta$ (as well as its trace part), 
the {\it shear} $\sigma_{ij}$ (as its symmetric, trace-free part) 
and the {\it twist} $\omega_{ij}$ (as its anti-symmetric part) of 
the congruence. 
Thus, $B_{ij}$ can be written as
\be
B_{ij}=\f13\theta h_{ij}+\sigma_{ij}+\omega_{ij} \, ,
\label{bi_def}
\ee
where the spatial metric is $h_{ij}=g_{ij}+\xi_i\xi_j$. 
Given a  deviation vector orthogonal to the vector field $\xi^i$,
the tensor $B_{ij}$ measures its  failure from 
 being parallelly transported
along  a 
geodesic in the congruence for this subfamily. \\
The rates of change of $\theta$, $\sigma_{ij}$ and $\omega_{ij}$ 
follow from the geodesic equation
\begin{multline}\label{eqgeB}
\xi^k\nabla_kB_{ij}=\xi^k\nabla_k\nabla_j\xi_i=
\xi^k(\nabla_j\nabla_k\xi_i+R_{kji}\,^l\xi_l)=\\
	=\nabla_j(\xi^k\nabla_k\xi_i)-(\nabla_j\xi^k)(\nabla_k\xi_i)+
	R_{kji}\,^l\xi^k\xi_l=-B^k_jB_{ik}+R_{kji}\,^l\xi^k\xi_l \, . 
\end{multline}
Taking the trace of  Eq. (\ref{eqgeB}), we obtain the 
Raychaudhuri equation
\be\label{riceq}
\dot\theta=\xi^k\nabla_k\theta=-\f13\theta^2-\sigma_{ij}\sigma^{ij}
	+\omega_{ij}\omega^{ij}-R_{kl}\xi^k\xi^l \, .
\ee  
This  result will be fundamental when proving the singularity theorems. 
Let us focus our attention on the right-hand side of it:
using Einstein equations, the last term can be written as
\be\label{eqt}
R_{kl}\xi^k\xi^l=\kappa \left(T_{ij}-\f12Tg_{ij}\right)\xi^i\xi^j=
	\kappa\left(T_{ij}\xi^i\xi^j+\f12T\right) \, .
\ee
Let us  state  the physical criterion preventing 
the stresses of matter from becoming  too large 
so that the right-hand side of (\ref{eqt}) is not negative, obtaining  
\be 
T_{ij}\xi^i\xi^j\geq-\f12T \, ,
\label{tgg}
\ee
which is known as the {\it strong energy condition} \cite{HE}. 
It is commonly accepted  that every reasonable kind of matter 
 would  satisfy 
the  condition (\ref{tgg}). 
Therefore, from the Raychaudhuri equation (\ref{riceq}) one can see that, 
if the congruence is non-rotating 
	\footnote{It is possible show that the congruence is orthogonal to the 
	hypersurface  if and only if $\omega_{ij}=0$},
which means $\omega_{ij}=0$, and the { strong energy condition} holds, 
$\theta$ always decreases along the geodesics. 
More precisely, we get 
$\dot\theta+\f13\theta^2\leq0$, whose integral implies 
\be
\theta^{-1}(\tau)\geq\theta_0^{-1}+\f13\tau,
\ee
where $\theta_0$ is the initial value of $\theta$. 
For negative values of  $\theta_0$ 
(i.e the congruence is initially converging), 
$\theta$ will diverge after a proper time no larger than 
$\tau\leq3/|\theta_0|$. 
By other words, the geodesics must intersect before such instant 
and form a \textit{caustic} (a focal point). 
Of course, the singularity of $\theta$ is nothing but a singularity in 
the congruence and not a space-time one, since 
the smooth  manifold is well-defined on caustics.

Let us briefly discuss the meaning of the { strong energy condition} 
in the simple case of a perfect fluid, for which it reads as 
\be
\rho+\sum_\alpha p_\alpha\geq 0, \qquad \rho+p_\alpha\geq 0\, ,
\ee
and is satisfied for $\rho\geq 0$ and for negative pressure components 
smaller  in magnitude
then $\rho$.

We need to introduce  some notions of differential geometry and topology
to translate the occurrence of caustics into space-time singularities. \\
Let $\gamma$ be a geodesic with tangent $v_i$ defined on a 
manifold $M$. 
A solution of the geodesic deviation equation is $\eta^i$ satisfying
\footnote{In this case, a minus sign appears in the right-hand side of (\ref{geodev})
due to the different signature of the metric.}
(\ref{geodev})
that constitutes a Jacobi field on $\gamma$. 
If it  is non-vanishing along $\gamma$, but is zero 
at both $p,q\in\gamma$, then $p,q$ are said to be conjugate. \\
It is possible to show that a point $q\in\gamma$ lying in the 
future of $p\in\gamma$ is conjugate to $p$ if and only if the 
expansion of all the  time-like geodesics congruence passing 
through $p$ approaches $-\infty$ at $q$ (i.e. a point is 
conjugate if and only if it is a caustic of such congruence). 
A necessary hypothesis in this statement is that the space-time 
manifold ($M,g_{ij}$) satisfy $R_{ij}\xi^i\xi^j\geq0$, 
for all time-like $\xi^i$. 
Moreover, the necessary and sufficient condition for a timelike 
curve $\gamma$, connecting $p,q\in M$, that locally maximizes 
the proper time between $p$ and $q$, is that $\gamma$ is a 
geodesic without any  point conjugate to $p$ between $p$ and $q$. 

An analogous analysis can be made for time-like geodesics and a 
smooth space-like hypersurface $\Sigma$. 
In particular, let $\theta=K=h^{ij}K_{ij}$ be the expansion of 
the geodesic congruence orthogonal to $\Sigma$, $K_{ij}$  being 
the extrinsic curvature of $\Sigma$. 
Then, for $K<0$ at the point $q\in\Sigma$, within a proper time 
$\tau\leq3/|K|$ a point $p$ conjugate to $\Sigma$ 
along the geodesic orthogonal to $\Sigma$ exists, for a space-time 
($M,g_{ij}$) satisfying $R_{ij}\xi^i\xi^j\geq0$. 
As above, a time-like curve that locally maximizes the proper time 
between $p$ and $\Sigma$ has to be a geodesic orthogonal to $\Sigma$ 
without  conjugate point  to $\Sigma$.

The last step toward the singularity theorems is to prove the existence 
of ma\-xi\-mum length curves in globally hyperbolic space-times. 
We recall that this is the case if they   posses 
Cauchy surfaces in accordance with the determinism of classical Physics. 
Without entering in  the details, in that case 
a curve $\gamma$ for which $\tau$ attains its maximum 
value exists, and a necessary condition  is that 
$\gamma$ be a geodesic without 
conjugate point.

\subsubsectionric{Singularity Theorems\label{sec:singtheo}}

In the previous subsection we have summarized 
the necessary concepts to analyze 
the  singularity theorems in some details, 
although we will discuss them  without any proofs. 

Let  a space-time manifold be globally hyperbolic, with 
$R_{ij}\xi^i\xi^j\geq0$ for all the time-like $\xi^i$. 
Suppose that the trace $K$ of the extrinsic curvature of a Cauchy surface 
 everywhere satisfies $K\leq C<0$, for a constant $C$. 
Therefore, no past-directed time-like curves $\lambda$ 
from 
$\Sigma$ can have a length 
greater than $3/|C|$. 
In fact, if there were any, a maximum length curve would exist
and  would  be a geodesic, 
thus contradicting the fact that no conjugate points  exist 
between $\Sigma$ and $p\in\lambda$: therefore such  curve cannot exist. 
In particular, all past-directed time-like geodesics are incomplete.

This theorem is valid in a cosmological context and expresses that, 
if the Universe is expanding everywhere at a certain instant of time, 
then it must have begun with a singular state at a finite time in the past.

It is also possible to show that the previous theorem remains valid 
also relaxing the hypothesis that the Universe is globally hyperbolic. 
The price to pay is the assumption that $\Sigma$ be a compact manifold 
(dealing with a closed Universe) and especially that only one 
incomplete geodesic is predicted. 

For the  theorems  proving null geodesic incompleteness in 
a gravitational collapse context, i.e. the existence of a singularity 
in a black hole space-time, 
we refer to the literature \reffcite{Penrose1964}. 
%

%

We conclude stating the most general theorem, 
which  entirely eliminates  the assumptions of a  Universe  
expanding everywhere and the global hyperbolicity 
of the manifold $(M,g_{ij})$. 
On the other hand, we loose any information about the 
nature of one 
incomplete geodesic at least, since it does not distinguish between 
a time-like and  a null geodesic. \\
The space-time is singular under the following hypotheses: 
\begin{enumerate}
\renewcommand{\labelenumi}{{\it\roman{enumi}})}
\renewcommand{\labelenumii}{{\it\alph{enumii}})}

\item 
	the condition $R_{ij}v^iv^j\geq0$ holds 
	for all time-like and null $v^i$, 

\item 
	no closed time-like curves exist and 

\item 
	  at least one of the following properties holds: 
	\begin{enumerate}

	\item  $(M,g_{ij})$ is a closed Universe, 
	\item  $(M,g_{ij})$ possesses a trapped 
		surface\footnote{A trapped surface is a  compact 
		smooth space-like manifold $T$, such that 
		the expansion $\theta$ of either 
		outgoing either ingoing future directed null geodesics 
		is everywhere negative. }, 
	\item there exists a point $p\in M$ such that the expansion $\theta$ of the 
		future or past directed null geodesics emanating from it 
		becomes negative along each geodesic in this congruence.

	\end{enumerate}
\end{enumerate}
In particular, our Universe must be singular. 
In fact, the conditions i)-ii) hold and $\theta$ for 
the past-directed null geodesics emanating from us at 
the present time becomes negative before the decoupling time, 
i.e. the time up to when the Universe is well 
described by the Friedmann-Robertson-Walker (FRW) model. 

The occurrence of a space-time singularity undoubtedly 
represents a breakdown of  the classical theory of gravity, 
i.e. the General Relativity. 
The removal of such singularities is a prerequisite for 
any fundamental theory, like the quantum theory of the 
gravitational field. 
Singularity theorems are very powerful instruments, 
although do not provide any  information about the nature of 
the predicted singularity. 
Unfortunately, we do not have a general classification of
singularities, 
i.e. many different types exist and the unbounded curvature 
is not the basic me\-cha\-nism behind such theorems. 

Let us conclude remarking how singularity theorems concern properties of 
differential geometry and topology. 
Einstein equations are used only with respect to the positive curvature case, 
being only used in the Raychaudhuri equation. 
Nevertheless, no general conditions for  non-singular solutions are known, 
and therefore it is not possible to disregard the singularity 
in General Relativity.


\sectionric{Homogeneous Universes\label{sec:homu}}

In this Section we introduce the homogeneous cosmological models, 
in order to describe the dynamics of the Universe  toward the cosmological 
singularity with a more realistic approach then the one offered by the 
homogeneous and isotropic Friedmann-Robertson-Walker one. 
We will summarize the derivation of the Bianchi classification, 
with particular attention to the dynamics of the  Bianchi types II and VII, 
while we will dedicate the entire Section  \ref{sec:bianchi89} 
 to types VIII and IX, 
the so-called \textit{Mixmaster} \cite{BKL1970,Misner1969PRL,Misner1969PR}.



\subsectionric{Homogeneous Spaces\label{sec:hspace}}

%
%
%
%
%
%
%
%
%
%
%
%
%

The study of homogeneous models arises from breaking the 
hypothesis of space isotropy.
We will focus in particular on space-times spatially homogeneous, without 
treating the non physical case of a space-time homogeneous manifold, 
i.e. where the metric is the same at all points of space and time, 
 because it represents a Universe not expanding at all. 
We will follow in particular the description as in 
Refs. \reffcite{wald,LF,Ry-Sh,MacCallexsol}.

When relaxing the   space isotropy hypothesis we gain a 
significantly larger arbitrariness  in the solution. 
In particular, the homogeneous and isotropic model possesses a 
single gravitational degree of freedom given by the scale factor
leaving free only the curvature  sign \cite{KT90}. 
This larger class of solutions has still a finite number of 
degrees of freedom and very general properties characterize 
the dynamics during the evolution toward the initial 
singularity \cite{BKL1970}.

A spatially homogeneous space-time is defined as 
a manifold with a group of 
isometries, i.e. a group of  transformations leaving invariant 
 the metric $g$.
The Killing vectors $\xi$ are the corresponding infinitesimal generators, 
with vanishing Lie derivative  $\mathcal L_\xi g=0$
 and  whose orbits are the 
space-like hypersurfaces which foliate the space-time.
We will discuss in which sense the metric properties are the same 
in all space points under homogeneity.

\subsubsectionric{Killing Vector Fields\label{sec:killvf}}

The Lie algebras of Killing vector fields  
generate the groups of motions via infinitesimal 
displacements, yielding conserved quantities and allowing 
a classification of  homogeneous spaces.
Consider a group of transformations
\be
\label{e1}
x^{\mu} \rightarrow \bar x^{\mu} = f^{\mu}\left(x,a\right)
\ee
on a space $M$ (eventually a manifold), 
where $\left\{ a^a \right\}_{a=1,..,r}$ 
are $r$ independent variables which parametrize the 
group and let $a_0$ correspond to the identity 
\be
\label{e2}
f^{\mu}\left(x,a_0 \right) = x^{\mu}\, .
\ee
Take an infinitesimal transformation 
 $a_0+\delta a$, i.e. one which is very close 
to the identity so that
\be
\label{e3}
x^{\mu} \rightarrow \bar x^{\mu} 
=f^{\mu}\left(x, a_0+\delta a \right) \approx \underbrace{ 
f^{\mu}\left(x, a_0 \right)}_{=x^{\mu}} 
+ \underbrace{\left( \frac{\partial f^{\mu}}{\partial 
a^a} \right) 
\left(x, a_0 \right) }_{\equiv\xi^{\mu}_a\left(x\right)}\delta a^a
\ee
i.e. 
\be
\label{e4}
x^{\mu} \rightarrow \bar x^{\mu} \approx x^{\mu} + \xi^{\mu}_a\left(x\right)\delta 
a^a=\left(1+\delta a^a \xi_a \right) x^{\mu} \, ,
\ee
where the $r$ first-order differential operators  
$\left\{\xi_a\right\}$ 
are defined by 
$\xi_a=\xi^{\mu}_a\frac{\partial}{\partial x^{\mu}}$ 
and correspond to the $r$ vector fields with components
$\left\{ \xi^{\mu}_a\right\}$.
These are the ``generating vector fields'' 
and when the group is a group of motions, 
they are called \textit{Killing vector fields}, 
satisfying 
$\mathcal L_{\xi}g=0$, 
%
so that, 
 under  infinitesimal transformations (\ref{e3})
all points of the space $M$ are translated by a distance 
$\delta x^{\mu}=\delta a^a\xi^{\mu}_a$ 
in the coordinates $\left\{x^{\mu}\right\}$ 
and 
\beq
\label{e5}
\bar x^{\mu} \approx \left( 1+ \delta a^a \xi_a \right)
x^{\mu} 
\approx e^{\delta a^a \xi_a} x^{\mu} \, .
\eeq
In fact, the finite transformations of the 
group may be represented as
\beq
\label{e6}
\bar x^{\mu} \rightarrow \bar x^{\mu} =e^{\theta^a \xi_a} x^{\mu}
\eeq
where $\left\{ \theta^a \right\}$ are $r$ new parameters 
on the group. \\
These generators 
form a Lie algebra\cite{Ry-Sh}, i.e. a real 
$r$-dimensional vector space with basis
$\left\{\xi_a\right\}$, which is closed under commutation, 
i.e. the commutators of the basis elements can be expressed
as constant linear combinations of themselves
\beq
\label{e7}
\left[ \xi_a ,\xi_b \right] \equiv \xi_a \xi_b 
- \xi_b \xi_a = \pm C^c_{~ab} \xi_c
\eeq 
where $C^c_{~ab} $ are the structure constants of the 
Lie algebra 
($(+)$ refers to left-invariant groups, while 
$(-)$ to right-invariant ones). 
Suppose $\left\{ e_a \right\}$ is a basis of the Lie
algebra $g$ of a group $G$ 
\beq
\label{e8}
\left[ e_a , e_b \right] = C^c_{~ab} e_c
\eeq
and define
\beq
\label{e9}
\gamma_{ab} =C^c_{~ad} C^d_{~bc} =\gamma_{ba}
\eeq
which is symmetric by definition,  
providing a natural inner product on the Lie algebra
\beq
\label{e10}
\gamma_{ab} \equiv e_a \cdot e_b 
= \gamma\left( e_a , e_b \right) \, ;
\eeq
when $\det\left(\gamma_{ab}\right)\not=0$, Eq. (\ref{e10})
is non-degenerate and the groups for which this is 
true are called semi-simple.
The $r$ vector fields $\left\{e_a \right\}$ 
may be used instead of the coordinate basis
 $\left\{ \frac{\partial}{\partial a^a} \right\}$ 
 as a basis in which to express an arbitrary vector field 
 on $G$. This basis is a frame.

%
%

\subsubsectionric{Definition of Homogeneity  \label{sec:homosp1}}

Suppose that the group acts on a manifold $M$
 as a group of transformations 
 \beq
 \label{e12}
 x^{\mu} \rightarrow f^{\mu}\left( x,a \right) 
 \equiv f^{\mu}_a \left(x\right) \, 
 \eeq
and 
let us define the orbit of $x$
\beq	
\label{e17}
f_G \left( x \right) = \left\{ f_a \left( x \right) \mid a \in G \right\}
\eeq
as the set of all points that can be reached 
from $x$ under the group of transformations. \\
Thus, the group of isotropy at $x$ is
\beq
\label{e18}
G_x = \left\{ a \in G \mid f_a \left(x \right) =x \right\}
\eeq
i.e., it is the subgroup of $G$ which leaves $x$ fixed.
Suppose $G_x=\left\{a_0\right\}$ and 
$f_G\left(x\right)=M$, i.e. every transformation
of $G$ moves the point $x$, and every point in $M$
can be reached from $x$ by a unique transformation.
Since $G|G_x=\{aa_0\,|a\in G\}=G$, 
$G$ is diffeomorphic to $M$ and one may identify 
the two spaces. \\
If $g$ is a metric on $M$ invariant under $G$, 
it corresponds to a left-invariant one on $G$, 
specified entirely by the inner products of the 
basis left-invariant vectors fields $e_a$. 
For three dimensions one obtains the family of spatially 
homogeneous space sections of the spatially homogeneous 
space-times. \\
Given a basis $\left\{e_a \right\}$ of the Lie algebra 
of a three dimensional Lie group $G$, with structure 
constants $C^c_{~ab}$, the spatial metric at each 
moment of time is specified by the spatially constant
inner products
\beq
\label{e19}
e_a \cdot e_b =g_{ab} \left(t \right) \, ,
\eeq
which are six functions of time. 
The Einstein equations, as we will apply, 
become ordinary differential equations for these 
six functions, plus whatever functions of time 
are necessary to describe the matter of the Universe.
For both homogeneous and spatially 
homogeneous space-times, one needs only to consider 
a representative group from each equivalence class 
of isomorphic Lie groups of dimension four and three,
respectively.
In three dimensions the classification 
of inequivalent three-dimensional Lie groups 
is called the Bianchi classification \cite{BIA97} 
and determines the various symmetry types possible 
for homogeneous three spaces, just as $(k=+1,0,-1)$ 
classify the possible symmetry types  for 
homogeneous and isotropic three-spaces (FRW). \\
%
%
%
%
After obtaining all the three-dimensional Lie groups
according to the Bianchi classification we
will  write down and discuss the corresponding Einstein equations.

\subsubsectionric{Application to Cosmology\label{sec:bianchispecc}}

A homogeneous space-time  is defined by space-like hypersurfaces 
$\Sigma$ such that for any points $p,q\in\Sigma$ there is a 
unique element $\tau\in G$ such that $\tau(p)=q$ 
(in this case the Lie group acts simply 
transitively on each $\Sigma$). 
Such uniqueness implies  $\dim G=\dim\Sigma=3$, 
and $G$ and $\Sigma$ can be identified
(for example, in the simplest case of translations 
group we have $G=\nR^3$); thus, the action of the isometries 
on $\Sigma$ is just the left multiplication on $G$
and tensor fields invariant under isometries are the 
left-invariant ones on $G$. 
In four dimensions one obtains the  
homogeneous space-times and 
the foliation of $M$ turns out to be $M=\nR\times G$. 
%
%
%
%
%
%
%
%
%
%
%
In order to preserve metric properties at all points, let us
consider the group of transformations
of coordinates which transform the space into itself, 
leaving the metric unchanged: if the line element 
before the transformation has the form
\beq
\label{f1}
dl^2 =h_{\alpha \beta} \left( x^1, x^2, x^3\right)
dx^{\alpha}dx^{\beta}  \, ,
\eeq 
then it becomes
\beq
\label{f2}
dl^2 =h_{\alpha \beta} \left( {x^{\prime1}}, {x^{\prime2}}, 
{x^{\prime3}}\right){dx^{\prime}}^{\alpha}{dx^{\prime}}^{\beta}  \, ,
\eeq 
where $h_{\alpha \beta}$ has the same form 
in the new coordinates. 

In the general case of a non Euclidean 
homogeneous three-di\-men\-sio\-nal space, there 
are three independent differential forms
which are invariant under the transformations 
of the group of  motions, however they do not
represent the total dif\-fe\-ren\-tial of any function
of the coordinates. We shall write them as $\omega^a = e^{a}_{\alpha} dx^{\alpha}$. 
Hence the metric (\ref{f2}) is re-expressed as
\beq
\label{f4}
dl^2 =\eta_{ab} \left( e^{a}_{\alpha} dx^{\alpha} \right) 
\left(e^{b}_{\beta} dx^{\beta} \right)
\eeq
so that the metric tensor reads as
\beq
\label{f5}
h_{\alpha \beta} =\eta_{ab}(t) e^{a}_{\alpha}(x^\gamma)e^{b}_{\beta}(x^\gamma),
\eeq
where $\eta_{ab}$ is a function of time only, 
symmetric in $ab$ and in contravariant
components we have
\beq
\label{f55}
h^{\alpha \beta} =
\eta^{ab}(t) e_{a}^{\alpha}(x^\gamma) e_{b}^{\beta}(x^\gamma) \, ,
\eeq
where $\eta^{ab}$ should be viewed as the 
components of the inverse matrix. 
All considerations developed in the previous Sections 
here apply straightforwardly.
%
%
%
%
%

The relationship between the covariant 
and contravariant expression for the three 
basis vectors is
\begin{equation}
\label{f6}
\mathbf{e}_{1} = 
\frac{1}{v} \left[ \mathbf{e}^{2} 
\wedge \mathbf{e}^{3} \right] \, , \qquad
\mathbf{e}_{2} = \frac{1}{v} 
\left[ \mathbf{e}^{3} \wedge \mathbf{e}^{1} \right] \, , \qquad
\mathbf{e}_{3} = \frac{1}{v} 
\left[ \mathbf{e}^{1}\wedge \mathbf{e}^{2} \right] \, , \qquad
\end{equation}
where $ \mathbf{e}^{a}$ and $ \mathbf{e}_{a}$
are to be understood formally as Cartesian vectors 
with components $e^a_{\alpha}$ and $e_a^{\alpha}$,
while $v$ represents 
\beq
\label{f7}
v= \mid e^{a}_{\alpha} \mid 
=e^{1} \cdot \left[e^{2}\wedge e^{3}\right] \, .
\eeq
The determinant of the metric tensor (\ref{f5}) 
is given by 
$\gamma=\eta v^2$
 where  $\eta$ is the determinant of the matrix $\eta_{ab}$. \\
The invariance of the differential form (\ref{f2}) 
means that
\beq
\label{f9}
e^{a}_{\alpha} \left(x \right) dx^{\alpha} =e^{a}_{\alpha} 
\left(x^{\prime} \right) dx^{\prime \alpha}
\eeq
and $e^{a}_{\alpha}$ on both 
sides of (\ref{f9}) are the same functions 
expressed in terms of the old and the new 
coordinates, respectively. \\
The algebra for the differential forms
permits to rewrite (\ref{f9}) as
\beq
\label{f10}
\frac{\partial x^{\prime \beta}}{\partial x^{\alpha}} = 
e^{\beta}_{a} \left(x^{\prime} \right) 
e^{a}_{\alpha} \left(x \right) 
\, .
\eeq
This is a system of differential 
equations which define the change of 
coordinates $x^{\prime \beta}(x)$ 
in terms of given basis vectors. \\
Integrability over the system (\ref{f10})
is rewritten in terms of the Schwartz condition
\beq
\label{f11}
\frac{\partial^2 x^{\prime \beta}}{\partial x^{\alpha}\partial x^{\gamma}} =
 \frac{\partial^2 x^{\prime \beta}}{\partial x^{\gamma}\partial x^{\alpha}} \, 
\eeq
which, explicitly, leads to 
\begin{align}
\label{f12}
\left[\frac{\partial e^{\beta}_{a} \left(x^{\prime} \right)}{\partial x^{\prime 
\delta}}  e^{\delta}_{b} \left(x^{\prime} \right) - \frac{\partial 
e^{\beta}_{b} \left(x^{\prime} \right)}{\partial x^{\prime \delta}}  
e^{\delta}_{a} \left(x^{\prime} \right) \right] &e^{b}_{\gamma} 
\left(x\right)e^{a}_{\alpha} \left(x\right)=   \nonumber \\
   &= 	e^{\beta}_{a} \left(x^{\prime} \right) \left[ \frac{\partial 
	e^{a}_{\gamma} \left(x\right)}{ \partial x^{\alpha}} - \frac{\partial 
	e^{a}_{\alpha} \left(x\right)}{ \partial x^{\gamma}}	\right].
\end{align}  
Multiplying both sides of (\ref{f12}) by 
$e^{\alpha}_{d}(x)
e^{\gamma}_{c}(x) e^{f}_{\beta}(x^{\prime})$ 
and differentiating, the left-hand side becomes
\begin{align}
\label{f13}
e^{f}_{\beta}\left(x^{\prime}\right) \left[\frac{\partial 
e^{\beta}_{d} \left(x^{\prime} \right)}{\partial x^{\prime \delta}}  
e^{\delta}_{c} \left(x^{\prime} \right) \right. &- \left. \frac{\partial e^{\beta}_{c} 
\left(x^{\prime} \right)}{\partial x^{\prime \delta}}  e^{\delta}_{d} 
\left(x^{\prime} \right) \right] = \nonumber \\
&=e^{\beta}_{c} \left(x^{\prime} \right) e^{\delta}_{d} 
\left(x^{\prime} \right)\left[ \frac{\partial e^{f}_{\beta} 
\left(x^{\prime}\right)}{ \partial x^{\prime \delta}} - \frac{\partial 
e^{f}_{\delta} \left(x^{\prime}\right)}{ \partial x^{\prime \gamma}}\right] \, 
\end{align}
and the right-hand side gives the same expression
but in terms of $x$. \\
Since $x$ and $x^{\prime}$ are arbitrary, 
both sides must be constant, 
and (\ref{f13})
reduces to
\beq
\label{f14}
\left( \frac{\partial e^{c}_{\alpha}}{\partial x^{\beta}} - \frac{\partial 
e^{c}_{\beta}}{\partial x^{\alpha}}\right) 
e^{\alpha}_{a}e^{\beta}_{b} =C^c_{~ab} \, , 
\eeq
which gives the relations of the vectors with 
 the group structure 
constants $C^c_{~ab}$. 
Multiplying (\ref{f14}) by 
$e^{\gamma}_{c}$, we finally have
\beq
\label{f15}
e^{\alpha}_{a} \frac{\partial e^{\gamma}_{b}}{\partial x^{\alpha}} - 
e^{\beta}_{b} \frac{\partial e^{\gamma}_{a}}{\partial x^{\beta}} 
=C^c_{~ab}e^{\gamma}_{c}\, .
\eeq

Similarly, such expression in the forms language 
is given by  
the left-invariant 1-form $\omega^a$ satisfying 
the Maurer-Cartan equation 
\be\label{maucar}
d\omega^a=\f12C^a_{bc}\omega^b\wedge\omega^c.
\ee

By construction, we have the antisymmetry 
property from (\ref{f13}) or (\ref{f14})
\beq
\label{f16}
C^c_{~ab}=-C^c_{~ba} \, .
\eeq
Defining 
\beq
\label{f17}
X_a =e^{\alpha}_{a}  \frac{\partial }{\partial x^{\alpha}} \, ,
\eeq
equation (\ref{f15}) rewrites  as
\beq
\label{f18}
\left[ X_a ,X_b \right] 
=C^c_{~ab}X_c \, .
\eeq
Homogeneity is expressed as the Jacobi identity
\beq	
\label{f19}
\left[ \left[ X_a,X_b \right] , X_c \right] + 
\left[ \left[ X_b,X_c \right] , X_a \right] 
+\left[ \left[ X_c,X_a \right] , X_b \right] =0  
\eeq
and explicitly 
\beq
\label{f20}
C^f_{~ab}C^d_{~cf}+C^f_{~bc}C^d_{~af}+C^f_{~ca}C^d_{~bf}=0 \, .
\eeq
With this formalism, the Einstein equations
for a homogeneous Universe can be written 
as a system of \textit{ordinary} differential 
equations which involve only functions of time, 
provided all three-dimensional vectors and
tensors are projected on the tetradic basis, 
while  the explicit coordinate dependence of the basis 
vectors
is not necessary
for the equations ruling the dynamics. 
In fact, such choice is not unique as $e_{a} = A_{ba} e^{b}$ yields again a set of basis vectors. \\
Introducing the two-index structure 
constants as $C^c_{~ab}= \varepsilon_{abd}C^{dc}$,
where $\varepsilon_{abc}=\varepsilon^{abc}$ 
is the Levi-Civita tensor ($\varepsilon_{123}=+1$),
the Jacobi identity (\ref{f20}) becomes
\beq
\label{f23}
\varepsilon_{bcd} C^{cd}C^{ba}=0.
\eeq

The problem of classifying  all homogeneous
spaces reduces to  finding all 
inequivalent sets of structure constants.


\subsectionric{Bianchi Classification and Line Element\label{sec:bianchile}}

The list of all three-dimensional Lie algebras was first 
accomplished by Bianchi \cite{Bianchi} and each algebra 
uniquely determines the local proprieties of a three-dimensional 
group. 
If a space is homogeneous and its Lie group is the ``Bianchi Type N'' 
(N=I,...,IX),
the subclassification of the Bianchi groups agrees with the one 
made by Ellis and MacCallum \cite{El-Ma,Ry-Sh,LF}. 

Any structure constant  
can be written also as 
\be\label{m+a}
C^a_{bc}=\varepsilon_{bcd}m^{da}+\delta^a_ca_b-\delta^a_ba_c \, ,
\ee
where the matrix $m^{ab}=m^{ba}$. 
The subclassification as class A and class B models
refers to the cases  $a_c=0$ or $a_c\neq 0$, respectively. 

The Jacobi identity (\ref{f20}) 
written for the structure constants like in 
(\ref{m+a}) reduces to the condition
\be\label{con1}
m^{ab}a_b=0.
\ee
Without loss of generality, we can put $a_c=(a,0,0)$ 
and the matrix $m^{ab}$ can be described by its 
principal values $n_1,n_2,n_3$. 
The condition (\ref{con1}) becomes  nothing but 
$an_1=0$, i.e. either $a$ or $n_1$ has to vanish.
Condition (\ref{f23}) rewrites explicitly as 
\bea
\label{f26}
\left[ X_1,X_2\right] &=&-aX_2 +n_3 X_3 \nonumber \\
\left[ X_2 ,X_3 \right] &=&n_1X_1 \\
\left[ X_3,X_1\right]&=& n_2X_2 +aX_3 \nonumber
\eea
where $a\geq 0$,  $(n_1, n_2, n_3)$ can be rescaled 
to unity, and 
we finally get the Bianchi classification
as in the Table \ref{tab:bianchi}.

Let us note that 
the Bianchi type I is isomorphic to the three-dimensional 
translation group $\mathcal\nR^3$, for which the flat 
FRW model is a particular case (once isotropy is restored)
and  analogously the Bianchi type V contains, as a particular case, 
the open FRW.

\begin{table}[htbp]

\begin{center}
\tbl{Inequivalent structure constants
corresponding to the Bianchi classification.}
{\begin{tabular}{l|c|c|c|r}

\toprule

\textbf{Type} 	&\quad \textbf{a} \quad	&\quad $\bf{n_1}$\quad 	& \quad$\bf{n_2}$ \quad	& 

\quad$\bf{n_3}$ \quad 	 \\
\hline \hline
I	&0	&0	&0	&0	\\

II	&0	&1	&0	&0	\\

VII	&0	&1	&1	&0	\\

VI	&0	&1	&-1	&0	\\

\hline \hline

\textbf{IX} &\bf{0}	&\bf{1}	&\bf{1}	&\bf{1}	\\

\textbf{VIII}	&\bf{0}	&\bf{1}	&\bf{1}	&\bf{-1}	\\

\hline \hline

V	&1	&0	&0	&0	\\

IV	&1	&0	&0	&1	\\

VII	&a	&0	&1	&1	\\

$ 
\begin{array}{ll} 
{\rm III} ~(a=1) \\
{\rm VI} ~(a \not=1)
\end{array}  \Biggl\}$ &a	&0	&1	&-1	\\
 \botrule
\end{tabular}
\label{tab:bianchi}}
\end{center}

\end{table}

Not all anisotropic dynamics are compatible with a 
satisfactory Standard Cosmological Model but, as shown 
in the early Seventies, some can be represented, 
under suitable conditions, 
as a FRW model plus a gravitational waves packet 
\cite{L74,Grishchuk1975ZETF}. \\
The interest in  
the Mixmaster \cite{Misner1969PR,Misner1969PRL} relies on having 
invariant geometry under the $SO(3)$ group, 
shared with the closed FRW Universe. 
From a cosmological point of view, the relevance of this model 
arises also from 
the decomposition of the  line element as
\begin{equation}
\label{ds}
ds^2={ds_0}^2 -\delta_{(a)(b)}G^{(a)(b)}_{ik}dx^i dx^k
\end{equation}
where $ds_0$ denotes the line element of 
an isotropic Universe having 
positive constant curvature, $G^{(a)(b)}_{ik}$ is a 
set of spatial tensors
and $\delta_{(a)(b)}(t)$ are amplitude functions, 
resulting small sufficiently far
from the singularity. The tensors introduced
in (\ref{ds})  satisfy the equations 
\beq
G^{(a)(b);l}_{~ik~~;l}= -(n^2-3)G^{(a)(b)}_{~ik}, \quad 
G^{(a)(b)k}_{~~;k}=0, \quad
G^{(a)(b)i}_{~~i}=0 \, , 
\eeq
in which the Laplacian is referred to the 
geometry of the sphere of unit radius.


Let us choose a basis of dual vector fields $\omega^a$, 
preserved under isometries. 
Therefore, recalling (\ref{f55}), the four-dimensional line element 
can be expressed as 
\be
ds^2= N^2(t) dt^2-\eta_{ab}(t)\,\omega^a\otimes\omega^b,
\ee
parametrized  by the proper time,
where $\omega^a$ obey the Maurer-Cartan equations (\ref{maucar}). 

The explicit expression for the $\omega^a$ 
for Bianchi I,  
whose structure constants are $C^a_{bc}=0$, is 
\be
\omega^1=dx^1 \, , \qquad \omega^2=dx^2 \, , \qquad \omega^3=dx^3 \, ,
\ee
while for Bianchi IX, being  $C^a_{bc}=\varepsilon_{abc}$, 
will be specified in (\ref{h27}).

In order to distinguish between expansion (change of volume) 
and anisotropy (change of shape) it is useful to parametrize 
the metric on the spatial slices as
\be
 \label{erreab}
\eta_{ab}=R_0^2e^{2\alpha}\left(e^{2\beta}\right)_{ab},
\ee
where $R_0$ is the initial radius of the Universe and all the 
other parameters are functions of time only. 
The matrix $\beta_{ab}$ satisfies the  condition Tr$\beta=0$,
ensuring  hypersurface three-volume  dependance  on the 
conformal factor $\alpha$ only as 
$V_{Univ}\sim R_0^3e^{3\alpha}$ \cite{MTW}.

For a review of works devoted to the study of the dynamics of 
Bianchi models in different cosmological paradigms see the 
collection of articles \reffcite{1989GReGr..21..211N, 
1989JMaPh..30..757R,1992GReGr..24..679F,1995IJPAM..26..169R,1996IJMPD...5...65C,1972PhLA...40..385M,1979IJTP...18..371T,1979PhR....56...65C,1986PhR...139....1B,1989GReGr..21..413V,1989grg..conf..351R,1990CQGra...7..611R,1990CQGra...7..625R,1990GReGr..22..595M,1990PhRvD..42..404U,1991CQGra...8.2191A,1992CQGra...9..651C,1992CQGra...9.1525M,1992CQGra...9.1923C,1993CQGra..10...99H,1993CQGra..10..703K,1993CQGra..10.1607C,1993PhRvD..47.1396R,1993PhRvD..48.5642O,1995CQGra..12.1099S,1997gr.qc.....9012R,1997IJMPD...6..491C,1999GReGr..31.1423K,2000GReGr..32.1319N,2000GReGr..32.1981N,2000JMP....41.4777M,
2002CMaPh.226..377C,2002PThPh.107..305K,2003CQGra..20.5425G,2003GReGr..35..475K,2004CQGra..21.1609F,2004CQGra..21.4193C,2004GReGr..36..799H,2004PhRvD..70h4040M,2006GReGr..38.1003E,2006MNRAS.369..598C,2007arXiv0710.5723T,2007CQGra..24..931V,1983NCimB..77...62S,0264-9381-20-15-307,0264-9381-22-2-006,0264-9381-22-21-002}.

\subsectionric{Field Equations\label{sec:bianchife}}

As we derived, the Einstein equations 
for a homogeneous 
Universe can be written in the form of 
a system of ordinary differential equations
which involve functions of time only, once introduced
 the tetradic basis
built as in the previous Section \ref{sec:tetrad}, whose projections 
in empty space take the form
\bseq
\bea
\label{g3a}
R^0_0 &=&-\frac{1}{2} {\dot{\kappa}}^{a}_{a} 
-\frac{1}{4}{\kappa}^{b}_{a}{\kappa}^{a}_{b} \\
\label{g3b}
R^0_{a}&=&-\frac{1}{2}{\kappa}^{c}_{b}\left(C^b_{~ca} 
-\delta^b_a C^d_{~dc}\right) \\
\label{g3c}
R^{a}_{b}&=&-\frac{1}{2 \sqrt{\eta}} 
\frac{\partial}{\partial t}
{\left(\sqrt{\eta} {\kappa}^{a}_{b}\right)} -P^{a}_{b} \, , 
\eea
\eseq
where the relations ${\kappa}_{ab} =\dot{\eta}_{ab}$ and ${\kappa}^{b}_{a}=\dot{\eta}_{ac}\eta^{cb}$
hold,  the dot denotes differentiation with 
respect to $t$, and the projection $P_{ab}=\eta_{bc}P^{c}_{a}$
of the three-dimensional Ricci tensor becomes
\begin{align}
\label{g6}
P_{ab}= -\frac{1}{2} \Big( C^{cd}_{~~b}C_{cda}~+~
C^{cd}_{~~b}C_{dca} ~-~\frac{1}{2} C_b^{~cd}C_{acd} 
   ~+~C^c_{~cd}C_{ab}^{~d} 
~+~C^c_{~cd}C_{ba}^{~~d}\Big) \, .
\end{align}
The Einstein equations have reduced to a much 
simpler differential system, involving only 
ordinary derivatives with respect to the 
temporal variable $t$.

In the following we will  discuss the 
Kasner solution which will be generalized to
 the dynamics of Bianchi types VIII
and IX.

\subsectionric{Kasner Solution\label{sec:kasnersol}}

\begin{figure}[ht]
\begin{center}
\includegraphics[
,width=0.6\textwidth]{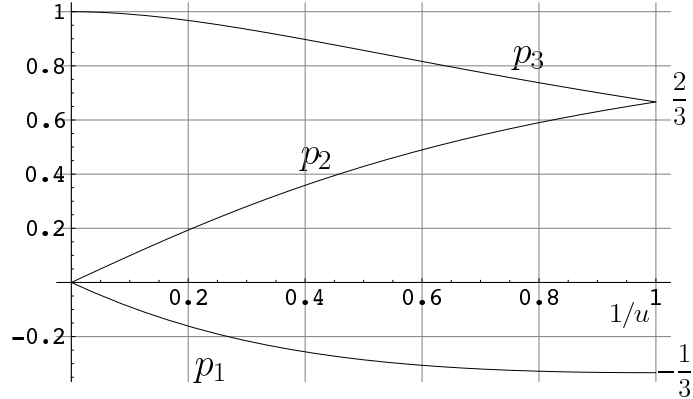}
\caption[Kasner indices in terms of the 
parameter $1/u$]{Evolution of Kasner indices in terms of the 
parameter $1/u$. The domain of  $u$ is $[1\,,\infty)$; 
for lower values of $u$ the inversion property 
(\ref{g20c}) holds.} 
\label{fig:kasner}
\end{center}
\end{figure}

The simplest and paradigmatic solution 
of the Einstein equations (\ref{g3a})-(\ref{g3c})
in the framework of the Bianchi classification
is the type I model, firstly obtained 
by  Kasner\cite{kasner} in 1921,
%
%
which is appropriate to describe the gravitational field 
in empty space. \\
The simultaneous vanishing of the three
structure constants and of the parameter $a$ implies
the vanishing  of the three-dimensional Ricci tensor as well
\beq
\label{g7}
\begin{array}{ll} 
e^a_{\alpha} =\delta^a_{\alpha} \\
C^c_{ab} \equiv 0
\end{array}  \Biggl\} \Longrightarrow P_{ab} =0 \, .
\eeq
Furthermore, since the three-dimensional metric tensor does not 
depend on space coordinates, also $R_{0\alpha}=0$, 
i.e. this model contains the standard Euclidean space as 
a particular case.
Then the system (\ref{g3a})-(\ref{g3c}) describes 
a uniform space and
reduces to 
\bseq
\bea
\label{g8a}
{\dot{\kappa}}^a_a +\frac{1}{2} {\kappa}^b_a{\kappa}^a_b &=&0  \\
\label{g8b}
\frac{1}{\sqrt{\eta}} \frac{\partial}{\partial t}
{\left( \sqrt{\eta} {\kappa}^b_a\right)} &=&0 \, .
\eea 
\eseq
%
%
%
%
From (\ref{g8b}) we get the first integral 
\beq
\label{g9}
\sqrt{\eta}\kappa^b_a =2 \lambda^b_a =\textrm{const.}\, ,
\eeq
and contraction of indices  $a$ and  $b$ leads to
\beq
\label{g10}
\kappa^a_a = \frac{\dot{\eta}}{\eta}=
\frac{2}{\sqrt{\eta}}\lambda^a_a \, ,
\eeq
and finally  
\beq
\label{g11}
\eta= G t^2 \qquad G=\textrm{const.} \, .
\eeq
Without loss of generality, a simple 
rescaling of the coordinates $x^{\alpha}$
allows to put such a constant equal to unity, thus 
providing 
\beq
\label{g12}
\lambda^a_a =1 \, .
\eeq
Substituting  (\ref{g9}) in 
(\ref{g8a}) one obtains the relations among the constants
$\lambda^b_a$
\beq
\label{g13}
\lambda^a_b\lambda^b_a =1 \, ,
\eeq 
and lowering index  $b$ in  (\ref{g9})  one gets
a system of ordinary differential equations with respect to
 $\gamma_{ab}$
\beq
\label{g14}
\dot{\eta}_{ab} =\frac{2}{t} 
\lambda^c_a\eta_{cb} \, .
\eeq
The set of coefficients $\lambda^c_a$ 
can be considered as the matrix of a certain 
linear transformation, reducible to its 
diagonal form.
In such a case, denoting its 
eigenvalues as  $p_1$, $p_2$, $p_3$ real
and not equal to each other, and  its eigenvectors as
${\bf n}^{\left(1\right)}$,
${\bf n}^{\left(2\right)}$, 
${\bf n}^{\left(3\right)}$, 
the solution of  (\ref{g14}) writes as
\begin{equation}
\label{solkasn}
\eta_{ab} = 
t^{2p_{1}} n_{a}^{(1)} n_{b}^{(1)} 
+ t^{2p_{2}} n_{a}^{(2)} n_{b}^{(2)} 
+ t^{2p_{3}} n_{a}^{(3)} n_{b}^{(3)} \, .
\end{equation}
If we choose the frame of the eigenvectors 
(recall that $e^a_\alpha= \delta^a_\alpha$) and denote 
them with $x^1, x^2, x^3$, then the spatial line element reduces to
\beq
\label{g16}
dl^2 =t^{2p_1} (dx^1)^2 +t^{2p_2} (dx^2)^2 +t^{2p_3} (dx^3)^2 \, .
\eeq
Here $p_1, p_2, p_3$ are 
the so-called Kasner indices, satisfying the relations
\beq
\label{g17}
p_1+p_2+p_3 =
p^2_1+p^2_2+p^2_3=1 \,  ,
\eeq
therefore only one of these numbers is 
independent. Except for the cases
$(0,0,1)$ and $(-\nicefrac{1}{3},\nicefrac{2}{3},\nicefrac{2}{3})$, 
such indices are never equal, one of them being 
negative and two positive;
in the peculiar case  $p_1=p_2=0,p_3=1$, the metric
is reducible to a Galilean form by the transformation
\beq
\label{g21}
t \sinh x^3 =\xi \, , \quad t \cosh x^3 =\tau \, ,
\eeq 
i.e. with a fictitious singularity in a 
flat space-time.\\
Once that Kasner indices have been ordered 
according to 
\beq
\label{g19}
p_1 <p_2<p_3 \, ,
\eeq
their corresponding variation ranges are 
\beq
\label{g20}
-\frac{1}{3} \leq  p_1 \leq 0 \, , \quad
0 \leq  p_2 \leq \frac{2}{3} \, , \quad
\frac{2}{3} \leq  p_3 \leq 1 \, .
\eeq
In parametric form we have the representation
\begin{align}
\label{g20a}
p_1\left(u\right) =\frac{-u}{1+u+u^2} \, \qquad 
p_2\left(u\right) =\frac{1+u}{1+u+u^2} \, \qquad
p_3\left(u\right) =\frac{u\left(1+u\right)}{1+u+u^2} 
\end{align}
as the parameter $u$ varies in the range (see Fig.\ref{fig:kasner})
\beq
\label{g20b}
1 \leq u < + \infty.
\eeq

The values $u<1$ lead to the same range following the 
inversion property 
\beq
\label{g20c}
p_1 \left(\frac{1}{u}\right) =p_1\left(u\right) \, , \quad
p_2 \left(\frac{1}{u}\right) =p_3\left(u\right) \, , \quad
p_3 \left(\frac{1}{u}\right) =p_2\left(u\right) \,. 
\eeq

The line element from (\ref{solkasn}) describes an 
anisotropic space where all the vo\-lu\-mes linearly grow 
with time, while  linear distances grow along two directions and 
decrease along the third, differently from 
 the 
Friedmann solution where all distances 
contract towards the singularity 
with the same behavior.
This metric has only one non-eliminable  singularity in  $t=0$
with the only exception of the case $p_{1} = p_{2} = 0,\,p_{3}=1$
mentioned above,
 corresponding to the standard Euclidean space.

For a discussion of the Bianchi I model in presence of several 
matter sources, see the following literature \reffcite{1977PThPh..57...67N,1978PhRvD..18..969H,1981PhRvD..24..305S,
1982AZh....59.1044R,1982PhRvD..26.3367C,1983JApA....4..295M,
1983PhRvD..28.1853A,1985A&A...151....7M,1985PhLB..152..171G,
1987Ap&SS.136...17R,1987PhLA..121....7C,1988PhLA..129..429H,
1990ApJ...358...23B,1991CQGra...8.1173M,1992CQGra...9S.183M,
1992PhRvD..46.1551B,1993Ap&SS.199..289G,1993PhRvD..47.3151M,
1994GReGr..26..307M,1994PhRvD..50.2431M,1995Ap&SS.225...57S,
1995PhRvD..52.5445C,1996JMP....37..438R,1997CQGra..14.2281L,
1997PhLB..408...47C,1998Ap&SS.262..145B,2000CQGra..17.2215T,
2000PhRvD..62j4017J,2000PhRvD..62l4016C,2001PhRvD..64j7301K,
2002IJMPD..11..447M,2003Ap&SS.288..523M,2005GReGr..37.1233F,
2005MPLA...20.2127S,2006Ap&SS.301..127P,2006CQGra..23.3463H,
2006JMP....47d2505C,2006PhyD..219..168S,2007Ap&SS.tmp..329C}.

\subsectionric{The role of matter}

\newcommand{\df}[2]{  \displaystyle\frac{#1}{#2}}
\newcommand{\DER}[3]{ \displaystyle\frac{\partial^{#3} #1}{\partial #2^{#3}}}
\newcommand{\DERT}[3]{\displaystyle\frac{d^{#3} #1}{d #2^{#3}}}
\newcommand{\DERO}[3]{\displaystyle\frac{\partial #1}{\partial #2^{#3}}}

Here we discuss the time evolution of a uniform distribution of 
matter in the Bianchi type I space near the singularity; 
it will result that it  behaves as a test fluid and thus 
it does not affect the properties of the solution.\\
Let us take a uniform distribution of matter and  assume that 
we can neglect its influence  on the gravitational field. 
The hydrodynamics equations describe the evolution \cite{LF} as
\bseq
\label{c1_idro}
\begin{align}
\df{1}{\sqrt{-g}}\DERO{}{x}{i} \left(\sqrt{-g} \sigma u^{i}\right) = 0\;,
\label{c1 idro2}
(p+\epsilon) u^{k} \left(\DERO{u_{i}}{x}{k}-\df{1}{2} u^{l} 
\DERO{g_{kl}}{x}{i}\right) = -\DERO{p}{x}{i}-u_{i} u^{k} \DERO{p}{x}{k}\;.
\end{align}
\eseq
Here $u^{i}$ is the four-velocity and $\sigma$ is the entropy density; 
in the neighbourhood of the singularity it is necessary to use the 
ultra-relativistic equation of state $p=\epsilon/3$, and then we get
$\sigma\sim\epsilon^{3/4}$.\\
As soon as all the quantities are functions of time, we have
\begin{equation}
\label{c1 id3}
\df{d}{d t} \left(a b c u_{0} \epsilon^{3/4}\right)=0\,,
\qquad  4\epsilon\df{d u_{\alpha}}{d t}+u_{\alpha} \df{d\epsilon}{d t}=0\;.
\end{equation}
From (\ref{c1 id3}), we obtain  the two integrals of motion
\begin{equation}
\label{c1 id4}
a b c u_{0}\epsilon^{3/4}=  \textrm{const.} \,,
\qquad u_{\alpha}\epsilon^{1/4}= \textrm{const.}\,.
\end{equation}
From (\ref{c1 id4}) we see that  all the covariant components 
$u_{\alpha}$ are  of the same order. 
Among the controvariant ones, the greatest as $t\to 0$ is $u^{3}=u_{3}$. 
Retaining only the dominant contribution in the identity $u_{i}u^{i}=1$, 
we have  $u_{0}^{2}\approx u_{3}^{2}$ and, from (\ref{c1 id4}), 
\begin{equation}
\label{c1 id5}
\epsilon \sim \df{1}{a^{2}b^{2}},\hspace{0.5cm}u_{\alpha}\sim \sqrt{a b}\,,
\end{equation}
or, equivalently, for the Kasner solution
\begin{equation}
\label{c1 id6}
\epsilon\sim t^{-2(p_{1}+p_{2})}=t^{-2(1-p_{3})}\;,
\hspace{0.5cm}u_{\alpha}\sim t^{(1-p_{3})/2}\;.
\end{equation}
As expected, $\epsilon$ diverges as $t\to 0$ for all the values 
of $p_{3}$, except $p_{3}=1$ (this is due to the non-physical 
character of the singularity in this case).\\
The validity of the test fluid approximation is verified from a direct 
evaluation of the components of the energy-momentum tensor $T^{k}_{i}$, 
whose dominant terms are 
\bseq
\begin{equation}
\label{c1 id7}
T^{0}_{0}\sim \epsilon u_{0}^{2}\sim t^{-(1+p_{3})}\;,
\hspace{0.5cm}T^{1}_{1}\sim \epsilon\sim t^{-2(1-p_{3})}\;,
\end{equation}
\begin{equation}
\label{c1 id8}
T^{2}_{2}\sim \epsilon u_{2} u^{2} \sim t^{-(1+2p_{2}-p_{3})}\;,\hspace{0.5cm}T^{3}_{3}\sim \epsilon u_{3}u^{3}\sim t^{-(1+p_{3})}\;.
\end{equation}
\eseq
As $t\to 0$, all the components grow slower than $t^{-2}$, 
which is the behavior of the do\-mi\-nant terms in the 
Kasner analysis. 
Thus the fluid contribution can be disregarded in the 
Einstein equations.

This test character of a perfect fluid on a Kasner background 
remains valid even in the following Mixmaster scenario, 
both in the homogeneous as well as in the inhomogeneous 
case \cite{BKL1970,BKL82}. 
The reason for the validity of such extension relies on 
the piece-wise Kasner behavior of the oscillatory regime.
For a discussion on the effects of ultra-relativistic matter, 
of scalar and electromagnetic  field on the quasi-isotropic solution, 
see \reffcite{Montani1999CQG,Montani2000CQGquasiisotropic,
Montani2000CQGultrarelativistic}.

\subsectionric{The Dynamics of the Bianchi Models\label{sec:bklpiece}}


The Kasner solution properly approximates those cases
when 
the Ricci tensor appearing in the Einstein 
equations $P_{\alpha \beta}$ is 
of higher order
in $1/t$ with respect to all other terms involved. 
However, since one of the Kasner exponents is 
negative, terms of order higher than $t^{-2}$
appear in the tensor $P_{\alpha \beta}$. In such a
case the discussion of solutions has to be 
extended to the general anisotropic case, in the search 
of a general behaviour of the Universe towards the 
initial singularity. 
In fact, the outlined Kasner regime

relies on a restriction over  the phase space of 
the solution (not discussed here in the details, 
see \reffcite{BKL1970}, \S 3)
which causes an instability with perturbations 
violating this condition.\\
A general solution,  by definition, 
\begin{itemize}
\item 
	is completely 
	stable, i.e. the effect of any perturbation 
	is equivalent to a change of the initial conditions
	at some moment of time and

\item
	must satisfy arbitrary initial conditions, 
	i.e. the perturbation
	cannot change the form of the solution. 

\end{itemize}
 
Nevertheless,
the cited restriction over the Kasner solution makes
it unstable with respect to perturbations 
destroying it: such perturbation promoting the 
transition to a new state cannot 
be considered small and lies outside the region 
of the infinitesimal ones.

Let us introduce three spatial vectors 
$e^{a}= l(x^\gamma), m(x^\gamma), n(x^\gamma)$ 
and take the matrix $h_{\alpha\beta}$ diagonal
in the form 
\begin{equation}
	\label{metricaix}
		h_{\alpha\beta}= a^{2}(t) l_{\alpha} l_{\beta} 
	+ b^{2}(t) m_{\alpha} m_{\beta} + c^{2}(t) n_{\alpha} n_{\beta}\;.
\end{equation}
Consequently, 
the Einstein equations in a synchronous 
reference system and for a generic homogeneous
cosmological model in empty space are given 
by the system 
\bseq
\begin{align}
-R^l_l &=\frac{{\left( \dot{a}  b c \right)}^{.}}{abc}
	+\frac{1}{2a^2b^2c^2} 
	\left[\lambda^2{a}^4 
	-{\left(\mu b^2-\nu c^2\right)}^2 \right]=0  \;\;\\
-R^m_m &= \frac{{( a \dot{b} c )}^{.}}{abc}
	+ \frac{1}{2a^2b^2c^2} 
	\left[\mu^2 {b}^4 -{\left(\lambda a^2- \nu c^2\right)}^2\right]  =0 \;\; \\
-R^n_n &= \frac{{\left(ab\dot{c} \right)}^{.}}{abc}
	+ \frac{1}{2a^2b^2c^2}
	\left[\nu^2 {c}^4 -{\left(\lambda a^2-\mu b^2\right)}^2 \right] =0 \;\;
\end{align}
\label{eeh}
\eseq
and 
\beq
\label{eeh0}
-R^0_0=\frac{\ddot{a}}{a}+\frac{\ddot{b}}{b}+\frac{\ddot{c}}{c}=0
\eeq
where 
the other 
off-diagonal components of the four-dimensional 
Ricci tensor  identically vanish as a consequence
of the choice of the diagonal form as in (\ref{metricaix}).
Eventually, the $0\alpha$ components of the Einstein equations
can be non-zero if some kind of matter is present, 
 leading to an effect of rotation on the Kasner axes\cite{BKL1970}.
The constants $\lambda,\mu,\nu$ correspond to the 
structure constants 
$C_{11},C_{22},C_{33}$  respectively, 
introduced earlier  
in Section \ref{sec:bianchispecc}.
In particular, we will study in details the
cases of $(\lambda, \mu, \nu)$ for Bianchi 
type II (1, 0, 0), VII (1, 1, 0), VIII (1, 1, -1) and IX (1, 1, 1). \\
All these equations are exact and contain 
 functions of time only,  without any 
restriction regarding the vicinity to the 
singular point $t=0$. 
Through the notation
\beq
\label{g28} 
\alpha =\ln a \, , \quad
\beta  = \ln b \, , \quad
\gamma = \ln c 
\eeq
and the new temporal variable $\tau$ defined by
\beq
\label{g29}
dt=abc ~d\tau \, , 
\eeq
Eqs. (\ref{eeh}) and (\ref{eeh0}) simplify to
\bseq
\bea
2\alpha_{\tau\tau}&=&{\left(\mu b^2-\nu c^2\right)}^2-\lambda^2 {a}^4  \\
2\beta_{\tau \tau}&=&{\left(\lambda a^2-\nu c^2 \right)}^2-\mu^2{b}^4 \\
2\gamma_{\tau \tau}&=&{\left(\lambda a^2-\mu b^2\right)}^2-\nu^2{c}^4 \, ,  
\eea
\label{g30}
\eseq
\beq
\frac{1}{2}\left(\alpha+\beta+\gamma\right)_{\tau\tau}
=\alpha_{\tau}\beta_{\tau}+
\alpha_{\tau}\gamma_{\tau}+\beta_{\tau}\gamma_{\tau} \, , 
\label{g30a}
\eeq
where  subscript $\tau$ denotes the  derivative with 
respect to $\tau$. 
Manipulating the system  (\ref{g30}) 
and using (\ref{g30a}), 
one obtains  the first integral
\begin{align}
\label{g31}
\alpha_{\tau}\beta_{\tau}+  \alpha_{\tau}\gamma_{\tau}
&+\beta_{\tau}\gamma_{\tau} = \nonumber \\
				= & 
	\frac{1}{4}
\Big(\lambda^2 a^4+\mu^2b^4+\nu^2c^4 
	- 	2\lambda \mu a^2b^2-
			2\lambda \nu a^2c^2-2\mu\nu b^2c^2\Big)
\end{align}
involving  first derivatives only.
The Kasner regime (\ref{g16}) discussed before is the solution of 
equations (\ref{g30}) corresponding to neglecting 
all terms on the right-hand side. 
However, such a situation cannot 
persist indefinitely as $t\rightarrow 0$ since
there are always some terms on the right-hand side
of (\ref{g30}) which are increasing.

\subsectionric{Application to Bianchi Types II and VII\label{sec:appl27}}

In this paragraph, the dynamics of the types II and VII is discussed 
in some detail: these spaces, in fact, present some features in 
common with the Mixmaster model, as it will be clear in the 
Sections \ref{sec:csol}-\ref{sec:bkl}.

\subsubsectionric{Bianchi type II\label{sec:appl2}}

Introducing the structure constants for the type II model,
the system (\ref{eeh}) reduces to
\bseq
\label{ein_ii}
\begin{align}
\label{ein_iia}
\displaystyle\frac{(\dot{a} b c)^{.}}{a b c} &= 
	-\displaystyle\frac{a^{2}}{2 b^{2} c^{2}}\,,\\
\label{ein_iib}
\displaystyle\frac{(a \dot{b} c)^{.}}{a b c} &= 
	~~\displaystyle\frac{a^{2}}{2 b^{2} c^{2}}\,,\\
\label{ein_iic}
\displaystyle\frac{(a b \dot{c})^{.}}{a b c} &= 
	~~\displaystyle\frac{a^{2}}{2 b^{2} c^{2}}\,,
\end{align}
\eseq
\begin{equation}
\label{ein eq bii r00}
\displaystyle\frac{\ddot a}{a} + \displaystyle\frac{\ddot b}{b} 
+ \displaystyle\frac{\ddot c}{c}  =0\,.
\end{equation}

In (\ref{ein_ii}) the right-hand sides 
play the 
role of a perturbation to the Kasner regime; if at a certain instant of time $t$ 
they could be neglected, then a Kasner dynamics would take place. 
This kind of evolution can be stable or not depending 
on the initial conditions; 
as shown earlier in Section \ref{sec:kasnersol}, 
the Kasner dynamics has a time evolution 
which differs along the three directions, growing along two 
of them and decreasing along the other. 
For example, for a perturbation growing as  $a^{4} \sim t^{4p_{a}}$,
toward the singularity, two scenarios are possible: 
if the perturbation is associated with one of the two positive indices, 
it will continue  decreasing till the singularity 
and the Kasner epoch is stable; 
on the other hand, if $p_{a}<0$,  the perturbation grows 
and cannot be indefinitely neglected. 
In this case, the analysis of the full dynamical system is 
required, and this can  be achieved 
with the  logarithmic variables (\ref{g28})-(\ref{g29})
and  the system (\ref{g30}) becomes
\bseq
\begin{align}
%
%
\label{g32}
\alpha_{\tau\tau}&=-\frac{1}{2}e^{4\alpha} \\
\label{g32b}
\beta_{\tau\tau}&=\gamma_{\tau\tau}=\frac{1}{2}e^{4\alpha} \, .
%
%
%
%
\end{align}
\label{g3232}
\eseq
The  equation (\ref{g32}) can be thought 
as the motion of a one-dimensional point-particle moving 
in an exponential potential: if the initial ``velocity'' 
$d\alpha/d\tau$ is equal to $p_{a}$, then the potential 
will slow down, stop and accelerate again the point up 
to a new ``velocity'' $-p_{a}$. 
From there on,  the potential will remain  negligible forever.\\
Furthermore, the second set of equations  
(\ref{g32b}) implies that the conditions 
\begin{equation}
\label{bii con vel}
\alpha_{\tau\tau}  + \beta_{\tau\tau} = 
	\alpha_{\tau\tau} + \gamma_{\tau\tau} = 0
\end{equation}
hold.
Considering the  explicit solutions of (\ref{bii con vel})
\bseq
\begin{align}
\label{bii sol}
\alpha (\tau) &= \displaystyle\frac{1}{2} \ln 
	\left(c_1 \text{sech}\left(\tau 
   		c_1+c_2\right)\right)\\
\beta (\tau) &= \displaystyle c_3 + \tau  c_4  
		-\frac{1}{2} \ln 
	\left(c_1 \text{sech}\left(\tau 
   		c_1+c_2\right)\right)\\
\gamma(\tau) &= \displaystyle c_5 + \tau  c_6
		-\frac{1}{2} \ln 	\left(c_1 \text{sech}\left(\tau 
   		c_1+c_2\right)\right)      \, , 
\end{align}
\eseq
where $c_1\, \ldots, c_6$ are integration constants and 
 we see how this dynamical scheme describes two connected 
Kasner epochs, where the perturbation has  the role 
of  changing the values of the Kasner indices. 
Let us assume that the Universe is initially 
described by a Kasner epoch for $\tau\to\infty$, 
with indices orderd as $p_{l}<p_{m}<p_{n}$; 
the perturbation starts growing and  the point bounces 
against the potential and a new Kasner epoch begins, 
where the old and the new indices (the primed ones) 
are related among them by the  so-called BKL map \cite{BKL1970}

\begin{equation}
\label{c1 mappa degli indici}
p'_{l} = \displaystyle\frac{|p_{l}|}{1 - 2 |p_{l}|} \;,
\qquad p'_{m} = -\displaystyle\frac{2 |p_{l}| - p_{m}}{1 - 2 |p_{l}|} \;,
\qquad p'_{n} = \displaystyle\frac{p_{n} - 2 |p_{l}|}{1 - 2 |p_{l}|} \;.
\end{equation}
In this new era, the negative  power 
is no longer related to 
 the $l$-direction so that the previously increasing 
perturbation is damped and eventually 
vanishes toward the singularity.
%
We will see how (\ref{c1 mappa degli indici}) will be valid in general.

For a detailed analysis on the main results concerning the Bianchi 
II model see the following literature 
\reffcite{1982GReGr..14..213B,1984GReGr..16.1189B,
Lorenz:1980fk,1987JMP....28.1382L,1997Ap&SS.253..205C,1997CQGra..14.2845R,
1997JMP....38.2611N,1998CQGra..15.1607L,2001GReGr..33...65H,
2001PhLB..501..264C}.

\subsubsectionric{Bianchi type VII\label{sec:appl7}}

The analysis of Bianchi type VII  can be performed
analogously, leading to  the Einstein equations
\bseq
\begin{align}
\label{ein eq bVii}
\displaystyle\frac{(\dot{a} b c)^{.}}{a b c} &= 
		~~\displaystyle\frac{-a^{4} + b^{4}}{2 a^{2} b^{2} c^{2}}\,,\\
\displaystyle\frac{(a \dot{b} c)^{.}}{a b c} &= 
		~~\displaystyle\frac{a^{4} -b^{4}}{2 	a^{2} b^{2} c^{2}}\,,\\
\displaystyle\frac{(a b \dot{c})^{.}}{a b c} &= 
		~~\displaystyle\frac{\left( a^{2}-b^{2} \right)^{2}}{2 b^{2} c^{2}}\,,
\end{align}
\eseq
and the constraint  (\ref{ein eq bii r00}) holding unchanged. 
Comparison of  (\ref{ein eq bVii}) with (\ref{ein_ii}) allows 
a similar qualitative analysis:
if the right-hand sides of (\ref{ein eq bVii}) are negligible at 
a certain instant of time, than the solution is Kasner-like
and  can be stable or unstable depending on initial conditions.
If the negative index is associated with the $n$ direction, 
than the perturbative terms $a^{4}$ and $b^{4}$, evolving  as 
$t^{4p_{l}}$ and $t^{4 p_{m}}$,  decrease up to the 
singularity and the Kasner solution turns out to be stable; 
in all  other cases, one and only one of the perturbation terms starts 
growing, blasting the initial Kasner evolution and ending as before 
in a new Kasner epoch. \\
The main difference between the types II and VII is that many 
other transitions can occur after the first one and 
this can happen, for example, if the new negative Kasner index 
is associated with the $m$ direction, i.e. with $b$. 
In this case,  the  $b^{4}$ term would  start growing and a new 
transition would occur with the same law (\ref{c1 mappa degli indici}). 
The problem of understanding if, when and how this mechanism can 
break up is unraveled considering the BKL map written in terms of 
the parameter $u$, i.e.
\beq
\label{g42}
\left.
\begin{array}{c}
p_l=p_1(u)  \\
p_m=p_2(u)  \\
p_n=p_3(u)  
\end{array}
\right\}\Rightarrow
\left\{
\begin{array}{c}
p^{\prime}_l=p_2\left(u-1\right)  \\
p^{\prime}_m=p_1\left(u-1\right)  \\
p^{\prime}_n=p_3\left(u-1\right) 
\end{array} \right.
\eeq
%
In this representation and from the properties (\ref{g20a}) and (\ref{g20c})
we see how  the exact number of exchanges among the $l$- and $m$-directions equates
the integer part $K$ of the initial value $u_{0}$ describing the  dynamics. 
In fact, for the first $K_{0}$ times, the negative index is exchanged among 
$l$ and $m$, then it passes to the $n$ direction,  a new and final 
(toward the singularity) Kasner epoch begins, and no more oscillations take place. 
The collection of the total of $K$ epochs is called a {\it Kasner era}: 
in this sense  we can say that in the general case the type VII dynamics is composed
by one era and  a final epoch. 

For additional informations about the Bianchi VII model 
(as well as some interesting features regarding the Bianchi VI 
Universes) see the 
articles \reffcite{1984CQGra...1...81R,1988JMaPh..29..449R,
1991Ap&SS.181...61R,1996CQGra..13.1273B,1998CQGra..15..331W,
2000CQGra..17.1435B,2006A&A...460..393J,2006ApJ...644..701J,
2006MNRAS.369.1858M,2007MNRAS.377.1473B,Apostolopoulos:2004uq}.


\sectionric{Chaotic Dynamics of the Bianchi Types VIII and IX Models \label{sec:bianchi89}}

In this Section we provide the detailed construction of 
the oscillatory-like regime of the Mixmaster model while 
approaching the initial singularity both in the field 
equations formalism and in the ADM one. A relevant part 
is centered around the chaotic properties of its dynamics. 
The cosmological implementation is also discussed, and in 
the last three Sections the effects on chaos of matter 
fields and of the number of space dimensions is reviewed.


\subsectionric{Construction of the solution\label{sec:csol}}

At this point we are going to address the solution of the 
system of equations (\ref{eeh}) for the cases of Bianchi types VIII and IX
cosmological models, following the standard approach of 
Belinsky,  Khalatnikov and Lifshitz (BKL) 
\cite{BKL1970,BKL82}\footnote{For a thorough discussion of 
works which concern the dynamics of such models see the 
literature \reffcite{1970JETPL..11..197Z,1971AnPhy..65..506R,
1971AnPhy..65..541R,1971Natur.230..112M,1971SvA....14..763D,
1971ZhETF..60.1201D,1972PhRvD...6.3390C,1972PhRvL..29.1616H,
1973PhRvD...8.1048H,1974PhRvD...9.3263H,1975PhRvD..12.1551H,
1978PhLA...67...19C,1980PhRvD..21..336B,1981JMaPh..22..623O,
1983ApJ...268..513S,1983PhRvD..28.....E,1984ApJ...286..379S,
1984GReGr..16.1119B,1990CQGra...7.1365U,1990PhRvD..41.2444X,
1991PhLA..160..123B,1994gr.qc.....5068M,1994JPhA...27.5357C,
1994PhRvD..49.2792G,1996PhRvL..76..857C,1997CQGra..14.2341R,
1997GrCos...3...85G,1997PhRvD..55.1896B,1998CSF.....9.1813A,
1999PhLB..465..101S,2000gr.qc.....6035R,2000PhRvD..62b3509B,
2000PhRvD..62l3501B,2003Ap&SS.283...67M,2003GReGr..35.2051A,
2004PhRvD..69f3514E,2005Ap&SS.298..419P}}.
Although the detailed discussion is devoted to the Bianchi IX model, 
it can be straightforwardly extended to the type VIII.\\
Explicitely, the Einstein equations (\ref{eeh}) reduce to 
\bseq
\begin{align}
\label{c1 equazioni di einstein in alpha, beta e gamma}
 2\alpha_{\tau \tau} &= (b^{2}-c^{2})^{2} - a^{4}\\
 2\beta_{\tau \tau}  &= (a^{2} - c^{2})^{2} - b^{4}\\
 2\gamma_{\tau \tau} &= (a^{2} - b^{2})^{2} - c^{4} \, ,
\end{align}
\eseq
togheter with the constraint (\ref{g30a}) unchanged, leading to the 
consequent constant of  motion 
\begin{equation}
\label{c1 costante del moto}
\alpha_{\tau} \beta_{\tau} + \alpha_{\tau}  \gamma_{\tau} 
+  \beta_{\tau} \gamma_{\tau} = \displaystyle\frac{1}{4} 
\left(a^{4} + b^{4} +c^{4} - 2 a^{2} b^{2} -2 a^{2} c^{2} -2 b^{2} c^{2}\right)\;.
\end{equation}
Let us therefore  consider again the case in which, for instance,
the negative power of the $p_i$'s exponents corresponds to the function 
$a(t)$ (that is to say $p_l=p_1$): the 
perturbation of the Kasner regime results from 
the terms $\lambda^2 a^4$ (remember that $\lambda=1$ for both models) 
while the other terms
decrease with decreasing $t$, in fact
\beq
p_1<0 \rightarrow p_1=-|p_1|\, , \quad 
\left\{
\begin{array}{l}
\alpha\sim-|p_1|\ln t \\
\displaystyle a \sim\frac{1}{t^{|p_1|}}
\end{array} \right.
\nearrow \quad \textnormal{for}\,t\rightarrow 0
\eeq
and along the other directions
\beq
\begin{array}{l}
p_2>0 \rightarrow p_2=|p_2|\, , \quad 
\beta\sim|p_2|\ln t \searrow \\
p_3>0 \rightarrow p_3=|p_3|\, , \quad 
\gamma\sim|p_3|\ln t \searrow 
\end{array}
\quad \textnormal{for}\,t\rightarrow 0 \, .
\eeq
Preserving only the increasing terms on the right-hand side of equations
(\ref{g30}),
we obtain a system identical to (\ref{g32}),
 whose solution describes 
the evolution of the metric from its initial state
(\ref{g16}). 
Let us fix, for instance,
\beq
\label{g33}
p_l=p_1 \, , \quad p_m=p_2 \, , \quad
p_n=p_3\, ,  
\eeq
so that  
\beq
\label{g33b}
a\sim  t^{p_1} \, , \quad
b\sim  t^{p_2} \, , \quad
c\sim  t^{p_3} \, , 
\eeq
and then 
\bea
\label{g34}
abc&=&\Lambda t \nonumber \\
\tau&=&\frac{1}{\Lambda}\ln t+\textnormal{const.}
\eea
where $\Lambda$ is a constant, so that  the initial conditions for
 (\ref{g32}) can be formulated as 
\begin{align}
\label{g35}
\alpha_{\tau}=\Lambda p_1 \, , \quad
\beta_{\tau} & =\Lambda p_2 \, , \quad
\gamma_{\tau}=\Lambda p_3 \, , \\
 & \textnormal{for}\qquad 
 \tau \rightarrow \infty \, . \nonumber
\end{align}
The system (\ref{g32}) with (\ref{g35})
is integrated to 
\bseq
\begin{align}
a^2 &=\frac{2\mid p_1\mid\Lambda}{\cosh\left(2\mid p_1\mid\Lambda\tau\right)} \\
b^2&={b_0}^2\exp\left[ 2\Lambda \left(p_2-\mid p_1\mid\right)\tau\right] \cosh\left(2\mid p_1\mid\Lambda \tau\right)    \\
c^2&={c_0}^2\exp\left[ 2\Lambda \left(p_3-\mid p_1\mid\right)\tau\right]\cosh\left(2\mid p_1\mid\Lambda \tau\right)
\end{align}
\label{g37}
\eseq
where $b_0$ and $c_0$ are integration constants.

\subsectionric{The BKL oscillatory approach\label{sec:bkl}}

Let us consider the solutions (\ref{g37}) 
in the limit $\tau\rightarrow\infty$: 
towards the  singularity they simplify to
\bseq
\begin{align}
a&\sim  \exp\left[-\Lambda p_1 \tau\right] \\
b&\sim  \exp\left[\Lambda \left( p_2 +2p_1\right) \tau\right] \\ 
c&\sim  \exp \left[\Lambda \left(p_3+2p_1\right)\tau\right] \\
t&\sim  \exp\left[ \Lambda \left(1+2p_1\right)\tau\right] 
\end{align}
\label{g38}
\eseq
that is to say, in terms of $t$,
\beq
\label{g39}
a\sim  t^{p^{\prime}_l} \,, \quad
b\sim  t^{p^{\prime}_m}  \,, \quad
c\sim  t^{p^{\prime}_n} \,, \quad
abc=\Lambda^{\prime}t  \, , 
\eeq
where the primed exponents are related 
to the un-primed ones by   
\bseq
\begin{align}
\label{g40}
& p^{\prime}_l =
\frac{\mid p_1\mid}{1-2\mid p_1\mid} \,, \quad
& p^{\prime}_m =
-\frac{2\mid p_1\mid -p_2}{1-2\mid p_1\mid} \,, \\
\label{g41}
& p^{\prime}_n=
\frac{p_3-2\mid p_1\mid}{1-2\mid p_1\mid} \, , \quad
& \Lambda^{\prime} =\left(1- 2\mid p_1\mid \right)\Lambda \, ,
\end{align}
\eseq
which, we note, is very similar to what found in the 
type II case with relations (\ref{c1 mappa degli indici})
plus the additional expression involving $\Lambda$.
Summarizing these results, we see the 
effect of the perturbation over the 
Kasner regime: 
a Kasner epoch is replaced by another one
so that the negative power of $t$ is transferred
from the $\mathbf{l}$ to the $\mathbf{m}$ direction, 
i.e. if in the original solution $p_l$ is 
negative, in the new solution $p^{\prime}_m<0$. 
The previously increasing perturbation 
$\lambda^2 a^4$ in (\ref{eeh}) is damped 
and eventually vanishes. The  
other terms involving $\mu^2$
instead of $\lambda^2$ will grow, therefore 
permitting the replacement of a Kasner 
epoch by another. Such rules of rotation 
in the perturbing property can be summarized 
with the rules (\ref{g42})
of the   BKL map, with the greater of 
the two positive powers remaining positive\cite{BKL1971}. \\ 
The following interchanges 
are characterized by a sequence of bounces,  
with a change 
of the negative power between the 
directions $\mathbf{l}$ and $\mathbf{m}$
continuing as long as the integral part 
of the initial value of $u$ is not exhausted, 
i.e. until $u$ becomes less than one. 
In terms of the parameter $u$, the map (\ref{c1 mappa degli indici})
takes the form 
\begin{equation}
\label{mapu}
u^{\prime} = u-1 \, \quad \textrm{for} \quad u>2 \, , \qquad 
u^{\prime} = \frac{1}{u-1} \, \quad \textrm{for} \quad u \leq 2 \, .
\end{equation}
At that point, according to (\ref{g20c}), 
the value $u<1$ is turned into $u>1$,  
thus  either the exponent 
$p_l$ or $p_m$ is negative and $p_n$
becomes the smaller one of the two positive
numbers, say $p_n=p_2$. The next sequence of
changes will bounce the negative power 
between the directions $\mathbf{n}$ and 
$\mathbf{l}$ or $\mathbf{n}$ and $\mathbf{m}$.

The phenomenon of increasing and decreasing of the various terms
with transition from a Kasner era to another is repeated infinitely 
many times up to the singularity.
Let us analyze the implications of the BKL map (\ref{g42}) and of the 
property (\ref{g20c}).

If we write  $u = k+x$ as the initial value of the 
parameter $u$, with  $k$ and $x$ being its integral and fractional part, respectively,
the continuous exchange of shrinking and enlarging directions (\ref{g42})
proceeds until $u < 1$, { i.e.} it lasts for $k^{0}$ epochs, 
thus leading a {\it Kasner era} to an end. The  new value of $u$ 
is $u' = 1/x > 1$ (\ref{g20c}) and the subsequent set of exchanges will be 
$l$-$n$ or $m$-$n$: for arbitrary initial values of $u$, the process will 
last forever and an infinite sequence of Kasner eras takes place.

%
%
%

For an arbitrary, irrational initial value of 
$u$ the changes (\ref{g42}) repeat indefinitely. 
In the case of an exact solution, the exponents 
$p_l$, $p_m$ and $p_n$ loose their literal 
meaning, thus in general, it has no sense to 
consider any well defined, for example
rational, value of $u$.

The evolution of the model towards the singularity
consists of successive periods, the 
\textit{eras}, in which distances along two
axes oscillate and along the third axis 
decrease monotonically while the volume 
decreases following a law approximately $\sim t$. 
The order 
in which the pairs of axes are interchanged and 
the order in which eras of different lengths
(number of Kasner epochs contained in it)
follow each other acquire a stochastic 
character.
Successive eras `condense' towards the singularity. 
%
%
%
%
%
Such general qualitative properties are 
not changed in the case of space filled in
with matter, however the 
meaning of the solution would change: the model 
so far discussed  would be considered as 
the principal terms of the limiting form 
of the metric as $t\rightarrow 0$.

\subsectionric{Stochastic properties and the Gaussian 
	distribution\label{sec:stochg}}

a decreasing
sequence of values of the parameter $u$
 corresponds to every $s$-th era there.
This sequence, from the starting era
 has the form
$u^{(s)}_{max},\,   
u^{(s)}_{max}-1,\,
u^{(s)}_{max}-2,\ldots,
u^{(s)}_{min}$.  
We can introduce the notation
\beq
\label{g45a}
u^{(s)}=k^{(s)}+x^{(s)}
\eeq
then
\beq
\label{g45b}
u^{(s)}_{min}=x^{(s)}<1  \, , \quad\quad
u^{(s)}_{max}=k^{(s)}+x^{(s)} \, ,
\eeq
where $u^{(s)}_{max}$ 
is the greatest value of $u$ for an assigned 
era and   $k^{(s)}=\left[u^{(s)}_{max}\right]$ 
(square brackets denote the greatest integer 
less or equal to $u^{(s)}_{max}$).
The number $k^{(s)}$denotes the era length, i.e. 
the number of Kasner epochs contained in it.
For the next era we obtain
\beq
\label{g46}
u^{(s+1)}_{max}=\frac{1}{x^{(s)}} \, ,\quad
k^{(s+1)}=\left[\frac{1}{x^{(s)}}\right] \, . 
\eeq
For large $u$, the Kasner exponents 
approach the values $(0,0,1)$
with the limiting form 
\beq
p_1 \approx -\frac{1}{u}\, , \quad
p_2 \approx \frac{1}{u}\, , \quad
p_3 \approx 1-\frac{1}{u^2}\, , 
\eeq
and the transition to the next era is governed 
by the fact that  not all  
terms in the Einstein equations  are negligible 
and some terms
are comparable: 
%
%
%
%
%
%
%
in such a case,  the transition is accompanied
by a long regime of small oscillations\cite{BKL1970} lasting 
 until the 
next era, whose details will be discussed in Section \ref{sec:oscill},
after which  
a new series of Kasner epochs begins. 
The probability $\lambda$ of all possible values 
of $x^{(0)}$ which lead to a dynamical 
evolution towards this specific case
is strongly converging to a number 
$\lambda \ll 1$\cite{LLK73}. If the initial value of 
$x^{(0)}$ is outside this special interval for
$\lambda$, the special case cannot occur; 
%
%
%
%
%
%
%
if $x^{(0)}$ lies in this interval, a peculiar
evolution in small oscillations take place, 
but after this period the model begins to 
regularly evolve with a new initial value 
$x^{(0)}$, which can only accidentally fall 
in this peculiar interval (with probability 
$\lambda$). The repetition of this situation 
can lead to these  cases only with 
probabilities $\lambda,\lambda^2,\ldots$, 
which asymptotically approach zero.

If the sequence begins with $k^{(0)}+x^{(0)}$, 
the lengths  $k^{(1)}, k^{(2)}, \ldots$ are
the numbers appearing in the expansion for 
$x^{(0)}$ in terms of the continuous fraction
\beq
\label{g47}
x^{(0)}=\frac{1}{k^{(1)}+\displaystyle \frac{1}{k^{(2)}
+\displaystyle \frac{1}{k^{(3)}+\ldots}	}} \, ,
\eeq
which is finite if related to a rational 
number, but in general it is an infinite one \cite{algebra}.

For the infinite sequence of 
positive numbers $u$ ordered as (\ref{g46}) and, 
admitting the expansion (\ref{g47}), it is possible
to note that
\begin{enumerate}\renewcommand{\labelenumi}
{\roman{enumi})}     	   
\item a rational number would have a finite expansion;
	\item periodic expansion represents quadratic
    irrational numbers (i.e. numbers which are roots of 
    quadratic equations with integral coefficients)
   \item irrational numbers have infinite expansion.
\end{enumerate}
 
All terms $k^{(1)},k^{(2)},k^{(3)},\ldots$ 
in the first two cases 
 having the exceptional 
property to be  bounded in magnitude 
are related to a set of numbers $x^{(0)}<1$ 
of zero measure
in the interval $(0,1)$.

An alternative to the numerical 
approach in terms of 
continuous fractions is the 
statistic distribution 
of the eras' sequence for the 
numbers $x^{(0)}$
over the interval $(0,1)$, governed by some 
probability law. For the series
$ x^{(s)}$ with increasing $s$ these 
distributions  tend to a stationary one 
$w(x)$, independent of $s$, 
 in which the initial conditions are 
 completely  forgotten
 \beq
\label{g53}
 w(x)=\frac{1}{(1+x)\ln 2} \, .
 \eeq
 In fact, instead of a well defined initial 
value as
in (\ref{g45a}) with $s=0$, 
 let us
consider a probability distribution 
for $x^{(0)}$ over the interval  $(0,1)$, 
$W_0(x)$ for $x^{(0)}=x$. 
Then also the numbers $x^{(s)}$ are distributed
with some probability. Let $w_s(x)dx$ 
be the probability that the last term in the
$s$-th series $x^{\left(s\right)}=x$ lies 
in the interval $dx$. 
The last  term  of the previous  
series must lie in the interval between 
 $1/(k+1)$ and $1/k$, in order for  
 the length of the $s$-th series to be $k$. \\
 The probability for the series 
 to have length $k$ is given by
\beq
\label{g50}
W_s(k)=\int^{\frac{1}{k}}_{\frac{1}{1+k}} w_{s-1}(x)dx \, .
\eeq
For each pair of subsequent series,
we get the recurrent 
formula relating the distribution 
 $w_{s+1}(x)$ to $w_{s}(x)$
\beq
\label{g51a}
w_{s+1}(x)dx =\sum^{\infty}_{k=1}
w_s\left(\frac{1}{k+x}\right) 
\left|d\frac{1}{(k+x)}\right|\, ,
\eeq
or, simplifying the differential interval,
\beq
\label{g51}
w_{s+1}(x) =\sum^{\infty}_{k=1}\frac{1}{{(k+x)}^2}
w_s\left(\frac{1}{k+x}\right) \, .
\eeq
If for increasing $n$ the $w_{s+n}$ 
distribution (\ref{g51}) tends to a stationary 
one, independent of $s$, $w(x)$ has to 
satisfy 
\beq
\label{g52}
w(x) =\sum^{\infty}_{k=1}\frac{1}{{\left(k+x\right)}^2}
w\left(\frac{1}{k+x}\right) \, .
\eeq
A normalized  solution to (\ref{g52}) is clearly given by (\ref{g53})\cite{BKL1970}; 
substituting it in (\ref{g50}) 
and evaluating the integral
\beq
\label{g54}
W\left(k\right)=\int^{\frac{1}{k}}_{\frac{1}{1+k}} w\left(x\right)dx=\frac{1}{\ln 2}\ln{\frac{{\left(k+1\right)}^2}{k\left(k+2\right)}} \, ,
\eeq
we get  the corresponding stationary distribution 
of the lengths of the series  $k$.
Finally, since $k$ and $x$ are not independent, 
they must admit a stationary joint  probability 
distribution 
\beq
\label{g55}
w\left(k,x\right)=\frac{1}{\left(k+x\right)\left(k+x+1\right)\ln2}
\eeq
which, for $u=k+x$, rewrites as 
\beq
\label{g56}
w\left(u\right) =\frac{1}{u\left(u+1\right)\ln2} 	\, , 
\eeq
i.e. a stationary 
distribution for the parameter $u$.

The existence of the Gauss map was firstly  demonstrated 
in the work of Belinskii, Khalatnikov and Lifshitz \cite{BKL1970},
showing how a statistical approach \cite{K8385}  describes the time 
evolution of the cosmological models near the singularity. 
These features opened the way to further investigations 
(see, for example \reffcite{BarrowMapPRL1981}), in view also of the 
peculiar properties of the discrete map $w$ 
%
%
leading to  final form for a measure of the full degrees of freedom of the discrete 
Mixmaster dynamics expressed as a not separable function 
\cite{barrow.cup84},
%
%
 and such map
%
%
\begin{itemize}
\item has positive metric- and topologic-entropy;
\item has the {\it weak Bernoulli properties} 
	(i.e., the map cannot be finitely approximated);
\item is {\it strongly mixing} and {\it ergodic}\cite{cit13barrow}.
\end{itemize}

Thus, we see that the approach to the initial singularity 
is characterized by an infinite series of Kasner epochs
described by the stochastic properties of the associated map.

For a discussion of the chaotic behavior inherent the homogeneous 
early cosmologies, 
see \reffcite{1986GReGr..18.1263B,1987GReGr..19...73H,
1989grg..conf..324B,1990PhLA..147..353R,1990STIN...9215945F,
1991GReGr..23.1385B,1991NYASA.631...15H,1991PhRvD..44.2369S,
1993JPhA...26.5795C,1993PhLB..299..223D,1993PhRvD..47.3222B,
1994AcC....20...17S,1994JPhA...27.1625C,1995JPhA...28..657C,
1995PhyD...87...70L,1995RpMP...36...75C,1996CQGra..13.1273B,
1996GrCos...2...92S,1996JPhA...29...59D,1996PhRvL..76..857C,
1997GReGr..29..185S,1997PhLB..399..207S,1997PThPh..98.1225A,
1998JPhA...31.2031M,1999magr.meet..616C,2000JMP....41.4777M,
2004GReGr..36.2635D,2005CQGra..22.1763C,2005GReGr..37.1097F,
2005PhRvD..71f4007J,2006PhRvD..73f9901S}.

\subsectionric{Small Oscillations\label{sec:oscill}}

Let us investigate a particular case of the general (homogeneous) 
solution constructed above. 
We analyze an era during which two of the three 
functions $a,b,c$ (for example $a$ and $b$) oscillate so that their 
absolute values remain close to each other and the third function 
(in such case $c$) monotonically decreases, so that 
$c$ can be neglected with respect to $a$ and $b$. 
As before, we will discuss only the Bianchi IX model, 
since for the Bianchi VIII case the argumentations and results 
are qualitatively the same \cite{BKL1970,BKL82}.
Let us consider the equations 
we can  obtain from (\ref{g30}) and (\ref{g31})
\bseq
\label{eqpopo}
\begin{align}
\label{eq1po}
\alpha_{\tau\tau}+\beta_{\tau\tau}&=0 \, , \\
\label{eq2po}
\alpha_{\tau\tau}-\beta_{\tau\tau}&=e^{4\beta}-e^{4\alpha} \, , \\
\label{eq3po}
\gamma_\tau(\alpha_\tau+\beta_\tau)&=-\alpha_\tau\beta_\tau+\f14(e^{2\alpha}-e^{2\beta})^2\, .
\end{align}
\eseq
The solution of  (\ref{eq1po}) is 
\be
\alpha+\beta=\f{2a_0^2}{\xi_0}(\tau-\tau_0)+2\ln(a_0),
\ee
where $a_0$ and $\xi_0$ are positive constants. 
For what follows we conveniently replace the time coordinate  $\tau$
with the new one $\xi$ defined as
\be\label{defchi}
\xi=\xi_0\exp\left(\f{2a_0^2}{\xi_0}(\tau-\tau_0)\right)
\ee
in terms of which the equations (\ref{eq2po}) and(\ref{eq3po}) rewrite as
\bseq
\label{chixixi}
\begin{align}
\label{chixi}
\chi_{\xi\xi} &+\f1\xi\chi_\xi+\f12\sinh(2\chi)=0 \, , \\
\label{gamm6axi}
\gamma_\xi &=-\f1{4\xi}+\f\xi8\left(2\chi^2_\xi+\cosh(2\chi)-1\right)\, ,
\end{align}
\eseq
where we have introduced the notation $\chi=\alpha-\beta$ 
and $A_\xi\equiv dA/d\xi$. 
Since $\tau$ is defined in the interval $[\tau_0,-\infty)$, 
from equation (\ref{defchi}) we note that $\xi\in[\xi_0,0)$. 
Since a general analytic solution of 
the system (\ref{chixixi}) is not available,
we shall  consider  the two limiting cases $\xi\gg1$ and $\xi\ll 1$ only.

Let us start with the $\xi\gg1$ region. 
In this approximation the solution of equation (\ref{chixi}) reads as
\be
\chi=\f{2A}{\sqrt\xi}\sin(\xi-\xi_0),
\ee
$A$ being  a constant and therefore obtaining 
  $\gamma\sim A^2(\xi-\xi_0)$. 
As we can see, the name ``small oscillations'' arises from 
the behavior of the function $\chi$. 
The  functions $a$ and $b$, i.e. the expressions for the scale factors,
are straightforwardly obtained as
\bseq
\begin{align}
\label{eqabpo}
a,b&=a_0\sqrt{\f\xi{\xi_0}}\left(1\pm\f A{\sqrt\xi}\sin(\xi-\xi_0)\right)\, , \\
\label{eqcpo}
c&=c_0\exp\left[-A^2(\xi_0-\xi)\right].
\end{align}
\eseq
The synchronous time coordinate $t$ can be re-obtained back 
from the relation $dt=abc\,d\tau$  as
\be
t=t_0\exp\left[-A^2(\xi_0-\xi)\right]\, .
\ee
Of course, these solutions only apply  when the 
condition $c_0\ll a_0$ is satisfied.

Let us discuss the region where $\xi\ll1$. 
In such a limit, the function $\chi$ reads as
\be
\chi=K\ln\xi+\theta, \qquad \theta=\textrm{const.}
\ee
where $K$ is a constant which, for consistency, is constrained 
in the interval $K\in(-1,1)$. 
We can therefore derive all other related  quantities, and in particular
\be
a\sim\xi^{(1+K)/2} \, , \quad b\sim\xi^{(1-K)/2} \, , \quad 
c\sim\xi^{-(1-K^2)/4} \, , \quad t\sim\xi^{(3+K^2)/2} \, .
\ee
This is again a Kasner solution, with the negative power 
of $t$ corresponding to $c$
and the evolution is the same as the general one. 
Moreover, we can easily note how,  for such a Kasner epoch, 
the parameter $u$ (introduced in (\ref{g20a})) becomes
\be
\textrm{for}  ~~K>0: \quad u=\f{1+K}{1-K} \, , \qquad 
\textrm{for} ~~K<0: \quad u=\f{1-K}{1+K} \, .
\ee 
Summarizing, the system initially crosses a long time interval 
during which the functions $a$ and $b$ satisfy $(a-b)/a<1/\xi$
and perform small oscillations
of constant period $\Delta\xi=2\pi$, while 
the function $c$ decreases with $t$  as $c=c_0t/t_0$. 
When $\xi\sim\mathcal O(1)$, equations (\ref{eqabpo}) 
and (\ref{eqcpo}) cease to be valid, thus after this period the $c$ 
starts increasing and $a$ and $b$ decreasing. 
At the end, when condition $c^2/(ab)^2\sim t^{-2}$ is realized, a 
new period of oscillations is reached (Kasner epochs) and the 
natural evolution of the system is restored.

Let us stress how the matching of the constants of the above 
limit regions is possible. 
In particular we can express the constants $(K,\theta)$ in 
terms of $(A,\xi_0)$. 
Such  procedure can be implemented also by  replacing $\sinh(2\chi)$ with 
$2\chi$ ($\chi\ll1$) in equation (\ref{chixi}), and finally 
the solution is expressed in terms of Bessel functions which can be 
asymptotically expanded and compared with the explicit mentioned 
solutions\cite{arfken}.


\subsectionric{Hamiltonian Formulation of the Dynamics\label{sec:hfd}}

So far we reviewed the formulation of General Relativity 
in terms of geometrodynamics, i.e. studying the equations 
governing the evolution of the metric tensor. 
In what follows we will consider 
a different interpretation of the terms involved, leading to 
a peculiar Hamiltonian formulation.
This paragraph is devoted to specialize it to the Mixmaster 
dynamics, i.e. to the Bianchi VIII and IX models 
\cite{Misner1969PR,Misner1969PRL,MTW}.

The space-time line element adapted to the 3+1 foliation can 
be written as in (\ref{linevn})\cite{MTW}
where, for a spatially homogeneous model, the three-metric tensor 
$h_{\alpha\beta}$ is parametrized by three functions $q^a=q^a(t)$, 
in particular as
\be\label{parq}
h_{\alpha\beta}=e^{q_a}\delta_{ab}e^a_\alpha(x) e^b_\beta(x) \,,
\ee
although such symmetry  is valid  in the case of absence of matter only. 
In fact,  the Einstein equations in  vacuum
$R_{0\alpha}=0$ allow the choice of a diagonal  metric 
$h_{\alpha\beta}$, thus leaving three degrees of freedom only.
The presence of matter induces  in general a non-diagonal metric
(which will be analyzed later)  while here  we focus
on the vacuum one. 

Given the parametrization (\ref{parq}), considering 
that all the canonical variables, including the lapse function 
and the shift vector, are independent of the spatial coordinates
 and that this model identically satisfies the super-momentum constraint,
we get the action as 
\be
S=\int dt\left(p_a\p_tq^a -NH\right),
\ee
$p_a=p_a(t)$ being the momenta canonically conjugate to $q^a$.
The super-Hamiltonian $H$ reads explicitly as
\be
\label{acca1}
H=\f{1}{\sqrt \eta}\left[\sum_a \left(p_a\right)^2 
	-\f{1}2\left(\sum_b p_b\right)^2-\eta {~}^{(3)}R\right], 
\ee
where $\eta\equiv\exp\sum_aq_a$ 
and
the Ricci three-scalar ${~}^{(3)}R$ plays the role of the potential in the dynamics.\\
Let us introduce the   ``anisotropy parameters'', which will be 
convenient in the following, as
\be
\label{qucona}
Q_a\equiv \f{q^a}{\sum_b q^b}, \qquad \sum_{a=1}^3Q_a=1 \, .
\ee

The functions defined as in (\ref{qucona}) allow a 
clearer interpetation of the last term 
in the right-hand side of equation (\ref{acca1}) as a potential 
form $U$ for the dynamics. 
In fact, it can be re-written in the more expressive form 
\be\label{bianpot}
U= \eta {~}^{(3)}R=\sum_a\lambda_a^2 \eta^{2Q_a}-\sum_{b\neq c}\lambda_b\lambda_c \eta^{Q_b+Q_c} \, ,
\ee
where the constants $\lambda_a$ specify the model under consideration, 
being  $\lambda_a=(1,1,-1)$ or $\lambda_a=(1,1,1)$ 
for Bianchi VIII and IX, respectively.

The main advantage of writing the potential as in (\ref{bianpot}), 
arises when investigating its proprieties in the asymptotical 
behavior toward the cosmological sin\-gu\-la\-ri\-ty ($\eta\rightarrow0$). 
In fact, the second term in (\ref{bianpot}) 
becomes negligible, while the value of  the first one results 
to be strongly sensitive  to the sign of $Q_a$s
and the potential 
can be modeled by an infinite well as
\be
\label{pottheta0}
U=\sum_a \Theta(Q_a)
\ee
where
\be
\Theta(x)=\begin{cases}+\infty,& \textrm{if}\ x<0 \cr 0,& \textrm{if}\  x>0 \, .\cr \end{cases}   
\ee
By (\ref{pottheta0}) we see how the  dynamics 
of the Universe resembles
that of a particle moving in a 
dynamically-closed domain $\Pi_Q$, 
defined as the one where all the anisotropy parameters $Q_a$ 
are simultaneously positive.

\subsectionric{The ADM Reduction of the Dynamics\label{sec:adm_3}}

The ADM reduction of the dynamics relies on the idea of identifying 
a temporal parameter as a functional of the geometric canonical 
variables, before applying any quantization procedure or, 
in other words, of solving the classical constraints before 
quantizing. 
For the moment, we will simply discuss the reduction of 
General Relativity to a pure canonical form focusing on 
the elimination of non-dynamical variables.

Let us count the degrees of freedom 
of the gravitational field. 
We have  twenty phase-space variables, given in the 3+1 formalism 
by the set  
$(N,\Pi;N^\alpha,\Pi_\alpha;h_{\alpha\beta},\Pi^{\alpha\beta})$ 
(see Section \ref{sec:hf}), 
subject to eight first-class constraints 
$(\Pi=0,\Pi_\alpha=0,H=0,H_\alpha=0)$. 
Since each constraint eliminates two 
phase-space variables, 
we remain with four of them,  
corresponding to  two physical degrees of freedom  of the gravitational field, 
i.e.  to the two independent polarizations 
of a gravitational wave in the 
weak field limit.

After eliminating $N$ and $N^\alpha$, 
we have $12\times\infty^3$ variables 
$(h_{\alpha\beta}(x),\Pi^{\alpha\beta}(x))$. 
Then, we can remove $4\times\infty^3$ variables thanks to 
the secondary constraints (\ref{seccon}). 
The remaining $4\times\infty^3$ non-physical degrees 
of freedom 
result from  the lapse function and the shift 
vector. 
In analogy with the Yang-Mills theory, also in this case it is 
necessary to impose some sort of gauge, 
fixing the lapse function and the shift 
vector.

In order to obtain a true canonical form for the canonical theory, 
we have to follow the key steps listed by Isham in \reffcite{ishamtime}. 
With respect to this formulation,
 we refer to a canonical description of the physical degrees of freedom only
and with the following procedure we are able to discard 
 the non-physical variables.

\renewcommand{\labelenumi}{\roman{enumi})}
\begin{enumerate}
\item Perform a canonical transformation
	\begin{align}
	(h_{\alpha\beta}(x), \Pi^{\alpha\beta}(x))
	\rightarrow &\left(\chi^A(x),P_A(x);\phi^r(x),\pi_r(x)\right) \,, \\
	 	    &A=1,2,3,4 \quad r=1,2 \nonumber
	\end{align}
	where $\chi^A(x)$ define a particular choice of the space 
	and time coordinates\footnote{These fields can be 
	interpreted as defining an embedding of $\Sigma$ in $M$ 
	via some parametric equations.}, 
	$P_A(x)$ are the corresponding canonically conjugate momenta 
and the four phase space variables $(\phi^r(x),\pi_r(x))$ represent 
the physical degrees of freedom of the system. 
We emphasize that these ``physical'' fields are not Dirac ob\-ser\-va\-bles, 
in the sense defined in Section \ref{sec:hf}.

Consequently,  the symplectic structure is determined by
	\bseq
	\begin{align}
	\{\chi^A(x),P_B(x')\}  &=\delta^A_B\delta^3(x-x') \,,\\
	\{\phi^r(x),\pi_s(x')\}&=\delta^r_s\delta^3(x-x')\,,
	\end{align}
	\eseq
	while all other Poisson brackets are vanishing.

\item 
Express the super-momentum and super-Hamiltonian in terms 
of the new fields and then write the Lagrangian density as 
\be\label{L'}
\mathcal L'(N,N^\alpha,\chi^A,P_A,\phi^r,\pi_r) =
	P_A\dot\chi^A 
	+\pi_r\dot\phi^r-NH'-N^\alpha H'_\alpha \,.
\ee

\item 
Remove $4\times\infty^3$ variables arising from 
the constraints $H=0$ and $H_\alpha=0$ once solved with respect to $P_A$
the equation
\be
\label{ha}
P_A(x)+h_A(x,\chi,\phi,\pi)=0
\ee
and re-inserting it in  (\ref{L'}). Removing the remaining 
$4\times\infty^3$ non-dynamical variables we yield  the so-called 
{\it true Lagrangian density} \cite{ishamtime}

\be
\label{ltrue}
\mathcal L_{true}=\pi_r\dot\phi^r-h_A\dot\chi^A \, ,
\ee
where the lapse function and the shift vector do not play any role, 
but spe\-ci\-fy\-ing the form of the functions $\dot\chi^A$. 
After solving the constraints, there is no longer information 
about the evolution of  $\chi^A$  in terms of  the parametric time $t$. 
Thus, in equation (\ref{ltrue}) we have chosen the conditions 
$\chi^A(x,t)=\chi^A_t(x)$ and the true Hamiltonian  is 
\be
\label{htrueg}
H_{true}=\int d^3x\dot\chi^A(x)h_A \left(x,\chi_t,\phi(t),\pi(t)\right) \, .
\ee
From (\ref{htrueg}) one can derive  the equations of 
motion as
\bseq
\begin{align} 
\p_t\phi^r &=\{\phi^r,H_{true}\}_{red} \\
\p_t\pi_s  &=\{\pi_s,H_{true}\}_{red},
\end{align}
\eseq
where the notation $\{...\}_{red}$ refers to the Poisson 
brackets evaluated in the reduced phase space with  coordinates
given by the physical modes $\phi^r$ and $\pi_s$. 

\end{enumerate}

This is an operative method to classically solve the constraints, 
i.e. pulling out all the gauges, and obtaining a canonical description 
for the physical degrees of freedom only. 

We also note that such procedure violates the geometrical structure 
of General Relativity, since it removes parts of the metric tensor.
Nevertheless, it avoids any obstacle at  a classical level, 
but there are several problems when implemented in the quantum framework 
\cite{ishamtime,Kuchar1980,Stachel}. 
In particular, they are related to the one of 
the reduced-phase space quantization, for more details see \reffcite{HT}.



\subsectionric{Misner variables and the Mixmaster model\label{sec:misner2ch}}


In order to implement the formalism to our purposes,
we start introducing a new set of variables, as mentioned 
in expression (\ref{erreab}). 
In fact, we
parametrize the tetradic projection of the metric as
\bseq
\beq
\label{h6}
\eta_{ab} =e^{2\alpha}{\left(e^{2\beta}\right)}_{ab}  
\leftrightarrow
%
\left(\ln \eta \right)_{ab} =2\alpha \delta_{ab} + 2\beta_{ab}
\eeq
\eseq
where $\beta_{ab}$ is a three-dimensional matrix with 
null trace of the kind $\textrm{diag}(\beta_{11}, \beta_{22},\beta_{33})$ and the exponential matrix has to 
be intended as a power series of matrices, so that  
\beq
\label{h8}
\textrm{det} \left(e^{2\beta}\right) =e^{2~\text{tr} \beta}=1 
\eeq
and  
\beq
\label{h9}
\eta=e^{6\alpha} \, .
\eeq
The BKL formalism used in Section \ref{sec:kasnersol} 
matches the present one (for the Bianchi I solution)
once considering the expression 
\beq
\label{h10}
p_{ab}=\frac{d\left(\ln \eta\right)_{ab}}{d\ln  \eta} \, .
\eeq
From equations (\ref{h6}) and (\ref{h10}) the relation 
\beq
\label{h11}
p_{ab} =\frac{1}{3} \left[\delta_{ab}+
 \left(\frac{d\beta_{ab}}{d\alpha}\right)\right]
\eeq
follows, so that the first Kasner condition 
(\ref{g17}) rewrites as
\beq
\label{h12}
1=\sum_a p_a \equiv \textrm{tr}~ p_{ab} =1 + 
\frac{1}{3}\textrm{tr}\left(\frac{d\beta}{d\alpha}\right)
\eeq
which is an identity since 
the trace $\textrm{tr} \beta_{ab}=0$. \\
The second Kasner relation (\ref{g17}) rewrites as
\beq
\label{h13}
\textrm{tr} \left(p^2\right)=1
\eeq
which, by virtue of (\ref{h11}) becomes 
\beq
\label{h13a}
\frac{1}{9}\tr\left(1+2\frac{d\beta_{ab}}{d\alpha}+
\left(\frac{d\beta_{ab}}{d\alpha}\right)^2\right)=
\frac{1}{3}+\frac{1}{9}
\left(\frac{d\beta_{ab}}{d\alpha}\right)^2 =1\, ,
\eeq
and then 
\beq
\label{h14}
\left(\frac{d\beta_{ab}}{d\alpha}\right)^2=6
\eeq
which is no longer an identity but a consequence 
of the Einstein equations in empty space.

\subsubsectionric{Misner Approach to the Mixmaster\label{sec:misnermix}} 			

Once seen how to perform the convenient change 
of variables for the diagonal case, we step 
further 
to approach the Mixmaster using this formalism.\\
The matrix $\beta_{ab}$ 
has only two independent components defined as 
\bseq
\bea
\beta_{11}&=&\beta_+ + \sqrt3 \beta_-  \\
\beta_{22}&=&\beta_+ - \sqrt3 \beta_- \\
\beta_{33}&=&-2 \beta_+  \, ,
\eea
\label{h21}
\eseq
and then the Kasner relation for Bianchi I  reads as 
\beq
\label{h22}
\left(\frac{d\beta_+}{d\alpha}\right)^2 + 
\left(\frac{d\beta_-}{d\alpha}\right)^2 =1 \, .
\eeq
The variables $\beta_{\pm}$ together with $\alpha$ 
are the \textit{Misner coordinates}. 
The relation (\ref{h22}) in terms of the Kasner 
exponents now is 
\bseq
\bea
\frac{d\beta_+}{d\alpha}&= &\frac{1}{2}\left(1-3p_3\right)  \\
\frac{d\beta_-}{d\alpha}&= &\frac{1}{2} \sqrt3 \left(p_1-p_2\right)
\eea
\label{h23}
\eseq
or equivalently with  the $u$ parameter
\bseq
\bea
\frac{d\beta_+}{d\alpha}&= &-1 + \frac{3}{2}\frac{1}{1+u+u^2} \\
\frac{d\beta_-}{d\alpha}&= &-\frac{1}{2} \sqrt3 \frac{1+2u}{1+u+u^2} \, .
\eea
\label{h23a}
\eseq
Such quantities represent the 
\textit{anisotropy velocity} 
$\boldsymbol{\beta}^{\prime}$
\beq
\label{h24}
\boldsymbol{\beta}^{\prime}\equiv \left(\frac{d\beta_+}{d\alpha}, \frac{d\beta_-}{d\alpha}\right) \, ,
\eeq
which measures the variation of the 
anisotropy amount with respect to the expansion 
as parametrized by the $\alpha$ parameter. The volume 
of the Universe behaves as $e^{3\alpha}$, 
tends to zero towards the singularity and 
 is directly related to the 
temporal parameter.\\
Eventually, the presence of matter as well the effects
of the spatial curvature can lead the  norm 
$||\boldsymbol{\beta}^{\prime}||$ to a deviation 
from the Kasnerian unity.

In order to develop a general metric for a 
homogeneous space-time we rewrite the 
line element in the general form 
\beq 
\label{h25}
ds^2={N^2(t)} d{t}^2  -
 e^{2\alpha}(e^{2\beta})_{ab}\omega^a \otimes \omega^b  \, .
\eeq
The cosmological problem reduces to the equations 
involving the functions $\alpha, N, \beta_{ab}$ 
in terms of the independent temporal parameter $t$, 
independently of the spatial coordinates.
Explicitly, the dual 1-forms associated to the Bianchi 
types VIII and IX  are, respectively,
\bseq
\bea 
\label{h26}
\omega ^1 &=&-\sinh \psi \sinh\theta d\phi~ 
			+ ~\cosh \psi d\theta \nonumber \\
\textrm{(VIII)}\quad \omega ^2 &=& -\cosh \psi \sinh\theta d\phi ~
			+~\sinh \psi d\theta \\
\omega ^3&=&~\cosh\theta d\phi ~+ ~d \psi  \nonumber  
\eea
\bea 
\label{h27}
		\omega ^1 &=& ~\sin \psi \sin\theta d\phi ~
		+~\cos \psi d\theta \nonumber\\
\textrm{(IX)}\quad \omega ^2 &=& -\cos \psi \sin\theta d\phi ~+~
			\sin \psi d\theta \\
	\omega ^3 &=& ~\cos\theta d\phi ~+~d \psi \, . \nonumber
\eea
\eseq
For example, the $0-0$ Einstein equation
with $N=1$ reads as 
\beq
\label{h28}
3\left(\dot{\alpha}^2 -\dot{\beta_+}^2
-\dot{\beta_-}^2\right) +
\frac{1}{2}\left({~}^{(3)}R_{B}\right)=  \kappa T^0_{~0} \, ,
\eeq
where ${~}^{(3)}R_{B}$ is the curvature scalar for the 
three-dimensional spatial surface corresponding 
to $t=\textrm{const.}$ and the index $B$ refers to the 
symmetry properties for the Bianchi cosmological 
models. 
Such  term contains  the peculiar 
difference between the nine types of the Bianchi 
classification, to be evaluated through 
expressions (\ref{g6}) in terms of the structure
constants. 

For the models referring to types VIII and IX
such curvature scalar reads as
\bseq
\bea
\label{h29}
{~}^{(3)}R_{VIII}&=& \frac{1}{2}e^{-2\alpha}\left(4e^{4 \beta_+}  
-2\tr~ e^{-2\beta}- \tr ~e^{4\beta}\right)\\
\label{h30}
{~}^{(3)}R_{IX}&=&\frac{1}{2}e^{-2\alpha}\tr\left(2e^{-2\beta}-e^{4\beta}\right) \, ,
\eea
\eseq
respectively, 
and the trace operation has to be intended 
over the exponential of diagonal matrices, 
without ambiguity. \\
Equation (\ref{h28}) with (\ref{h30}) or
(\ref{h29}) can be interpreted as a contribution 
of the anisotropy energy, connected to the term $T^0_{~0}$,
to the volume expansion $\dot{\alpha}^2$, 
so that it appears as a potential term together 
with the kinetic ones  $\dot{\beta}^2_{\pm}$. 
Close to the singularity, this term becomes negligible 
for small values of the 
anisotropy variables $\beta_{\pm}$. \\
Finally, equation (\ref{h28}) has to be regarded as a 
 constraint over the field equations.

\subsubsectionric{Lagrangian Formulation in the Misner Variables\label{sec:hamilfo}}

Despite the variables $q_a$ introduced in Section \ref{sec:hfd}
are linked to the Misner ones by the linear transformations
\bseq
\begin{align}
q_1 & = 2 \alpha + 2 \beta_+ + 2 \sqrt{3} \beta_- \\
q_2 & = 2 \alpha + 2 \beta_+ - 2 \sqrt{3} \beta_- \\
q_3 & = 2 \alpha 		- 4 \beta_-  \, , 
\end{align}
\eseq
nevertheless we conveniently restate \textit{ab initio}
the Lagrangian approach.
From the general expression of the action (\ref{eh_act}) in vacuum, 
we get the variational principle 
\beq
\label{h36}
\delta\int_{t_1}^{t_2} L\, dt=0 \, ,
\eeq
in which $t_1 $ and $t_2$ 
$(t_2>t_1)$ are two values 
of the temporal coordinate and integration
is performed  with respect to the 
spatial coordinates. \\
In particular, 
integration for the Bianchi 
type VIII is considered over a spatial vo\-lu\-me 
$(4\pi)^2$, in order to have the same integration
constant used for the type IX (and to keep 
a uniform formalism), having
\beq
\label{h37}
\int \omega^1 \wedge\omega^2 \wedge\omega^3 = 
\int \sin\theta d\phi \wedge d\theta\wedge d\psi = 
\left(4\pi\right)^2 \, .
\eeq
This way, the  Lagrangian $L$ is 
written as\footnote{Here we choose the Einstein constant $\kappa=16\pi$, 
i.e. $G=1$.}
\beq
\label{h38}
L=-\frac{6\pi}{N}e^{3\alpha}
\left[{{\alpha}^{\prime}}^2-{{\beta_+}^{\prime}}^2
-{{\beta_-}^{\prime}}^2\right]+ N \frac{\pi}{2} 
e^{\alpha}U^{\left(B\right)}\left(\beta_+ , \beta_-\right)
\eeq
where ${()}^{\prime}=\displaystyle\frac{d}{dt}$ and $U^{(B)}$ 
is a function linear in ${~}^{(3)}R_B$ having  a potential role. 
The variational principle rewrites explicitely as
\beq
\label{h39}
\delta S = \delta \int \Big( p_{\alpha} {\alpha}^{\prime} 
+ p_+ {\beta_+}^{\prime} +p_- {\beta_-}^{\prime} - 
N{ H} \Big)dt = 0
\eeq
in which ${ H}$ is given  by
\beq
\label{h40}
{ H} = \frac{e^{-3\alpha}}{24 \pi} 
\Big( -p^2_{\alpha} +p^2_+ +p^2_- +{\cal V}   \Big)
\eeq
and the potential  ${\cal V}$ as 
\beq
\label{h41}
{\cal V}=-12 \pi ^2 e^{4\alpha} 
U^{(B)}\left(\beta_+ , \beta_-\right)
\eeq
where $U^{(B)}$ is specified for the two Bianchi models
under study as
\bseq
\begin{align} 
\label{9p2a}
&U^{\textrm{VIII}}= e^{-8\beta_+}
         +4e^{-2\beta_+}\cosh(2\sqrt3 \beta_-)   
	+2e^{4\beta_+}\Big(\cosh(4\sqrt3 \beta_-)-1\Big)  \\
&U^{\textrm{IX}}=e^{-8\beta_+} -4e^{-2\beta_+}\cosh(2\sqrt3 \beta_-)
	+2e^{4\beta_+}\Big(\cosh(4\sqrt3 \beta_-)-1\Big) \, .
\label{9p2}
\end{align}
\label{9pp}
\eseq
From Lagrangian (\ref{h38}) it is standard to derive the
conjugate momenta 
\bseq
\bea
\label{h41a}
p_{\alpha} &=& \frac{\partial L}{\partial {\alpha}^{\prime}}= 
- \frac {12 \pi}{N}e^{3\alpha} {\alpha}^{\prime}  \\
p_{\pm} &=& \frac{\partial L}{\partial {\beta_{\pm}}^{\prime}}= 
\frac{12 \pi}{N} e^{3 \alpha} {\beta_{\pm}}^{\prime} \, .
\eea
\eseq

\subsubsectionric{Reduced ADM Hamiltonian\label{sec:adm}}

In order to obtain Einstein equations, the variational 
principle requires  $\delta S$ 
to be null for arbitrary and independent variations of 
$p_{\pm}, p_{\alpha}, \beta_{\pm}, \alpha, N$. Variation with 
respect  to $N$ leads to the super-Hamiltonian constraint 
$ H=0$.
As we have seen in Section \ref{sec:adm_3}, 
the procedure prescribes the choice of one of the field 
variables, or one of the momenta, as the temporal coordinate
and subsequently one can solve the constraint (\ref{h40}) with 
respect to the corresponding conjugate quantity.\\
It is customary, as in this general approach, to set
$t=\alpha$ and solve ${H}=0$ obtaining 
\beq
\label{h43}
{\cal H}_{ADM}=-p_{\alpha}=\sqrt{p^2_+ +p^2_- +{\cal V}} \, .
\eeq
Within this equation a relation between 
the temporal gauge described by the function $N$ and 
the dynamical quantity ${\cal H}_{ADM}$ is defined.

Through (\ref{h43}) we explicit $p_{\alpha}$ in the 
action integral, so that the reduced variational principle
in a canonical form reads as
\beq
\label{h44}
\delta S_{\textrm{red}} =0
\eeq  
$S_{\textrm{red}} $ being written as
\beq
\label{h45}
S_{\textrm{red}} =\int \left( p_+ d\beta_+ + p_- d\beta_- 
- {\cal H}_{ADM}d\alpha \right)
\eeq
together with the equation defining the temporal gauge for $\dot{\alpha}=1$
\begin{equation}
\label{nadm1}
N_{ADM} = \frac{12 \pi e^{3 \alpha}}{{\cal H_{ADM}}} \, .
\end{equation}

\subsubsectionric{Mixmaster Dynamics\label{sec:mix}}

In the present Section we will write in general the 
approach to the Mixmaster dynamics and later on it 
will be applied to prove specific properties, such 
as chaoti\-ci\-ty in a covariant approach, with respect to the 
temporal gauge, and subsequent statistical effects. 

The Hamiltonian introduced so far differs from 
the typical expression of classical mechanics for the 
non positive definiteness of the kinetic term, i.e. 
the sign in front of $p_{\alpha}^2$, and for the peculiar 
form of the potential  as a function of $\alpha$ 
(say \textit{time})
and $\beta_{\pm}$, reduced to the study of a function 
of the kind $V\left(\beta_+, \beta_-\right)$.

%
%
\begin{figure}[ht] 
\begin{center} 
\includegraphics[angle=-90, width=0.6\textwidth,clip]{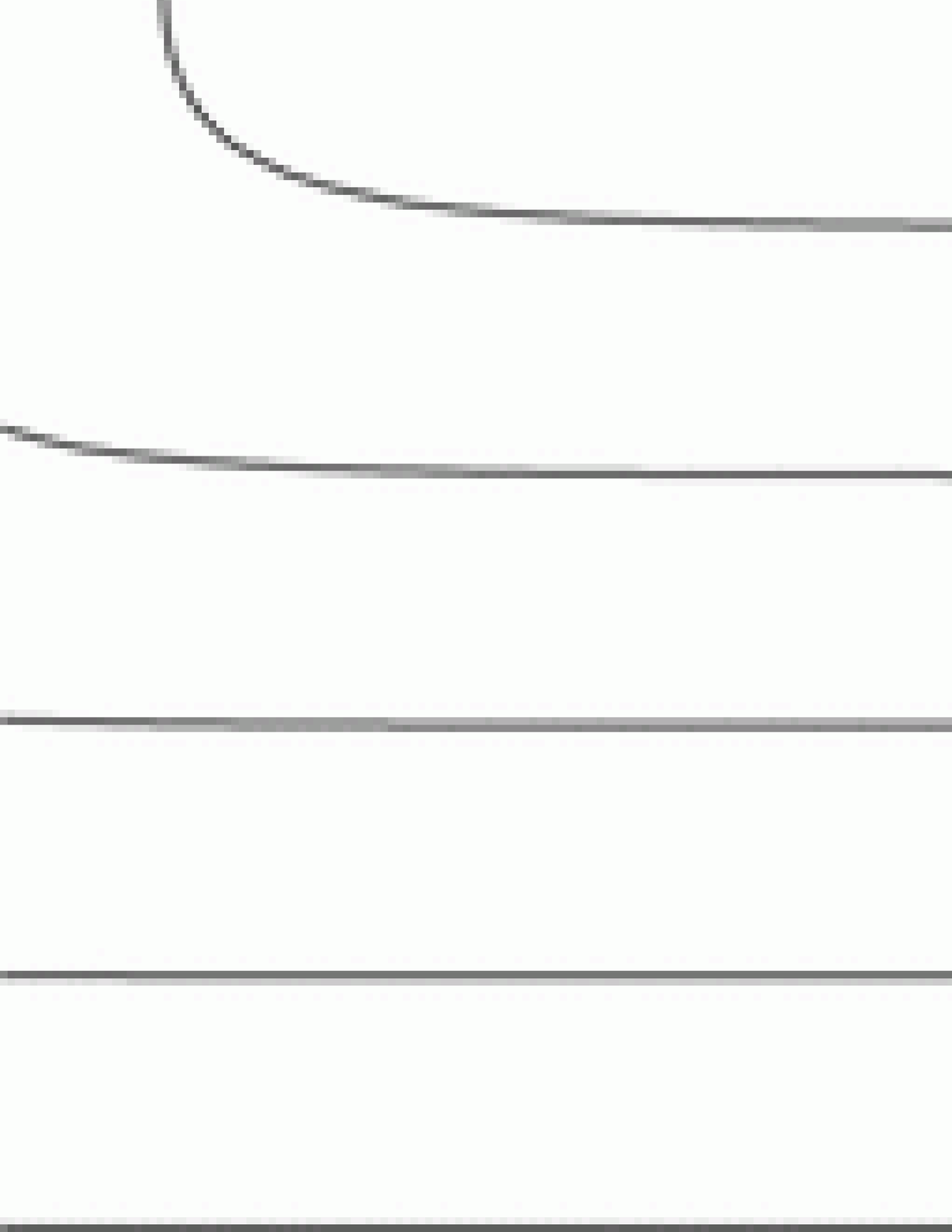}
\caption{Equipotential lines of the Bianchi
 type VIII model in the $\beta_+,\beta_-$
 plane.}  
\label{fig:B8} 
\end{center}
\end{figure}
\begin{figure}[ht] 
\begin{center} 
\includegraphics[angle=-90, width=0.6\textwidth,clip]{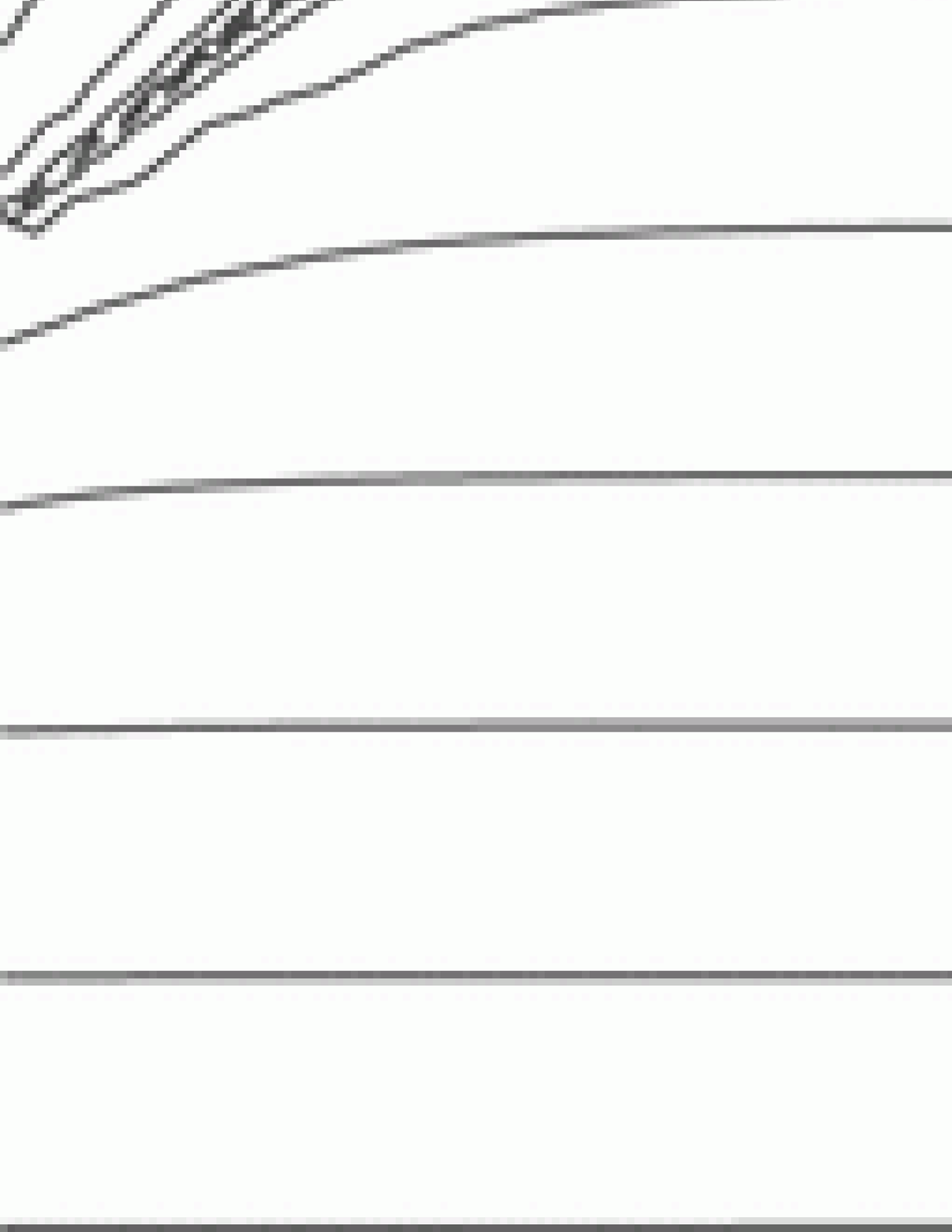}
\caption{Equipotential lines of the Bianchi
 type IX model in the $\beta_+,\beta_-$
 plane.}  
 \label{fig:B9}
\end{center}
\end{figure}

A Hamiltonian approach permits to derive 
the equations of motion as
\bseq
\bea
\label{h46}
\alpha^{\prime}&=&
	\frac{\partial { H}}{\partial p_{\alpha}} \, ,\quad 
p_{\alpha}^{\prime}=
	-\frac{\partial {H}}{\partial {\alpha}} \, ,  \\
\beta_{\pm}^{\prime}&=&
	\frac{\partial { H}}{\partial p_{\pm}}  \, , 
\quad p_{\pm}^{\prime}=
	-\frac{\partial { H}}{\partial \beta_{\pm}} \, .  
\eea
\eseq
This set considered  with the explicit 
form of the potential (see Figure \ref{fig:B8} 
and \ref{fig:B9}), 
can be interpreted as the motion of 
a ``point-particle'' in a potential. The term 
${\cal V}$ is proportional to the curvature scalar 
and describes the anisotropy of the Universe, i.e.
in the regions of the configuration space where it
can be negligible the dynamics resembles the pure
Kasner behaviour, corresponding to 
$|\boldsymbol{ \beta}^{\prime}|=1$. 

In general, it is necessary a detailed 
study of the potential form 
which behaves
as a potential well 
in the plane  
$\beta_+ ,\beta_-$.\\ 
Asymptotically close to the origin, i.e. 
$\beta_{\pm}=0$, equipotential lines for 
the Bianchi type IX are circles
\beq
\label{h49}
U^{IX}\left(\beta_+, \beta_-\right)\simeq -3 
+\left({\beta_+}^2 +{\beta_-}^2\right) +o\left(\beta^3\right),
\eeq
while for Bianchi type VIII are ellipses 
\begin{align} 
\label{h50}
\! \! \! U^{VIII}\left(\beta_+, \beta_-\right)
\simeq \left(40{\beta_+}^2 +24{\beta_-}^2\right)-
8\left(\beta_+ + \beta_-\right)+5 +o\left(\beta^3\right).
\end{align}
The expressions for large values of $\beta$ are 
the same  for both types
\beq 
\label{h51a} 
U\left(\beta\right)\simeq e^{-8\beta_+} \, ,\qquad  
\beta_+\longrightarrow -\infty 
\eeq
or
\beq
\label{h51b}
U\left(\beta\right)\simeq 48 {\beta_-}^2e^{4\beta_+} \, , \qquad \beta_+\longrightarrow +\infty  
\eeq
when 
\beq
\label{h51ab}
\mid \beta_-\mid  \ll  1 \, . 
\eeq
In the figures \ref{fig:B8} and \ref{fig:B9}
are represented some of the 
equipotential lines $U\left(\beta\right)=\textrm{const.}$, 
for which  the potential value 
has an increment of a factor $e^8\sim3\times10^3$ 
for 
$\Delta \beta \sim 1$. \\
The Universe evolution is described as the motion 
of a point-like particle under the influence of 
such potentials and it corresponds to 
bounces on the potential walls when  evolving 
towards the singularity. 
The behaviour of 
the anisotropy variables  $\beta_{\pm}$ 
in this regime consists of a Kasner epoch followed
by a bounce and then a new epoch with different 
Kasner parameters, in correspondence with the 
description given in Section \ref{sec:bkl} 
according to the BKL approach.



Let us describe in more detail the bounces performed 
by the billiard ball  representing the Universe. 
From the asymptotic form (\ref{h51a}) for the Bianchi IX potential, 
we get  the equipotential line $\beta_{wall}$ 
cutting  the region where the potential terms are significant. 
In particular we have
\be\label{bwall}
\beta_{wall}=\f\alpha2-\f18\ln(3H^2),
\ee
where the super-Hamiltonian $H$ is defined as in (\ref{h40}). 
Since we have  $H=\textrm{const.}$ inside the potential well, 
from  (\ref{bwall}) one gets  $|\beta^\prime_{wall}|=1/2$, 
i.e. the $\beta$-point moves twice as fast as the receding 
potential wall, i.e. the particle will collide against the 
wall and will be reflected from one straight-line motion (Bianchi I) 
to another one.

A relation of reflection-type  lays for the bounces \cite{Misner1969PR}. 
In fact, 
considering the super-Hamiltonian (\ref{h40}) in the (\ref{h51a})-case,
if  $\theta_i$ and $\theta_f$ are  the angles of incidence 
and of reflection of the particle off the potential wall, respectively, 
the relation
\be\label{reflay}
\sin\theta_f-\sin\theta_i=\f12\sin(\theta_i+\theta_f) \, , 
\ee
holds and in terms of the parameter $u$ introduced in section \ref{sec:appl7}, 
it  is nothing but $u_f=u_i-1$.

\subsectionric{Misner-Chitre--like variables\label{sec:mcl}}

A valuable framework of analysis of the Mixmaster 
evolution, able to join together the two  
points of view of the map approach and of the
continuous dynamics evolution, relies on a 
Hamiltonian treatment of 
the equations in terms of Misner-Chitre 
variables \cite{C72}. This formulation  allows to 
individualize the existence of an asymptotic 
(energy-like) constant of motion once  performed an 
ADM reduction. 
By this result, the stochasticity of 
the Mixmaster can be treated either in terms of the 
statistical mechanics (by the microcanonical ensemble)
\cite{ImponenteMontani2002JKPS}, 
either by its characterization as isomorphic to a 
billiard on a two-dimensional Lobachevsky 
space \cite{A89} and such  scheme can be 
constructed independently of the choice of a time 
variable, simply providing very general 
Misner-Chitre--like (MCl) coordinates 
\cite{ImponenteMontani2001PRD,ImponenteMontani2001IJMPD}. 

To this purpose, let us here re-define the anisotropy parameters 
$Q_a $ (following \reffcite{Kir93,KirillovMelnikov1995}, 
\reffcite{Montani1995CQG,KirillovMontani1997PRD}) as the functions 
\bseq
\label{ani}
\begin{eqnarray} 
Q_1 &=& \frac{1}{3}+ \frac{\beta_+ + 
\sqrt{3} \beta_-}{3 \alpha}  \\ 
Q_2 &=& \frac{1}{3}+ \frac{\beta_+ - 
\sqrt{3} \beta_-}{3 \alpha}   \\ 
Q_3 &=&\frac{1}{3}- \frac{2\beta_+}{3 \alpha}  \, ,
\label{c} 
\end{eqnarray} 
\eseq
excluding the pathological cases when two or three of them coincide.

We then introduce the Misner-Chitre variables 
$\{\tau, \zeta,\theta\}$ as
\bseq
\label{dmc}
\bea
\alpha &=& - e^{\tau} \cosh \zeta  \\ 
\beta_+ &=& ~~e^{\tau}   \sinh \zeta  \cos \theta \\ 
\beta_- &=& ~~e^{\tau}  \sinh \zeta  \sin \theta 
\eea
\eseq
where $0\leq \zeta<\infty $, $0\leq\theta<2\pi$, and  
$\tau$ plays the role of a ``radial'' coordinate 
coming out from the origin of the $\beta_{\pm}$
space \cite{MTW}. 
In terms of (\ref{dmc}), it is possible to study 
the first interesting approximation of the 
potential (\ref{9pp}) as 
independent of $\tau$ towards the singularity, 
i.e. for $\alpha\rightarrow - \infty$.

To discuss the contrasting 
results concerning chaoticity and dynamical properties
which arose from numerics
\cite{FranciscoMatsas1988,RU94,1990CQGra...7..203B,HobillNum1991}, it is necessary to 
introduce a slight modification to the set (\ref{dmc})
via the MCl coordinates 
$\{\Gamma(\tau),\xi,\theta\}$
through the transformations 
\bseq
\label{d}
\begin{eqnarray} 
\alpha &=& - e^{\Gamma \left(\tau\right)}\xi  \\ 
\beta_+ &=& ~~e^{\Gamma \left(\tau\right)}\sqrt{\xi^2 -1}\cos \theta \\ 
\beta_- &=& ~~e^{\Gamma \left(\tau\right)}\sqrt{\xi^2 -1}\sin \theta   \end{eqnarray} 
\eseq
where $1\leq\xi<\infty$, and
$\Gamma(\tau)$ stands for a \textit{generic} 
function of $\tau$: in the original work, 
Chitre took simply $\Gamma(\tau)\equiv \tau$
and set  $\xi=\cosh{\zeta}$. \\
This modified set of variables permits 
to write the anisotropy parameters (\ref{ani})
$Q_a$ as \textit{independent} of the 
variable $\Gamma$ as
\bseq
\label{sh} 
\begin{eqnarray} 
Q_1 &=& \frac{1}{3} - \frac{\sqrt{\xi ^2 - 1}}{3\xi }
\left(\cos\theta + \sqrt{3}\sin\theta \right) \\ 
Q_2 &=& \frac{1}{3} - \frac{\sqrt{\xi ^2 - 1}}{3\xi }
\left(\cos\theta - \sqrt{3}\sin\theta \right)  \\ 
Q_3 &=& \frac{1}{3} + 2\frac{\sqrt{\xi ^2 - 1}}{3\xi } \cos\theta \, .
\end{eqnarray}
\eseq
\textit{All dynamical quantities, if expressed in terms 
of (\ref{sh}) will be independent of $\tau$ too}.

\begin{figure}[ht] 
\begin{center} 
\includegraphics[width=8cm]{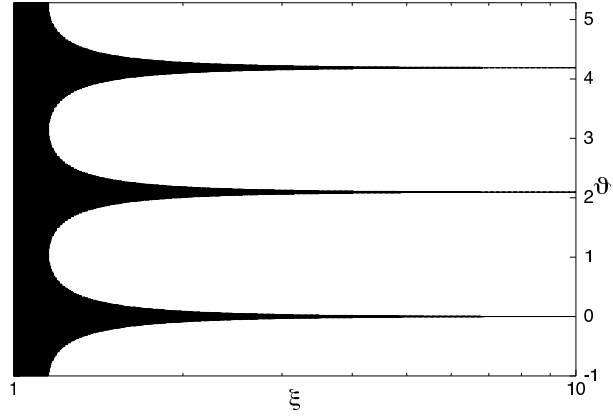} 
\caption[The reduced configurational space 
		$\Gamma_Q(\xi,\theta)$]{The 
		limited region $\Pi_Q(\xi,\theta)$ of the 
		configuration space where the dynamics 
		of the point Universe is restricted by mean 
		of the curvature term which corresponds to an 
		infinite potential well. In this  region the 
		conditions $Q_{a}\geq0$ are fulfilled.}
\label{DOMINIO XI THETA}
\end{center}
\end{figure}

\subsubsectionric{The Hamilton Equations\label{sec:hameqa}}

The main advantage relying on the 
reformulation of the dynamics as a chaotic 
scattering process consists of replacing the 
discrete BKL map by a geodesic flow in a space of 
continuous variables 
\cite{ChernoffBarrow1983PRL,KirillovMontani1997PRD}--
\cite{Barrow1981PR,Montani2000CQGultrarelativistic}, 
\cite{ImponenteMontani2003IJMPD}.   

%

In terms of (\ref{d}), 
the Lagrangian (\ref{h38}) becomes
\begin{align}
\label{e} 
L=\frac{6 D}{N} \Bigg[ \frac{{\left(e^{\Gamma } 
{\xi}^{\prime}\right)}^2}{\xi ^2 -1} + {\left(e^{\Gamma}
{\theta}^{\prime}\right)}^2\left(\xi ^2 -1\right) 
-             {{\left(e^{\Gamma }\right)}^{\prime}}^2 \Bigg] 
-\frac{N}{D}V \big( \Gamma \left(\tau\right), \xi, \theta \big) \, ,
\end{align} 
while  $D$ expresses  as 
\begin{equation} 
\label{f}
D = \exp\left\{ -3 \xi e^{\Gamma \left(\tau\right)} \right\} \, ,
\end{equation} 
and since it vanishes  towards the 
singularity independently of its particular 
form, the only property required for 
$\Gamma$ is to approach infinity in this limit.\\
The Lagrangian (\ref{h38})
leads to the conjugate momenta 
\bseq
\label{io}
\begin{align}
\label{ioa}
p_{\tau}&= -\frac{12 D}{N}{\left(e^{\Gamma} 
\frac{d\Gamma}{d\tau}\right)}^2  {\tau}^{\prime} \\
\label{iob}
p_{\xi}&= \frac{12 D}{N}
\frac{e^{2\Gamma} }{{\xi}^2 -1}{\xi}^{\prime}  \\
\label{ioc}
p_{\theta}&= \frac{12 D}{N}e^{2\Gamma} 
\left({\xi}^2 -1\right) {\theta}^{\prime} 
\end{align}
\eseq
which by a Legendre transformation 
make the variational principle assume the form
\cite{ImponenteMontani2001PRD}
\begin{equation} 
\label{g} 
\delta \int \left(p_{\xi} {\xi}^{\prime} +  
p_{\theta} {\theta}^{\prime}+
p_{\tau}  {\tau}^{\prime} - 
\frac{Ne^{-2\Gamma }}{24 D} { H}
\right) dt =0 ,
\end{equation}
where 
\begin{equation} 
{ H} = -\frac{{p_{\tau}}^2}{\left(
\displaystyle\frac{d\Gamma }{d\tau}\right)^2} 
+ {p_{\xi}}^2\left(\xi ^2 -1\right)
+\frac{{p_{\theta}}^2}{\xi ^2 -1} +24 V e^{2\Gamma } . 
\label{h} 
\end{equation}

\subsubsectionric{Dynamics in the Reduced 	Phase Space\label{sec:dynredph}}

Solving the super-Hamiltonian constraint we get 
 the expression for
${\cal H}_{ADM}$
\begin{equation} 
\label{l} 
- p_{\tau}\equiv \frac{d\Gamma }{d\tau} 
{\cal H}_{ADM} = \frac{d\Gamma }{d\tau}
\sqrt{\varepsilon ^2 +24 V e^{2\Gamma }} \, ,
\end{equation}
where
\begin{equation}
\varepsilon ^2 \equiv  \left({\xi}^2 -1\right){p_{\xi}}^2 
+\frac{{p_{\theta}}^2}{{\xi}^2 -1} \, .
\label{m} 
\end{equation} 
In terms of this constraint, the principle 
(\ref{g}) reduces to the simpler form
\begin{equation}
\label{ee}  
\delta \int \Big(   p_{\xi} {\xi}^{\prime} +  
p_{\theta} {\theta}^{\prime} 
- {\Gamma }^{\prime}{\cal H}_{ADM} \Big) dt = 0 \, ,
\end{equation} 
whose variation  provides  
the Hamilton equations for ${\xi}^{\prime}$ 
and ${\theta}^{\prime}$ \cite{ImponenteMontani2001PRD}
\bseq
\begin{eqnarray}
{\xi}^{\prime}&=& \frac{\Gamma^{\prime}}{{\cal H}_{ADM}}
\left(\xi ^2 -1\right)p_{\xi}     \\
{\theta}^{\prime}&=& \frac{\Gamma^{\prime}}{{\cal H}_{ADM}} 
\frac{p_{\theta}}{\left(\xi ^2 -1\right)} \, .
\end{eqnarray}
\label{zsz}
\eseq
From (\ref{ioa}) we find the time-gauge relation 
\begin{equation} 
\label{o} 
N_{ADM}(t) = \frac{12 D}{{\cal H}_{ADM}} 
e^{2\Gamma } \frac{d\Gamma }{d\tau} 
{\tau }^{\prime} \, ,
\end{equation} 
thus our analysis remains fully independent of the 
choice of the 
time variable until the form of $\Gamma$ and 
${\tau }^{\prime}$ is  fixed. \\
By choosing $d\Gamma/d\tau=1$, 
the principle (\ref{ee}) reduces to the 
two-dimensional one \cite{ImponenteMontani2001PRD}
\begin{equation}
\label{qqw} 
\delta \int \left(   p_{\xi} {\xi}^{\prime} 
+  p_{\theta} {\theta}^{\prime} 
- {\cal H}_{ADM} \right) dt =0 \, ,
\end{equation}
where, remebering (\ref{bianpot}) 
\begin{equation} 
\label{n2}
{\cal H}_{ADM} = \sqrt{\varepsilon ^2 +{ \cal U}}\, ,\qquad 
{ \cal U}\equiv 24 V e^{2\tau}\, ;
\end{equation}
moreover, the choice $\tau^{\prime}=1$ for 
the temporal gauge provides
the lapse function as
\begin{equation} 
N_{ADM}\left(\tau\right)= 
\frac{12 D}{{\cal H}_{ADM}} e^{2\tau} \, .
\label{rs} 
\end{equation} 
The reduced principle (\ref{qqw}) finally  gives  
the Hamilton 
equations 
\bseq
\begin{align}
\label{s}
{\xi}^{\prime}	&= \frac{\left(\xi ^2 -1\right)}
			{{\cal H}_{ADM}}p_{\xi}  \, , \\
{\theta}^{\prime}&= \frac{1}{{\cal H}_{ADM}} 			 
			 \frac{p_{\theta}}{\left(\xi ^2 -1\right)} \, , \\
p^{\prime}_{\xi} &= -\frac{\xi}{{\cal H}_{ADM}}
			\left[ {p_{\xi}}^2-\frac{{p_{\theta}}^2}
			{\left(\xi^2-1\right)^2} \right]
			-\frac{1}{2{\cal H}_{ADM}}
			\frac{\partial U}{\partial \xi}   \, , \\
p^{\prime}_{\theta}& = -\frac{1}{2{\cal H}_{ADM}}
			\frac{\partial U}{\partial \theta} \, ,
\end{align}
\eseq
where, because of the choice of the time 
gauge, $\displaystyle (~)^{\prime}=\frac{d}{d\tau}$.

\subsubsectionric{Billiard Induced from the 
	Asymptotic Potential\label{sec:billinduced}}

The Hamilton equations 
are equivalently viewed through the two 
time variables $\Gamma$ and $\tau$ and then, 
for this Section only, we choose the natural 
time gauge $\tau^{\prime }=1$, in order to write  the 
variational principle (\ref{ee}) in terms of 
the time variable $\Gamma$
 as
\begin{equation} 
\delta \int \left( p_{\xi} \frac{d\xi }{d\Gamma } +
 p_{\theta} 
\frac{d\theta}{d\Gamma } 
- {\cal H}_{ADM} \right) d\Gamma = 0  \,.
\label{p} 
\end{equation} 
Nevertheless, for any choice of time variable 
$\tau$ (i.e.\ $\tau=t$), there  
exists a corresponding function 
$\Gamma \left(\tau \right)$ (i.e. a set of 
MCl variables leading to the scheme (\ref{p})) 
defined by the (invertible) relation 
\begin{equation} 
\label{q} 
\frac{d\Gamma }{d\tau} = 
\frac{{\cal H}_{ADM}}{12 D } 
N\left(\tau \right)e^{-2\Gamma } \, . 
\end{equation} 


The asymptotically vanishing of  $D$ 
is ensured by 
the Landau-Raichaudhury theorem
near the initial singularity (which occurs by convention 
at $T=0$, where $T$ now denotes the synchronous time, i.e. 
$dT=- N\left(\tau\right)d\tau$), for $T\rightarrow 0$ we 
have $d \ln D/d T > 0$. 
In terms of the adopted variable $\tau$ we have
\beq
D\rightarrow 0 \quad \Rightarrow \quad \Gamma\left(\tau\right) 
\rightarrow \infty \, ,
\eeq
consequently, by (\ref{f}) and (\ref{q}), also 
\begin{equation} 
\frac{d \ln D}{d \tau } = 
\frac{d \ln D}{d T}\frac{dT}{d\tau } = 
- \frac{d \ln D}{d T}N\left(\tau \right)  
\label{x1} 
\end{equation}
and therefore $D$ monotonically vanishes even in the generic 
time gauge as soon as $d\Gamma / d\tau >0$ for 
increasing $\tau$, according to (\ref{q}). 

Approaching the initial singularity, 
the limit $D\rightarrow 0$ for the Mixmaster 
potential (\ref{bianpot}) implies 
an infinite potential well behavior, as discussed in Section \ref{sec:hfd}

Furthermore, by (\ref{l}) the important relation 
\begin{align} 
\frac{d\left({\cal H}_{ADM}\Gamma^{\prime}\right)}{dt} =
 \frac{\partial \left({\cal H}_{ADM}\Gamma^{\prime}\right)}{\partial t} 
\Longrightarrow  
\frac{d\left({\cal H}_{ADM}\Gamma^{\prime}\right)}{d\Gamma} = 
\frac{\partial \left({\cal H}_{ADM}
\Gamma^{\prime}\right)}{\partial \Gamma} \, ,
\label{t} 
\end{align} 
 holds, i.e. explicitly
 \begin{equation} 
\label{uh}
\frac{\partial{\cal H}_{ADM}}{\partial \Gamma}=
\frac{e^{2\Gamma}}{2 {\cal H}_{ADM}} 24
\left( 2U+ \frac{\partial U}{\partial \Gamma} \right) \, .
\end{equation} 
In this reduced Hamiltonian formulation, 
the functional $\Gamma(t)$ simply  plays
the role of a parametric 
function of time and we recall how actually the 
anisotropy pa\-ra\-me\-ters $Q_a$  are functions 
of the variables $\xi, \theta$ only (see (\ref{sh})).

Therefore in the dinamically allowed domain  $\Pi_Q$ 
(see Fig.\ref{DOMINIO XI THETA})
the ADM Hamiltonian becomes 
(asymptotically) an integral of motion
\begin{align}
\!\!\!\! \!\!\! \forall \{\xi, \theta\}\in{ \Pi_Q} \;\;
\left\{ 
\begin{array}{lll} \displaystyle
\frac{\partial {\cal H}_{ADM}}{\partial \Gamma} 
=0= \frac{\partial E}{\partial  \Gamma}  \\ 
{\cal H}_{ADM}= \sqrt{\varepsilon ^2 +
24~ U} \cong \varepsilon =E ={\rm const.}\, .
\end{array}
\right.
\label{bb}
\end{align}
%
%

%
%


\subsubsectionric{The Jacobi Metric and the 
Billiard Representation\label{sec:billiardr}}

Since above we have shown that asymptotically to 
the singularity ($\Gamma  \rightarrow \infty$, 
i.e. $\alpha \rightarrow  -\infty$) 
$d{\cal H}_{ADM}/d\Gamma =0$, i.e. 
${\cal H}_{ADM} =\epsilon =E=\textrm{const.}$, 
the variational principle (\ref{p}) reduces to
\begin{equation}
\delta \int \big( p_{\xi} d\xi + 
p_{\theta} d\theta -Ed\Gamma \big) =
\delta \int \big(  p_{\xi} d\xi + 
p_{\theta} d\theta \big)=0 \, ,
\label{cc}
\end{equation}
where we dropped the third term on the left-hand side 
since it behaves as an exact 
differential.

By following the standard Jacobi procedure 
\cite{A89} to reduce our variational 
principle to a geodesic one in terms of the configuration 
variables $x^a$, we set 
${x^a}^{\prime} \equiv g^{ab}p_b$, and by 
the Hamilton equation (\ref{zsz}) we 
obtain the metric \cite{ImponenteMontani2001PRD,ImponenteMontani2002JKPS}
\beq
g^{\xi \xi} =\frac{\Gamma^{\prime}}{E}
\left({\xi}^2 -1\right) \, , \qquad 
g^{\theta \theta} =\frac{\Gamma^{\prime}}{E} 
\frac{1}{{\xi}^2 -1} \, .
\label{dd}
\eeq
By (\ref{dd}) and by the fundamental constraint 
relation obtained rewriting (\ref{m}) as
 \begin{equation}
\left({\xi}^2 -1\right){p_{\xi}}^2 
+\frac{{p_{\theta}}^2}{{\xi}^2 -1} =E^2 \, ,
\label{eew}
\end{equation}
we get 
\begin{equation}
g_{ab}{x^a}^{\prime} {x^b}^{\prime} 
=\frac{\Gamma^{\prime}}{E} 
\left[ \left({\xi}^2 -1\right){p_{\xi}}^2 + \frac{{p_{\theta}}^2}{{\xi}^2 -1}\right]=
\Gamma^{\prime}  E \, .
\label{ff}
\end{equation}
Using the definition 
\beq
\label{defban}
{x^a}^{ \prime}= \frac{dx^a}{ds} 
\frac{ds}{dt}\equiv u^a \frac{ds}{dt}\, ,
\eeq
equation (\ref{ff}) is rewritten as
\begin{equation}
g_{ab}u^a u^b \left( \frac{ds}{dt} \right) ^2  
= \Gamma^{\prime} E \, ,
\label{gg}
\end{equation}
which leads to the key relation
\begin{equation}
dt = \sqrt{ 
\frac{g_{ab}u^a u^b}{{\Gamma}^{\prime} E }}~ds \, .
\label{hh}
\end{equation}
Indeed the expression (\ref{hh}) together with 
$p_{\xi} \xi^{\prime} +p_{\theta} \theta^{\prime}=E{\Gamma}^{\prime}$ 
allows us to put the variational principle (\ref{cc}) 
in the geodesic form
\begin{align}
\delta \int  \Gamma^{\prime} E ~dt      = 
\delta \int \sqrt{ g_{ab}u^a u^b 
			\Gamma^{\prime} E} ~ds  
 =\delta \int  \sqrt{ G_{ab}u^a u^b}~ds =0 \, ,
\label{ii}
\end{align}
where the metric $G_{ab} \equiv \Gamma^{\prime} E g_{ab}$ 
satisfies the normalization condition 
$G_{ab}u^a u^b =1$ and therefore 
\begin{equation}
\frac{ds}{dt}=E\Gamma^{\prime}\Rightarrow 
\frac{ds}{d\Gamma} =E \, ,
\label{ll}
\end{equation}
where we take the positive root since 
we require that the curvilinear coordinate 
$s$ increases monotonically with increasing 
values of $\Gamma$, i.e. approaching the initial 
cosmological singularity. 


Summarizing, the 
 dynamical problem in the region $\Pi_Q$ reduces to a 
geodesic flow on a two-dimensional Riemannian 
manifold described by the 
line element \cite{AN67,ImponenteMontani2001PRD}
\begin{equation}
ds^2 =E^2 \left[ 	\frac{d{\xi }^2}{{\xi}^2 -1}+ 
 \left(\xi^2 -1\right) d {\theta }^2 \right] \, .
\label{mm}
\end{equation}
The above metric 
has negative curvature, since the associated 
curvature scalar is 
$ R=-\nicefrac{2}{E^2}$;
therefore the point-Universe moves over a 
negatively curved bidimensional space on which 
the potential wall (\ref{pottheta0}) cuts the region 
$\Pi_{Q}$, depicted in Figure \ref{DOMINIO XI THETA}. \\
By a way completely independent 
of the temporal gauge we provided a satisfactory 
representation of the system as isomorphic to a 
billiard on a Lobachevsky plane \cite{A89}.


From a geometrical point of view, the domain 
defined by the potential walls is 
not strictly closed, since there are three 
directions corresponding to the three corners 
in the traditional Misner picture \cite{Misner1969PR,Misner1969PRL} from which 
the point universe could in principle escape 
(see Fig.\ref{DOMINIO XI THETA}).

However, as discussed in Section \ref{sec:bkl} 
for the Bianchi models, the only case in which an 
asymptotic solution of the field equations  has  
this behaviour corresponds to having two scale factors equal
to each other (i.e. $\theta=0$); but, as shown by \reffcite{BK69}, 
these cases are dynamically unstable and  correspond 
to sets of zero measure in the space of the initial conditions. 
Thus, it has no sense 
to speak of a probability to reach certain configurations 
and the domain is {\it de facto} dynamically closed.

The bounces (billiard configuration) against 
the potential walls together with the geodesic flow 
instability on a 
closed domain of the Lobachevsky plane imply the 
point-Universe to have stochastic features,
 with a formalism true for any 
Bianchi type model.
Indeed the types VIII and IX are the only 
Bianchi models having a compact configuration 
space, hence the claimed compactness   
of the domain bounded 
by the potential walls guarantees that the 
geodesic instability is upgraded to a real 
stochastic behaviour \cite{arlond1968,sinai,MPVb}.
On the other hand, the possibility to deal with  
a stochastic scattering is justified
by the constant negative curvature of the 
Lobachevsky plane and therefore
these two notions (compactness and curvature) are both
necessary for our considerations \cite{Woj,BeniniMontani2004PRD}.

\subsectionric{The Invariant  Liouville Measure \label{sec:invmeasure}}

Here we show how the derivation of an 
invariant measure for the Mixmaster model 
(performed by \reffcite{ChernoffBarrow1983PRL,KirillovMontani1997PRD,Montani2001NCB} 
within the framework of the statistical mechanics) 
can be extended to a generic time gauge \cite{ImponenteMontani2001IJMPD}
(more directly than in previous approaches
relying on fractal methods as in \reffcite{CornishLevin1997PRL,CornishLevin1997PRD}) 
provided the  Misner-Chitre-like 
variables used so far. 
We have seen how the  (ADM) reduction of the variational 
problem asymptotically close to the cosmological singularity 
permits to modelize the Mixmaster dynamics by a two-dimensional
point-Universe randomizing in a closed domain with fixed 
``energy'' (just the ADM kinetic energy) (\ref{bb}); 
the key point addressed here is that we consider 
an approximation dynamically induced by the asymptotic 
vanishing of the metric determinant.   

From the statistical mechanics point of view\cite{ImponenteMontani2004PA}, 
such a stochastic motion within the closed domain $\Pi_Q$
of the phase-space, induces
a suitable ensemble representation which, 
in view of the existence of the ``energy-like''
constant of motion, has to have the natural features of 
a {\it microcanonical} one.
Therefore, the stochasticity of this system can be 
described in terms of the Liouville invariant
measure
\begin{equation} 
\label{u}
d\varrho = {\rm const} \times \delta \left(E - \varepsilon \right)d\xi d\theta dp_{\xi }dp_{\theta }  
\end{equation} 
characterizing the {\it microcanonical ensemble},
having denoted by $\delta\left(x\right)$ the 
Dirac function. \\
The particular value taken by the constant 
$\varepsilon$ $(\varepsilon=E)$ cannot influence 
the sto\-cha\-sti\-ci\-ty property of the system
and must be fixed by the initial conditions. 
This useless information from the statistical 
dynamics is removable by integrating over all admissible 
values of $\varepsilon$. 
Introducing the natural variables 
$(\varepsilon,\phi)$ in place of $(p_\xi,p_\theta)$ by
\beq
p_{\xi } = \frac{\varepsilon}{\sqrt{\xi ^2 - 1}}
\cos\phi  \, , \qquad  
p_{\theta } =  \varepsilon \sqrt{\xi ^2 - 1}\sin\phi \, , \qquad
0 \leq \phi < 2\pi 
\label{vh}
\eeq
the integration removes the Dirac function, 
leading to the uniform and  normalized
invariant measure \cite{ImponenteMontani2001IJMPD}
\begin{equation} 
d\mu = d\xi d\theta d\phi \frac{1}{8\pi^2} \, . 
\label{x} 
\end{equation} 
The approximation on which our analysis is based 
(i.e. the potential wall model) 
is reliable since it is dynamically induced, no matter 
what time variable $\tau$ is adopted. 
Furthermore, such invariant measure turns out to be 
independent on the choice of the temporal gauge, as shown 
in \reffcite{ImponenteMontani2001IJMPD}.


The use of the invariant measure  and of the 
Artins theorem \cite{artin1965msq} provides  the complete equivalence 
between the BKL piece-wise description and the Misner-Chitre 
continuous one\cite{KirillovMontani1997PRD}.


According to the analysis presented in \reffcite{Montani2001NCB}, 
by virtue of (\ref{zsz}) and (\ref{ll}), the a\-sympto\-tic 
functions $\xi\left(\Gamma\right), 
\theta\left(\Gamma\right), \phi\left(\Gamma\right)$ 
during the free geodesic motion are governed by the equations
\bseq
\begin{align}
\frac{d\xi}{d\Gamma}&=\sqrt{\xi^2-1}\cos\phi \\
\frac{d\theta}{d\Gamma}&=\frac{\sin\phi}{\sqrt{\xi^2-1}} \\
\frac{d\phi}{d\Gamma}&=
-\frac{\xi\sin\phi}{\sqrt{\xi^2-1}} \, ,
\end{align}
\eseq
whose parametric solution  $\xi\left(\Gamma\right)$ 
has the form
\bseq
\begin{eqnarray}
\label{xx1}
\xi \left(\phi\right)&=&\frac{\rho}{{\sin^2\phi}}  \\
\Gamma \left( \phi \right)&=& \frac{1}{2} {\rm arctanh}
 \left(\frac{1}{2} \frac{{\rho^2} +a^2 {\cos^2\phi}}{a\rho 
 \cos \phi}\right) +b \\
&~&\rho \equiv\sqrt{a^2 +\sin^2\phi} \, \quad 
a,b={\rm const.}\in \nR  \, .\nonumber 
\end{eqnarray}
\eseq

However, the global behaviour 
of $\xi$ along the whole geodesic flow 
is described by the invariant measure (\ref{x}) 
and therefore the temporal behavior of 
$\Gamma\left(\tau\right)$  acquires 
a stochastic character: if we assign one of the two 
functions $\Gamma \left(\tau\right)$ or 
$N\left(\tau\right)$ 
with an \textit{arbitrary} analytic functional form, then 
the other one will exhibit a stochastic behaviour.
Finally, by retaining the same dynamical scheme 
adopted in the construction of the invariant measure, 
we see how the one-to-one correspondence between 
any lapse function $N\left(\tau\right)$ and the 
associated set of MCl variables (\ref{d}) 
guarantees the covariance with respect to the time-gauge.

\subsectionric{Chaos covariance\label{sec:chaos}}


We have discussed the oscillatory regime whose 
properties characterize the behavior of the Bianchi 
types VIII and IX cosmological models 
in the BKL for\-ma\-lism 
 \cite{BKL1970,BKL82,Misner1969PRL} 
 near a physical singularity, in which 
it is outlined the appearance of chaotic
properties\cite{Ott}: firstly, the dynamics evolution 
of the Kasner exponents characterized the 
sequence of the Kasner epochs, each one described
by its own line element, with the epochs sequence
nested in multiple eras. Secondly, the use of the
pa\-ra\-me\-ter $u$ and its relation to dynamical 
functions offered the statistical 
treatment connected to each Kasner era, finding 
an appropriate expression for the distribution 
over the space of variation: the entire evolution 
has been decomposed in a discrete mapping in terms 
of the rational/irrational initial values attributed
to $u$. \\
The limits of this approach essentially reside 
in the non-continuous evolution 
 toward the initial singularity
and the lack of an assessment of chaoticity 
in accordance with the indicators commonly 
used in the theory of dynamical systems, 
say in terms of the 
estimate of Lyapunov exponents. \\
A wide literature faced over the years this subject 
in order to provide the best possible understanding 
of the resulting chaotic dynamics 
\cite{B00,1990CQGra...7..203B}. 

The research activity developed overall in two 
different, but related, directions: 
\begin{enumerate}
\renewcommand{\labelenumi}{({\it\roman{enumi}})}
	\item 
on one hand, 
the dynamical analysis was devoted to remove the 
limits of the BKL approach due to its discrete 
nature (by analytical treatments 
\reffcite{Barrow1981PR,ChernoffBarrow1983PRL,BBT,BI95,
CornishLevin1997PRL,CornishLevin1997PRD,KirillovMontani1997PRD,M00a}
and by numerical simulations \reffcite{RU94,
1990CQGra...7..203B,1994PhRvD..49.1120B,BGS,BGS},
%
\item 
on the other one, to get a better characterization of the 
Mixmaster chaos (especially in view of its properties 
of covariance \cite{FFM91,FF92,SK93,Hobill1994}.
%
%
%
\end{enumerate}
The first line of investigation provided 
satisfactory representations of the Mixmaster 
dynamics in terms of continuous variables \cite{BN73}, 
mainly studying the pro\-per\-ties of the BKL map 
and its reformulation as a Poincar\'e one 
\cite{BarrowMapPRL1981}.

In parallel to these studies,
detailed numerical descriptions  have been performed
with the aim  to test the  precise validity of the 
analytical results \cite{RU94,1990CQGra...7..203B,berger1993nic,berger2002nas,Cot97cm}.

The efforts \cite{BI95,CGR94,RU94} 
to develop a precise characterization 
of the chaoticity observed in the Mixmaster dynamics 
found non-trivial difficulties due to the impossibility, 
or in the best cases the ambiguity, to apply the standard 
chaos indicators to relativistic systems. 
However, the chaotic properties summarized so 
far were questioned when numerical evolution 
of the Mixmaster equations yielded zero Lyapunov 
exponents \cite{1990GReGr..22..349B,Hobill1994,FranciscoMatsas1988}.
Nevertheless, exponential divergence of 
initially nearby trajectories was  
found by other numerical studies yielding 
positive Lyapunov numbers. 
 This 
issue was understood when in  \reffcite{1990CQGra...7..203B}
and \reffcite{FFM91}  numerically and 
analytically  was shown how such calculations
depend on the choice of the time 
variable and parallely to the failure of  the 
conservation of the Hamiltonian constraint 
in the numerical simulations by \reffcite{Z83}, although 
was still debated by \reffcite{Hobill1994}.

In particular, the first clear distinction 
between the direct numerical study of the 
dynamics and the map approximation, stating 
the appearance of chaos and its relation 
with the increase of entropy, has been introduced 
by \reffcite{BBT}.
The puzzle consisted of simulations  
providing even in the following years zero (see for example \reffcite{FF92})
Lyapunov numbers, claiming that the Mixmaster 
Universe is non-chaotic with respect to the intrinsic 
time (associated with the function $\alpha$ introduced
for the Hamiltonian formalism) 
but chaotic with respect to the synchronous time
(i.e. the temporal parameter $t$).
The non-zero claims \cite{SK93} about Lyapunov exponents,
using different time variables, have been obtained 
reducing the Universe dynamics to a geodesic flow on 
a pseudo-Riemannian manifold: on average, local
instability has been discussed for the BKL approximations. 
Moreover, a geometrized model of dynamics defining 
an average rate of separation of nearby trajectories 
in terms of a geodesic deviation equation in a Fermi 
 basis has been interpreted for detection of chaotic 
 behavior as a principal Lyapunov exponent.  
A non-definitive result was given: the principal 
Lyapunov exponents result always positive in the 
BKL approximations but, if the period of oscillations 
in the long phase (the evolution of long 
oscillations, i.e. when the particle enters 
the corners of the potential) is infinite, 
the principal Lyapunov exponent tends to zero.

For example, the author of \reffcite{1990CQGra...7..203B} reports the dependence 
of the Lyapunov exponent on the choice of the  time variable. 
Through numerical simulations,  the  Lyapunov exponents were  evaluated
along some trajectories in the $(\beta_{+},\;\beta_{-})$ plane for different 
choices of the time variable,  more precisely  
$\tau$ (BKL), $\;\Omega$ (Misner) and $\lambda$, 
the ``mini-superspace'' one, i.e.  
$d\lambda = |-p_{\Omega}^{2}+p_{+}^{2}+p_{-}^{2}|^{1/2}d\tau$
.
It is shown that the same trajectory giving 
zero Lyapunov exponent for $\tau$ 
or $\Omega$-time,  fails  for $\lambda$. 

Such contrasting results can  find a clear
explanation realizing the non-covariant nature of 
these indicators and their inapplicability to 
hyperbolic manifolds. The existence of 
such difficulties prevented, up to now, to say a 
definitive word about the general picture concerning 
the covariance of the 
Mixmaster chaos, with particular reference to the 
possibility of removing the observed chaotic features 
by a suitable choice of the time variable, apart from 
the indications provided by \reffcite{CornishLevin1997PRL,CornishLevin1997PRD}.

Interest in these 
covariance aspects has increased in recent years 
in view of the contradictory and often dubious results 
that  emerged. \\
%
%
%
%
%
The confusion which arises regarding the effect of a 
change of the time variable in this problem depends 
on some special properties of the Mixmaster 
model when represented as a dynamical system, 
in particular the vanishing 
of the Hamiltonian and its non-positive definite kinetic 
term (typical of a gravitational system). 
These peculiar features 
prevent the direct application of the most common 
criteria provided by the 
theory of dynamical systems for characterizing 
chaotic behavior 
(for a review, see \reffcite{Hobill1994}).

Although a whole line of research opened up \cite{PU91,SS94}, 
the first widely accepted indications 
in favor of covariance were derived 
with a fractal formalism by \reffcite{CornishLevin1997PRL,CornishLevin1997PRD}
(see also \reffcite{MotterLetelier2001PLA}).
Indeed, the requirement of a complete 
covariant description of the Mixmaster 
chaoticity when viewed in terms of continuous 
dynamical variables, due to the discrete 
nature of the fractal approach, 
leaves this subtle question open and prevents 
a general consensus in this sense from 
being reached.

\subsubsectionric{Invariant Lyapunov Exponent\label{sec:lyapinv}}

In order to characterize the dynamical instability 
of the billiard in terms of an invariant treatment 
(with respect to the choice of the coordinates 
$\xi,\theta$), let us introduce the following 
(orthonormal) tetradic basis \cite{ImponenteMontani2001PRD}
\bseq
\begin{eqnarray} 
v^i &=&\left(\frac{1}{E}\sqrt{{\xi}^2-1}\cos{\phi},\, 
\frac{1}{E}\frac{\sin{\phi}}{\sqrt{{\xi}^2-1}}\right)  \\
w^i &=&\left(-\frac{1}{E}\sqrt{{\xi}^2-1}\sin{\phi},\,
\frac{1}{E}\frac{\cos{\phi}}{\sqrt{{\xi}^2-1}} \right)  \, .
\label{nn}
\end{eqnarray} 
\eseq
Indeed, the vector $v^i$ is nothing else than the 
geodesic field, i.e. it satisfies
\begin{equation}
\frac{Dv^i}{ds}=\frac{dv^i}{ds}+\Gamma^i_{kl}v^k v^l =0 \, ,
\label{oo}
\end{equation}
while the vector $w^i$ is parallely transported 
along the geodesic, according to the equation
\begin{equation}
\frac{Dw^i}{ds}=\frac{dw^i}{ds}+\Gamma^i_{kl}v^k w^l =0 \, ,
\label{pp}
\end{equation}
where  $\Gamma^i_{kl}$ are the Christoffel 
symbols constructed by the reduced metric (\ref{mm}).
Projecting the geodesic deviation equation along 
the vector $w^i$ (its component along the geodesic 
field $v^i$ does not provide any physical information 
about the system instability), the corresponding 
connecting vector (tetradic) component $Z$ satisfies 
the following equivalent equation 
\begin{equation}
\frac{d^2 Z}{ds^2}=\frac{Z}{E^2} \, .
\label{qqz}
\end{equation}
This expression, as a projection on the tetradic 
basis, is a scalar one and therefore completely 
independent of the choice of the variables.
Its general solution reads as
\begin{equation}
Z\left(s\right)=c_1 e^{\frac{s}{E}}+
c_2 e^{-\frac{s}{E}} \, , \qquad c_{1,2}=\textrm{const.} \,  ,
\label{rr}
\end{equation}
and the corresponding  invariant Lyapunov exponent \cite{LYAP07}
is defined as \cite{PE77} 
\begin{equation}
\lambda_v =\sup \lim_{s\rightarrow \infty} 
\frac{\ln\left(Z^2+ \left(\frac{dZ}{ds} \right)^2\right)}{2s} \, ,
\label{ssz}
\end{equation}
which, in terms of (\ref{rr}),  takes the value 
\cite{ImponenteMontani2001PRD}
\begin{equation}
\lambda_v =\frac{1}{E} > 0 \, .
\label{tt}
\end{equation}

The limit (\ref{ssz}) is well defined as soon as the 
curvilinear coordinate $s$ approaches $\infty$.
In fact, from (\ref{ll}) we see that  the singularity 
corresponds to the limit  $\Gamma\to\infty$, and this implies $s\to\infty$.


When the point-Universe bounces against the 
potential walls, it is reflected from a geodesic 
to another one, thus making each of them unstable. 
Though with  the limit of our potential wall 
approximation, this result shows
 that, independently of the choice of 
the temporal gauge, the Mixmaster dynamics is 
isomorphic to a well described chaotic system. 
Equivalently, in terms of the BKL representation, 
the free geodesic motion corresponds to the 
evolution during a Kasner epoch and the bounces 
against the potential walls to the transition 
between two of them. By itself, the positive 
Lyapunov number (\ref{tt}) is not enough to 
ensure the system chaoticity, since its derivation 
remains valid for any Bianchi type model; the 
crucial point is that for the Mixmaster (types 
VIII and IX) the potential walls reduce the 
configuration space to a compact region ($\Pi_Q$), 
ensuring that the positivity of $\lambda_v$ 
implies a real chaotic behaviour (i.e. the geodesic 
motion fills the entire configuration space).

Furthermore, it can be shown that the Mixmaster asymptotic 
dynamics and the structure of the potential walls fulfill 
the hypotheses at the basis of the Wojtkowsky 
theorem \cite{Woj,BeniniMontani2004PRD}; this result ensures that 
the largest Lyapunov exponent has a positive sign almost 
everywhere\footnote{In \reffcite{Motter2003PRL} it is shown 
that, given a dynamical system of the form
$	d{\bf x}/dt ={\bf F}(x)$,
the positiveness of the associated Lyapunov exponents are 
invariant under the following diffeomorphism: 
${\bf y}=\phi({\bf x},t), d\tau=\lambda({\bf x},t)dt$, as soon 
as several requirements hold, fulfilled bu the  
Mixmaster \cite{BeniniMontani2004PRD}}.

Generalizing,  for any 
choice of the time variable, we are able to 
give  a stochastic representation of 
the Mixmaster model, provided the factorized 
coordinate transformation in the configuration space
\bseq
\begin{eqnarray}
&\alpha &= -e^{\Gamma\left(\tau\right)} 
a\left(\theta , \xi\right) \\
&\beta_+ &=~e^{\Gamma\left(\tau\right)} 
b_+ \left(\theta , \xi\right) \\
&\beta_- &=~e^{\Gamma\left(\tau\right)} 
b_- \left(\theta , \xi\right) \, , 
\label{uu}
\end{eqnarray}
\eseq
where   $\Gamma,a,b_{\pm}$ denote generic functional 
forms of the variables $\tau, \theta, \xi$.

It is worth noting that this analysis
relies on the use of a standard ADM 
reduction of the variational principle 
(which reduces the system by one degree 
of freedom) and overall because, adopting 
MCl variables, the 
asymptotic potential walls are fixed 
in time. 
This effectively is  the difference between the above  
approach and the one presented in 
\reffcite{SL90,SZ93} (see also for a critical 
analysis \reffcite{BurdTavakol1993PRD}), though in those works 
the problem of the Mixmaster 
chaos covariance is faced even with respect to the choice 
of generic configuration variables.

\subsubsectionric{On the Occurrence of Fractal Boundaries\label{sec:fractal}} 

In order to give an in\-va\-riant cha\-rac\-te\-ri\-zation
of the dy\-na\-mics chaoticity,   many methods along the years 
have been proposed, but not all 
approaches have  reached an undoubtable
consensus. A very interesting one, relying on 
techniques considering fractality of the basin 
of initial conditions  evolution has been proposed
in 1997 in \reffcite{CornishLevin1997PRL} and  opened a whole 
line of debate. The conflict among different 
approaches has been tackled by using an 
observer-independent fractal method, 
though leaving some questions open about the conjectures
lying at the basis of it. \\
The asymptotic behaviour towards the initial 
singularity of a Bianchi type IX trajectory depends
on whether or not we have a rational or irrational 
initial condition for the parameter $u$ in the BKL map.
In such a scheme, the effect 
of the Gauss map has been considered together with the evolution of the 
equations of motion, in order to ``uncover'' dynamical 
properties about the possible outcoming configurations
with the varying of the corresponding initial  
conditions. \\   
Nevertheless, such approach led to some doubts
regarding the reliability of the method itself.
In fact, let us observe that 
rational numbers initial conditions are dense and yet 
constitute a set of zero  measure and
moreover correspond to 
\textit{fictitious} singularities 
\cite{BKL1970,Misner1969PRL,Misner1969PR}.
The nature of this initial set needs to be compared
with the one regarding the \textit{complete} set of initial 
conditions, with finite measure over a finite interval:  
the conclusions obtained after the dynamical evolution 
are not necessarily complementary between the two 
initial assumptions.

\begin{figure}[ht]
	\begin{center}
		\includegraphics[width=7cm]{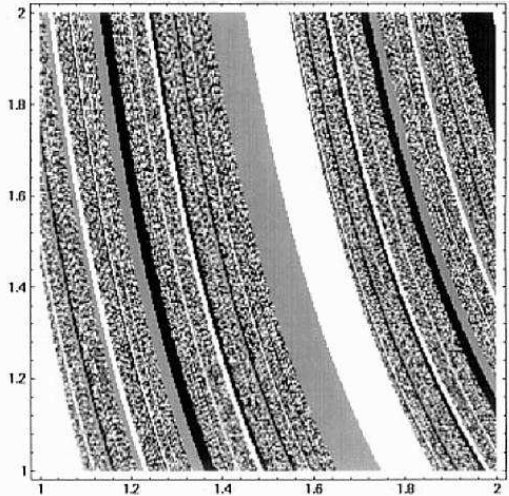}
		\caption[Fractal structure in Mixmaster dynamics]{The numerically 
		generated basin boundaries in  the $(u,\;v)$ plane are built of 
		Universes which ride the repellor for many orbits before being 
		thrown off. 
		Similar fractal basins can be found by viewing alternative 
		slices through the phase space, such as the 
		$(\beta,\;\dot{\beta})$ plane. 
		The overall morphology of the basins is altered little 
		by demanding more strongly anisotropic outcomes. (From \reffcite{CornishLevin1997PRL})}
		\label{c2 CL repellor}
	\end{center}
\end{figure}

In \reffcite{CornishLevin1997PRD,CornishLevin1997PRL}, 
Cornish and Levin used a coordinate-independent fractal method
to show that the Mixmaster Universe is indeed chaotic.  
By exploiting techniques originally developed to study chaotic scattering, 
they gained a new perspective on the evolution of the Mixmaster cosmology, 
finding  a fractal structure, namely {\it the strange repellor} 
(see Fig.\ref{c2 CL repellor}), that well describes  chaos. 
A strange repellor is the collection of all Universes periodic in $(u,\;v)$,
and an aperiodic one  will tipically experience a transient age of chaos 
if it brushes against the repellor. 
The fractal pattern was exposed in both the exact Einstein equations and in the 
discrete map used to approximate the solution. 
The most important feature gained is a fractal approach independent of 
the adopted time coordinate and the chaos reflected in the fractal weave 
of Mixmaster Universes is unambiguous.  

The approach used in \reffcite{CornishLevin1997PRL,CornishLevin1997PRD} is based on the method 
firstly stated in \reffcite{BGOB88} where it is shown how 
fractal boundaries can occur for some solutions involving 
chaotic systems. The space of initial conditions is 
spanned giving rise to different exit behaviours whose
borders have fractal properties: this constitutes a 
\textit{conjecture} as a typical property of chaotic 
Hamiltonian dynamics with multiple exit modes. \\
For the case of the Bianchi IX model potential 
(see Figure \ref{fig:B9}) the openings are obtained 
widening the three corners, on the basis that the 
point representing the evolution spends much of the 
time there nearby. \\
This method has three essential fallacies:

\begin{enumerate}

	\item  the case-points chosen as representatives 
				within this framework are the ones whose dynamics
				proceeds never reaching the singularity;
				
	\item  the ``most frequent'' dynamical evolution is the 
				one in which the point enters the corner with the 
				velocity \textit{not parallelly} oriented towards 
				the corner's bisecting line and, after some 
				oscillations, it is sent back in the middle of the 
				potential. This effect is altered when opening the potential 
				corners;
				
	\item the artificial opening up of the potential corners 
				adopted in the basin boundary approach could be
				creating the fractal nature of it.	
	
\end{enumerate}

In particular, the third observation is supported by the 
existence of strange attractors that \textit{are not} 
chaotic, as counter-exampled by \reffcite{GOPJ84} 
and discussed by 
\reffcite{HH94}. The choice of the method adopted 
to characterize the property of chaos or its absence
is very relevant, especially when based on  the 
presence of fractal boundaries in the dynamics underlying 
Bianchi IX models. 
  This is important to be checked,
first of all, because the result of
  \reffcite{CornishLevin1997PRL,CornishLevin1997PRD}  relies on the conjecture as in \reffcite{BGOB88}
  that opening gates in a chaotic Hamiltonian system
  can result in the presence of fractal basin boundaries
  (which needs in principle to be checked in the case of 
  Bianchi IX), not satisfying the necessity of 
  a general statement concerning chaos: 
  even the opening of the corners does not solve 
  the question about  what happens when taking the 
  limit of closing them and if there is 
  an universal behaviour (for general systems).\\
 Secondly, it is needed to 
%
%
  integrate the Bianchi IX flow and this operation  
  is not necessarily commuting with the statement 
  regarding the remaining (and equally relevant) part 
  of the set of initial conditions constituted by the 
  irrational 
  numbers, which needs to be checked.

Motter and Letelier, in \reffcite{MotterLetelier2001PLA} 
in the criticism
to the paper of  \reffcite{CornishLevin1997PRL,CornishLevin1997PRD}, 
 claim the same results with more accurate comprehension
of the global chaotic transient and afford calculations
involving a more stable constraint check and a higher
order integrator.
Again  the same criteria used by \reffcite{CornishLevin1997PRL,CornishLevin1997PRD} is followed
to get the same results.
The informations obtained
following the Farey map approach are not relevant
(only rational values of $u$ are led to the
three peculiar outcomes) because the corresponding
invariant set contains almost every point of phase
space. But they claim it is possible
to get strict indications of chaos with the
Hamiltonian exit method \cite{BGOB88,STN02,MG02,MG01,BOG89}. 

Firstly one has to fix the width 
$(\longleftrightarrow u_{exit})$
of the open corners, then let the
system evolve.
The future invariant set leads to
a box-counting dimension $D_0$ (estimated from the
uncertainty exponent method \cite{Ott})
coherent with previous results, which is, by
construction, a function of the width itself.
The value of $D_0$ found, equal numerically to 
$1.87$, is dependent
on the change done to the original potential,
and converges to the value of 2, which is an
indicator of {\it non}-chaoticity \cite{OAST02}.
Any of such fundamental properties, if outlined in a
specific case, must be jointed through a limit
procedure to the general one.

Pianigiani and Yorke in \reffcite{PY79} 
study the evolution of a ball
on a billiard table with smooth obstacles so that all
trajectories are unstable with respect to initial data.
This is a system energy conserving and then they open a small
hole on such table in order to allow the ball to go through.
Such two differences have not been taken account of in the 
Bianchi IX analysis.

In the work by \reffcite{STN02} Schneider et al. it is supposed
to show the existence of a chaotic saddle, whose
signature is the chaotic basin.

The paper by \reffcite{OAST02} 
declares
the absence of such points, in a model with $\Lambda=0$, 
hence we infer the inapplicability of that method to
discover a supposed unknown feature of a dynamical system.\\
They stress too that the limit (not unnatural)
for $\Lambda\rightarrow 0$ doesn't matches: it
doesn't permit to characterize the chaos in mixmaster
vacuum model: a continuous change in a parameter of
the theory heavily affects the method's applicability, 
mainly while the study of Bianchi IX dynamics is of interest
towards the initial singularity, where the BKL
approach applies: in such approximation, the domain
walls close to a circle. \\
Hence there are objections which are subject for 
interesting further investigation

\begin{enumerate}

\renewcommand{\labelenumi}{({\it\roman{enumi}} )} 

	\item has the opening of a polygonal domain the same
				effect as the opening a circle (which has curvature)?
	
	\item Is the system truly independent from temporal
				reparameterization?			
				
	\item Is the opening independent of temporal
(either spatial) reparameterization?			

 \item  Could exist a temporal reparameterization
			whose effect is to close the artificial openings?

 \item How to interpret this eventuality?

\end{enumerate}
Even if 
this is not relevant for dynamical system in
classical mechanics, in General Relativity it would; 
for example, Cornish and Levin claim that they open a non compact domain, 
differently from Misner.

For all the criticism here outlined we consider
an analytical approach crucial to distinguish 
among chaos indicators relying on numerical 
properties not well-manageable via numerical 
simulations.

\subsectionric{Isotropization Mechanisms\label{sec:isomech}}

The isotropic FRW model can accurately describe the evolution 
of the Universe until the decoupling time, i.e. $10^{-3}-10^{-2}$ 
seconds after the Big-Bang\cite{KT90}. 
On the other hand, the description of its very early stages requires 
more general models, like at least the homogeneous ones. 
Therefore we are interested to investigate  some mechanisms  allowing 
a transition  between these two cosmological epochs. 
When the anisotropy of the Universe is sufficiently suppressed, 
we can speak of a quasi-isotropization of the model \cite{Star83,Coleman,CW73,CH73} (for a detailed discussion of the isotropization mechanism see \reffcite{1971SvA....14..763D,1971ZhETF..60.1201D,1980A&A....87..236S,1985A&A...151....7M,1987A&A...179...11F,1993Ap&SS.199..289G,1993Ap&SS.203..169B,1995PhRvD..51..928I,1996IJTP...35..419G,1997PhLB..408...47C,1998CQGra..15..331W,1999PhRvD..59d3501C,1999PhRvD..59j7302V,2001GReGr..33..767C,2004CQGra..21.1609F,2006A&A...460..393J,2006MNRAS.369..598C,2006MNRAS.369.1858M,2007MNRAS.380.1387P}) and such 
mechanism can  be regarded as a ``bridge`` between the two stages. 

In this paragraph we will discuss the origin of a background 
space\footnote{The chaotic nature of the evolution toward the 
singularity implies that the geometry, and therefore  all the 
geometric quantities, should be described in an average sense only. 
With this respect, the Universe does not possess a stable background 
near the singularity \cite{Kir93,KirillovMontani1997JL}.} 
when a real self-interacting scalar field $\phi$ is taken into account, 
following the work \reffcite{KirillovMontani2002PRD}. 
Let us extend the Misner-like variables \cite{KirillovMelnikov1995} (see (\ref{parq}))
\be
q^a=A^a_j\beta^j+\alpha, \qquad \beta^3=\f \phi {\sqrt 3},
\ee
where $j=1,2$, $a=1,2,3$ and $A^a_j$ satisfy
\be
\sum_a A^a_j=0, \qquad \sum_a A^a_jA^a_k=6\delta_{jk} \, .
\ee
As usual, $\alpha$ parametrizes the isotropic change of the metric 
with the singularity appearing as $\alpha\rightarrow-\infty$ and 
the $\beta^j$  the anisotropies of the model. 
When these variables approach some constants as a limit
we can speak of a quasi-isotropization of the model (for a discussion on the quasi-isotropic solution see \reffcite{LK1960,KHA02,KHA03,IM03}).

The action expressed in terms of them writes as 
\be\label{azionekimo}%
S=\int dt\left[P_r\p_t\beta^r+P_\alpha\p_t\alpha 
- \f N {6\sqrt h}\left(\sum_rP_r^2+6U-P_\alpha^2\right)\right]\, , 
\ee
 where $r=1,2,3$,  $h=\exp(3\alpha)$ and the potential term 
(\ref{n2}) is now 
defined as $U=h(W-   {~}^{(3)}R)$ with 
$W(\phi)=\nicefrac{1}{2}[h^{ij}\p_i\phi\p_j\phi+V(\phi)]$. 
From the  action (\ref{azionekimo}) 
an inflationary solution comes out imposing the constraint 
\be
\f1 h U\simeq V(\phi)\simeq \textrm{const.}\gg {~}^{(3)}R
\ee
which can be realized by an appropriate process of spontaneous symmetry 
breaking, exhaustively studied in \reffcite{KirillovMontani2002PRD,Montani2000CQGultrarelativistic}. 
Let us consider the situation where $U=h\Lambda$, where $\Lambda=\textrm{const}$. \\
The Hamilton-Jacobi equation is 
\be  
\sum_r\left(\f{\delta S}{\delta\beta^r}\right)^2
-\left(\f{\delta S}{\delta\alpha}\right)^2+6\exp(3\alpha)\Lambda=0 \, ,
\ee
whose solution can be expressed as
\be
S(\beta^r,\alpha)\sim K_r\beta^r+\f 23K_\alpha+\f K 3
\ln \left| \f{K_\alpha-K}{K_\alpha+K}\right|\, ,
\ee
where $K_\alpha(K_r,\alpha)=\pm\sqrt{\sum_rK_r^2+6\Lambda\exp(3\alpha)}$,  
with some generic constants $K=\sqrt{\sum_rK_r^2}$ and $K_r$. 
The equation of motion for $\alpha$ is readily  obtained  as
from (\ref{azionekimo})
\be
\f{\p\alpha}{\p t}=-\f{NP_\alpha}{3\exp(3\alpha/2)} \, .
\ee
Choosing  $\alpha$ as the  time coordinate, i.e. $\p_t\alpha=1$, 
the time gauge condition becomes $N=-3\exp(3\alpha/2)/P_\alpha$.
Since the lapse function is positive defined  we must also have $P_\alpha<0$. 

Accordingly to the Hamilton-Jacobi method, one has firstly to 
differentiate with respect to $K^r$ and then, equating 
the results  to arbitrary constants, one  finds the solutions 
describing the trajectories of the system as
\be\label{solbeta}%
\f{\delta S}{\delta K_r}=\beta_0^r \quad \Rightarrow \quad \beta^r(\alpha)=\beta^r_0+\f{K_r}{3|K|}\ln\left|\f{K_\alpha-K}{K_\alpha+K}\right| \, ,
\ee  
where $\beta^r_0$ are new arbitrary constants. 
Let us investigate the two limits of interest. 
First of all, let us note  that for $h\rightarrow\infty$ 
(i.e. $\alpha\rightarrow\infty$, $K_\alpha\rightarrow\infty$) 
the solution (\ref{solbeta}) transforms into the inflationary one 
obtained in \reffcite{Star83}, corresponding to the quasi-isotropization 
of the model as the  functions $\beta^r$ approach 
the constants $\beta^r_0$. 
On the opposite limit, i.e. for $h\rightarrow0$ 
($\alpha\rightarrow-\infty$, $K_\alpha\rightarrow K$), 
the solution (\ref{solbeta}) provides the generalized Kasner one 
as expected, simply modified by the presence of the scalar field \cite{BK73}
\be
\beta^r(\alpha)=\beta^r_0-\f{K_r}K(\alpha-\alpha_0),
\ee 
$\alpha_0$ being the remaining constants.

The existence of the solution (\ref{solbeta}) shows how the inflationary 
scenario\cite{G81,GP82,Linde1,ST87} can provide the necessary dynamical ``bridge'' between the fully 
anisotropic and the quasi-isotropic epochs of the Universe evolution. 
In fact, during that time the anisotropies $\beta_\pm$ 
are dumped  and the only effective dynamical variable is $\alpha$, 
i.e. the one related to the isotropic volume of the Universe. 
This shows how the dominant term during the inflation is 
$\Lambda e^{3\alpha}$ and any  term in the spatial curvature
becomes more and more negligible
although increasing like (at most) $e^{2\alpha}$.

For a sample of works dealing with the Bianchi models dynamics involving a cosmological term, see \reffcite{1982GReGr..14..751S,1987JMaPh..28.1658W,1997Ap&SS.253..205C,2002CQGra..19.1013C,2005Ap&SS.298..419P,2006Ap&SS.301..127P,2006PhyD..219..168S,2007Ap&SS.tmp..329C,Podolsky:2007vu}.

\subsectionric{Cosmological Implementation of the Bianchi Models\label{sec:cosmob}}

In this paragraph we shall examine the cosmological 
issues of the Bianchi models. 
In particular, we will focus attention on the 
question about the corresponding 
theoretical predictions regarding  the relic
Cosmological Background Radiation (CMB) 
anisotropy \cite{deBernardis2000Na,MAXIMA}. 
Confirmed observations \cite{DASI1,DASI2,Spergel:2003cb,LU03,BOOM} show that the large-scale relic 
radiation anisotropy $\Delta T/T$ is about 
\be
\Delta T/T\leq 2\cdot10^{-5} \, ,
\ee
in terms of the equivalent black body radiation temperature $T$.
Therefore the Universe becomes transparent to the  relic radiation 
at an epoch when the expansion anisotropy is ``small''. 
A comparison between the theory and the experimental observations
requires the analysis of a quasi-isotropic stage of the Bianchi models.

The discussion which follows is mainly based on the 
textbook \cite{Zel-Nov} and references therein (see also \reffcite{STA79}). 
Initially we approach the Bianchi I model and then 
we focus attention on the more general Bianchi IX model.

\subsubsectionric{Expected Anisotropy of CMB for Bianchi I\label{sec:ani1}}

Let us analyze the Bianchi I model,  characterized by 
a flat, comoving three-dimensional space. 

We will include in the dynamics an ordered magnetic field having oriented fluxes
of relativistic particles since they inevitably develop as a consequence of the 
processes during the early stages of anisotropic expansion. 
Let the magnetic field be oriented along the $z$-axis, 
$W$ be its energy density and $p_x, p_y, p_z$ 
components of the pressure of the free particles along the corresponding axes. 
Let us also  assume $\epsilon\gg W$ and  $\epsilon\gg p_x, p_y, p_z$ 
and moreover  $a^{-1}da/dt\sim b^{-1}db/dt\sim c^{-1}dc/dt$, 
$\epsilon$ being the energy density of  ordinary matter. 
Such relations ensure that, to first approximation, 
the expansion isotropically takes place  and thus  the equations of motion 
read as
\bseq
\begin{align}
\f1{abc}\f d{dt}\left[\left(a^{-1}\f{da}{dt}-b^{-1}\f{db}{dt}\right)abc\right] &=\kappa (p_x-p_y) \\
\f1{abc}\f d{dt}\left[\left(a^{-1}\f{da}{dt}-c^{-1}\f{dc}{dt}\right)abc\right] &=\kappa [(p_x-p_z)+2W] \,.
\end{align}
\label{equapippi}
\eseq
Suppose that $P$ is the pressure associated to $\epsilon$ and 
consider the particular case of ultra-relativistic matter 
$P=\epsilon/3$ (corresponding to $a\sim b\sim c\sim t^{1/2}$). 
In this case, we can derive the behavior of the deformation anisotropies
(defined as the left-hand sides of the following  equations (\ref{eqddd})) 
when the expansion is almost periodic and (\ref{equapippi}) rewrite as
\bseq
\label{eqddd}
\begin{align}\label{eqdefani}
\f{a^{-1}da/dt-b^{-1}db/dt}{a^{-1}da/dt} &=\f{3(p_x-p_y)}\epsilon+2\left(\f{t^\star}t\right)^{1/2} \\
\label{eqdefani1}
\f{a^{-1}da/dt-c^{-1}dc/dt}{a^{-1}da/dt} &=\f{3(p_x-p_z)}\epsilon
	+\f{6W}\epsilon+2\left(\f{t^\star}t\right)^{1/2} \,,
\end{align}
\eseq
$t^\star$ being a constant. 
In  absence of matter the anisotropy decays as $t^{-1/2}$. 
On the other hand, when the expansion is almost periodic, 
the ratios $(p_x-p_y)/\epsilon$ as well as $W/\epsilon$ remain constant. 
Therefore, in  presence of a magnetic field or of a flux of relativistic 
particles, the deformation anisotropy is conserved during the era 
when $P=\epsilon/3$. 
In conclusion, the anisotropy of the stress-energy 
tensor slows down the isotropization of solution. 
This statement can be generalized for all (reasonable) kinds of matter
and  in particular for the case $P=0$.

Let us discuss the equation for the  anisotropy of the relic radiation 
temperature and how it is related to the expansion anisotropy, 
and consider an observer receiving the radiation from the direction 
of the two axes with scale factors $a$ and $b$. 
The observed difference of  temperature in those two directions is
\be
\left. \f{\Delta T}T\right|_{t\rightarrow\infty}=\left. \f{a-b}a\right|_{z_1} \, ,
\ee
where $z_1$ is the redshift associated to the time when the Universe 
is transparent to the radiation and the radiation field is 
isotropic\footnote{After this time the radiation freely propagates.}. 

If the sharply anisotropic stage ends before the time when the radiation 
density $\rho_r$ equates the baryon one $\rho_m$, then the deformation 
anisotropy (\ref{eqdefani})  
is conserved. 
After this stage, in the $P=0$ epoch  the anisotropy decreases  according to
\be
\f{a-b}a=\f{6\rho_{anis}}{\rho_m}\sim t^{-2/3} \, ,
\ee
$\rho_{anis}$ being the density of the anisotropic neutrino flux 
($\rho_{anis}/\rho_r\sim0.1-1$). 
Assuming the transparency redshift as $z_1=8$, 
one obtains
\be\label{dtsut}
\left. \f{\Delta T}T\right|_{t\rightarrow\infty}=
\left. \f{a-b}a\right|_{z_1=8}
=\left. \f{6\rho_{anis}}{\rho_m}\right|_{z_1=8}\simeq 10^{-4}-10^{-3}.
\ee
Hence,  expression (\ref{dtsut}) gives the expected value for 
the relic radiation anisotropy although it is valid only in 
the case $\rho_m=\rho_{crit}$  (flat space).

\subsubsectionric{Expected Anisotropy of CMB for Bianchi IX\label{sec:ani9}}

In the homogeneous cosmological models with curved comoving space, 
as Bianchi IX is, the deformation anisotropy behaves in a way similar 
to the one described above. 
In fact, it is constant during the radiation dominated era up to the 
time when $\rho_r=\rho_m$. 
Nevertheless, the difference relies in the role played by the non-interacting 
particles in Bianchi I, which actually is taken by the spatial curvature. 
Then, after this epoch, the anisotropy decreases as $t^{-2/3}$, while 
during the time when  $\rho_r=\rho_m$  the matter becomes transparent 
to the background radiation and the photons freely propagate.

Therefore, let us underline two main aspects of the relic radiation 
anisotropy in the anisotropic models. 
The first feature is related to the conservation of the deformation 
anisotropy during the radiation dominated era, and it is responsible 
for the amplitude of $\Delta T/T$. 
The second characteristic is connected to the curvature of the 
comoving space and the consequent motion of the photons which gives rise 
to the different angular distributions on the sky of the background anisotropy. 

The anisotropy of the relic radiation temperature  
can now be calculated with the 
same arguments given above for Bianchi I and it depends on three 
characteristic temporal steps: 
\begin{itemize}
\item[i)] $t_\phi$, the time of the beginning of the isotropic stage, 
when $a\simeq b$; 
\item[ii)] $t_c$, when the equation of the state changes, i.e. when it effectively 
becomes $P=0$ and after this time all the deviations from exact isotropy 
decay  as a power law; 
\item[iii)] $t_e$, when matter becomes transparent. 
\end{itemize}
Explicitly, the anisotropy of the temperature background reads as \cite{Zel-Nov}
\be
\label{tempphi}
\f{\Delta T}T\simeq\f8{\ln[(t_c/t_\phi)e^8]}\left(\f{t_c}{t_e}\right)^{2/3} \, .
\ee
It is worth noting that the above formula (\ref{tempphi}) is alarmingly unstable to 
small numerical changes in underlying parameters. This feature is widely discussed in \reffcite{Barrow:1997sy} where an analysis of the cosmological evolution of matter sources with small anisotropic pressures is performed (see also \reffcite{Barrow:1997mj} for applications to the magnetic case and \reffcite{Barrow:1995fn} in which a more general analysis is shown). Furthermore the dependence of (\ref{tempphi}) on the 
isotropization time $t_\phi$ is extremely weak. 
In fact, taking $t_c\simeq t_e$ and assuming 
$\Delta T/T\sim\mathcal O(10^{-3})$,  one has $t_\phi\sim\mathcal O(t_P)$, 
where $t_P\sim\mathcal O(10^{-44}s)$ is the Planck time. 
Therefore isotropization would take place in the region when 
the applicability of classical General Relativity is  expected to fail. 
On the other hand, $\Delta T/T$ is of order of the values experimentally observed 
only if the factor $(t_c/t_e)^{2/3}$ is small, i.e. if 
the matter becomes transparent much later than the end of the radiation 
dominated epoch ($t_e\gg t_c$) and this is possible only if the amount 
of ionized intergalactic gas is so relevant that the critical parameter 
is $\Omega\sim\mathcal O(1)$.

With respect to the angular distribution, we stress that the light 
propagates along the principal direction of the deformation tensor 
and therefore the anisotropy has a quadrupole character.

As last point, we have to stress that in above analysis no mention to inflation is made. Indeed it would a major damping effect on any primordial anisotropy, massively reducing it below (\ref{tempphi}) though. About this point we refer to \reffcite{Barrow:1997sy,Barrow:1997mj,Barrow:1995fn}.

\subsectionric{The Role of a Scalar Field\label{sec:scal}}

In this Section we face the influence of a scalar field 
when  approaching the cosmological singularity. 
As shown in the initial studies \cite{BK72,BK73},
such field can suppress the Mixmaster oscillations during the 
evolution toward the singularity. \\
Following  the approach given in \reffcite{Berger1999PRD}, let us consider the 
Mixmaster Universe in the presence of a self-interacting scalar 
field $\phi$. 
As we have seen  previously, the Einstein equations are obtained
from the variation of the Hamiltonian constraint $H=0$, 
where ($N \propto e^{3\alpha}$)
\be
H=H_K+H_V \, ,
\ee
$H_K$ and $H_V$ being the kinematic and potential part of 
the Hamiltonian, respectively. In particular we have
\be
H_K=-p_\alpha^2+p_+^2+p_-^2+p_\phi^2 \, ,
\ee
and
\begin{multline}\label{pot}
H_V=e^{4\alpha}\left[e^{-8\beta_+}+e^{4\beta_++4\sqrt 3\beta_-}+e^{4\beta_+-4\sqrt 3\beta_-}\right.+\\-\left.
2\left(e^{4\beta_+}+e^{-2\beta_+-2\sqrt 3\beta_-}+e^{-2\beta_++2\sqrt 3\beta_-}\right)\right]+e^{6\alpha}V(\phi),
\end{multline}
where with $V(\phi)$ we denote a generic potential of the scalar field. 
Working with such Misner variables  the cosmological singularity appears 
as  $\alpha\rightarrow-\infty$. 
Therefore, unless $V(\phi)$ contains terms exponentially growing with $\alpha$, 
the very last term in (\ref{pot}) can be neglected at early times, 
i.e. $e^{6\alpha}V(\phi)\rightarrow0$ as $\alpha\rightarrow-\infty$.

Let us consider the kinematic part $H_K$. 
Its variation yields equations whose solution reads as
\be\label{betaphi}
\begin{cases}
\beta_\pm=\beta_\pm^0+v_\pm|\alpha| \cr \phi=\phi^0+v_\phi|\alpha|,
\end{cases}
\ee
where $v_\pm=p_\pm/|p_\alpha|$ and $v_\phi=p_\phi/|p_\alpha|$. 
Therefore the constraint $H_K=0$ becomes
\be\label{phi=0}
v_+^2+v_-^2+v_\phi^2=1.
\ee
Let us firstly analyze the case without the scalar field, i.e. 
$\phi\equiv 0$. 
Introducing  polar coordinates in the anisotropy planes as 
$v_+=\cos\theta$ and $v_-=\sin\theta$, 
through equations (\ref{betaphi})  the potential (\ref{pot}) rewritee as
\be\label{pottheta}
H_V\sim e^{-4|\alpha|(1+2\cos\theta)}+ 
e^{-4|\alpha|(1-\cos\theta-\sqrt 3\sin\theta)}
+e^{-4|\alpha|(1-\cos\theta+\sqrt 3\sin\theta)},
\ee 
where we maintained the dominant terms only, i.e. the first three 
terms in (\ref{pot}). 
Except for the set of zero measure  of values 
$\theta=(0,2\pi/3,4\pi/3)$, any generic value of $\theta$ will cause 
the growth of one of the terms on the r.h.s. of (\ref{pottheta}).

Let us consider the case $\phi\neq0$ and hence $v_\phi^2>0$. 
Equation (\ref{phi=0}) is replaced by
\be
v_+^2+v_-^2=1-v_\phi^2<1 \, ,
\ee
thus none of the terms in (\ref{pottheta}) will grow if the following conditions
are satisfied
\be\label{vxx}%
\begin{cases}
1+2v_+>0 \cr 1-v_+-\sqrt 3v_->0 \cr 1-v_++\sqrt 3v_->0 \, ,
\end{cases}
\ee 
situation which is realized if $v_+^2<1/2$ and $v_-^2<1/12$, 
which occur if $2/3<v_\phi^2<1$. 
In \reffcite{BGS} it is described  how $p_\alpha$ decreases at each bounce, 
and therefore for any initial value of $p_\phi$ it is always  $v_\phi^2>2/3$.

As we have seen, the approach to the singularity of the Bianchi IX model
 is described by a particle moving in a potential with exponentially
closed  walls bounding a triangular domain. 
During the evolution, the particle bounces against the walls providing  
an infinite number of oscillations toward the singularity. 
The scalar field influences such dynamics so that for values 
of $v_\pm$ satisfying (\ref{vxx}), there are not further bounces 
and the solution approaches (\ref{betaphi}). 
In other words there will be an instant of time 
after which the point-Universe 
does not ever reach the potential walls 
and no more oscillations appears. 
In this sense the scalar field can suppress the chaotic Mixmaster dynamics 
toward the classical cosmological singularity.

\subsectionric{Multidimensional Homogeneous Universes\label{sec:multidh}}

When the number of spatial dimensions is greater than three, 
homogeneous models can loose their chaotic dynamics, as already 
arises in the four-dimensional case.\\
The question of chaos in higher dimensional cosmologies has been 
widely investigated, and many 
authors  \cite{1985PhRvD..32.1595B,Furusawa1985PTP,Halpern2003GRG,Halpern2002PRD} 
showed that none of higher-dimensional extensions of the Bianchi IX model 
possesses proper chaotic features and the crucial 
difference is given by the finite number of oscillations characterizing the 
dynamics near the singularity. 
Without loss of generality we will follow the analysis proposed 
by Halpern\cite{Halpern2003GRG,Halpern2002PRD} and limit our discussion 
to the case of a homogeneous model with four spatial dimensions.

The work of Fee\cite{Fee} classifies the four-dimensional homogeneous 
spaces in 15 types, named  $G0 - G14$, and   it is based on the analysis 
of the corresponding  Lie groups.
The line element  can be written using the Cartan 
basis of left-invariant forms and explicitly reads as ($N=1$)
\begin{equation}
\label{c1 elemento di linea 5 dim}
ds^{2} = dt^{2} - 
{~}^4\eta_{rs}(t)\omega^r \otimes \omega^s \, . 
\end{equation}
The 1-forms $\omega^{r}$ obey the relation 
$d \omega^{r} =\displaystyle\frac{1}{2} C^{r}_{\phantom{c}pq} \omega^{p} \wedge \omega^{q}$, 
where the $C^{r}_{\phantom{c}pq} $ are the four-dimensional  
structure constants. 
The discussion remains quite general even  limiting 
it to the case of a  diagonal matrix ${~}^4\eta_{rs}$
\begin{equation}
\label{c1 metrica diagonale}
{~}^4\eta_{rs} = \textrm{diag}(a^{2}, b^{2}, c^{2}, d^{2})\;.
\end{equation}
The Einstein equations are  obtained with the standard procedure as
\bseq
\begin{align}
R^{0}_{0} &= \displaystyle\displaystyle\frac{\ddot{a}}{a} + 
	\displaystyle\frac{\ddot{b}}{b} + \displaystyle\frac{\ddot{c}}{c} +
	\displaystyle\frac{\ddot{d}}{d} = 0\;,\label{c1 r00 5 dim}\\
R^{1}_{1} &= \displaystyle\displaystyle\frac{(\dot{a}bcd)\dot{}}{abcd} + 
	S^{1}_{\phantom{1}1} = 0\;,\label{c1 r11 5 dim}\\
R^{2}_{2} &= \displaystyle\displaystyle\frac{(a\dot{b}cd)\dot{}}{abcd} + 
	S^{2}_{\phantom{2}2} = 0\;,\label{c1 r22 5 dim}\\
R^{3}_{3} &= \displaystyle\displaystyle\frac{(ab\dot{c}d)\dot{}}{abcd} + 
	S^{3}_{\phantom{3}3} = 0\;,\label{c1 r33 5 dim}\\
R^{4}_{4} &= \displaystyle\displaystyle\frac{(abc\dot{d})\dot{}}{abcd} + 
	S^{4}_{\phantom{4}4} = 0\;,\label{c1 r44 5 dim}\\
R^{0}_{n} &= \left( \displaystyle\displaystyle\frac{\dot{x}_{n}}{x_{n}} -
	 \displaystyle\frac{\dot{x}_{m}}{x_{m}}\right) C^{m}_{\phantom{n}mn} = 0\;;\label{c1 r0n 5 dim}
\end{align}
\eseq
where $x_{n}$ $(n =1,2,3,4)$ denote the scale factors $a, b, c, d$, 
respectively, and the $S^{n}_{\phantom{n}n}$ are functions of them 
and of  the structure constants.\\
The analysis developed for the standard Bianchi type IX  can be 
straightforwardly generalized to the five-dimensional case, 
obtaining the system 
\bseq
\label{eee5}
\begin{align}
\label{c1 eq di einstein alpha beta gamma 5 dim}
\alpha_{\tau\tau} &= -\Lambda^{2} S^{1}_{\phantom{1}1}\;,  \qquad 
\beta_{\tau\tau} = -\Lambda^{2} S^{2}_{\phantom{1}2}\;,    \\  
\gamma_{\tau\tau} &= -\Lambda^{2} S^{3}_{\phantom{1}3}\;, \qquad   
\delta_{\tau\tau} = -\Lambda^{2} S^{4}_{\phantom{1}4}\;,\nonumber
\end{align}
\begin{align}
\alpha_{\tau\tau} + \beta_{\tau\tau} &+ \gamma_{\tau\tau} + \delta_{\tau\tau} = \nonumber \\
&= 2 \alpha_{\tau} \beta_{\tau} + 2 \alpha_{\tau} \gamma_{\tau} 
 + 
2 \alpha_{\tau} \delta_{\tau} + 2 \beta_{\tau} \gamma_{\tau} + 2\beta_{\tau} \delta_{\tau} 
	+ 2 \gamma_{\tau}\delta_{\tau}\;,\label{c1 integrale primo 5 dim}
\end{align}
and
\begin{equation}
\label{c1 r0n 5 dim in alpha}
R^{0}_{n} = 0\;.
\end{equation}
\eseq
The dynamical scheme (\ref{eee5})
is valid for any of the 15 models using the corresponding 
$S^{n}_{\phantom{n}n}$.

\begin{figure}[ht]
\begin{center} 
\includegraphics[width=6.5cm]{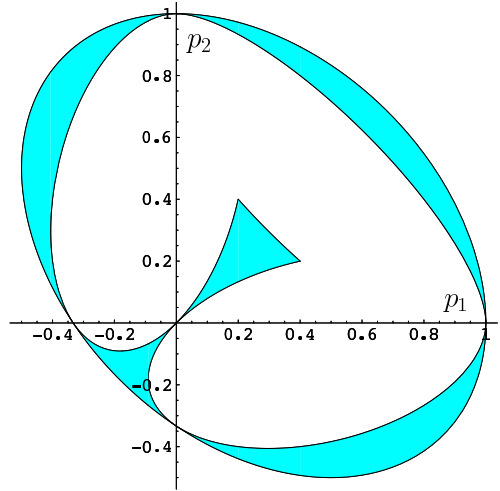} 
\end{center}
\caption[The region where the 5-d BKL mechanism breaks down]{The shaded region corresponds to  all of the couples 
$(p_{1},\,p_{2})$ not satisfying (\ref{c1 condizione di kasner 5 dim 2}): 
as soon as a sey $(p_{1},\,p_{2})$ takes values in that portion,  
the BKL mechanism breaks down  and the Universe experiences 
the last Kasner epoch till the singular point.}
\label{c1 cond 5 dim kasner}
\end{figure}

Among the five-dimensional homogeneous space-times, $
G13$ is the analogous of the Bianchi type IX, having the same set of  
structure constants.\\
The Einstein equations can be written as
\bseq
\label{c1 eq einstein mixmaster 5 dim}
\begin{align}
2 \alpha_{\tau \tau}& =  \left[(b^{2} - c^{2})^{2} - a^{4} \right] d^{2}\;,\\
2  \beta_{\tau \tau}& = \left[(a^{2} - c^{2})^{2} - b^{4} \right] d^{2}\;,\\
2 \gamma_{\tau \tau}& = \left[(b^{2} - a^{2})^{2} - c^{4} \right] d^{2}\;,\\
  \delta_{\tau \tau}& = 0\;,
\end{align}
\eseq
togheter with (\ref{c1 integrale primo 5 dim}).
If we assume that the BKL approximation is valid, 
{ i.e.}  that the right-hand sides of equations 
(\ref{c1 eq einstein mixmaster 5 dim}) are negligible, 
then the asymptotic solution for $\tau\to  -\infty$                
is the five-dimensional and Kasner-like line element
\begin{equation}
ds^{2}  = dt^{2}-\sum_{r=1}^{4} t^{2 p_{r}} (dx^{r})^{2}\;,
\end{equation}
with the Kasner exponents $p_{r}$ satisfying the generalized Kasner relations
\begin{equation}
\label{c1 vincoli di kasner 5 dim}
\sum_{r=1}^{4} p_{r} = \sum_{r=1}^{4} p_{r}^{2} = 1\;.
\end{equation}
This regime can only hold  until the BKL approximation works; 
however, as $\tau$ approaches the singularity, 
one or more of the terms may increase. 
Let us assume $p_{1}$ as the smallest index; 
then $a = \exp(\alpha)$ is the largest contribution and we can 
neglect all other terms, obtaining
\begin{eqnarray}
\alpha_{\tau \tau}  &=& -\displaystyle\displaystyle\frac{1}{2} \exp (4\alpha + 2\delta)\nonumber\;,\\
\beta_{\tau \tau}   &=& \gamma_{\tau \tau} = 
\displaystyle\displaystyle\frac{1}{2} \exp (4\alpha + 2\delta)\label{c1 eq mixmaster 5 dim con potenziale}\;,\\
\delta_{\tau \tau}  &=& 0\nonumber\;.
\end{eqnarray}
As soon as the asymptotic limits for $\tau\to\pm\infty$ are considered, 
from the  solution to (\ref{c1 eq mixmaster 5 dim con potenziale}) we 
obtain the map  
\begin{equation}
p_{1}' = -\displaystyle\displaystyle\frac{p_{1}+p_{4}}{1+2p_{1}+p_{4}}\;,\hspace{0.5cm} p_{2}' = \displaystyle\displaystyle\frac{p_{2}+2p_{1}+p_{4}}{1+2p_{1}+p_{4}}\;,\nonumber
\end{equation}
\begin{equation}
p_{3}' = \displaystyle\displaystyle\frac{p_{3}+2p_{1}+p_{4}}{1+2p_{1}+p_{4}}\;,\hspace{0.5cm} p_{4}' = \displaystyle\displaystyle\frac{p_{4}}{1+2p_{1}+p_{4}}\label{c1 mappa bkl 5 dim}\;;
\end{equation}
\begin{equation}
abcd =\Lambda' t\;,\hspace{0.5cm} \Lambda' = (1+2p_{1}+p_{4}) \Lambda  \;.
\end{equation}
The difference of  this dynamical scheme with the four-dimensional 
case relies in  the conditions needed to undergo a transition: 
analysing the behavior of the potential terms in 
(\ref{c1 eq einstein mixmaster 5 dim}) we see how  two of the 
four parameters must satisfy the inequality
\begin{equation}
\label{c1 condizione di kasner 5 dim 1}
1-3p_{1}^{2} - 3p_{2}^{2} -2p_{1}p_{2}+ 2p_{1} + 2p_{2}\geq0\;,
\end{equation}
and one of the following ones
\begin{eqnarray}
&&3p_{1}^{2}+p_{2}^{2}+p_{1} -p_{2}-p_{1}p_{2}<0\;,\nonumber\\
&&3p_{2}^{2}+p_{1}^{2}+p_{2} -p_{1}-p_{1}p_{2}<0\;,\label{c1 condizione di kasner 5 dim 2}\\
&&3p_{1}^{2}+p_{2}^{2}-5p_{1} -5p_{2}+5p_{1}p_{2}+2<0\;. \nonumber
\end{eqnarray}
Figure  \ref{c1 cond 5 dim kasner} shows the existence of a region 
where condition (\ref{c1 condizione di kasner 5 dim 1}) is satisfied 
but none of (\ref{c1 condizione di kasner 5 dim 2}) is,
thus  the Universe undergoes a certain number of transitions 
and Kasner epochs and eras; as soon as the Kasner indices 
$p_{1}, p_{2}$ assume  values in the shaded region, 
then no more transitions can take place and the evolution 
remains Kasner-like until  the singular point is reached.

Type $G14$ case is quite similar to $G13$: for this model, 
the structure constants are the same as Bianchi type VIII 
and, under the same hypotheses, only a finite sequence of epochs occurs.

The results of this analysis for the five-dimensional homogeneous 
space-times can be extended to higher dimensions and reveal how chaos is a 
{\it dimensional phenomenon}, for the homogeneous case, 
 limited to the four-dimensional space-time.

\subsectionric{The Role of a Vector Field\label{sec:vec}}

In this section we investigate the effects of 
an Abelian vector field on the dynamics of a generic $(n+1)$-dimensional
homogeneous model in the BKL scheme; 
the chaos is restored for any number of dimensions, 
and a BKL-like map, exhibiting a peculiar  dependence on the 
dimension number, is worked out\cite{BeniniMontani2005CQG}. 
These results have also been inserted in  more general treatment by 
Damour and Hennaux\cite{cosmbill}.

A generic $(n+1)$-dimensional space-time coupled to an 
Abelian vector field $A_{\mu }=( \varphi ,A_{\alpha })$, 
with $\alpha =(1,2,\ldots ,n)$ in the ADM framework is described
 by the  action
\begin{equation}
 S=\int d^{n}xdt\left( \Pi ^{\alpha \beta }\displaystyle\frac{\partial }{\partial t} 
h_{\alpha \beta }+\Pi ^{\alpha }\displaystyle\frac{\partial }{\partial t}A_{\alpha }
+\varphi D _{\alpha }\Pi ^{\alpha }-N H-N^{\alpha } H_{\alpha
}\right)\; ,  \label{c4 act}
\end{equation}
where
\bseq
\begin{align}
 H&=\displaystyle\frac{1}{\sqrt{h}}\left[ \Pi _{\beta }^{\alpha }
\Pi _{\alpha }^{\beta}-\displaystyle\frac{1}{n-1}
\left( \Pi _{\alpha }^{\alpha }\right) ^{2}+
\displaystyle\frac{1}{2} 
h_{\alpha \beta }\Pi ^{\alpha }\Pi ^{\beta }+
h\left( \displaystyle\frac{1}{4}F_{\alpha
\beta }F^{\alpha \beta }-{~}^{(n)} R\right) \right]  \label{c4 hamcn}\;, \\
H_{\alpha }&=-\nabla _{\beta }\Pi _{\alpha }^{\beta }+\Pi ^{\beta }F_{\alpha
\beta }\;,  \label{c4 momcn}
\end{align}
\eseq
denote the super-Hamiltonian and the super-momentum respectively, 
while 
$F_{\alpha \beta }$ is the spatial electromagnetic tensor,  
and the relation $D_\alpha\equiv \partial_\alpha+A_\alpha$ holds.
Moreover, $\Pi^\alpha$ and $\Pi^{\alpha\beta}$ are the conjugate momenta 
to the electromagnetic field and to the $n$-metric, respectively, 
which result to be a vector and a tensorial density of weight $1/2$, 
since their explicit  expressions contain the square root of the spatial 
metric determinant.
The variation with respect to the lapse function $N$ 
yields the super-Hamiltonian constraint 
$H = 0$, 
 while  with respect to $\varphi$ it provides the constraint 
$\partial _{\alpha }\Pi ^{\alpha }=0$.\\
We will deal with a source-less Abelian vector field and in this case 
one can  consider the transverse (or Lorentz) components 
for $A_{\alpha }$ and $\Pi ^{\alpha }$  only. 
Therefore, we choose the gauge conditions 
$\varphi =0$ and $D _{\alpha }\Pi ^{\alpha }=0$, 
enough to prevent 
the longitudinal parts of the vector field from
taking part to the action.\\
It is worth noting how, 
in the general case, {\it i.e.}
either in presence of the sources, or in the case of
non-Abelian vector fields,
this simplification can no longer take place 
in such explicit form and the terms 
$\varphi (\partial _{\alpha }+A_\alpha)\Pi ^{\alpha }$ must be considered
in the action principle.

A BKL-like analysis can be developed\cite{BeniniMontani2005CQG} 
as well as done previously, 
following some steps: after introducing a set of Kasner vectors 
$\vec{l}_{a}$ and the Kasner-like expanding factors $\exp(q^{a})$, 
the dynamics is dominated by a potential of the form 
$\sum e^{q_{a}}\widetilde{\lambda } _{a}^{2}$, where  
$\widetilde{\lambda } _{a}$ are the projection of the momenta of 
the Abelian field along the Kasner vectors. 
With  the same spirit of the Mixmaster analysis, an unstable 
$n$-dimensional Kasner-like evolution arises, nevertheless 
the potential term inhibits the solution to last 
up to the singularity and, as usual, induces the  BKL-like transition 
to another epoch. 
Given the relation $\exp(q^{a})=t^{p_{a}}$, 
the map that links two consecutive epochs is 
\bseq
\label{c4 234}
\begin{align}
p_{1}^{\prime } &=\frac{-p_{1}}{1+\frac{2}{n-2}p_{1}}\;, \qquad
p_{a}^{\prime }= \frac{p_{a}+\frac{2}{n-2}p_{1}}{1+\frac{2}{n-2}p_{1}}\;, \\
\widetilde{\lambda }_{1}^{\prime }&=\widetilde{\lambda }_{1}\,,\qquad \qquad \qquad
\widetilde{\lambda }_{a}^{\prime } =\widetilde{\lambda }_{a}
\left( 1-2\frac{\left(n-1\right) p_{1}}{\left( n-2\right) p_{a}+np_{1}}\right)\;.
\end{align}
\eseq
An interesting new feature, resembling that of the inhomogeneous Mixmaster 
(as we will discuss later), is the \textit{rotation of the Kasner vectors},
\bseq
\begin{align}
\label{rotax}
\vec{\ell}_{a}^{\prime } &=\vec{\ell}_{a}+\sigma _{a}\vec{\ell}_{1}\;,\\
\sigma_{a} &=\displaystyle\frac{\widetilde{\lambda }_{a}^{\prime } 
-\widetilde{\lambda }_{a}}{\widetilde{\lambda }_{1}} 
=-2\displaystyle\frac{\left( n-1\right) p_{1}}{\left( n-2\right)
p_{a}+np_{1}}\displaystyle\frac{\widetilde{\lambda }_{a}}{\widetilde{\lambda }_{1}}\;.
\label{c4 40}
\end{align}
\eseq
which completes our dynamical scheme.

The homogeneous Universe in this case  approaches the 
initial singularity  described by a metric tensor with 
oscillating scale factors and rotating Kasner vectors.
Passing from one Kasner epoch to another, the negative Kasner index 
$p_1$ is exchanged between different directions 
(for istance $\vec\ell_1$ and $\vec\ell_2$) and, at the same time, 
these directions rotate in the space according to the rule (\ref{c4 40}).
The presence of a vector field is crucial because, 
independently of the considered model, it induces a dynamically closed 
domain on the configuration space\cite{BeniniMontani2004PRD}.\\
In correspondence to these oscillations of the scale factors, 
the Kasner vectors $\vec{\ell}_{a}$ rotate and
the quantities $\sigma_{a}$ remain constant during a Kasner epoch
to lowest order in $q^a$; thus, the vanishing  of the determinant
$h$ approaching the singularity does not significantly affect the 
rotation law (\ref{c4 40}).\\
There are two  most interesting features of the resulting dynamics: 
the map exhibits a \textit{dimensional-dependence}, 
and it reduces to the standard BKL one  for the four-dimensional case.


%
\sectionric{Quantum dynamics of the Mixmaster\label{sec:qdm}}

This Section faces the question of the quantum evolution of the Mixmaster model. In particular, after a discussion on the Wheeler-DeWitt approach and the so-called problem of time, the first quantization of the model is achieved in both Misner variables and MCl ones. Furthermore, the model is also analyzed in the Loop Quantum Gravity framework (with particular attention to the disappearance of chaos) and in the Generalized Uncertainty Principle one. A brief discussion of the so-called Quantum Chaos closes the Section.\\
In the rest of the paper, we set $\hbar = 1$.

\subsectionric{The Wheeler-DeWitt Equation\label{sect:wdweq}}

In this Section we will briefly review 
the Wheeler-DeWitt approach to the quantum gravity formulation, 
in the metric formalism. 
For the detailed literature about this topic we refer 
to \cite{ishamtime,Kuchar1980,Stachel,Espo94,har88}.

This scheme relies on the Dirac approach to a first-class constrained 
system \cite{HT,Dirac64}, i.e. the quantum theory is constructed 
without solving the constraints. 
Of course this method carries some un-physical information that 
will be removed imposing some conditions to select the physical states. 
In particular, if  $G_a$ is a first class constraint, 
a physical state must remain unchanged when performing a 
transformation generated by $G_a$. 
Thus, the physical states are the ones annihilated by the 
quantum operator constraints, i.e. $\hat G_a\vert\Psi\rangle=0$.

The quantization of General Relativity in a 
canonical formalism prescribes to implement the Poisson 
algebra (\ref{alg1}-\ref{alg3}) in the form of the 
canonical commutation relations
\bseq
\begin{align}
\label{canc1}
\left[\widehat{h}_{\alpha\beta}(x,t),\widehat{h}_{\gamma\delta}(x',t)\right] & =0 \\
\label{canc2}
\left[\widehat{\Pi}^{\alpha\beta}(x,t),\widehat{\Pi}^{\gamma\delta}(x',t)\right] & =0 \\
\label{canc3}
\left[\widehat{h}_{\gamma\delta}(x,t),\widehat{\Pi}^{\alpha\beta}(x',t)\right] & = 
	i\kappa\delta^\alpha_{(\gamma}\delta^\beta_{\delta)}\delta^3(x-x') \, .
\end{align}
\eseq 
We note that equation (\ref{canc1}) is a kind of microcausality 
condition for the three-metric field, though  the functional form of the constraint is 
independent of any foliation of space-time: thus this confirms 
that the points of the three-manifold $\Sigma$ are space-like separated.

In the next step, we  impose the constraint equations 
(\ref{seccon}) as operators to select the physically allowed states
\bseq
\begin{align}
\label{hhat}
\widehat{H}\left(  x;\widehat{h},\widehat{\Pi}\right)  \Psi &=0 \\
\label{hahat}
\widehat{H}^\alpha\left(  x;\widehat{h},\widehat{\Pi}\right)  \Psi &=0 \, .
\end{align}
\eseq 
Since the dynamics of General Relativity is fully contained in 
the classical constraints, we do not have to analyze the dynamical 
equations, whose meaning is readily obtained. 
As we have seen, the Hamiltonian for GR (\ref{hamilgener}) reads as
\be
\label{HGR}
\mathcal H\equiv\int_\Sigma d^3x(N H + N^\alpha H_\alpha),
\ee
and therefore, considering (\ref{hhat}) and (\ref{hahat}), 
in a putative Schr\"{o}dinger-like equation  
\be
i\f d{dt}\Psi_t=\hat{\mathcal H}\Psi_t=0
\ee
the wave functional $\Psi$ (also known as "the wave function of the Universe" \cite{HartleHawking1983PRD}) turns out to be independent of ``time''. 
This is the so-called ``frozen formalism'', because it apparently 
implies that nothing evolves in a quantum theory of gravity. 
By other words, we have an identification of the quantum Hamiltonian 
constraint as the zero-energy Schr\"{o}dinger equation 
$\hat{\mathcal H}\Psi=0$.  
This is known as the problem of time and has deep consequences 
on the interpretation of the wave function of the Universe and deserves 
to be treated in some details in  Section \ref{sec:ptime}. 

When writing the expression (\ref{HGR}), we 
assumed  the primary constraints
\be
C(x,t)\equiv\Pi(x,t)=0 \, , \qquad C^\alpha(x,t)\equiv\Pi^\alpha(x,t)=0\, ,
\ee 
implemented at a quantum level and therefore the wave functional $\Psi=\Psi(h_{\alpha\beta},N,N^\alpha)$ becomes 
function of the three-metric only, i.e. $\Psi=\Psi(h_{\alpha\beta})$.

Let us now explicitly discuss the meaning of  constraints 
(\ref{hhat}) and (\ref{hahat}). 
First of all, a representation of the canonical algebra can be chosen to be
\begin{equation}
\widehat{h}_{\alpha\beta}  \Psi =h_{\alpha\beta}\Psi, 
\qquad  \widehat{\Pi}^{\alpha\beta}  \Psi = -i \kappa\dfrac{\delta\Psi}{\delta h_{\alpha\beta}  }.
\end{equation}
This is the widely used representation of the canonical approach 
to quantum gravity in the metric formalism. 
However, the above equations do not define proper self-adjoint operators 
because of the absence of any Lebesgue measure on $\Sigma$ \cite{ishamtime}.

Let us firstly address the  constraint (\ref{hahat}), which 
is the so-called diffeomorphisms one, because 
the wave functional $\Psi(h_{\alpha\beta})$ depends on 
a whole class of three-geometries $\{h_{\alpha\beta}\}$ 
(invariant under three-diffeomorphisms) and not only 
on the three-metric, i.e. $\Psi=\Psi(\{h_{\alpha\beta}\})$. 
Therefore the configuration space for quantum gravity will be 
the Wheeler {superspace}\cite{MTW}. 
In   literature this is referred to as  the ``kinematical constraint''. 
The dynamics is generated via the scalar constraint (\ref{hhat}), 
providing the famous Wheeler-DeWitt equation\cite{DeWitt1967I,DeWitt1967II,DeWitt1967III}, which explicitly reads as
\begin{equation}\label{WDW}
\widehat{H}(x) \Psi=-\mathcal G_{\alpha\beta\gamma\delta}(x)
\f{\delta^{2}\Psi}{\delta h_{\alpha\beta}(x)\delta h_{\gamma\delta}(x)}
-\sqrt h {~}^{(3)}R\Psi=0 \, ,
\end{equation}
where $\mathcal G_{\alpha\beta\gamma\delta}$ is the supermetric (\ref{supermetrica}). 
This equation is at  the heart of the Dirac constraint 
quantization approach and the key aspects of the canonical quantum 
gravity are all connected to it.

There are several problems in the WDW approach to quantum gravity,
both mathematical and conceptual ones, for which we refer to a wide literature \cite{Kuchar1980,ishamtime,Thi,Thirev} on the topic. 
We will discuss the two most relevant among them: 
\begin{enumerate}
\item 
since equation (\ref{WDW}) contains products of functional 
differential operators evaluated at the same spatial point, 
it is therefore hopelessly divergent; 
moreover the  distributions in the denominator are not clearly defined. 

\item
Understanding the physical meaning of the WDW equation requires a clarification 
of the notion of time at a quantum level, and eventually this is 
related to the classical ($3+1$)-splitting performed before the 
quantization procedure \cite{Montani2002NPB,Mercuri,BattistiMontani2006PLB}.  

\end{enumerate}

\subsectionric{The Problem of Time\label{sec:ptime}}

A major conceptual problem in quantum gravity is the issue of what 
time is and how it has to be treated once a formalism is adopted (for a detailed discussion see \reffcite{ishamtime}, while for the role of conformal three-geometries we refer to \reffcite{Y72}). 
This task is deeply connected with the special role assigned to 
temporal concepts in all theories of physics different from GR. 
For example, in Newtonian physics, as well as in non relativistic 
quantum mechanics, time is an external parameter to the system itself and 
is treated as a background degree of freedom. 
In ordinary quantum field theory 
the situation is similar since the Minkowski background is fixed 
and the Newtonian time is replaced by the time measured in a set 
of relativistic inertial frames. Such notion of  ``time'' plays a crucial 
role in the conceptual foundations of the quantum theory. 
In fact, in the conventional Copenhagen interpretation 
of quantum mechanics, an observable is a quantity whose value can 
be measured at fixed time. 
Moreover, the scalar product is conserved under the time evolution 
and the quantum fields have to satisfy the microcausality conditions. 
Finally, it can be shown \cite{Un-Wa89} how a perfect clock, in the 
sense of a quantum observable $T$ whose  values monotonically grow with 
abstract time $t$, is not compatible with the physical requirement 
of a positive spectrum of the energy, and 
this is a peculiar feature of the quantum theory.  
 
The problem of time arises also in the canonical formulation of 
the quantum theory of gravity, as happens in any diffeomorphism-invariant 
quantum field theory, and the Schr\"{o}dinger equation is replaced 
by a Wheeler-DeWitt one, where  the time coordinate is not present 
in  the formalism.

The proposals to address  this fundamental problem are often related 
with the introduction of a reference system. 
In fact, this can be achieved in two different ways:
the first one consists in adding a dynamical fluid or fields to the vacuum 
gravitational picture (see \cite{BK94,R91a,R91b,BattistiMontani2006PLB}), while the other 
in fixing the frame in a geometrical way (\cite{KT91,BK94,Montani2002NPB,Mercuri}). 
However, both of them lead to an equivalent evolutionary quantum 
dynamics \cite{R91a,R91b,BK94,Mercuri2}. Below we will discuss the Brown and 
Kucha$\breve r$ mechanism, the so-called evolutionary quantum gravity and, finally,
the multi-time approach.

\subsubsectionric{The Brown and Kucha$\breve{r}$ mechanism\label{sec:bkmech}}

This  approach\cite{BK94} is devoted to  find a medium leading  to a 
Schr\"{o}dinger equation when applying the Dirac quantization 
to a constrained system. 
In particular,  an incoherent dust, i.e. one  with the gravitational interaction 
only, is included in the dynamics. 
This procedure leads to the  new constraints $H_\uparrow(X)$ 
and $H_{\uparrow\alpha}(X)$ in which the dust plays the role of 
time and the true Hamiltonian  does not depend on the dust variables.

Let us introduce the variables $T,Z^\alpha$ and the corresponding 
conjugate momenta $M,W_\alpha$, so that the values of $Z^\alpha$ be  
the comoving coordinates of the dust particles and $T$ be 
the proper time along their worldlines. 
In this scheme the new constraints read as 
\bseq
\begin{align}
\label{BKH}
H_\uparrow & =P(X)+h(X,h_{\alpha\beta},\Pi^{\alpha\beta})=0  \\
H_{\uparrow\alpha}& =P_\alpha(X)+h_\alpha(X,T,z^\alpha,h_{\alpha\beta},\Pi^{\alpha\beta})=0  
\end{align}
\eseq
where
\bseq
\begin{align}
h  & =-\sqrt{G(X)}  \, , \qquad 
	G(X)=H_{(grav)}^2-h^{\alpha\beta}H_{(grav)\alpha} H_{(grav)\beta} \, ,  \\
h_\alpha & =Z^\beta_\alpha H_\beta+\sqrt{G(X)}\p_\beta T Z^\beta_\alpha \, ,
\end{align}
\eseq
 $H_{(grav)}$ and $H_{(grav)\alpha}$ being the usual scalar and momentum 
constraints, respectively,  $P$ the projection of the rest mass 
current of the dust into the four velocity of the observers, 
and $P_\alpha=-PW_\alpha$. 
This way the Hamiltonian $h$ does not depend on the dust.

The quantization of this model is performed in the canonical way 
and yields, from equation (\ref{BKH}), the Hamiltonian constraint operator
\be
\hat H_\uparrow=\hat P(X)+\hat h(X,\hat h_{\alpha\beta},\hat \Pi^{\alpha\beta})=0 \,.
\ee
This mechanism leads to a Schr\"{o}dinger equation for the wave 
functional $\Psi=\Psi(\hat T,\hat h)$ 
\be
i\f{\delta\Psi}{\delta T}=\hat h\Psi.
\ee
The central point of this  procedure is the independence of the effective 
Hamiltonian  $h(X)$ on the dust; this allows a well posed 
spectral analysis formulation because $h$ commutes with itself. Furthermore, the Schr\"{o}dinger equation can be split  into a  
dust- (time-) dependent part and a truly gravitational one.

\subsubsectionric{Evolutionary Quantum Gravity\label{sec:evol}}

In this section we remark some of the fundamental aspects 
of the evolutionary quantum gravity as presented in \reffcite{Montani2002NPB,Mercuri}. 
As a first step we analyze the implication of a Schr\"{o}dinger 
formulation of the quantum dynamics for the gravitational field\cite{H91c}, and 
then we establish a dualism between time evolution and matter fields.

Let us assume that the quantum evolution of the gravitational field 
is governed by the smeared Schr\"{o}dinger equation
\be
i\p_t \Psi=\hat{\mathcal{H}}\Psi\equiv\int_{\Sigma}d^3x\left(N\hat{H}\right)\Psi \, ,
\ee
where 
the wave functional $\Psi$ is defined on the  superspace, 
i.e. it is annihilated by the super-momentum operator $\hat H_\alpha$. 
Let us take the following expansion for the wave functional
\be
\Psi=\int D\epsilon ~ \chi(\epsilon,\left\{h_{\alpha\beta}\right\}) \exp\left[-i\int_{t_0}^tdt'\int_{\Sigma}d^3x(N\epsilon)\right],
\ee
where $D\epsilon$ is  the Lebesgue measure on the space 
of the functions $\epsilon(x)$. 
This expansion reduces the Schr\"{o}dinger dynamics to an eigenvalue problem of the form
\be\label{eipro}%
\hat{H}\chi=\epsilon\chi,\qquad \hat{H_\alpha}\chi=0,
\ee
which outlines the appearance of a non-zero super-Hamiltonian eigenvalue.

In order to reconstruct the classical limit of the dynamical 
constraints (\ref{eipro}), 
we  replace the wave 
functional $\chi$ by its corresponding zero-order WKB 
approximation $\chi\sim e^{iS}$. 
In this case the eigenvalue problem (\ref{eipro}) reduces to 
its classical counterpart
\be
\widehat{HJ}S=\epsilon\equiv-2\sqrt h T_{00},\qquad \widehat{HJ_\alpha}S=0
\ee
where $\widehat{HJ}$ and $\widehat{HJ_\alpha}$ denote operators which, 
when applied to the phase $S$, reproduce the super-Hamiltonian 
and super-momentum Hamilton-Jacobi terms, respectively. 
We see that the classical limit of the Schr\"{o}dinger quantum 
dynamics is characterized by the appearance of a new matter 
contribution (associated with the non-zero eigenvalue $\epsilon$) 
whose energy density reads as
\be\label{enden}%
\rho\equiv T_{00}=-\f{\epsilon(x)}{2\sqrt h},
\ee  
where by $T_{ij}$ we refer to the new matter energy-momentum tensor.

Since the spectrum of the super-Hamiltonian has,
in general, a negative component, then we can  infer that,
when the gravitational field is in its ground state,
this matter comes out in the classical limit with a positive
energy density. 
The explicit form of (\ref{enden}) is that
of a dust fluid co-moving with the slicing of the three-hypersurfaces,
i.e. the normal field $n^{i}$ becomes the four-velocity 
of the appearing fluid (in other words,
we deal with an energy-momentum tensor
$T_{ij}=\rho n_i n_j$).

We stress how the space of the solutions can
be turned  into the  Hilbert one and therefore a
notion of probability density naturally arises, from the squared modulus of
the wave functional.

Let us now consider the opposite sector, i.e. a gravitational 
system in the presence of
a macroscopic matter source. 
In particular, we chose  a perfect fluid with  a
generic equation of state $p = (\gamma - 1 )\rho$ ($p$ being the 
pressure and $\gamma$ the polytropic index). 
The energy-momentum
tensor, associated to this system reads as
\begin{equation}
T_{ij} =
\gamma \rho u_{i}u_{j} - (\gamma -1)\rho g_{ij}
\, .
\label{tens}
\end{equation} 
To fix the constraints when matter is included in the
dynamics, let us make use of the relations
\bseq
\begin{align}
\label{ten2}%
G_{ij} n^{i}n^{j} & = - \kappa \frac{H}{2\sqrt{h}} \\
\label{due}%
G_{ij} n^{i}\partial _\alpha y^{j} & = \kappa \frac{H_i}{2\sqrt{h}} \, , 
\end{align}
\eseq
where $\partial _\alpha y^{i}$ are the tangent vectors to the
three-hypersurfaces, i.e. $n_{i}\partial _\alpha y^{i} = 0$. 
Equations (\ref{ten2}) and (\ref{due}), by (\ref{tens}) and 
identifying $u_{i}$ with $n_{i}$
(i.e. the physical space is filled by the fluid),
rewrite as
\be
\label{ten3}
\rho = -\frac{H}{2\sqrt{h}} \, , \qquad H_i = 0  \, , 
\ee
and furthermore, we get the equations
\begin{equation}
G_{ij}\partial _\alpha y^{i}\partial _\beta y^{j}
\equiv G_{\alpha\beta} = \kappa (\gamma -1)\rho h_{\alpha\beta} \, .
\label{ij}
\end{equation}
The conservation law
$\nabla_jT_i^{j} = 0$ implies the additional  two conditions
\bseq
\begin{align}
\label{ten4}%
\gamma \nabla_i\left( \rho u^{i}\right) & =
(\gamma - 1) u^{i}\partial _{i}\rho  \\
\label{ten5}%
u^{j}\nabla_j u_{i} &= 
\left( 1 - \frac{1}{\gamma }\right)
\left( \partial _{i} \ln \rho - u_{i}u^{j}
\partial _{j}\ln \rho \right)
\, .
\end{align}
\eseq  
With  the space-time slicing,   
looking at the dynamics into the fluid frame
(i.e. $n^{i} = \delta _0^{i}$), by the relation
$n^{i} = (1/N, \; -N^\alpha/N)$, the co-moving
constraint implies the synchronous nature of the reference frame.
Since  a synchronous reference is also a geodesic one, 
the right-hand side of equation 
(\ref{ten5}) must  identically vanish and, for a generic
inhomogeneous case, this implies 
$\gamma \equiv 1$. Hence, equations (\ref{ten4})
yields $\rho = -\bar{\epsilon }(x)/2\sqrt{h}$ and 
substituted  into  (\ref{ten3}), we get
the same Hamiltonian constraints associated to the Evolutionary
Quantum Gravity given above in eq.(\ref{eipro}),   as soon as the function
$\bar{\epsilon}$ is turned into the eigenvalue $\epsilon$.
In this respect, while
$\bar{\epsilon}$ is positive by definition, the corresponding
eigenvalue can also take negative values because of
the structure of $H$.

Thus, we conclude that a dust fluid is a good choice to realize
a clock in Quantum Gravity, because it induces a non-zero
super-Hamiltonian eigenvalue into the dynamics;
furthermore, for vanishing pressure ($\gamma = 1$), the
equation (\ref{ij}) reduces to the proper  vacuum evolution equation 
for $h_{\alpha\beta}$, thus outlining  
a real dualism between time evolution and the presence of a dust fluid.

For the cosmological applications of the above approach in the isotropic sector see \reffcite{Corvino,Montani2003IJMPD} 
 and in a generic cosmological model see \reffcite{Montani2004IJMPDminisuperspace} and  \reffcite{BattistiMontani2006PLB} 
where it is shown how, from a phenomenological point of view, 
an evolutionary quantum cosmology overlaps the Wheeler-DeWitt
approach.

\subsubsectionric{The Multi-Time Approach\label{sec:multit}}

The multi-time approach \cite{kuchar1972,ishamtime}  represents an alternative 
interesting way to get a Schr\"{o}dinger quantum dynamics, 
even if deeply  different from the one described above, 
since it is based on the ADM reduction of the dynamics. 
In fact, starting from equation (\ref{ha}),
\be
P_A(x)+h_A(x,\chi,\phi,\pi)=0 \, ,
\ee 
and performing a  canonical quantization of the model, 
we obtain a  Schr\"{o}dinger-like equation
\be
i\f{\delta\Psi}{\delta\chi^A}=\hat h_A\Psi \, ,
\ee
where $\hat h_A$ is the operatorial version of the classical 
Hamiltonian density. 
Even if this approach and the evolutionary one (discussed in the previous subsection) seem to overlap each other when solving the problem of the frozen formalism, this is not 
the case. 
As matter of fact, the evolutionary quantum dynamics approach 
is based on a full quantization of the system. 
By the other hand, the multi-time approach relies on a quantization 
of only some degrees of freedom. 
In fact, the constraints are classically solved before implementing 
the quantization procedure, thus violating the geometrical nature of 
the gravitational field in view of real physical degrees of freedom. 
This fundamental difference between the two approaches is evident, 
for example, in a cosmological context. 
In fact, when we quantize a minisuperspace model 
 in the ADM formalism, 
the scale factor of the Universe is usually chosen as a ``time'' 
coordinate, and therefore  the dynamics is consequently expressed.
By the other hand, in an evolutionary approach the scale factor 
is treated on the same footing of other variables, 
i.e. the anisotropies, and the evolution of the system is considered with 
respect to a privileged reference frame.

\subsectionric{The Minisuperspace Representation\label{sec:minisup}}

Although the full theory (``Quantum General Relativity'') 
is far from being reached, 
many approaches are properly treated in the context of the so-called minisuperspace,  
as for example  quantum cosmology. 
In fact, only a finite number of gravitational degrees of 
freedom are invoked in the quantum theory, and the remaining ones are frozen out  
imposing some symmetries on the spatial metric. 
These space-times are, for instance, the homogeneous cosmological models. 

In this sense quantum cosmology is a  toy model for quantum gravity
(with finite degrees of freedom) which is a simple arena to test 
ideas and constructions introduced in the full theory 
(a genuine quantum field theory). 
In particular, since on a classical level the Universe dynamics 
is described by such symmetric models, their  quantization 
 is required to answer the fundamental questions like the fate of the 
classical singularity, the inflationary expansion and the chaotic 
behavior of the Universe toward the singularity. 

Moreover, as we will see in Section \ref{sec:generic}, in the general 
context of inhomogeneous cosmology, the spatial derivatives in the Ricci scalar are 
negligible with respect to the temporal ones, toward the 
singularity. 
This is the well-known BKL
scenario \cite{BKL1970,BKL82} where, as the singularity is approached 
(on a classical level), the spatial geometry can be viewed as a 
collection of small independent patches, in general  Bianchi IX like models. 
Therefore a minisuperspace reduction of the dynamics is important also in 
the description of a generic Universe toward the classical singularity when 
restricted to each cosmological horizon.

\subsectionric{On the Scalar Field as a Relational Time\label{sec:sfrtime}}

Let us now discuss in some details the role of a matter field, 
in particular that of a scalar field $\phi$, used as a definition of time 
for the quantum dynamics of the gravitational field. 
As we have seen, when the canonical quantization procedure is applied,
 the usual Schr\"{o}dinger equation is 
replaced by a Wheeler-DeWitt one in which the time coordinate is dropped 
from the formalism. 
One possible solution to this fundamental problem is adding a matter field 
in the dynamics and then let evolve the physical degrees of freedom of the 
gravitational system with respect to it. 
This way the dynamics is  described on a relational point of view, 
i.e. the matter field behaves as a relational time. 
Such an idea is essentially based on the absence of  time at a fundamental 
level and therefore a field can evolve  with respect to another one  only
(for more details on this approach see \cite{rovelli2004qg}). 

In particular, in quantum cosmology, such choice 
appears as the most natural one. 
In fact,  near the classical singularity, a monotonic behavior of 
a massless $\phi$ 
 always appears as a function of the scale 
factor\footnote{In fact, such behavior well approximates 
the one of an inflaton field when its potential 
is negligible at enough high temperature\cite{CW73}. For a detailed discussion of the inflationary scenario within the framework of homogeneous cosmologies, see \reffcite{1985PhLB..159..256R,1985PhRvD..31..718Z,1986PhRvD..33.1204G,1986PhRvD..34.2535H,1988PhRvD..38.2392C,1991PhLB..256..359H,1992PhLB..287...61S,1993CQGra..10.1147F,1997PhRvD..55.1896B,2001PhRvD..64j7301K,2001PhRvD..64l7502C,2004PhRvD..70h4040M}.}
(more precisely the variable which describes the isotropic expansion 
of the Universe). \\
Let us consider the case of the Bianchi IX model in  presence of 
a massless scalar field, whose  Hamiltonian constraint in the Misner 
variables $(a\equiv e^{\alpha},\beta_\pm)$ has the form 
\be\label{hamgrasca}
H_{IX}+H_{\phi}=\kappa\left[-\f{p_a^2}a+\f1{a^3} 
\left(p_+^2+p_-^2\right)\right]-\f a{4\kappa}V(\beta_\pm)
+\f{p_{\phi}^2}{a^3}\approx 0 \, ,
\ee
$V(\beta_\pm)$ being the relative potential term.  \\
When this system is canonically quantized, the associated 
 Wheeler-DeWitt equation describes how the wave function 
$\Psi=\Psi(a,\beta_\pm,\phi)$ evolves with  $\phi$. 
More precisely, from (\ref{hamgrasca}) 
we obtain\footnote{The choice of the normal ordering is not important 
for the following discussion, thus  we adopt the simplest one.}
\be
-\p_\phi^2\Psi=\Xi\Psi, \qquad \Xi\equiv 
\kappa\left[-a^2\p_a^2+\p_+^2+\p_-^2+
	\f{a^4}{4\kappa^2}V(\beta_\pm)\right] \, ,
\ee
behaving as  a Klein-Gordon equation with  $\phi$ playing 
the role of (relational) time and $\Xi$ of the spatial Laplacian. 
An explicit Hilbert space arises after  performing  the natural decomposition 
of the solution into positive and negative frequencies parts. 
In particular,  the positive frequency sector offers the 
Sch\"odinger-like equation $-i\p_\phi\Psi=\sqrt{\Xi}\Psi$.

We can analyze in which sense the scalar field can be regarded as 
a good time parameter for the dynamics. 
First of all, near the cosmological singularity ($a\rightarrow0$) 
the potential term $V(\beta_\pm)$ can be 
neglected\footnote{For a discussion on the consistency condition 
	ensuring that the quasi-classical limit of the Universe dynamics is 
	reached before the potential term becomes important, see \cite{BattistiMontani2006PLB}}. 
In such approximation the (classical) equations of motion obtained 
from  (\ref{hamgrasca}) read as
\be
\label{cicciodot}
\dot a=\f{a^2p_a}{\sqrt{a^2p_a^2-p_\beta^2}}, \qquad \dot p_a=-\f{ap_a^2}{\sqrt{a^2p_a^2-p_\beta^2}}, \qquad \dot p_+=\dot p_-=0,
\ee
where $\dot{(...)}\equiv d(...)/d\phi$ and $p_\beta^2\equiv p_+^2+p_-^2$. 
The solutions of  system (\ref{cicciodot}) have the form
\be\label{monrel}
a(\phi)=B\exp\left(\f{A\phi}{\sqrt{A^2-p_\beta^2}}\right), 
\qquad p_a(\phi)=\f AB\exp\left(-\f{A\phi}{\sqrt{A^2-p_\beta^2}}\right) \, ,
\ee
$A$ and $B$ being integration constants and $p_\beta^2=\textrm{const}$.

We have recovered a monotonic dependence of the scalar field $\phi$ 
with respect to the isotropic variable of the Universe $a$ and 
therefore the massless $\phi$  shows to be  a good (relational) 
time for the gravitational dynamics. 

As we will see in Section \ref{sec:dryt}, the dynamics toward the 
cosmological singularity of a generic inhomogeneous Universe is 
described, point by point, by the one of a Bianchi IX model. 
More precisely, the spatial points  dynamically decouple toward the singularity 
and the spatial geometry can be viewed as a collection of small patches, 
each one  independently evolving as a Bianchi IX model. 
From this point of view, the monotonic relation (\ref{monrel}) is a 
proper general feature of the gravitational field, better clarifying 
the choice of the  scalar field as a relational time.

\subsectionric{Interpretation of the Universe Wave Function\label{sec:uwf}}

In this Section we face the problem of  the probabilistic interpretation 
of the Universe wave function, in agreement with the analysis 
developed in \reffcite{Vilkenin89,Vilkenin88} (for a different point of view see \reffcite{Kirillov1992JETPL,ImponenteMontani2001NPPS,grahamquantumbirth}). 
In fact, in quantum cosmology, the Universe is described by a single wave 
function $\Psi$ providing  puzzling interpretations when 
analyzing  the differences between ordinary quantum mechanics 
and quantum cosmology. 

In quantum mechanics, given a wave function $\Psi(q_i,t)$ describing 
a system, the probability to find the system in a configuration-space 
element $d\Omega_q$ at time $t$ is given by
\be
dP=|\Psi(q_i,t)|^2d\Omega_q,
\ee
and it is positive semi-definite, i.e. $dP\geq0$. 
On the other hand, the Universe wave function (which is a solution 
of the Wheeler-DeWitt equation) depends on the three-geometries, 
on the possible matter fields and no dependence on time appears. 
Therefore, the associated probability (here we denote whole the set of 
the superspace variables by $h$)
\be
dP=|\Psi(h)|^2\sqrt \mathcal{G}d^nh \, ,\hspace{5mm}\mathcal{G} = \det(\mathcal{G}_{\alpha \beta \gamma \delta})\, ,
\ee
($\mathcal{G}$ being the supermetric) is not normalizable, because its integral over the whole super-space is 
diverging. 
Such  behavior can be considered as the analogue of the quantum mechanical feature
\be
\int|\Psi(q_i,t)|d\Omega_qdt=\infty \, .
\ee
In fact, in quantum cosmology, the ``time'' is included among the set of variables $h$. 

An alternative  definition of the Universe probability can be given in 
terms of conserved\footnote{This idea relies on the consideration that 
	the WDW equation is nothing but  a Klein-Gordon equation with 
	variable mass.} 
current \cite{DeWitt1967I} (for notation see Vilenkin\cite{Vilkenin89})
\be
\nabla_ij^i=0, \qquad j^i=-\f i2\mathcal{G}^{ij}(\Psi^\star\nabla_j\Psi-\Psi\nabla_j\Psi^\star) \, .
\ee 
This approach is limited as well since the corresponding probability 
to find the Universe in a surface element $d\Sigma_\alpha$ is 
\be\label{provil}
dP=j^\alpha d\Sigma_\alpha
\ee
and it can be negative, similarly to the problem of negative 
probabilities in the Klein-Gordon equation. 

In order to correctly  define a probability, i.e. a positive 
semi-definite one, we can proceed in two different ways: 
first we can consider a  Universe, in which all the 
variables are treated semiclassically, then we can analyze a Universe 
where a quantum subsystem is taken into account, i.e. 
when some variables are pure quantum ones.  

In the case of semiclassical variables, the wave function $\Psi$ is given by
\be
\Psi=A(h)e^{iS(h)},
\ee
which admits a WKB expansion and leads to a  conserved current 
$j^i=|A|^2\nabla^iS$. 
The classical action $S(h)$ describes a congruence of classical trajectories 
and we shall define probability distribution on the $(n-1)$-dimensional equal-time 
surfaces. 
Requiring only single crossing between the trajectories and these equal-time surfaces, 
the probability (\ref{provil}) results to be positive semi-definite.

Let us now consider the case in which not all the superspace variables 
are semiclassical. 
We shall assume that the quantum variables are labeled by $q$ and 
their effect on the dynamics of the semiclassical variables $h$ can be neglected. 
Therefore, the WDW equation can be decomposed in a semiclassical part $H_0$, 
which corresponds to the one of the previous case, and a ``quantum'' one  $H_q$. 
The fact that the quantum subsystem is small is shown by the existence 
of a parameter $\lambda$ (proportional to $\hbar$) such that 
$H_q\Psi/H_0\Psi=\mathcal O(\lambda)$. 
Thus, since  $H_0=\mathcal O(\lambda^{-2})$,  therefore  $H_q=\mathcal O(\lambda^{-1})$. 
The superspace metric can then be expanded in terms of $\lambda$ as
\be 
\mathcal{G}_{ij}(h,q)=\mathcal{G}_{ij}^0(h)+\mathcal O(\lambda) \, ,
\ee
and  the Universe wave function can be written as
\be
\Psi=A(h)e^{iS(h)}\chi(h,q),
\ee
where the function $\chi$ has to satisfy 
\be
\left[\nabla_0^2+2[\nabla_0(\ln A)]\nabla_0+ 2i(\nabla_0S)\nabla_0-H_q\right]\chi =0 \, ,
\ee
where the operator $\nabla_0$ is built using the metric $\mathcal{G}^0(h)$. 
The first two terms are of higher order in $\lambda$ with respect the 
third and can be neglected, resulting into 
\be
2i(\nabla_0S)\nabla_0\chi=H_q\chi \, ,
\ee
which can be written as a Schr\"{o}dinger equation as $i\p_t\chi=NH_q\chi$ (we recall that $\nabla_0S$ coincides with the classical conjugate momentum to the semiclassical variable). 
Thus, we can get two different currents, one for the components in 
the classical subspace and one for those in the quantum one with 
the corresponding  probability distribution written as
\be\label{prodivil}
\rho(h,q,t)=\rho_0(h,t)\left|\chi(q,h(t),t)\right|^2,
\ee
where $\rho_0(h,t)$ and $|\chi|^2$  are the  probability 
distribution  for the classical and the quantum variables, respectively. 
Considering the surface element on  equal-time surfaces 
$d\Sigma=d\Sigma_0d\Omega_q$, for   $d\Sigma_0$ 
defined from the metric $\mathcal{G}^0(h)$, 
the probability distribution (\ref{prodivil}) results to be normalizable.

By other words, we have recovered the standard interpretation of the wave 
function  for a small subsystem of the Universe (only). 
This result agrees with the intrinsic  approximate interpretation 
of the Universe wave function. 
In fact, in the interpretation of quantum mechanics,
all realistic measuring  devices have some quantum uncertainty. 
In particular, the bigger is the apparatus the smaller are the quantum 
fluctuations. 
In this sense, we are able to give a  physical interpretation of the wave 
function of the Universe only in a domain in which some variables are semiclassical. 

Finally, we recall two assumptions underlying this model:

\begin{enumerate}
\renewcommand{\labelenumi}{({\it\roman{enumi}})}

\item
the analysis has been developed in the minisuperspace homogeneous models only;

\item 
the fundamental requirement of existence of a family of 
equal-time surfaces is  taken as a  general feature.

\end{enumerate}

\subsectionric{Quantization in the Misner Picture\label{sec:qmisner}}

In this Section  we provide  a quantum representation 
of the dynamics,  relying  on the adiabatic approximation
ensured by the potential term, reduced
to an infinite well and, according to 
C. W. Misner \cite{Misner1969PR}, we model the potential 
as an infinite square \textit{box} with the same measure as in the 
original triangular picture. 
The volume-dependence of the wave function acquires 
increasing amplitude and frequency of oscillations as the 
Big Bang is approached and the occupation number grows, 
respectively.

For a review of the canonical quantization of Bianchi cosmologies in the WDW framework see \reffcite{1977PThPh..58..842N,1982GReGr..14..549P,1985PhLA..107...33P,1985PhRvD..31..742C,1986PhRvD..33.3581B,1987PhRvD..35.3760L,1989IJTP...28..415P,1989PhRvD..40.3982K,1990PhRvD..41.1047Y,1991PhRvD..44.2375M,1993PhLB..300...44D,1993PhRvD..48.3704M,1995CQGra..12.1441M,1996gr.qc.....7020M,1996IJTP...35.1381O,1998GrCos...4...23K,1998PhRvD..58h4007H,1999IJMPA..14.4473S,2000CQGra..17.2765H,2000GrCos...6...19F,2000GReGr..32.1255F,2000PhRvD..62d3515B,2001PThPh.106..323Y,2002CQGra..19.1013C,2004CQGra..21.4087B,2004PhRvD..70h4040M,2006PhRvD..74f3508V}. For a discussion on the quantization in a supersymmetric form of the Bianchi IX model, see \reffcite{grahamsuper,grahamhartlehawk}.

In  a canonical framework, 
by replacing the canonical variables with the corresponding 
operators and implementing the  Hamiltonian constraints
we get the state function describing the system 
$\psi=\psi(\alpha, \beta_+, \beta_-)$.


Adopting the canonical representation in the configuration space
we address the WDW equation 
\beq
\label{wdw}
\hat{H}\psi=  e^{-3 \alpha}
\left[ - \frac{\delta^2}{\delta \alpha^2}
	+\frac{\delta^2}{\delta \beta_+^2}
	+\frac{\delta^2}{\delta \beta_-^2}	  \right]\psi
-e^{\alpha}V\psi =0 \, . 
\eeq
We can 
find a solution in the form 
\beq
\psi= \sum_n \Gamma_n(\alpha)\varphi_n(\alpha, \beta_+, \beta_-) \, ,
\eeq
where the coefficients $\Gamma_n$ 
are $\alpha$-dependent amplitudes.\\
Thus, equation (\ref{wdw}) is reduced to the ADM eigenvalue problem
\beq
\label{eeii}
\left[-\frac{\partial^2}{\partial \beta_+^2} 
	-\frac{\partial^2}{\partial \beta_-^2}	+
e^{4\alpha}V  \right]\varphi_n
 = E_n^2(\alpha)\varphi_n  \,.
\eeq
According to \cite{Misner1969PR},
we approximate the triangular infinite walls of the 
potential by a box having the same measure 
to find  the eigenvalues $E_n$
\beq
\label{eig_1}
E_n(\alpha)= \pi \left(\frac{4}{3^{3/2}}\right)^{1/2} 
\frac{\mid n \mid}{\alpha} \, ,
\eeq
where $n^2\equiv n_+^2+ n_-^2$, and  
$n_+,n_- \in \mathbb{N}$ are the two independent quantum 
numbers corresponding to the  variables $\beta_+, \beta_-$, 
respectively.

\begin{figure}[ht]
\begin{center}
  \includegraphics[height=.3\textheight]{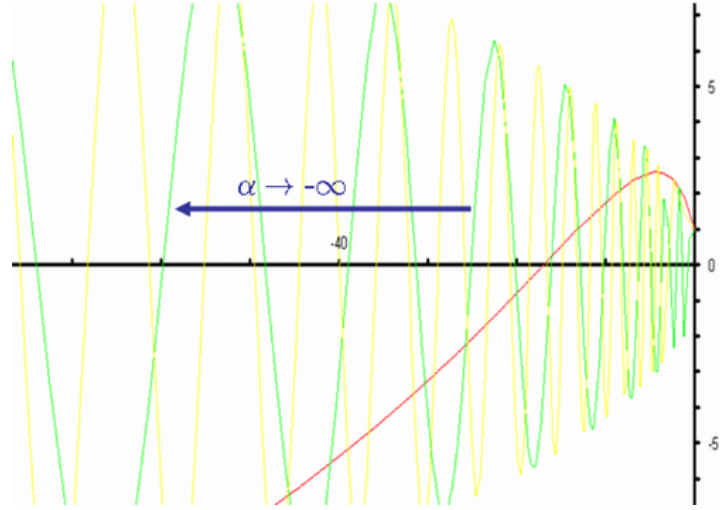}
 \caption[$\Gamma_n(\alpha)$ for three
 different values of the parameter $k_n=1, 15,30$]{Behavior of the solution $\Gamma_n(\alpha)$ for three
 different values of the parameter $k_n=1, 15,30$. 
 The bigger $k_n$, the higher the frequency of oscillation. \label{fig:c_alpha}}

\end{center}
\end{figure}

Substituting the expression for 
$\psi$ in equation (\ref{wdw}) 
we get the differential equation for $\Gamma_n$
\beq
\sum_n(\partial^2_{\alpha} \Gamma_n) \varphi_n + 
\sum_n \Gamma_n (\partial^2_{\alpha} \varphi_n)  +
2\sum_n (\partial_{\alpha} \Gamma_n )(\partial_{\alpha}\varphi_n)+
\sum_n E_n^2 \Gamma_n \varphi_n =0 \, , 
\eeq
which, in the limit of the Misner adiabatic approximation 
of neglecting $\partial_{\alpha}\phi_n$ 
(i.e. $\phi \sim \phi(\beta_+, \beta_-)$), simplifies to
%
\beq
\label{e_diff2}
\frac{d^2\Gamma_n}{d\alpha^2} +\frac{k^2}{\alpha^2}\Gamma_n=0
\eeq
where 
\beq
k_n^2=\left(\frac{2 \pi}{3}\right)^{3/2}\mid n\mid^2  \, .
\eeq
The above equation is solved by $\Gamma_n(\alpha)$ in the form
\beq
\label{gamman}
\Gamma_n(\alpha) =C_1\sqrt{\alpha} \sin\left(\frac{1}{2} \sqrt{p_n} 
\ln \alpha\right)+
C_2\sqrt{\alpha} \cos\left(\frac{1}{2} \sqrt{p_n} \ln \alpha\right)\, , 
\eeq
where $\sqrt{p_n}=\sqrt{k_n^2-1}$. 
From (\ref{gamman}) the self-consistence of the adiabatic 
approximation is ensured.
Figure \ref{fig:c_alpha} shows
the behavior of $\Gamma_n(\alpha)$ for various values of the 
parameter $k_n$. 
Such wave function 
behaves like an oscillating profile whose frequency 
increases with occupation number $n$ and approaching the Big Bang,
while the amplitude  depends on the $\alpha$ variable only.

By this treatment, one finds \cite{Misner1969PR} the interesting 
result that  $n$ on average is constant  toward the
singularity and then if the initial state of evolution of the Universe
is classical, extrapolating backwards it maintains a semiclassical character.

\subsectionric{The quantum Universe in the Poincar\'e half-plane\label{sec:poinc}}

The Misner representation  provides a good insight 
in some qualitative aspects of the Mixmaster model  quantum 
dynamics, and allows some physical considerations on the 
evolution toward the singularity. 
In this picture the potential walls move with time, and this is 
an obstacle toward a full implementation of a Schr\"odinger 
like quantization scheme. 
These difficulties can be by-passed as soon as MCl variables are 
adopted, characterized by {static potential walls}; in particular, 
we will choose the so-called Poincar\'e variables $(u,\,v)$, defined as
\bseq
\label{c4 poincare variable}
\begin{align}
\xi   &=\displaystyle\frac{1+u+u^2+v^2}{\sqrt{3} v}\;,\\
\theta &=-\arctan\displaystyle\frac{\sqrt{3} (1+2 u)}{-1+2 u+ 2u^2+2 v^2}\;.
\end{align}
\eseq
In the vicinity of the initial singularity, we have seen that the potential 
term behaves as a potential well and as soon as we restric the dynamics 
to $\Pi_Q$, ${\mathcal H}_{ADM}= \epsilon$ and we can rewrite 
(\ref{qqw}) and  (\ref{n2}) as
\bseq
\begin{align}
	\label{c4 hamiltoniana ridotta}
	\delta S_{\Pi_Q} &= \delta\int d\tau  (p_\xi\dot\xi+p_\theta\dot\theta-H_{ADM})=0 \\
\label{c4 Hamiltoniana uv}
H_{ADM} &= v \sqrt{p_u^2+p_v^2}\;.
\end{align}
\eseq
The asymptotic dynamics is  defined in a portion $\Pi_{Q}$ of 
the Lobatchevsky plane, delimited by inequalities 
\bseq
\label{anisotropie geodetiche}
\begin{align}
	Q_1(u,v) &= -u/d  \geq 0\\
	Q_2(u,v)& = (1+u)/d \geq 0\\
	Q_3(u,v)& = (u(u+1)+v^2)/d \geq 0\\
	d &= 1+ u+ u^2 +v^2 \, , 
\end{align} 
\eseq
whose boundaries  are composed by geodesics of the plane, 
{i.e.} two vertical lines and one semicircle centred on the absolute $v=0$.

\begin{figure}[ht]
	\begin{center} 
		\includegraphics[width=8cm]{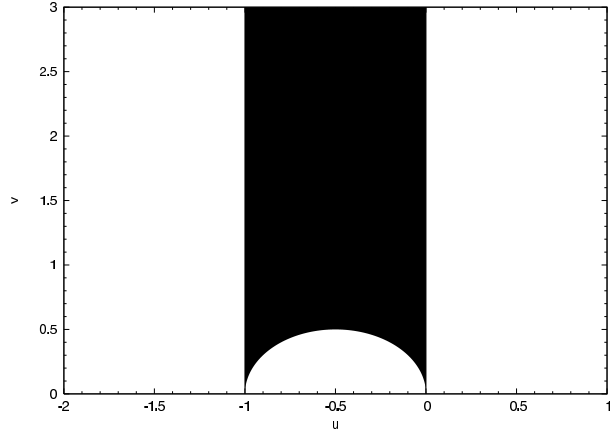} 
		\caption[The triangular domain $\Gamma_Q(u,v)$ in the Poincar\'e plane]{$\Gamma_Q(u,v)$ 
	is the available portion of the configuration space in the Poincar\'e upper half-plane. 
	It is bounded by three geodesic $u=0, u=-1$, and $(u+1/2)^2+v^2=1/4$, it has a finite 
	measure $\mu=\pi$ (it is  ``dynamically'' bounded).}
		\label{c2 fig:cuspidi2 cap51} 
	\end{center}
\end{figure}

The billiard has a finite measure, and its open region at infinity 
together with the two points on the absolute $(0,0)$ and $(-1,0)$ 
correspond to the three cuspids of the potential in Fig. \ref{DOMINIO XI THETA}.\\
It is easy to show that, in the $u,\,v$ plane, \ref{x} becomes
\begin{equation}
\label{c4 misura uv}
d\mu = \displaystyle\frac{1}{\pi}
\frac{du\,dv}{v^{2}}\displaystyle\frac{d\phi}{2\pi} \, .
\end{equation}



\subsectionric{Continuity Equation and the Liouville theorem\label{sec:celth}}

Since  the Mixmaster provides  an energy-like constant of motion 
toward the singularity, the-point Universe randomizes within 
a closed domain and  we can characterize the dynamics as a 
microcanonical ensemble, as discussed in Section \ref{sec:mcl}.

The physical properties of a stationary ensemble are described 
by a distribution function $\rho=\rho(u,v,p_u,p_v)$ \cite{Huang}, 
representing the probability of finding the system within an 
infinitesimal interval of the phase-space $(u,v,p_u,p_v)$, 
and it obeys  the continuity equation
\begin{equation}
\label{c3 eq di continuita def}
\frac{\partial (\dot{u} \rho)}{\partial u}+\frac{\partial (\dot{v} \rho)}{\partial v}+\frac{\partial (\dot{p_u} \rho)}{\partial p_u}
+\frac{\partial (\dot{p_v} \rho)}{\partial p_v}=0 \;, 
\end{equation}
were the dot denotes the time derivative and the Hamilton 
equations associated to (\ref{c4 Hamiltoniana uv}) read as 
\bseq
\label{c3 equazioni di hamilton}
\begin{align}
\dot{u} &= \frac{v^2}{\epsilon} p_u\;,   \qquad  \dot{p}_u =0\;, \\
\dot{v} &= \frac{v^2}{\epsilon } p_v\;, \qquad \dot{p}_v = -\frac{\epsilon }{v}\;.
\end{align}
\eseq
From (\ref{c3 eq di continuita def}) and  
(\ref{c3 equazioni di hamilton}) we obtain
\begin{equation}
\label{c3 eq di continuita esplicitata}
\frac{v^{2}p_{u}}{\epsilon} \DER{\rho}{u}{}+
\frac{v^{2}p_{v}}{\epsilon} \DER{\rho}{v}{}- 
\frac{\epsilon}{v}\DER{\rho}{p_{v}}{} =0\;.
\end{equation}
The continuity equation provides an appropriate representation 
 when we are sufficiently close to the initial singularity  only, 
and the infinite-potential-wall approximation works. 
Such model  corresponds to deal with the energy-like constant of 
motion, and fixes the microcanonical nature of the ensemble. 
From a dynamical point of view, this picture naturally arises 
because the Universe volume element monotonically vanishes 
(for non stationary corrections to this scheme in the MCl variables, 
see \cite{Montani2001NCB}).\\
Since we are interested to  the distribution function in the 
$(u,v)$ space, we will reduce the dependence on the momenta by 
integrating $\rho(u,v,p_u,p_v)$ in the momentum space.
Assuming $\rho$ to be a regular, vanishing at infinity in the phase-space, 
limited function, we can integrate over 
(\ref{c3 eq di continuita esplicitata}) getting  the  equation for 
$\tilde{w}=\tilde{w}(u,v;k)$  
\begin{equation}
\label{c3 eq w}
\frac{\partial \tilde{w}}{\partial u}+
\sqrt{\left(\frac{E}{C v}\right)^2-1}~
\frac{\partial \tilde{w}}{\partial v}+\frac{E^2-2 C^2v^2}{C v^2} 
\frac{\tilde{w}}{\sqrt{E^2-(C v)^2}}=0\;,
\end{equation}
where the constant $C$ appears,  due to the analytic expression of 
the HJ solution, fixed by  the initial conditions. 
However, we  deal  with a distribution function that cannot consider 
these initial conditions, and must be ruled out  from the final result.
We obtain the following solution in terms of the generic function  $g$
\begin{equation}
	\label{c3 soluzione eq w}
	\tilde{w}(u,v ;C)=
\frac{g\left(u+v\sqrt{\frac{E^2}{C^2 v^2}-1}\right)}{v \sqrt{E^2-C^2 v^2}}\;.
\end{equation}
The distribution function cannot contain the constant $C$, 
and the final result is obtained after the integration over it. 
We define the reduced distribution $w(u,v)$ as
\begin{equation}
w(u,v)\equiv\int_A \tilde{w}(u,v;k)dk\;,
\end{equation}
where the integration is taken over the classical available 
domain for $p_u\equiv C$
\begin{equation}
\label{c3 dominio k}
	A\equiv \left[-\frac{E}{v},\frac{E}{v}\right]\;.
\end{equation}
In (\ref{c4 misura uv}) we proved demonstrated   that the 
measure associated to it is the Liouville one; 
the  measure  $w_{mc}$ (after integration over the 
admissible values of $\phi$) corresponds to the case $g=\textrm{const.}$
\begin{equation}
	\label{c3 misura microcanonica}
	w_{mc}(u,v)=\int_{-\frac{E}{v}}^{\frac{E}{v}}
\frac{1}{C v^2 \sqrt{\frac{E^2}{C^2 v^2}-1}}dC=\frac{\pi}{v^2}\;.
\end{equation}
Summarizing, we have derived the generic expression of the 
distribution function fixing its form for the microcanonical 
ensemble. This choice, in view of the energy-like constant of 
motion $\mathcal{H}_{ADM}$, is appropriate to describe the 
Mixmaster system restricted to the configuration space.
This analysis reproduces in the Poincar\'e half-plane the 
same result as the stationary invariant measure described in 
Section \ref{sec:invmeasure}.


For completeness we report the explicit solution to 
(\ref{SHJE})  fo this model in the restricted domain $\Pi_Q$
as the Hamilton-Jacobi function for the point-Universe
\begin{equation}
\label{c3 S}
{\mathcal S}_0(u,v)=C u + \sqrt{\epsilon^2-C^2 v^2} 
-\epsilon \ln \left(2\frac{\epsilon+\sqrt{\epsilon^2- C^2 v^2}}{\epsilon^2 v}\right)+ D\;,
\end{equation}
where $C$  is the separation constant, and $D$ is an integration one.

\subsectionric{Schr\"{o}dinger dynamics\label{sec:schrdy}}

The Schr\"odinger quantum picture is obtained in the standard way, 
{i.e.} by promoting the classical variables to operators and imposing 
some boundary condition to the wave function. \\
The first step is achieved as
\begin{align}
\nonumber
\hat{v} &\mid\rangle =  v \mid\rangle \;,\ \qquad  \hat u \mid\rangle = u \mid\rangle\;, \\
\hat{p}_v & \rightarrow-\imath \frac{\partial}{\partial v}\;,
\qquad \hat{p}_u\rightarrow-\imath \frac{\partial}{\partial u}\;,
\qquad \hat{p}_\tau\rightarrow-\imath\frac{\partial}{\partial \tau}\;,
\end{align}
while for the second one  we will require the Dirichlet boundary conditions 
\begin{equation}
\Phi(\partial \Pi_{Q}) = 0\; .
\end{equation}
The quantum dynamics for the state function $\Phi=\Phi(u,v,\tau)$ 
obeys the Schr\"odinger equation
\begin{align}
i \frac{\partial \Phi}{\partial \tau}&=\hat{H}_{ADM}\Phi =
\sqrt{-v^2\frac{\partial^2}{\partial u^2}
-v^{2-a}\frac{\partial}{\partial v}\left(v^a 
\frac{\partial}{\partial v}\right)} ~\Phi \;.
\label{c3 scroedinger}
\end{align}
Here we have adopted a generic operator-ordering for the 
position and momentum parametrized by the constant 
$a$\cite{KT90}, as soon as we have no indication 
on it providing a first  problem. 
The other one is linked to the multi-time approach, {i.e.} to the 
non-locality of the Hamiltonian operator: when solving the 
Hamilton constraint with respect to one of the momenta, 
the ADM Hamiltonian  contains a square root  and consequently 
it might define a non-local dynamics.

The question of the correct operator-ordering is addressed the next 
Section comparing the classic evolution versus the WKB limit of the 
quantum-dynamics and requiring the overlapping of the two. 
On the other hand,  we will assume  the operators 
$\hat{\mathcal H}_{ADM}$ 
and $\hat{\mathcal H}_{ADM}^2$ having the same set of eigenfunctions 
with eigenvalues $E$ and $E^2$, respectively.\footnote{The problems 
discussed  by \cite{Kheyfets2006CQG} do not arise here because 
in the domain $\Pi_Q$  the  ADM Hamiltonian has a positive sign 
(the potential vanishes asymptotically).} 

Under these assumptions, we will solve the eigenvalue problem for 
the squared ADM Hamiltonian given by 
\begin{equation}
\label{c3 hquadro}
\hat {\mathcal H}^2 \Psi = \left[-v^2\frac{\partial^2}{\partial u^2}
-v^{2-a}\frac{\partial}{\partial v}\left(v^a \frac{\partial}{\partial v}
\right)\right]\Psi = E^2 \Psi\;,
\end{equation}
where $\Psi = \Psi(u,v,E)$.

\subsectionric{Semiclassical WKB limit\label{sec:semiwkb}}

In order to study the WKB limit of equation (\ref{c3 hquadro}), 
we separate the wave function into its phase and 
modulus\footnote{In this Section we restore $\hbar$ in the notation 
because we deal with the semiclassical limit.}
\begin{equation}
\label{c3 forma funzionale psi}
\Psi(u,v,E)=\sqrt{r(u,v,E)} e^{\imath \sigma(u,v,E)/\hbar}\;.
\end{equation}
In Eq. (\ref{c3 forma funzionale psi}) the function $r(u,v)$ 
represents the probability density, and the quasi-classical 
regime appears in the limit  $\hbar\to 0$; substituting 
(\ref{c3 forma funzionale psi}) in (\ref{c3 hquadro}) 
and retaining only the lowest order in $\hbar$, we obtain the system
\bseq
\label{c3 hj system}
\begin{align}
\label{c3 hj systema}
&v^2 \left[\left( \frac{\partial \sigma}{\partial u}\right)^2
+\left(\frac{\partial \sigma}{\partial v}\right)^2\right]=E^2\;, \\
\label{c3 hj systemb}
&\frac{\partial r}{\partial u} \frac{\partial \sigma}{\partial u} +
	\frac{\partial r}{\partial v} \frac{\partial \sigma}{\partial v}
	+r\left(\frac{a}{v} \frac{\partial \sigma}{\partial v} +
	\frac{\partial^2 \sigma}{\partial v^2} + 
	\frac{\partial^2 \sigma}{\partial u^2}\right)=0\;.
\end{align}
\eseq
In view of the HJ equation and of Hamiltonian (\ref{c4 Hamiltoniana uv}), 
we can identify the phase $\sigma$ as the functional $S_0$.\\
Taking (\ref{c3 S}) into account, Eq. (\ref{c3 hj systemb}) 
reduces to
\begin{equation}
\label{c3 eq per r}
C \frac{\partial r}{\partial u}+\sqrt{\left(\frac{E}{v}\right)^2-C^2} 
\frac{\partial r}{\partial v}+\frac{a(E^2-C^2 v^2)-E^2}{v^2 \sqrt{E^2-C^2 v^2}} r=0\;.
\end{equation}
Comparing (\ref{c3 eq per r}) with (\ref{c3 eq w}), we see 
that they coincide  for  $a=2$ only\cite{BeniniMontani2007CQG}.
This correspondence is expectable for a suitable  choice of the 
configurational variables; 
however, it is remarkable that it arises for the chosen  
operator-ordering only. 
Here arises  the importance of the correspondence, wich fixes a 
particular quantum dynamics for the system.\\
Summarizing, we have demonstrated from our study that it is 
possible to get a WKB correspondence between the quasi-classical 
regime and the ensemble dynamics in the configuration space, and 
we provided the operator-ordering when quantizing the  Mixmaster model  
\begin{equation}
	\label{c3 ordinamento operatoriale}
	\hat{v}^2 \hat{p}_v^2\,\rightarrow\,-\hbar^2 
	\frac{\partial}{\partial v}\left( v^2\frac{\partial}{\partial v}\right)\,.
\end{equation}

\subsectionric{The Spectrum of the Mixmaster\label{sec:mixsp}}

\subsubsectionric{Eigenfunctions and the vacuum state\label{sec:mixeig}}

\begin{figure}[ht]
\begin{center} 
\includegraphics[width=8.0cm]{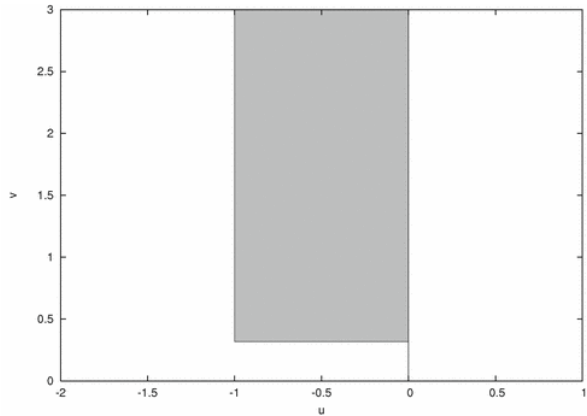} 
\caption[The approximate domain for the quantum dynamics]{The approximate 
domain where we impose the boundary conditions.
The choice $v=1/\pi$ for the straight line preserves the measure $\mu=\pi$ (from \reffcite{BeniniMontani2007CQG}).}
\label{c3 pot approx}
\end{center}
\end{figure}

Once fixed the operator ordering $a=2$, the eigenvalue equation 
(\ref{c3 hquadro})  rewrites as
 \begin{equation}
	\left[v^2 \frac{\partial^2}{\partial u^2}  
+v^2 \frac{\partial^2}{\partial v^2} +2 v \frac{\partial}{\partial v}
+\left({E}\right)^2\right] \Psi(u,v,E)=0\;.
\label{c3 autoval} 
\end{equation}
By redefining $\Psi(u,v,E) =\psi(u,v,E)/v$, we can reduce 
(\ref{c3 autoval}) to the eigenvalue problem for the Laplace-Beltrami 
operator in the Poincar\'e plane \cite{Terras}
\begin{equation}
	\label{c3 lb equation}
	\nabla_{LB} \psi(u,v,E)\equiv v^2 
	\left(\frac{\partial^2}{\partial u^2}+
	\frac{\partial^2}{\partial v^2} \right) \psi(u,v,E)=E_s\psi(u,v,E)\;,
\end{equation}
which is central in the harmonic analysis on symmetric spaces and has been widely investigated in terms of its invariance under $SL(2,\mathbb{C})$ and 
its eigenstates and eigenvalues are
\begin{align}
	\label{c3 forma generica delle autofunzioni}
 &\psi_s(u,v)= a v^s+ b v^{1-s}+ 
 \sqrt{v}\sum_{n\neq0}a_n K_{s-1/2}(2\pi |n| v) e^{2\pi {\mathit i} n u} \, ,
\qquad  a,b,a_n \in \mathbb{C}\;, \\
 	\label{c3 condizioni al bordo}
  &\nabla_{LB} \psi_s(u,v)=s(s-1) \psi_s(u,v)\;,
\end{align}
where  $K_{s-\nicefrac{1}{2}}(2\pi n v)$ are the modified Bessel 
functions of the third kind\cite{abramowitz} and $s$ denotes the 
index of the eigenfunction.
This is a continuous spectrum and the sum runs over every real value of $n$.\\
The eigenfunctions for our model, then, read as
\begin{equation}
	\label{c3 autofunzione bianchi}
	\Psi(u,v,E)=a v^{s-1}+ b v^{-s} +\sum_{n\neq 0} a_n 
	\frac{K_{s-1/2}(2\pi|n|v)}{\sqrt{v}} e^{2 \pi{\mathit i} n u}\;,
\end{equation}
with eigenvalues
\begin{equation}
	\label{c3 autovalore bianchi}
	E^2=s(1-s)\;.
\end{equation}
To impose Dirichlet boundary conditions for the wave functions 
we will require a vanishing behaviour on the edges of the geodesic 
triangle of Fig. \ref{c2 fig:cuspidi2 cap51}. 
Let us  approximate\cite{BeniniMontani2007CQG} the domain with the  one in Fig. \ref{c3 pot approx}; 
the value of  the horizontal line $v=1/\pi$ provides the same measure 
for the exact as well as for  the approximate domain
\begin{equation}
\label{c3 equiv misura dei domini}
\int_{\Pi_{Q}}\frac{du dv}{v^{2}} = \int_{\text{\tiny  Approx domain}}\frac{du dv}{v^{2}} = \pi\;.
\end{equation}

The difficulty to deal with the exact boundary conditions relies in the 
sophisticated number theory is linked to these functions while the circle, 
that bounds from below the domain, furthermore mixes solutions with different 
indices $s$.\\
The Laplace-Beltrami operator and the exact boundary conditions are 
invariant under parity transformation $u \to -u-1$; 
however, the full symmetry group is $C_{3v}$, and this can be seen in 
the disk representation of the Lobachewsky plane. 
$C_{3v}$ has two one-dimensional irreducible representations and 
one two-dimensional representation\cite{Graham1991PRA1,Graham1991PRA2}. 

The eigenstates transforming according to one of the two-dimensional 
representations are twofold degenerate, while the others are  non-degenerate. 
The latters can be divided in two classes, either satisfying  Neumann 
boundary conditions, or Dirichlet ones. 
We  focus our attention to second case. 
The choice of the line $v=1/\pi$  approximates symmetry lines of 
the original billiard and corresponds to one-dimensional 
irreducible representations\cite{Puzio1994CQG,Graham1991PRA1,Graham1991PRA2}.\\
The conditions on the vertical lines $u=0$, $u=-1$ require to 
disregard the first two terms in (\ref{c3 autofunzione bianchi}); 
furthermore, we get the condition on the last term
\begin{equation}
\sum_{n\neq 0} e^{2\pi \imath  n u} \to \sum^{\infty}_{n=1} \sin(\pi n u)\;,
\end{equation}
for  integer  $n$. 
As soon as we restrict to only one of the two one-dimensional 
representations, we get
\begin{equation}
\sum_{n\neq 0} e^{2\pi \imath n u} \to \sum^{\infty}_{n=1} \sin(2\pi n u)\;,
\end{equation}
 while the condition on the horizontal line implies
\begin{equation}
	\label{c3 relazione sulle k}
	\sum_{n>0}a_n K_{s-1/2}(2n) \sin (2n\pi u)=0\;,\quad \forall u \in [-1,0]\;,
\end{equation}
which in general is satisfied by requiring $K_{s-\nicefrac{1}{2}}(2n)=0$ 
only, for every $n$. 
This last condition, together with the form of the spectrum 
(\ref{c3 autovalore bianchi}), ensure the discreteness of the energy 
levels, thanks to discreteness of the zeros of the Bessel functions. \\
The functions $K_\nu(x)$ are real and positive for real argument and real 
index, therefore the index must be imaginary, { i.e.}  
$s=\displaystyle\nicefrac{1}{2}+\imath t$. 
In this case, these functions have (only) real zeros, and the corresponding eigenvalues turn out to be real and positive. 

\begin{figure}[ht]
	\begin{center}
	\includegraphics[width=8cm]{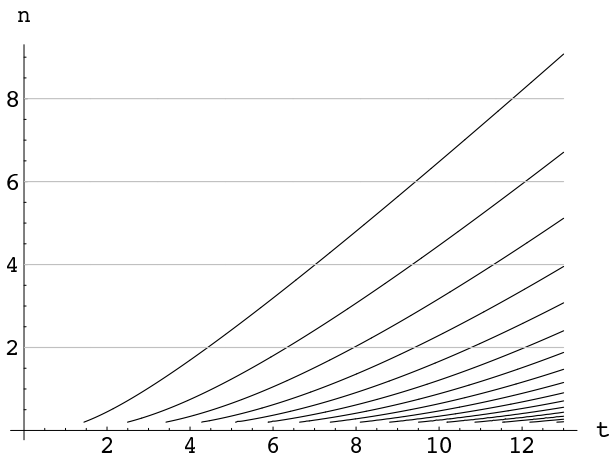} 
      	\end{center}
\caption[Roots of the eq. $K_{i t}(n)=0$]{The intersections between the straight lines and the curves represents the roots of the equation $K_{i t}(n)=0$, where $K$ is the modified Bessel function (from \reffcite{BeniniMontani2007CQG})..}
\label{c3 zeri}
\end{figure}

\begin{equation}
	\label{c3 spettro finale}
	E^2=t^2+\frac{1}{4}\,.
\end{equation}
The  eigenfunctions (\ref{c3 autofunzione bianchi}) exponentially 
vanish as infinite values of $v$ are approached.\\
The conditions (\ref{c3 relazione sulle k}) cannot be analytically 
solved for all the values of $n$ and $t$, and the roots must be numerically 
worked out for each $n$. 
There are several results on their distribution that allow one to 
find at least the first levels: a theorem\cite{Palm} on the zeros of 
these functions states that 
$K_{\imath \nu}(\nu x)=0$ $\Leftrightarrow$ $0<x<1$; 
furthermore,  the energy levels (\ref{c3 spettro finale}) monotonically 
depend on the values of the zeros. 
Thus,  one can   search the lowest levels by solving 
Eq. (\ref{c3 relazione sulle k}) 
for the firsts $n$; in the next Section we will discuss some 
properties of the spectrum, while now we will discuss the 
ground state only.

\begin{figure}[ht]
	\begin{minipage}[b]{0.45\textwidth}
		\centering 
			\includegraphics[width=0.9\textwidth]{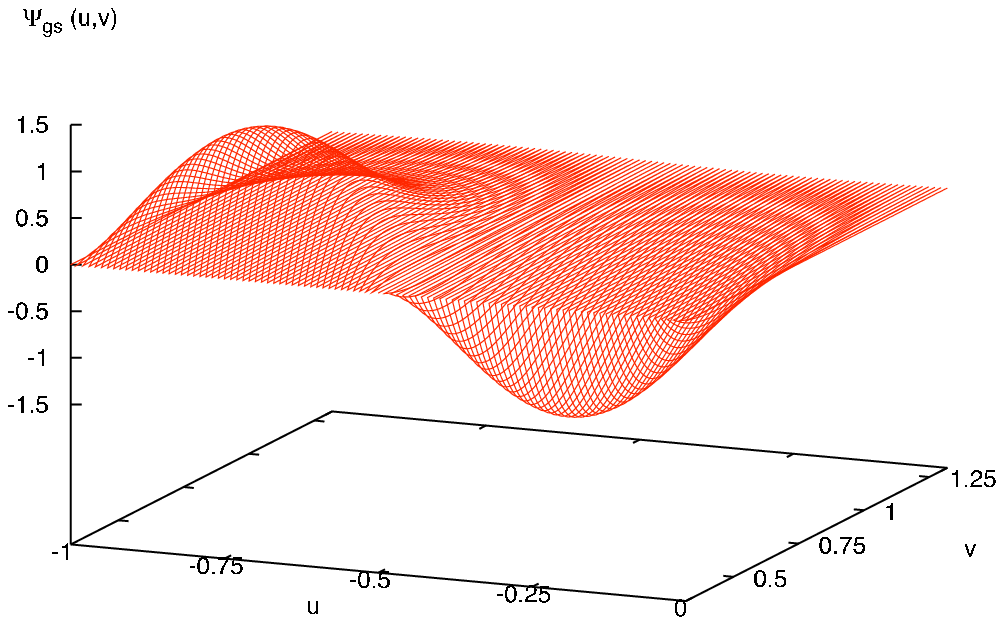} 
			\caption[The wave function of the ground-state]{The ground state wave function of the Mixmaster model is sketched (from \reffcite{BeniniMontani2007CQG}).}
			\label{c3 groundstate}
			\vspace{4mm}
	\end{minipage}%
	\begin{minipage}[b]{0.1\textwidth}
	\phantom{stron}
	\end{minipage}%
	\begin{minipage}[b]{0.45\textwidth}
			\centering 
			\includegraphics[width=0.9\textwidth]{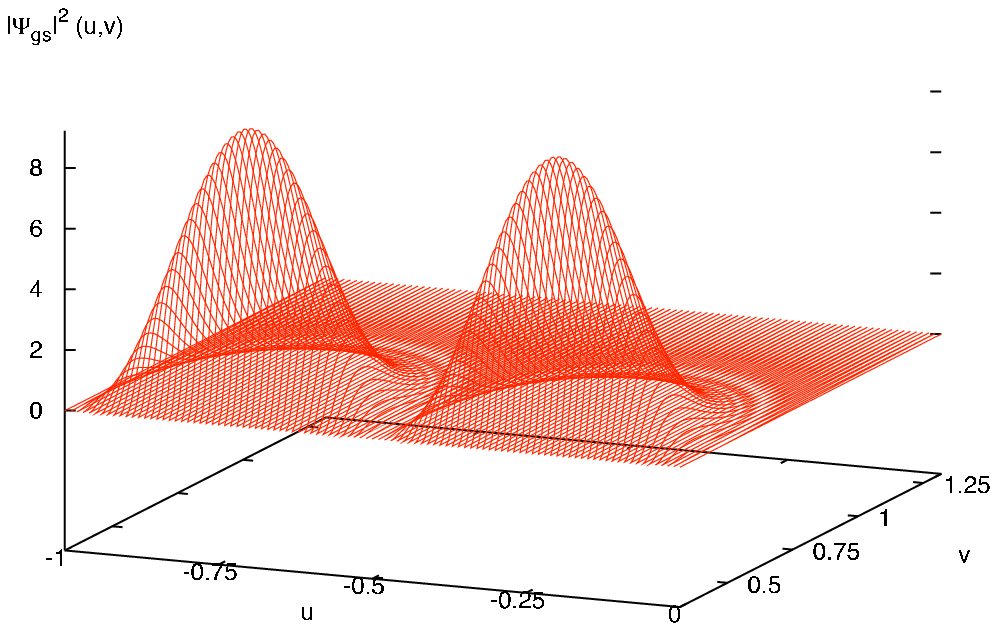}
			\caption[The probability distribution associated to the ground state]{The probability distribution in correspondence to the ground state of the theory, according to the boundary conditions (from \reffcite{BeniniMontani2007CQG}).}
			\label{c3 probabilita}
	\end{minipage}
\end{figure}

A minimum energy exists, as follows from the quadratic structure 
of the spectrum and from the properties of the Bessel zeros, and its value is 
 $E^{2}_{0} = 19.831 \hbar^{2} $, 
and correspondingly the eigenfunction is  plotted in Fig. \ref{c3 groundstate}, 
together with the probability distribution in Fig. \ref{c3 probabilita}; 
the eigenstate is normalized through the normalization constant $N=739.466$.
The existence of such a ground state has been numerically derived, 
but it can be inferred on the basis of general considerations about 
the Hamiltonian structure; the Hamiltonian, indeed, contains a term 
$\widehat{v^2}\widehat{p^2_v}$ which has positive definite spectrum 
and does not admit vanishing eigenvalues.

\subsubsectionric{Distribution of the Energy Levels\label{sec:mixnrg}}

In Table \ref{c3 primi livelli} we report the first ten ``energy'' 
levels evaluated solving (\ref{c3 relazione sulle k}).\footnote{
For a detailed numerical investigation of the energy spectrum 
of the standard Laplace-Beltrami operator, especially with 
respect to the high-energy levels,   
see\cite{Graham1991PRA1,Graham1991PRA2}, where the effects on the 
level spacing of deforming the circular boundary condition towards 
the straight line are numerically analyzed. }
%

\begin{table}[ht]
\tbl{The first ten energy eigenvalues, ordered from the lowest one.}
{\begin{minipage}{0.8\textwidth}
\begin{center}
\begin{tabular}{@{}c@{}}
	\toprule
	 \textbf{\mbox{${E}^2=t^2+\frac{1}{4}$}} \\
    \colrule
   			\phantom{1}19.831\\
   			\phantom{1}40.357\\
   			\phantom{1}49.474\\
   			\phantom{1}63.405\\
			\phantom{1}87.729\\
			\phantom{1}89.250\\
			116.329\\
			128.234\\
			138.739\\
			146.080\\
		\botrule
\end{tabular}
\end{center}
\end{minipage}
\label{c3 primi livelli}
}
\end{table}

To study the distribution of the highest energy levels, 
we need to consider  the asymptotic behavior of the 
zeros for the modified Bessel functions of the third kind.
We will discuss the asymptotic regions of the $(t,n)$ plane 
 in the two cases $t\gg n$ and  $t\simeq n\gg1$.

\begin{enumerate}
\renewcommand{\labelenumi}{({\it\roman{enumi}})}

\item 
	 For $t\gg n$, the Bessel functions admit the  representation
	\begin{align}
	K_{i t}(n)=\frac{\sqrt{2\pi}e^{-t\pi/2}}{(t^2-n^2)^{1/4}}
	& \left[ \sin a\sum_{k=0}^\infty \frac{(-1)^k}{t^{2k}} 
	u_{2k}\left(\frac{1}{\sqrt{1-p^2}}\right)+ \right. \nonumber \\
	+&  \left. \cos a \sum_{k=0}^\infty \frac{(-1)^k}{t^{2k+1}} 
	u_{2k+1}\left(\frac{1}{\sqrt{1-p^2}}\right)\right] \, , 
	\label{c3 sviluppo asintotico tutto a destra}
	\end{align}
where $a=\nicefrac{\pi}{4}-\sqrt{t^2-n^2}+t \textrm{acosh}(t/n)$, 
$p\equiv n/t$ and $u_k$ are the  polynomials
\begin{equation}
	\left\{ \begin{array}{ll}
	u_0(t)=1\;,\\
	u_{k+1}(t)=\displaystyle\frac{1}{2} t^2(1-t^2) u_k'(t)
	+\frac{1}{8}\int_0^1(1-5t^2)u_k(t)dt\;.
	\end{array}\right.
\end{equation}
Retaining in the expression above only  terms of order 
$\mathsf{o} (\nicefrac{n}{t})$, the zeros are fixed by the  relation
\begin{equation}
	\label{c3 arcotangente}
	\sin\left[\frac{\pi}{4}-t+t \left(\log(2)-\log(p)\right)\right]
	-\frac{1}{12 t} \cos\left[\frac{\pi}{4}-t+t \left(\ln(2)-\ln(p)\right)\right]=0\;.
\end{equation}
In the limit $n/t\ll 1$, Eq. (\ref{c3 arcotangente}) can be recast as 
\begin{equation}
	\label{c3 arcotangente2}
	t \log(t/n)=l \pi\;\Rightarrow\; 	
	t= \frac{l \pi}{\mathrm{productlog}\left(\frac{l \pi}{n}\right)}\;,
\end{equation}
where productlog(z) is a generalized function giving the solution 
of the equation $z=w e^w$ and, for real and positive domain, 
is a monotonic function. 
In (\ref{c3 arcotangente2}) $l$ is an integer number much greater
 than 1 in order to verify  $n/t\ll 1$.

\item 
 In case the difference between $2n$ and $t$ is $\mathsf{o}(n^{1/3})$ 
for $t,n\gg 1$, we can evaluate the first zeros $k_{s, \nu}$ 
by the  relations\cite{Balogh}
\begin{equation}
\label{c3 9 balogh}
k_{s,\nu}\sim \nu +\sum_{r=0}^{\infty}(-1)^{r} s_{r}(a_{s}) \left(\displaystyle\frac{\nu}{2}\right)^{-(2r-1)/3}\;,
\end{equation}
where $a_{s}$ is the $s$-th zero of Ai$\left((2/z)^{1/3}\right)$, 
$Ai(x)$ is the Airy function\cite{arfken} and $s_{i}$ are some 
polynomials. 
From this expansion it results that, to lowest order
\begin{equation}
  \label{c3 zeri asintotici}
  t= 2n + 0.030 n^{1/3}\,.
\end{equation}
Eq. (\ref{c3 zeri asintotici}) provides the lowest zero 
(and therefore the energy) for a fixed  value of $n$ and also
the relation for the eigenvalues for high occupation numbers
\begin{equation}
\label{c3 alti numeri}
 {E}^2\sim 4 n^2 + 0.12 n^{4/3} \;.
\end{equation}
\end{enumerate}

Let us  to discuss {i.e.} the completeness of the spectrum and the 
definition  of a scalar product.\\
The problem of completeness can be faced by studying firstly the sine 
functions  and then  the Bessel ones.
On the interval $[-1,0]$, the set $\sin(2\pi n u)$ is not a complete 
basis, but  as soon as we request the wave function to satisfy the 
symmetry of the problem, it becomes complete.\\
Let us take a value $n > 0$, thus  functions (\ref{c3 autofunzione bianchi}) 
have the form $\Phi(u, v) = \sin(2\pi n u)g(v)$, 
which substituted in (\ref{c3 autoval}) provides 
$v^{2} \left(\partial_{v}^{2}+ (2\pi n)^{2})\right) g(v) = s(1 - s)g(v)$, 
whose  solutions are exactly the Bessel functions.\cite{Puzio1994CQG}\\
This property togheter with the condition on the line $v=1/\pi$ 
form a Sturm-Liouville problem and we deal with a complete of eigenfunctions.

Therefore, such eigenfunctions define a space of functions where we can 
introduce a scalar product,  naturally induced by the metric of the 
Poincar\'e plane\cite{Puzio1994CQG}
\begin{equation}
\label{c3 prodotto scalare}
(\psi,\phi) = \int \psi(x,y)\phi^{*}(x,y)\displaystyle\frac{dx dy}{y^{2}} \;,
\end{equation}
where $^{*}$ denotes complex conjugation.


Now we briefly discuss  if the presence of a non-local function, 
like the square-root of a differential operator, can give rise 
to non-local phenomena. 
Following the work of Puzio\cite{Puzio1994CQG},  
a wavepacket which is non-zero in a finite region of the domain 
($v < M$) and far from  infinity fails to run  to infinity 
in a finite time, {i.e.}, the probability $P(v>M)$ to find the 
packet far away exponentially vanishes.
\begin{align}
P(v>M)&=\int_{-1}^{0}\int_{M}^{\infty}|v \sqrt{\partial_{u}^{2}
	+\partial_{v}^{2}}\Psi(u,v)|^{2}\frac{d v\, d u}{v^{2}}<\nonumber\\
&< 4\, M^{2}\sqrt{\frac{\pi}{2}}\, 
\left(\text{sup }\Psi\right)^{2}\int_{-1}^{0}\int_{M}^{\infty}e^{-2v} 
	\frac{d v\, d u}{v^{2}}=\nonumber\\
&= 4\, M^{2}\sqrt{\frac{\pi}{2}}\,\left(\text{sup }\Psi\right)^{2} \left(\frac{e^{-2M}}{M}+\text{Ei}(-2M)\right)<\nonumber\\
&< 4\sqrt{\frac{\pi}{2}}\,\left(\text{sup }\Psi\right)^{2}\,M e^{-2M}
\end{align}
where sup$(\Psi)$ is the maximum value of the wavepacket in the domain $v < M$ 
and $\text{Ei}(z)=-\int_{-z}^{\infty}e^{-t}/t dt$ is the exponential integral function.
We can thus  conclude that nevertheless the square root is a 
non-local function,
non-local phenomena don't appear (like the case of a wavepacket starting 
from a localized zone and falling out to infinity).

\subsectionric{Basic Elements of Loop Quantum Gravity\label{sec:blqg}}

In this Section we will review some of the basic aspects of Loop Quantum 
Gravity (LQG) in a pedagogical manner for a non-expert reader. 
For a detailed and deeper approach, we refer to the textbooks \reffcite{rovelli2004qg,Thi}
and to some dedicated reviews \reffcite{Ro97,Thirev,ashtekar2004,Smo}, 
while for a critical point of view  we recommend \reffcite{Nic}. 

LQG is an attempt to rigorously quantize General Relativity in a background 
independent manner, trying to define a quantum field theory just on a 
differential manifold $M$ and not on a background space-time $(M,g_0)$, 
i.e. independently of the choice of a fixed background metric $g_0$. 
However, a way to give a formulation of the theory independently of 
the possible topological changes of the underling manifold  has not been 
achieved so far in LQG.

In order to see how fundamental fixed background metric is, 
let us consider ordinary quantum field theory, whose 
whole framework, namely the Wightman axioms\cite{Haag}, 
breaks down as soon as the metric is no longer  considered fixed
but with a  dynamical structure according to General Relativity. 
In fact we could only  construct a rigorous quantum theory  in the Minkowski 
background\footnote{The only quantum fields in four 
	dimensions fully understood
	to-date are the free, or perturbatively interacting, 
	fields.}, 
which implies a preferred notion of causality (locality) and a 
symmetry group, i.e. the Poincar$\acute e$ one.

LQG works in a Hamiltonian 
approach\footnote{For the covariant approach to  LQG, i.e. the 
	spin foams models, see \reffcite{O01,Per}.} 
and it is able to overcome some of the problems of the older 
geometro-dynamical approach 
(described in the previous Sections) using new conceptual and technical ingredients. 

Let us  rewrite GR in the form of a $SU(2)$ Yang Mills theory, so that 
the phase space is 
endowed with coordinates \footnote{Here and in the following paragraphs on 
	the LQG/C 
	we will adopt 
$i,j,k,...=1,2,3$ as $SU(2)$ indices.} 
by a $SU(2)$ connection $A^j_{\alpha}$ and by 
an electric field $E^{\alpha}_j$. 
These are related to the ADM variables $(h,K)$ by
\be
\label{varAE}
A^j_{\alpha}=\omega^j_{\alpha}+\gamma K_{\alpha\beta}e^{\beta j}, 
\qquad E^{\alpha}_j=\sqrt{h} e^{\alpha}_j,
\ee
where $\omega^j_{\alpha}= \nicefrac{1}{2} \epsilon_{\alpha\beta\gamma} \omega^j_{\beta\gamma}$ 
is the spin connection, $K_{\alpha\beta}$ the extrinsic curvature 
and $\gamma>0$ 
the so-called Immirzi parameter which does not affect the classical 
dynamics but brings to inequivalent quantum 
predictions \cite{RovThi}. 
The symplectic geometry is determined by the only non-trivial Poisson brackets
\be
\{A^i_{\alpha}(x),E^{\beta}_j(x')\}=\kappa\gamma\delta^i_j\delta^{\beta}_{\alpha}\delta^3(x-x') \, .
\ee
Let us rewrite the constraints (\ref{seccon}) in terms of these new variables, 
which in the  phase phase satisfy  the usual spatial diffeomorphism constraint
\bseq
\be
\label{scacon0}
H_{\alpha}=F^i_{\alpha\beta}E^{\beta}_i\approx0 \, , 
\ee
where  $F=dA+A\wedge A$ is the curvature of $A$,  and the scalar one 
\be\label{scacon}
H=\f{\epsilon_{ijk}E^{\alpha}_jE^{\beta}_k}{\sqrt{|\det(E)|}}
\left[F^i_{\alpha\beta} -\left(1+\gamma^2 \right) 
\epsilon_{imn}K^m_{\alpha}K^n_{\beta} \right]\approx0 \, ,
\ee
\eseq
where $\gamma K^i_{\alpha}=A^i_{\alpha}-\omega^i_{\alpha}$.
In the connection formalism, 
with respect to the metric approach, we have the additional 
 Gauss constraint
\be
\label{scagau}
\mathcal{G}_i=\mathcal{D}_{\alpha}E^{\alpha}_i=
\p_{\alpha}E^{\alpha}_i+\epsilon_{ijk}A^j_{\alpha}E^{\alpha}_k\approx 0
\ee
which gets rid of the $SU(2)$ degrees of freedom. 
Since  an observable is a gauge invariant function, 
it has to commute with all constraints.  

As a  second step we have to  smear \cite{rovelli2004qg,Thi} the fields 
$A^i_\alpha$ and $E^\alpha_j$ 
to overcome some of the problems arising in the direct approach. 
In fact, we switch from the connections $A^i_\alpha$ to the {\it holonomies} 
as the basic variables. 
Given a  curve on $\Sigma$, i.e. an edge $\ell$, 
a holonomy is defined as
\be
h_{\ell}[A]=\mathcal P\exp\left(\int_{\ell} A\right)
	=\mathcal P\exp\left(\int_{\ell} A^i_{\alpha}\tau_idx^{\alpha}\right) \, ,
\ee 
where $\mathcal P$ denotes the path order and the $\tau_i=\sigma_i/2i$ 
form a basis of $SU(2)$ and $\sigma_i$ are the Pauli matrices. 
The holonomies $h_{\ell}$ are elements of $SU(2)$ 
and define the parallel transport 
of the connection $A^i_\alpha$ along the edge $\ell$. 
They are  gauge invariant  and have a one-dimensional support 
(rather than all $\Sigma$). 
The  variable conjugate to the $h_{\ell}[A]$ is the {\it flux} vector
\be
P^i_S[E]=\int_S\star E^i=\int_S\epsilon_{\alpha\beta\gamma}E^{i\alpha}dx^{\beta}\wedge dx^\gamma
\ee
through any two-dimensional surface $S\subset\Sigma$, whose 
support  is a two-dimensional submanifold of $\Sigma$. 
In order to compute the Poisson brackets between these variables, 
let us consider an edge $\ell$ that intersects a surface 
$S$ in one point, 
thus obtaining 
\be\label{pbhf}
\{h_{\ell}[A],P^i_S[E]\} = 
\f{\kappa\gamma}4  \alpha(e,S)\tau^i h_{\ell}[A] \, ,
\ee
where $\alpha=0$ in the case of the edge not intersecting  the surface
and $\alpha = \pm 1$  
when the orientation of the edge and surface are the same or the opposite, 
respectively. 
We note that the  commutation relation (\ref{pbhf}) is non-canonical 
for the presence of $h_{\ell}[A]$.

The quantum kinematics can be constructed promoting such variables to 
quantum operators obeying  appropriate commutation relations. 
The essential feature of LQG is promoting the holonomies 
$h_{\ell}[A]$ to operators 
rather than the connections $A^i_\alpha$ themselves. 

Let us investigate the kinematical Hilbert space $\mathcal H_{kin}$ of LQG, 
i.e. that of the spin networks. 
These are  defined as  a graph $\Gamma$ consisting of a finite 
number of edges and vertices with a given collection of spin quantum numbers 
$j_{\ell}=\nicefrac{1}{2},1, \nicefrac{3}{2}, \ldots$, 
one for each edge, and of other quantum 
numbers $I$, the intertwiners, one for each vertex. 
The wave function of the spin network, the so-called cylindric function, 
can thus  be written as
\be\label{cylfun}
\Psi_{\Gamma,\psi}[A]=\psi(h_{{\ell}_1}[A],...,h_{{\ell}_n}[A]).
\ee
If the wave functions $\psi$ are $SU(2)$-gauge invariant, they satisfy the 
Gauss constraint and {\it viceversa}. 
These functions are called cylindrical because they have a one-dimensional 
support, i.e. they probe the gauge connection on one-dimensional networks only. 
When promoted to quantum operators, the fundamental variables of the theory 
$(h_{\ell}[A], P^i_S[E])$ act on the wave functions (\ref{cylfun}) as
\bseq
\begin{align}
\hat h_{\ell}[A]\Psi_{\Gamma,\psi}(A) &=  h_{\ell}[A]\Psi_{\Gamma,\psi}(A)  \\
\hat P^i_S[E]\Psi_{\Gamma,\psi}(A)          &=  i\{P^i_S[E],\Psi_{\Gamma,\psi}(A)\} \, .
\end{align}
\eseq
A key point in the LQG approach to the quantum gravity problem is 
the kinematical scalar product between two cylindric functions, since 
the main results of  discretization of areas and volumes are based on it, 
which is  defined as
\be\label{kscpr}
\langle\Psi_\Gamma\vert\Psi_{\Gamma'}\rangle= 
	\bcas 
		0 & \textrm{if}\, \Gamma\neq\Gamma' \\ 
		\int\prod_{{\ell}\in\Gamma}dh_{\ell} 
			\bar\psi_\Gamma(h_{{\ell}_1},...) \psi_{\Gamma'}(h_{{\ell}_1},...) & 
			\textrm{if}\, \Gamma=\Gamma' 	, ,
	\ecas
\ee
where the integrals $\int h_{\ell}$ are performed with the $SU(2)$ 
Haar measure\cite{Stanco}. 
The definition is based on a strong uniqueness theorem \cite{LOST}. 
The inner product vanishes if the graphs $\Gamma$ and $\Gamma'$ do not 
coincide and it is invariant under spatial diffeomorphisms, even if the states 
$\Psi_1$ and $\Psi_2$ themselves are not, because the information that the 
two graphs coincide is  diffeomorphism invariant. 
The information on the position of the graphs, carried by the wave function,  
disappears in the scalar product (\ref{kscpr}). 

Let us stress three relevant aspects: 

\begin{enumerate}
\renewcommand{\labelenumi}{({\it\roman{enumi}})}

\item 
in contrast with the lattice gauge theory, where all 
quantities depend on the scale parameter, 
the ``discretuum'' of LQG is built via the construction of the 
scalar product (\ref{kscpr}). 
This feature is responsible for the  failure of the Stone-von Neumann 
theorem \cite{Haag}:  LQG and  WDW  lead to different results, since 
the two quantizations procedures are not equivalent. 

\item

The obtained Hilbert space is not-separable, as it does not 
admit a countable basis, because the set of all spin networks is 
not numerable and two non-coincident spin networks are orthogonal with 
respect to (\ref{kscpr}). 

\item 

States with  negative norm  are absent without imposing the 
constraints (\ref{scacon0}), (\ref{scacon}), (\ref{scagau}), 
in contrast with the usual gauge theories, where the negative norm states 
can be eliminated only after imposing the constraints. 
However, this kinematical Hilbert space is not relevant to solve the 
quantum constraints of LQG. 
Anyway, the area and volume operators (which we do  not discuss here) 
are computed at this level and the discretization of the corresponding 
spectrum is related to this  Hilbert 
space properties.\footnote{The spectrum of the area operator $\hat A_S$ is 
computed applying it to a wave function $\Psi$ defined in a given graph. 
Such a graph is refined in a way that the elementary surface $S_I$ is 
pierced only by the  edge of the network, therefore obtaining 
\beq
\label{fnote}
\hat A_S\Psi = \kappa\gamma\sum_p\sqrt{j_p(j_p+1)}\Psi \, .
\eeq
The minimal accessible length appears to be of order of the Planck one  
and the spectrum discreteness  is related 
to the structure of the spin networks. 
In fact, a further refinement of the area operator 
does not contribute to the result.}

\end{enumerate}

The last step is to  impose the constraint at a quantum level to  obtain 
the physical states. 
These constraints are implemented  firstly to express them in terms 
of holonomies and fluxes, and secondly to  investigate their properties. 

The Gauss constraint  (\ref{scagau}) imposes several  restrictions on the quantum 
numbers $I$ of the cylindric functions. 
The diffeomorphism constraint (\ref{scacon0}) is more difficult to deal with and will not  
be treated in terms of an operatorial one. 
In fact, a diffeomorphism generator does not exist as an operator 
and diffeomorphism invariant states 
(with the exception of the empty spin network state $\Psi=1$) do not 
exist in $\mathcal H_{kin}$. 
This constraint is imposed implementing a ``group average method'' 
in a  way properly adapted to the scalar product (\ref{kscpr})\cite{marolf}. 
A key step arises from solving the constraints on a larger space, 
i.e. the dual  $Cyl^\ast$ of the cylindric functions $Cyl$, such that 
\be
Cyl\subset\mathcal H_{kin}\subset Cyl^\ast \, , 
\ee  
which  is the so-called Gel'fand triple\cite{Haag}. 
An important feature of the new Hilbert space $\mathcal H_{diff}$ 
is of being separable, differently from $\mathcal H_{kin}$. 
The final challenge will be finding  a space annihilated by all the 
constraints and defining a physical inner product which yields 
the final physical Hilbert space.

As mentioned in the WDW formalism, the main problem of all quantum 
gravity theories is to impose at a quantum level the scalar constraint. 
This difficulty appears also in the LQG approach and will be reflected 
also in the minisuperspace theory, i.e. the Loop Quantum Cosmology (LQC). 

Let us rewrite the constraint (\ref{scacon}) in terms of variables 
corresponding to a well defined quantum operator. 
This can be done using classical identities from which we can express the triads, 
the extrinsic curvature and the field strength in terms of holonomies and well-defined 
operators as the volume one. 
Then, similarly to quantum field theory, the Hamiltonian has to be 
regularized with a  parameter $\epsilon$ shrunk to zero at 
the end of the computation. 
This  is  a highly non trivial step  and poses serious challenges. 
In fact, the regularisation  $\epsilon$ enters in the spin networks 
picture via a plaquette $P(\epsilon)$, attached to a vertex and thus 
it modifies the underlying graph, but two wave functions supported on 
different networks are orthogonal by the scalar product (\ref{kscpr}). 
In other words, for any cylindric function $\Psi$, the limit 
$\epsilon\rightarrow0$ of $\hat H_\epsilon\Psi$ does not exist on $Cyl$. 
Usually one  transfers the action of the scalar constraint to the dual 
space, adopting a weaker notion of limit. 
More specifically, the limit $\epsilon\rightarrow0$ is defined by
\be\label{welim}
\langle\hat H^\ast\chi\vert\Psi\rangle=
\lim_{\epsilon\rightarrow0}\langle\chi\vert\hat H_\epsilon\Psi\rangle
\ee
for all $\Psi\in Cyl$ and $\chi\in V^\ast\subset Cyl^\ast$. 
The space $V^\ast$ must  be selected taking into account physical motivations. 
A natural choice is $V^\ast=\mathcal H_{diff}$, 
but this is not unique and furthermore poses several problems\cite{Nic}.

The necessary use of the weaker limit (\ref{welim}) is reflected 
on the quantum constraints algebra. 
In fact, in LQG a weaker notion of algebra closure 
(with respect to the so-called off-shell closure) is formulated, 
requiring that equation 
\be
\left[\hat H^\ast,\hat H'^\ast\right]\chi=0
\ee  
holds. 
The closure of the algebra is required only after the imposition of 
the constraints, and in this sense it is weaker. 
This is relevant when addressing the quantum space-time covariance, 
because only the off-shell closure exhibits this quantum gravity 
property and, relaxing the algebra closure, some crucial information may 
get lost.

Let us conclude noting that all the main result of LQG and LQC are related
to the discretization of the spectrum of area and volume operators. 
The nature of  this cut-off could be ambiguous\cite{Alex,samuel}, 
since it can be  dynamically 
generated or imposed by hand when dealing with the compact $SU(2)$  rather 
than the true gauge group of GR, i.e. the Lorentz one. 
In fact, in a covariant formalism,  without imposing a time-gauge, 
the area spectrum is continuous \cite{Alexb}, but for a discussion on 
this point of view  we demand to \reffcite{livine} and 
references therein.
For the recent interesting approaches to solve the scalar constraints
problems mentioned above, see \reffcite{ThiAGQ1,ThiAGQ2,ThiAGQ3,Thimascon,Han}.

\subsectionric{Isotropic Loop Quantum Cosmology\label{sec:isolqc}}

The Loop Quantum Cosmology is a minisuperspace model, i.e. a truncation 
of the phase space of classical GR, which is quantized according to 
the methods of LQG. 
Therefore LQC is not the cosmological sector of LQG, i.e. the 
inhomogeneous fluctuations are switched off by hand rather than being suppressed 
quantum mechanically. 
However the minisuperspace is an important arena to test a theory and 
in LQC the most spectacular results appear: the absence of the classical 
singularity \cite{Bo01}, replaced by a Big-Bounce \cite{ashtekar2006,A06} in the 
isotropic settings, a geometrical inflation \cite{Bo03} and the suppression 
of the Mixmaster chaotic behavior  toward the singularity \cite{Bo04}.

In this Section  we will analyze some basic aspects of the so-called 
isotropic Loop Quantum Cosmology and the fate of the classical singularity. 
For exhaustive discussions on this argument see \reffcite{Bojoiso,ABL}.

In this case, the phase space of GR is two-dimensional, since the scalar factor 
$a=a(t)$ is the only degree of freedom of an isotropic model. 
Therefore, the variables (\ref{varAE}) are reduced  to
\be\label{cpvar}
A=c\, ~^o\omega^i\tau_i, \qquad E=p\sqrt{^oh}~\,~^oe_i ~\tau^i
\ee
where a fiducial metric in $\Sigma$ is fixed by $^oh_{ab}$ and thus 
by the triad $^oe_i$ and co-triad $^o\omega^i$. 
Then, 
the phase space has coordinates ($c,p$), which are conjugate variables satisfying 
$\{c,p\} = \f{\kappa \gamma}3 $. 
This connection formalism is related to the metric one via the relations
\be
\label{cpvar9}
|p| =  a^2, \qquad c = \f12 (k+\gamma\dot a),
\ee
where $k=0,\pm1$ is the usual curvature parameter. 
Since the Gauss and the diffeomorphism constraints are already satisfied 
(with (\ref{cpvar}) they identically vanish), the scalar constraint
\be
\label{isocon}
H^{(c,p)} = -\f3\kappa  \sqrt{|p|}\left(\f1{\gamma^2}(c-\Gamma)^2+\Gamma^2\right)\approx0 \, ,
\ee
 where $\Gamma\propto k$, is the only remaining. 
This  is nothing but the Friedmann equation and reduces to a  
simple form in the flat case: $H^{(c,p)}_{k=0} = -\f3{\kappa \gamma^2} c^2\sqrt{|p|}$. 
We stress that  $p\in(-\infty,+\infty)$ 
and the classical singularity appears for $p=0$. 
Furthermore the changes in the sign of $p$ correspond to 
the changes in  the orientation of the physical triad $e^{\alpha}_i$, 
related to $^oe^{\alpha}_i$ via $\textrm{sign}(p)/\sqrt{|p|}$.

The quantization of this model follows the lines of LQG and 
therefore we have to construct $SU(2)$ holonomies and fluxes, 
in order to promote these variables to quantum operators, 
which become 
\be
h_{\ell}(c)=\cos\left(\f{\mu c}2\right) + 2\left(\dot e^{\alpha}\,^o\omega^i_{\alpha} \right)
\tau_i\sin\left(\f{\mu c}2\right) \, , \qquad P_S(p)=A_Sp \, ,
\ee
where $A_S$ is a factor determined by the background metric and 
$\mu\in(-\infty,+\infty)$
is a real continuous parameter, along which the holonomies are computed. 
The elements of the holonomies can be recovered from the almost periodic 
functions $N_\mu(c)=e^{i\mu c/2}$ and then  the cylindirc functions 
of this reduced model are given by
\be
g(c)=\sum_j\xi_je^{i\nicefrac{\mu_j c}{2}}\quad\in Cyl_S,
\ee
$Cyl_S$ being the space of the symmetric cylindric functions. 
The holonomy flux algebra is generated by $e^{i\mu c/2}$ and by $p$, 
i.e. we are in a hybrid representation between the Heisenberg ($p$) and 
Weyl ($e^{i\alpha x}$) ones.

The Hilbert kinematical space is now obtained requiring 
that the $N_\mu(c)$ form an orthoromal basis, i.e. 
$\langle N_\mu\vert N_{\mu'}\rangle=\delta_{\mu,\mu'}$, 
in analogy\footnote{Note that this is a Kronecker-delta rather 
than the usual Dirac one.} 
with the scalar product (\ref{kscpr}) of the full theory. 
From general theoretical considerations, the Hilbert space 
is necessarily $\mathcal H_S=L^2(\bar R_{Bohr},d\mu)$, 
where $\bar R_{Bohr}$ is a compact Abelian group 
(the Bohr compactification of the real line). 
This is the space of the almost periodic function. 

Let us stress that the construction of the Hilbert space is the 
key point of all the theory and of its main results. 
In fact, using $\bar R_{Bohr}$ one can introduce a new representation 
of the Weyl algebra. 
This is inequivalent to the standard Schr\"{o}dinger 
representation and therefore leads to different results from the WDW theory.
 
In this Hilbert space the operators $\hat N_\mu$ and $\hat p$ act 
by multiplication and derivation, respectively. 
As usual, let us introduce the bra-ket notation 
$N_\mu(c)=\langle c\vert\mu\rangle$. 
The volume operator ($V=|p|^{3/2}$) has a continuous spectrum 
$\hat V\vert\mu\rangle = V_\mu\vert\mu\rangle \propto |\mu|^{3/2} l_P^3\vert\mu\rangle$, 
in contrast to  
LQG\footnote{This feature can be attributed to the high degree of symmetry. 
In fact, in LQG the spin networks are  characterized by a pair (${\ell},j$) 
consisting of a continuous edge ${\ell}$ and a discrete spin $j$. 
Due to the symmetry, such  pair now collapses to a single continuous 
label $\mu$.}. 
In the reduced theory the spectrum is discrete in a weaker sense: 
all the eigenvectors are normalizable. 
Hence the  Hilbert space can be expanded as a direct sum, 
rather than as a direct integral, of the one-dimensional eigenspaces of $\hat p$.

In view of the analysis of the singularity, we have to address 
the inverse scale factor operator which is a fundamental one because, at a 
classical level, the inverse scale factor 
$\textrm{sign}(p)/\sqrt{|p|}$ diverges toward  the singularity. 
Let us  express it in terms of holonomies and fluxes and then 
proceed to the quantization. 
We note that the following classical identity 
\be
\label{invscfac}
\f{\textrm{sign}(p)}{ \sqrt{|p|}} 
= \f4{\kappa \gamma  }  
\textrm{Tr} \left(\sum_i\tau^ih_i\left\{h_i^{-1},V^{1/3}\right\}\right) \, ,
\ee
holds, where the holonomy $h_i$ is evaluated along any given edge. 
In fact, since we have $h_ih_i^{-1}$,  the  choice of the  edge  is not 
important, i.e. we do not introduce a regulator and the expression 
(\ref{invscfac}) is exact. 
With the scalar constraint the situation will be different.

We can proceed with the quantization of  operator (\ref{invscfac}) 
in a canonical way. 
The eigenvalues are given by
\be\label{volop}
\widehat{\left(\f{ \textrm{sign}(p)}{\sqrt{|p|}}\right)}
\vert\mu\rangle \propto \f1{\gamma l_P^2}\left(V_{\mu + 1}^{1/3} 
	- V_{\mu-1}^{1/3} \right) \vert\mu\rangle,
\ee
where $V_\mu$ is the volume operator eigenvalue defined above, 
whose  fundamental properties  are   to be  bounded from above 
and to  coincide with the operator $1/\sqrt{|\hat p|}$ 
for $|\mu|\gg1$. 
The upper bound is obtained for  the value $\mu=1$ 
and 
therefore\footnote{We recall that in this model the 
classical Ricci scalar curvature is given by 
$R\sim 1/a^2$ and therefore, from equation (\ref{volop}), 
at the classical singularity it doe not  diverge, 
assuming the value $R\sim 1/l_P^2$.} 
$|p|^{-1/2}_{max}\sim l_P^{-1}$.

The physical situation emerging is intriguing. 
Although the volume operator admits a continuous spectrum and a  zero volume 
eigenstate (the $\vert\mu=0\rangle$ state),  the inverse scalar 
factor is non diverging at the classical singularity, but is 
bounded from above. 
This can indicate that,  at a kinematical level, 
the classical singularity is avoided in  a quantum approach. 
The semiclassical picture, i.e.  the WDW behavior of the inverse scalar factor, 
appears for $|\mu|\gg1$ and therefore far from the fully quantum regime. 
Such a behavior contrasts with  the WDW formalism where the inverse scale factor 
is unbounded from above and the differences reside in the non-standard Hilbert
 space adopted. 
In fact, differently from the WDW theory,  all 
eigenvectors of $p$ are normalizable in LQC, 
including the one with zero eigenvalue. 
Therefore, in order to define an inverse scale factor operator, one has 
to formulate the  alternative procedure  discussed above.

Let us stress that a key ingredient  is the absence of the inhomogeneous 
fluctations which conversely are present in the full theory. 
In fact \reffcite{BruThi05} computed in full LQG 
that the analogue of  the inverse scale factor is unbounded from 
above on zero volume eigenstates. 
The boundedness of the inverse scale factor is 
neither necessary nor sufficient for the curvature singularity avoidance. 

As a  last point we have to impose the scalar constraint at a quantum level
to discuss the fate of the singularity from a dynamical point of view.

In order to follow the lines of the the full theory, 
the starting point will be the constraint (\ref{scacon}) 
and not the reduced one (\ref{isocon}). 
In the expression (\ref{isocon}), the connection $c$ itself 
is present rather than the holonomies and the operator 
$\hat H^{(c,p)}$ is not well defined at a quantum level. 
Let us investigate, for simplicity, the flat  case. 
The two parts of (\ref{isocon}) are then proportional to each other. 
Mimicking the procedure followed in the full theory \cite{Thi}
 we rewrite the constraint in terms of holonomies and fluxes as
\be
\label{accamu5}
H^{(h,p)}_{\mu_0} = -\f4{\kappa\gamma^3\mu_0^3}\sum_{ijk}
\epsilon^{ijk} \textrm{Tr} \left(h_{ij}^{\mu_0}
	h_k^{\mu_0}\{(h_k^{\mu_0})^{-1},V\} \right) +\mathcal O(c^3\mu_0) \,,
\ee
where $h_{ij}^{\mu_0}$ denotes the holonomy computed around a 
square $\alpha_{ij}$ with each side having length $\mu_0$, 
i.e. $h_{ij}=h_ih_jh_i^{-1}h_j^{-1}$. 
Differently from  expression (\ref{invscfac}), the dependence on 
$\mu_0$ does not drop out and now  plays the role of a regulator. 
However, at a classical level, we can take the limit $\mu_0\rightarrow0$ 
and verify that the resulting expression coincides with the 
classical Hamiltonian (\ref{isocon}). 
As in the full theory, the problems arise  at a quantum level. 
Let us investigate the action of the quantum operator 
$\hat H^{(h,p)}_{\mu_0}$ on the eigenstates which is 
\be
\label{accamu6}
\hat H^{(h,p)}_{\mu_0}\vert\mu\rangle
= \f3{\kappa\gamma^3\mu_0^3} (V_{\mu+\mu_0}-V_{\mu-\mu_0})
(\vert\mu+4\mu_0\rangle-2\vert\mu\rangle+\vert\mu-
4\mu_0\rangle) \, .
\ee
The limit $\mu_0\rightarrow0$ fails to exist and  the 
classical regulated version of equation (\ref{isocon})
behaves as $H^{(h,p)}_{\mu_0}\rightarrow H^{(c,p)}$. 
Since it contains $c$,  a limit at a quantum level does not 
exist because the operator $\hat c$ itself does not exist in the 
Hilbert space $\mathcal H_S$. 
As a matter of fact, in this reduced theory, there is no way to 
remove the regulator. 
This is resolved in  comparison with the 
full theory. 
If we assume that the predictions of LQG are true, then the regulator 
$\mu_0$ can be shrunk until the minimal admissible length given 
by the area operator spectrum. 
In this sense, the $\mu_0\rightarrow0$ limit is physically meaningless. 
However,  how this reduced theory (LQC) can see a minimal length coming 
out from LQG is not fully understood, since it is not the cosmological 
sector, but the usual cosmological mi\-ni\-su\-per\-space phase space 
quantized through the LQG methods.

Let us investigate the physical states. 
As in the full theory,  the physical states are those  annihilated by 
all the constraints and live in some bigger space $Cyl_S^\ast$ 
and they do not  need to be normalizable. 
A generic state\footnote{The notation $(\Psi\vert$ is adopted to the 
eventuality of non-renormalizable states.} 
$(\Psi\vert\in Cyl_S^\ast$ can be expanded as
\be
(\Psi\vert=\sum_\mu\psi(\mu,\phi)\langle\mu\vert,
\ee
where $\phi$ represents a generic matter field and 
satisfies the constraint equation
\be
(\Psi\vert\left(\hat H^{(h,p)}_{\mu_0}+\hat H^{\phi}_{\mu_0}\right)^\dag=0
\ee
and therefore the  equation for $\psi(\mu,\phi)$ 
\begin{align}
\label{difeq}
(V_{\mu+5\mu_0} &-V_{\mu+3\mu_0}) \psi(\mu+4\mu_0,\phi) 
	-2(V_{\mu+\mu_0}-V_{\mu-\mu_0})\psi(\mu,\phi)+\\ \nonumber
&+(V_{\mu-3\mu_0}-V_{\mu-5\mu_0})\psi(\mu-4\mu_0,\phi)
=-\f\kappa3(\gamma^3\mu_0^3l_P^2)\hat H^{\phi}_{\mu_0}\psi(\mu,\phi) \, .
\end{align}
has to hold.
This is nothing but a recurrence relation for the 
coefficients $\psi(\mu,\phi)$ which ensures that 
$(\Psi\vert$ is a physical state. 
Even though $\mu$ is a continuous variable, 
it is given by (\ref{difeq}) as an algebraic rather then by 
 a differential equation as, for example, in the WDW theory.

Let us now briefly discuss what happens to the classical singularity 
at a dynamical level. 
It corresponds to the state $\vert\mu=0\rangle$ and 
as we can see from equation (\ref{difeq}), 
starting at $\mu=-4N\mu_0$ we can  compute the coefficients 
$\psi(4\mu_0(n-N),\phi)$ for $n>1$, all but 
$\psi(\mu=0,\phi)$, because the generic coefficient vanishes 
if and only if $n=N$. 
Although the quantum evolution seems to break down 
right at the classical singularity,  this is not the case, 
since the coefficient $\psi(\mu=0,\phi)$ is decoupled from the others
thanks to $V_{\mu_0}-V_{-\mu_0}=0$  and to  
$\hat H^{\phi}_{\mu_0}\psi(\mu=0,\phi)=0$ realizes. 
Therefore the coefficients in (\ref{difeq}) are such that one can 
unambiguosly evolve the states through the singularity even 
though $\psi(\mu=0,\phi)$ is not determined and 
the classical singularity is solved in the LQC framework.

We conclude reporting that the previous results have been significantly 
extended \cite{A06} with a rigorous formulation of the physical Hilbert
 space, of the Dirac observables and of the semi-classical states. 
Furthermore, taking a massless scalar field as a relational clock 
for the system, the classical Big-Bang is replaced by a quantum 
Big-Bounce displaying  in detail an intuitive picture of the 
Universe evolution in the 
Planckian era is displayed.

\subsectionric{Mixmaster Universe in LQC\label{sec:mixlqc}}

We review the basic aspects of 
the Mixmaster Universe in the LQC framework in this Section, while 
for the detailed analysis we demand to the literature \reffcite{Bo04,Bo04a,2004CQGra..21.3541B}.

As we have seen, the Bianchi IX evolution toward the singularity sees 
  infinite sequences of Kasner epochs characterized by a series of 
permutations as well as 
by possible rotations of the expanding and contracting spatial directions. 
However, this infinite number of bounces within the potential, 
at the basis of chaos, is a consequence of an unbounded growth of the spatial 
curvature. 
When the theory offers a cut-off length and the curvature is bounded, 
the Bianchi IX model naturally shows a finite number of  oscillations
and in  LQC  a quantum suppression of the chaotic behavior appears
 close to the singularity. 

Let us  formulate  in the connection formalism the vacuum Bianchi IX model. 
The spatial metric $dl^2=a_I^2(\omega^I)^2$ can be taken diagonal, leaving 
three degrees of freedom only. 
The basic variables for a homogeneous model are
\be 
A^i_\alpha=c_{(I)}\Lambda^i_I\omega^I_\alpha \, , \qquad E^\alpha_i=p^{(I)}\Lambda^I_iX^\alpha_I \, ,
\ee
where  $\omega^I$ are the left-invariant 1-forms, $X_I$ are the left-invariant 
fields dual to $\omega^I$ ($\omega^I(X_J)=\delta^I_J$), the $SO(3)$-matrix 
$\Lambda$ contains the pure gauge degrees of freedom and $I,J,K,\ldots,$ 
run as $1,2,3$. 
The physical information are expressed  in terms of the gauge invariants
 $c_{(I)}$ and $p^{(I)}$ which satisfies the Poisson brackets 
$\{c_I,p^J\}=\gamma\kappa\delta_I^J$. 
These variables are related to the scale factors $a_I$, 
the spin connections $\Gamma_I$ and the extrinsic curvature 
$K_I=-\nicefrac{1}{2}\dot a_I$ by  the relations
\be
p^I= \textrm{sign}(a_I)~ |a_Ja_K| \textrm{sign}(a_I), \qquad c_I=\Gamma_I-\gamma K_I  \, , 
\ee
where
\be
\Gamma_I=\f12\left(\f{a_J}{a_K}+\f{a_K}{a_J}-
\f{a_I^2}{a_Ja_K}\right)=\f12\left(\f{p^K}{p^J}+\f{p^J}{p^K}
-\f{p^Jp^K}{(p^I)^2}\right).
\ee
The classical dynamics is governed by the scalar constraint (\ref{scacon}) which reads as
\be\label{bixham}
H=\f2\kappa\left[\left(\Gamma_J\Gamma_K-\Gamma_I \right)a_I
-\f14a_I\dot a_J\dot a_K+ \textrm{cyclic} \right]\approx0 \, .
\ee
The  particular case of the isotropic  model, the open ($k=1$) 
FRW, can be  recovered setting $a_1=a_2=a_3=a$. 
On the other hand, the Bianchi I model can be obtained for $\Gamma_I=0$. 
The potential term  from (\ref{bixham}) is given by
\be\label{potinp}
W(p)=2\left(p^Ip^J\left(\Gamma_I\Gamma_J-\Gamma_K\right)+\textrm{cyclic} \right)
\ee
which has infinite walls at small $p^I$ due to the divergence of 
the spin connection components.

The loop quantization is performed straightforwardly as in the isotropic 
case. 
In fact, an orthonormal basis is given by the $\hat p^I$-eigenstates 
$\vert \mu_1,\mu_2,\mu_3\rangle= 
\vert\mu_1\rangle\otimes\vert\mu_2\rangle\otimes\vert\mu_3\rangle$
and the Hilbert space is taken as a direct product of the isotropic ones, 
it is separable and is a subspace of the kinematical non-separable Hilbert space. 
As usual, the cylindric functions are given by a superposition of the functions $\langle c\vert\mu\rangle\sim\exp(\nicefrac{i\mu c}{2})$ and 
the basic quantum operators are the gauge invariant 
triad operators $\hat p^I$ (fluxes) and the holonomies 
\be
h_I(c)=\cos\left(\f{c_I}2\right)+2\Lambda^i_I\tau_i\sin\left(\f{c_I}2\right) \,.
\ee
The $\hat p^I$ and $\hat h_I$  act as derivative and multiplication 
operators, respectively. 
In par\-ti\-cu\-lar, the volume operator defined from $\hat p^I$ 
as $\hat V=\sqrt{|\hat p^1\hat p^2\hat p^3|}$, 
acts on the eigenstates $\vert \mu_1,\mu_2,\mu_3\rangle$ as
\be
\label{veap}
\hat V\vert \mu_1,\mu_2,\mu_3\rangle
= (\f12\kappa\gamma)^{3/2}
\sqrt{|\mu_1\mu_2\mu_3|}\,\vert \mu_1,\mu_2,\mu_3\rangle \, , 
\ee
and it has a continuous spectrum.

From these basic operators we can obtain, similarly to the isotropic case, 
the inverse triad operator. 
Since it is  diverging toward the classical singularity, we are interested to 
 its behavior at a quantum level. 
Conceptually its construction in the homogeneous cosmological sector 
is  the same as in  the isotropic one, 
the only difference residing in  the  computations and in 
the appearance of some quantization ambiguities, as in particular 
the half-integer $j$ and a continuous parameter $l\in(0,1)$. 
Nevertheless,  all the results are independent on them. 
This behavior mimics the quantization ambiguities present either in LQG 
or in the isotropic sector of LQC. 
While  in the full theory we find an  ambiguous  choice of the spin number 
$j$ associated to a given edge of the spin network, 
in the isotropic LQC it is reflected on the choice of the 
Hamiltonian regulator $\mu_0$. 
Nevertheless, this is but  a parameter that we can neither shrunk 
to zero, neither fix in some way in the context of LQC theory itself
but it outcomes from keeping some prediction of another theory 
(usually LQG). 
In the inverse scale factor of isotropic LQC none quantization ambiguity
appears.

Since  we can express $|p^I|^{-1}$ in terms of holonomies and volumes
 via a classical identity, it can be canonically quantized
as follows
\be
\label{oppla}
\widehat{|p^I|^{-1}}\vert \mu_1,\mu_2,\mu_3\rangle=
A_{j,l} f_{j,l}(\mu_I)\vert \mu_1,\mu_2,\mu_3\rangle \, ,
\ee
where
\be
f_{j,l}(\mu_I)=\left(\sum_{k=-j}^jk|\mu_I+2k|^l\right)^{\nicefrac{1}{1-l}}
\ee    
and $A_{j,l}$ is a function of the quantization ambiguities, 
i.e. $j$ and $l$. 
The values $f_{j,l}(\mu)$ decrease for $\mu<2j$ and we have 
$f_{j,l}(0)=0$ and therefore 
$\widehat{|p^I|^{-1}}\vert\mu_I=0\rangle=0$. 
Thus, the inverse triad operator annihilates the state corresponding 
to the classical singularity $\vert\mu_I=0\rangle$.

Fundamental properties of the eigenvalues of the operator
(\ref{oppla}) can be extracted from the asymptotic expansions 
of  $F_{j,l}(\mu_I)\equiv A_{j,l} f_{j,l}(\mu_I)$. 
For large $j$, $F_{j,l}(\mu_I)=F_l(\nu_I)$, with  $\nu_I=\mu_I/2j$, and no
 dependence on $j$ appears, in particular
\be\label{inteopex}
F_l(\nu_I)=
	\bcas 
		1/\nu_I & \textrm{if}\, \nu_I\gg1\, 	\quad  (\mu_I\gg j)\\ 
		\displaystyle \left(\f{\nu_I}{l+1}\right)^{\nicefrac{1}{1-l}} 
		& \textrm{if}\, \nu_I\ll1\, \quad (\mu_I\ll j)\, .
	\ecas
\ee 
The classical behavior for the inverse triad components is obtained 
for $\nu_I\gg1$ and the loop quantum modifications arise for $\nu_I\ll1$. 
Rewriting the spin connection in the triad representation, the potential 
(\ref{potinp}) is given by 
\be
\label{pottrirep}
W_{j,l}(\nu)= 2(\gamma\kappa j)^2
\left(\nu_I\nu_J\left(\Gamma_I\Gamma_J-\Gamma_K \right)+\textrm{cyclic} \right) \, ,
\ee
where
\be
\Gamma_I(\nu_I)=\f12\left[\nu_K \textrm{sign}(\nu_I)F_l(\nu_J)
+\nu_J \textrm{sign}(\nu_K)F_l(\nu_K)-\nu_J\nu_KF^2_l(\nu_I)\right].
\ee

The loop quantization of the scalar constraint is 
different in the homogeneous case. 
In fact, unlike the full theory, 
the spin connections are tensors due to the homogeneity symmetry and
cannot vanish as in LQG, thus the holonomies will depend also on them
and  the  quantum scalar operator leads to a partial difference equation, 
like in isotropic LQC, for which we refer to \reffcite{Bo04a}.

Let us  discuss the Mixmaster Universe in the LQC framework, 
 analyzing its behavior at a semiclassical level, considering the 
modifications induced in the dynamics by the loop quantization. 
Since  an {\it effective} Hamiltonian must arise  from the underlying
 quantum evolution, we will proceed in two steps.
\begin{enumerate}
\renewcommand{\labelenumi}{{\it\roman{enumi}})}

\item
For a slow varying solution, 
the differences equation is specialized, 
in the continuum regime $\mu_I\sim p^I/\gamma\kappa\gg1$,
\footnote{The eigenvalues of the triad operator $\hat p^I$ 
are given by 
$\hat p^I\vert \mu_1,\mu_2,\mu_3\rangle= 
p^I\vert \mu_1,\mu_2,\mu_3\rangle$, 
where $p^I\sim \gamma\kappa\mu_I$.}
thus  obtaining the Wheeler-DeWitt equation
\begin{multline}\label{WDWloop}
\left(\kappa^2 p^Ip^J\f{\p^2\sqrt{|p^1p^2p^3|}S(p)}{\p p^I\p p^J}
+\textrm{cyclic} \right)  + W_{j,l}(p)\sqrt{|p^1p^2p^3|}S(p)=\\
=-\kappa|p^1p^2p^3|^{3/2}\hat\rho_\phi S(p) \, ,
\end{multline}
where $\phi$ denotes a generic matter field.

\item
We take the WKB limit of the wave function 
$T(p)=\sqrt{|p^1p^2p^3|}S(p)$, i.e. $T\sim e^{iA/\hbar}$, 
leading to  the Hamilton -Jacobi equation for the phase $A$ 
to zero-th $\hbar$ order.

\end{enumerate}

We have obtained the classical dynamics plus the quantum loop 
corrections. 
The key point for the classical analysis of the effective Hamiltonian
 is that the classical region $\mu_I\gg1$ can be separated in two 
subregions. 
In fact, remembering  the $p$-dependence in the WDW equation 
(\ref{WDWloop}) given by $\mu_I/2j\sim p^I/j\gamma\kappa$, 
we obtain the condition $\mu_I\gg1$  for 
$p^I\ll j\gamma\kappa$ ($j\gg\mu_I\gg1$) and $p^I\gg j\gamma\kappa$ ($\mu_I\gg j\gg1$). 

The second subregion ($\mu_I\gg j\gg1$) is the purely classical one, 
i.e. where the Misner picture is still valid. 
From expansion (\ref{inteopex}) the eigenvalues of the inverse triad 
operator correspond  to the classical values. 
On the other hand, the region where  $j\gg\mu_I\gg1$ 
is characterized by loop quantum modifications. 
In fact, the inverse triad operator eigenvalues have a power 
law dependence. 
The quantum modifications of the classical dynamics are controlled 
by the parameter $j$: if it is large enough, one can move the quantum 
effects within the effective potential into the semi-classical domain. 
On the other hand, the WKB limit ($\hbar\rightarrow0$) is strictly 
valid in the Misner region ($\mu_I\gg j\gg1$), 
because in the first region a dependence on $(\gamma\kappa)^{-1}$, 
appears in the potential term. 
We assume  the validity of such approximation in the 
($j\gg\mu_I\gg1$)-region,  because the inverse triads vanish as $p^I\rightarrow0$.

A qualitative study of the modified classical dynamics, arises 
from analyzing the potential term, whose  explicit expression
is more complex then the original one, making the analysis of the 
dynamics tricky. 
The volume variable is regarded as a time variable and in general 
the dependence on 
it does not factorize. 
Nonetheless, in the second region, the Misner potential is restored. 

Considering the particular case $q_-=0$ (the Taub Universe\footnote{The Taub model is nothing but a particular case of 
the Bianchi IX one in which $\beta_-=0$ \cite{Ry-Sh}.} ), 
we can qualitatively study the effective (loop) potential. 
In this case we get $\nu_2=\nu_3\equiv\nu$ and 
$\nu_1\equiv\rho=V^2/((2j)^3\nu)$ and therefore $\Gamma_2=\Gamma_3$. 
The wall is seen for $\nu\gg1$ so that $F_l(\nu)\simeq\nu^{-1}$, 
but $\rho$ not negligible, i.e. the relation $F_l(\rho)\simeq\rho^2$ holds
and  the potential wall (\ref{pottrirep}) becomes
\be
\label{wvj7}
W_{j,l}\simeq\f{V^4}{j^4\rho^2}F^2_l{\rho} \left(3-2\rho F_l(\rho)\right)\, , 
\ee
where $V$ denotes the volume. 
For $\rho\gg1$, $F_l(\rho)\simeq \rho^{-1}$ and the classical wall 
$e^{4\alpha-8\beta_+}$ is restored. 
The key difference with the Misner case relies in the wall finite height. 
As the volume decreases, the wall moves inwards and its height decreases
as well. 
In the following evolution the wall completely disappears. 
In particular it gets its  maximum  for $\beta_+=-\alpha$ and 
vanishes as $e^{12\alpha}\propto V^4$ toward the classical 
singularity ($\alpha\rightarrow-\infty$).

This peculiar behavior shows that the Mixmaster evolution stops at 
a given time and therefore  chaos  disappears. 
In fact, the point-Universe, when the volume is so small that the 
quantum modifications arise, will never bounce against the potential 
wall and the  Kasner epochs will continue without any replacement. 
This behavior predicted by the LQC framework produces (qualitatively) 
the same results of the influence of a scalar filed on the Universe 
dynamics: in fact, also in that case, at a given time the particle 
performs  the last bounce and then it freely moves 
(see Section \ref{sec:scal}).   

For a discussion of the Bianchi cosmologies in this new canonical first order formalism see \reffcite{1988PThPh..80.1024K,1989grg..conf..613K,1991CQGra...8.2191A,1993CQGra..10..559M,1995CQGra..12.1287G,1996PhRvD..54.2589G,2002PThPh.107..305K,2004CQGra..21.3541B,2005PhRvD..72f7301D,2005PhRvD..72h4008C,2005PhRvL..94k1302C,2007PhRvD..75b4029C,2007PhRvD..76h4015C}.

\subsectionric{On the GUP and the Minisuperspace Dynamics\label{sec:gup}}

This paragraph is devoted to explain some results obtained in a 
recent approach \cite{BM07a,BM07b} (see also \reffcite{2007CQGra..24..931V,2007gr.qc.....3101O,2007PhLB..651...79V}) to quantum cosmology, 
in which the notion of a minimal length naturally appears. 
In particular, the purpose is to quantize a cosmological model by 
using a modified Heisenberg algebra, which reproduces a Generalized Uncertainty Principle (GUP) 
\be\label{gup}
\Delta q \Delta p\geq \f 1 2\left(1+\beta \left(\Delta p\right)^2
+\beta \langle{\bf p}\rangle^2\right),
\ee
where $\beta$ is a ``deformation'' parameter. 
The  uncertainty principle (\ref{gup}) can be obtained by considering 
an algebra generated by $\bf q$ and $\bf p$ obeying the commutation relation   
\be\label{modal}
[{\bf q},{\bf p}]=i\left(1+\beta{\bf {p}}^2\right) \, . 
\ee 
Such deformed Heisenberg uncertainty principle  appeared in 
studies on string theory \cite{gross1988,witten1997dsa,ACamelia97} and leads to a fundamental 
minimal scale. 
From this point of view, a minimal observable length is
a consequence of the limiting scale given by the string one.
However, we stress that the minimal scale predicted by the GUP 
is different from the  one  predicted by other approaches. 
In fact,  equation (\ref{gup}) implies a finite minimal 
uncertainty in position $\Delta q_{min}=\sqrt\beta$\cite{Kempf95}. 
This way, we will introduce a minimal scale in the quantum 
dynamics of a cosmological model.

The appearance of a non-zero uncertainty in position poses 
some difficulty in the construction of a Hilbert space, as 
position eigenstates cannot be  physically constructed. 
An eigenstate of an observable necessarily has to have a vanishing
uncertainty. 
Although it is possible to construct position eigenvectors, 
they are  formal but not physical states. 
In order to recover information on position, we 
consider the so-called {\it quasiposition wave functions}
\be\label{qwf} 
\psi(\zeta) = N \int^{+\infty}_{-\infty}\f{dp}{(1+\beta p^2)^{3/2}} 
\exp\left(i\f{\zeta}{\sqrt{\beta}} 
\arctan \left(\sqrt{\beta}p\right)\right)\psi(p) \, ,
\ee
where $\zeta$ is the {\it quasiposition} defined by the main 
value of the position $\bf q$ on appropriate functions, 
i.e. $\langle\bf q\rangle=\zeta$ and $N$ is a normalization factor. 
The quasiposition wave function (\ref{qwf}) represents the probability 
amplitude to find a particle maximally localized around the position 
$\zeta$, i.e. with standard deviation $\Delta q_{min}$. 
For more details on quantum mechanics in this framework we 
recommend \reffcite{Kempf95,Kempf97}.

The GUP approach relies on a modification of the canonical 
prescription for quantization and  can be  applied to any 
dynamical system. 
Moreover,  such formalism allows us to analyze some peculiar 
features of string theory in the minisuperspace dynamics.

Let us investigate the consequences of a Heisenberg deformed 
algebra (\ref{modal}) of the quantum dynamics of the flat ($k=0$) 
FRW model in presence of a massless scalar field $\phi$. 
We are interested on  the fate of the classical singularity. 
In the  following we summarize the discussion and results reported in 
\reffcite{BM07a,BM07b}. \\
The Hamiltonian constraint for this model has the form
\be\label{con}
H_{grav}+H_{\phi}= -9\kappa p_x^2x
	+\f{p_{\phi}^2}{x}\approx 0\, ,  \qquad x\equiv a^3 \, ,
\ee      
where $a$ is the scale factor. In the classical theory, the phase space 
is four-dimensional, with coordinates $(x,p_x;\phi,p_{\phi})$,
 and the physical volume of the Universe vanishes  at $x=0$ and the 
singularity appears. 
Moreover each classical trajectory can be specified in the 
$(x,\phi)$-plane, i.e. $\phi$ can be considered as a relational 
time. 
The dynamical trajectories read as
\be\label{clastra}
\phi=\pm\f 1 {3\sqrt{\kappa}}\ln\left|\f x {x_0}\right|+\phi_0 \, ,
\ee
where $x_0$ and $\phi_0$ are integration constants. 
In  equation (\ref{clastra}), the plus sign corresponds to 
a  Universe expanding from the Big-Bang, while the minus sign to a 
contracting one into the Big-Crunch. 
The classical cosmological singularity is for $\phi=\pm\infty$ 
and every classical solution, in this model, reaches it.

The canonical approach (the WDW theory) does not solve the 
singularity problem. In fact, we can  construct a state localized at 
some initial time. 
Then, in the backward evolution, the  peak of the wave packet  
will move along the classical trajectory (\ref{clastra}) 
falling in the classical singularity\cite{BlyIsh}, so that 
the latter  is not tamed by quantum effects.

This picture is radically changed in the GUP framework and 
the modifications can be realized in two steps. 
Firstly, we can  show how the probability density $\vert\Psi(\zeta,t)\vert^2$ to find the Universe around $\zeta\simeq0$ (around the Planckian region) can be expanded as 
\be
\vert\Psi(\zeta,t)\vert^2\simeq\vert A(t)\vert^2+\zeta^2\vert B(t)\vert^2 \, ,
\ee 
where $t$ is a dimensionless time $t=3\sqrt{\kappa}\phi$ and the wave packets
\be
\Psi(\zeta,t)=\int_0^\infty d\epsilon ~ g(\epsilon)\Psi_\epsilon(\zeta)e^{i\epsilon t}
\ee
are such that the state is initially peaked at late time, i.e. the weight 
function $g(\epsilon)$ is a Gaussian distribution centered  at some 
$\epsilon^\ast\ll1$ (at energy much less then the 
Planck one $1/l_P$). 
Of course, $\Psi_\epsilon(\zeta)$ rapresents the 
{\it quasiposition eigenfunctions} (\ref{qwf}) of this problem.

Near the Planckian region, the probability density to find the Universe 
is $\vert A(t)\vert^2$, which is very well approximated by a Lorentzian 
function peaked in $t=0$, which corresponds to the classical time when 
$x(t)=x_0$. 
For $x_0\sim\mathcal{O}(l_P^3)$, the probability density to find the 
Universe in a Planck volume has a maximum 
around the corresponding classical values and vanishes for 
$t\rightarrow-\infty$, where the classical singularity appears. 
In this sense the classical cosmological singularity is solved.  

\begin{figure}[ht]
\begin{center}
\includegraphics[width=0.7\textwidth]{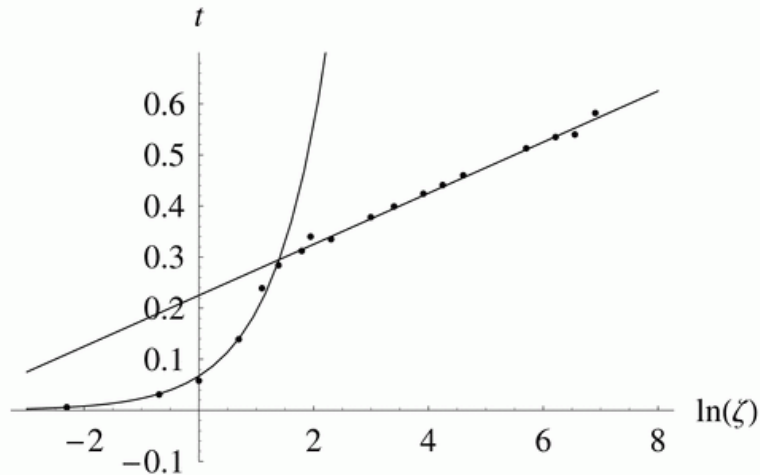}
\caption[The peaks of the probability density $\vert\Psi(\zeta,t)\vert^2$]{The peaks of the probability density $\vert\Psi(\zeta,t)\vert^2$ are plotted as functions of $t$ and $\ln(\zeta)$. The points (resulting from numerical computation) are fitted by a logarithm $0.050\ln(\zeta)+0.225$ for $\zeta\geq4$ and by a power law $0.067\zeta^{1.060}$ for $\zeta\in[0,4]$ (from \reffcite{BM07a}).} \label{fig:vale}
\end{center}
\end{figure}

The more interesting differences between the WDW and the GUP 
approaches relies on the wave packets dynamics. 
If we consider a wave packet initially peaked at late time, 
once numerically  evolved  backward,
the integration results in a  probability density, 
at different fixed values of $\zeta$, still  approximated 
by a Lorentzian function. 
The width of this function remains the same as the states 
evolve from large $\zeta$ ($\sim 10^3$) to $\zeta=0$, 
while its  peaks  move along the classically expanding 
trajectory (\ref{clastra}) for values of $\zeta$ larger than $\sim4$. 
Near the Planck region, i.e. when $\zeta\in[0,4]$, we observe a 
modification of the trajectory of the peaks since 
they follow a power-law up to $\zeta=0$, reached in a finite time 
interval and they escape from the classical trajectory toward the 
singularity (see Figure \ref{fig:vale}). 
The peaks of the Lorentzian at fixed time $t$ slowly evolve 
remaining close to the Planck region. 
Such behavior outlines how  the Universe has a stationary approach 
to the cutoff volume, accordingly to Figure \ref{fig:vale}, and 
is different from other approaches.  

In fact, it was shown how the classical Big-Bang is replaced by a 
Big-Bounce in the framework of LQC \cite{A06}. 
Qualitatively, one expects that the bounce (and the consequently 
repulsive features of the gravitational field in the Planck regime) 
are consequence of a Planckian cut-off length, but  this is not the case. 
We can observe from Figure \ref{fig:vale} that there is not a bounce for the 
quantum Universe. 
The main difference between the two approaches resides in the quantum 
modification of the classical trajectory. 
In fact, in LQC we observe a ``quantum bridge'' between the expanding and 
contracting Universes, while  in our approach the probability density of 
finding the Universe reaches the Planck region in a stationary way.

Finally, we emphasize two points:
i) the above result on the singularity-free behavior of the GUP 
(FRW) Universe is confirmed by the more general Taub 
model
(for a discussion about  the Taub Universe in the GUP framework 
we refer to \reffcite{BM07b}). 
Also  in this case the singularity is probabilistically suppressed. 
ii) In the GUP Taub Universe, the wave packets favour the establishment 
 of a quantum isotropic Universe  with respect to  those in the WDW theory.

\subsectionric{Quantum Chaos}

In the previous section we have widely discussed the chaotic 
features of the Mixmaster dynamics; 
as usual in chaotic systems, we expect that this classical 
behavior is maintained also at a quantum level. 
Several attempts were made over the years to characterize the 
``quantum chaos'' of the Bianchi IX model, analyzing the 
evolution of the shape of the wave functions \reffcite{1986PThPh..75...59F,Furusawa1985PTP,Berger1989PRD} 
or the distribution of the energy 
levels \reffcite{Graham1991PRA1,Graham1991PRA2,Graham1994Chaos}. 

The numerical simulation performed 
in \reffcite{1986PThPh..75...59F,Furusawa1985PTP} is focused on 
a wave packet initially peaked in the center of the 
$(\beta_+,\,\beta_-)$ plane and is a solution of the WDW equation. 
The evolution is  numerically computed as a multiple-reflection 
problem of the wave-packet within moving potential walls. 
The quantized Mixmaster model possessing a quantum mixing property
can also be seen as a quantum chaotic system. 
The mixing operates on the expectation values of the scale factors, 
which tend to an equilibrium value without any help of the matter 
field or of the cosmological term. 
A self-similar structure of the distribution of probability density
emerges, strongly suggesting the existence of a fractal structure.

From \reffcite{Berger1989PRD} one finds another feature of the 
classical chaos emerging at a  quantum level. 
Adopting a path-integral representation of the dynamics, 
a Monte Carlo simulation of the dynamics shows how the 
anisotropy potential at the origin of the classical chaos yields 
a quantum mechanical correlation between large universe volume 
and high anisotropy, although the peak at zero anisotropy is 
significant as well. 
This result suggests  a possible failure of the quantized 
Mixmaster cosmology to represent the very Early Universe, 
or it might be a consequence of neglecting non gravitational 
fields.

A different point of view is addressed in 
\reffcite{Graham1991PRA1,Graham1991PRA2}, where the quantum 
chaos is outlined when studying the distribution of the energy 
levels of the quantized dynamics, assuming Dirichlet boundary 
conditions. 
In the Poincar\'e upper half-plane representation, the authors 
evaluate approximately the first $10^3$ eigenvalues, 
theie  distribution and how the energy levels are influenced 
by the deformation of the circular side of the bounding triangle.

\sectionric{Inhomogeneous Mixmaster Model\label{sec:inhomix}}

This Section is devoted to the generalization of the Mixmaster dynamics to the inhomogeneous case and to its multidimensional extensions. We firstly provide the construction of the generic cosmological solution of the Einstein equations in the neighborhood of the Big Bang and then we discuss the so-called fragmentation process. The Hamiltonian formulation of the inhomogeneous Mixmaster is reviewed with particular attention to the role played by the Ricci scalar and the quantum picture is then briefly analyzed.  We close this review showing how the chaos is a dimensional phenomenon summarizing the most important results obtained in the multidimensional extensions of GR.

\subsectionric{Formulation of the generic cosmological problem\label{sec:generic}}

We will focus our attention on the generic behavior of the Einstein
 equations near the Big Bang, {i.e.}, we will 
 discuss the properties of their general cosmological solution, 
meaning with it a solution that possesses the right number of 
physical arbitrary functions allowing to specify arbitrary initial 
conditions on a generic non-singular space-like hypersurface at a 
fixed instant $t$, thus  four arbitrary functions in vacuum, 
or eight if a perfect fluid is a source \cite{LF}(for a discussion in presence of inflationary matter see \reffcite{DT94}).

Belinskii, Khalatnikov ans Lifshitz  in the 70's derived such a solution and how  its dynamics resembles the one  of 
the homogeneous indicess of types VIII and IX \reffcite{BelinskiiKhalatnikov1971gsgSPJ,BKL1972,BKL82} (see also \reffcite{Berger98MPLA,Berger01wd}). 
The construction can be achieved firstly considering 
the generic  solution for the individual Kasner epoch, and then 
providing a general description of the alternation of two 
successive epochs.
The answer to the first question is given by the so-called  generalized 
Kasner solution, while  to the latter is found to be in close 
analogy to the replacement rule in the homogeneous 
indices

For a general view on inhomogeneous cosmologies, see \reffcite{1980PhLA...75..333C,1982AN....303..227S,1982ApJ...253....1A,1983AnPhy.150..392C,1984CQGra...1..291H,1985Ap&SS.108..195R,1986Ap&SS.128..447R,1991Ap&SS.181...61R,1995IJPAM..26..169R,1998GrCos...4...23K,2004CQGra..21S..81B,0264-9381-7-4-008}.

\subsubsectionric{The Generalized Kasner  Solution\label{sec:genk}}

The homogeneous indices of types VIII and IX provide the prototypes for 
the construction of the generic (non-homogeneous) solution of the 
Einstein equations in the neighbourhood of the singularity.  
In \reffcite{LK1963}, it is shown that the Kasner solution can be generalized 
to the inhomogeneous case and near the singularity, as
\begin{equation}
\label{c1 kasner gen}
    \begin{cases}
    dl^{2}=h_{\alpha\beta}dx^{\alpha} dx^{\beta}\;,\\
    h_{\alpha\beta}=a^{2}l_{\alpha} l_{\beta}+b^{2}m_{\alpha} m_{\beta} + c^{2} n_{\alpha} n_{\beta}\;,
    \end{cases}
\end{equation}
where
\begin{equation}
\label{c1 and coeff kasn}
a\sim t^{p_{l}}\;,\hspace{ 0.5cm}b\sim t^{p_{m}}\;,\hspace{0.5cm}c\sim t^{p_{n}}\;,
\end{equation}
and $p_{l,},\,p_{m},\,p_{n}$ are functions of spatial coordinates subjected to the conditions
\begin{equation}
\label{c1 vinc kasn gen}
p_{l}(x^{\gamma})+p_{m}(x^{\gamma})+p_{n}(x^{\gamma})=
p_{l}^{2}(x^{\gamma})+p_{m}^{2}(x^{\gamma})+p_{n}^{2}(x^{\gamma})=1\;.
\end{equation}
Differently from  the homogeneous indicess, the frame vectors $l,\,m,\,n$ 
are now arbitrary functions of the coordinates (subject to the conditions 
imposed by the  Einstein equations  $0\alpha$\cite{LK1963}).\\
The behaviour (\ref{c1 kasner gen}) cannot last up to  the singularity, 
unless a further condition is imposed on the vector $l$ ({ i.e.}, the one 
corresponding to the negative index $p_{1}$)
\begin{equation}
\label{c1 cond rot}
l\cdot\nabla\wedge l=0\;.
\end{equation}
This condition reduces to three the number of arbitrary functions, 
i.e. one less than the generic solution: in fact,  
(\ref{c1 kasner gen}) possesses twelve arbitrary functions of the 
coordinates (nine components of the Kasner axes and three indexes 
$p_{i}(x^{\gamma})$), and satisfies the two Kasner relations 
(\ref{c1 vinc kasn gen}), the three $0\alpha$ Einstein equations, 
the three conditions from the invariance under three-dimensional 
coordinate transformations, and (\ref{c1 cond rot}). 

The generalized Kasner solution (\ref{c1 kasner gen}) does not require
any additional condition, since the individual Kasner epoch lasts only 
for a finite interval of time, and  contains, for example, four  
arbitrary functions in vacuum.

\subsubsectionric{Inhomogeneous BKL indices\label{sec:inhobkl}}

Let us now generalize our scheme by investigating
the implications of the remaining conditions (\ref{c1 cond rot}). 
This analysis leads to the inhomogeneous BKL indices.

We  assume that the factors that determine the order of magnitude 
of the components  of the spatial metric tensor (\ref{c1 kasner gen})
can be included in the functions $a,\,b,\,c$, which  change with time 
according to (\ref{c1 and coeff kasn}), { i.e.} the vectors $l,\,m,\,n$ 
define the directions of the Kasner axes. 
For non homogeneous spaces there is no reason to introduce a fixed 
set of frame vectors, which would be independent of Kasner axes.\\
The time interval of applicability of solution (\ref{c1 kasner gen}) 
is determined by conditions which follow from the Einstein equations. 
Near the singularity, the matter energy-momentum tensor in the 
$00$- and $\alpha\beta$-components  may be neglected
\bseq
\begin{align}
\label{c1 einstein 00}
- R^{0}_{0} &=\frac{1}{2}\dot{\chi}^{\alpha}_{\alpha}
	+\frac{1}{4}\chi^{\alpha}_{\beta} \chi^{\beta}_{\alpha} = 0\;, \\
\label{c1 einstein alphabeta}
- R^{\beta}_{\alpha} &=\frac{1}{2\sqrt{h}}
\partial_t\left(\sqrt{h} \chi^{\beta}_{\alpha} \right) + P^{\beta}_{\alpha}=0 \;.
\end{align}
\eseq
Solution (\ref{c1 kasner gen})-(\ref{c1 vinc kasn gen}) is obtained 
neglecting the three-dimensional Ricci tensor $P^{\beta}_{\alpha}$ 
in (\ref{c1 einstein alphabeta}) and its  validity is easily 
formulated in terms of the projections of the tensors along the 
directions $l,\,m,\,n$ \cite{LK1963}, which  must satisfy the conditions
\begin{equation}
\label{c1 condizioni diagonali}
P^{l}_{l},\,P^{m}_{m},\, P^{n}_{n}\ll t^{-2}\;,
\qquad P_{l}^{l}\gg P_{m}^{m},\,P^{n}_{n\;.}
\end{equation}
The off-diagonal projections of (\ref{c1 einstein alphabeta}) 
determine the off-diagonal projections of the metric tensor 
$(\eta_{lm},\,\eta_{ln},\,\eta_{mn})$, which should only be 
small corrections to the leading terms of the metric, as 
given by (\ref{c1 kasner gen}). 
In the latter, the only non-vanishing projections are the diagonal 
$(\eta_{ll},\,\eta_{mm},\,\eta_{nn})$, and satisfy\cite{BKL82}
\begin{equation}
\label{c1 cond mista}
\eta_{lm}\ll \sqrt{\eta_{ll} \eta_{mm}}\;,\qquad 
\eta_{ln}\ll \sqrt{\eta_{ll} \eta_{nn}}\;,\qquad
\eta_{mn}\ll \sqrt{\eta_{mm} \eta_{nn}}\;,
\end{equation}
and consequently 
\begin{equation}
\label{c1 cond on tens ricci}
P_{lm}\ll ab/t^{2}\,,\qquad P_{ln}\ll ac/t^{2}\,,\qquad P_{mn}\ll bc/t^{2}\;.
\end{equation}
In this case,  the off-diagonal components of (\ref{c1 einstein alphabeta}) 
can be disregarded to  leading order. 
The Ricci tensor $P^{\beta}_{\alpha}$ for the metric (\ref{c1 kasner gen})
 is given in Appendix D of \reffcite{LK1963}.\\
 The diagonal projections $P^{l}_{l},\,P^{m}_{m},\,P^{n}_{n}$ contain the terms
\begin{equation}
\label{c1 term nelle proiezioni diagonali}
\frac{1}{2}\left(\frac{a\, l \nabla\wedge (a\, l)}{a\,b\,c 
\left(l\cdot[m\times n]\right)}\right)
\;\sim\;\frac{k^{2} a^{2}}{b^{2} c^{2}}=\frac{k^{2}a^{4}}{\Lambda^{2}t^{2}}\;,
\end{equation}
and analogous terms with $a\,l$ replaced by $b\,m$ and $c\,n$,  
where $1/k$ denotes the order of magnitude of spatial distances over which 
the metric significantly changes, and $\Lambda$ is the same as in 
(\ref{g29}). 
In (\ref{c1 term nelle proiezioni diagonali}), all the vectorial 
operations are performed as in the Euclidean case.\\
According to (\ref{c1 condizioni diagonali}), we get the inequalities
\begin{equation}
\label{c1 ancora}
a\sqrt{k/\Lambda}\ll 1\;,\qquad b\sqrt{k/\Lambda}\ll 1\;,\qquad c\sqrt{k/\Lambda}\ll 1\;,
\end{equation}
which  are not only necessary, but also sufficient conditions for 
the existence of the generalized Kasner solution. 
After conditions (\ref{c1 ancora}) are satisfied, all other terms 
in $P^{l}_{l},\,P^{m}_{m},\,P^{n}_{n}$, as well as $P_{lm},\,P_{ln},\,P_{mn}$, 
automatically satisfy  (\ref{c1 condizioni diagonali}) and 
(\ref{c1 cond on tens ricci}) as well.\\
 An estimate of these terms leads to the conditions
\begin{equation}
\label{c1 ultima stima}
\frac{k^{2}}{\Lambda^{2}} \left(a^{2}b^{2}\,,\ldots,\,
	a^{3}b\,,\ldots,\,a^{2}bc\,,\dots\right)\ll 1\;, 
\end{equation}
containing on the left-hand side the products of powers of two 
or three of the quantities which enter in (\ref{c1 ancora}), 
and therefore are \emph{a fortiori} true if the latter are satisfied.
Inequalities (\ref{c1 ultima stima}) moreover represent a natural 
generalization of those which already appearing in the  oscillatory regime 
in the homogeneous case.

As $t$ decreases, an instant $t_{tr}$  may eventually occur when one 
of the conditions (\ref{c1 ancora}) is  violated 
(the case when two of these are simultaneously violated can happen when 
the exponents $p_{1}$ and $p_{2}$ are close to zero, corresponding to the 
case of  small oscillations). 
Thus, if during a given Kasner epoch the negative exponent refers to 
the function $a(t)$, { i.e.}, $p_{l}=p_{1}$, then  at  $t_{tr}$, 
we have
\begin{equation}
\label{c1 cond di trans}
a_{tr}=\sqrt{\frac{k}{\Lambda}} \sim 1\,.
\end{equation}
Since during that epoch the functions $b(t)$ and $c(t)$ decrease 
with $t$, the other two inequalities in (\ref{c1 ancora}) remain valid and 
at $t\sim t_{tr}$ we shall have
\begin{equation}
\label{c1 cond di trans 2}
b_{tr} \ll a_{tr}\,,\hspace{0.5cm}c_{tr}\ll a_{tr}\,.
\end{equation}
At the same time all  conditions (\ref{c1 ultima stima}) continue to hold, 
and  all off-diagonal projections of (\ref{c1 einstein alphabeta}) may be 
disregarded. 
In the diagonal projections (\ref{c1 term nelle proiezioni diagonali}), 
only terms  containing $a^{4}/t^2$ become relevant. 
In such remaining terms we have
\begin{equation}
\label{c1 term import}
\left(a\,l\,\cdot\nabla\wedge\,(a\,l)\right) 
= a \left(l\cdot [\nabla a \times l]\right)+ 
a^{2}\left(l\cdot \nabla\wedge\, l\right) = 
a^{2} \left(l\cdot\nabla\wedge\, l\right)\;,
\end{equation}
{ i.e.}, the spatial derivatives of $a$ drop out.\\
As a result, we obtain the following equations for the 
replacement of two Kasner epochs
\bseq
\begin{align}
-R^{l}_{l} &=\frac{(\dot{a}bc)^{.}}{abc} + \lambda^{2} \frac{a^{2}}{2b^{2}c^{2}}=0\;,\\
-R^{m}_{m} &=\frac{(a\dot{b}c)^{.}}{abc} - \lambda^{2} \frac{a^{2}}{2b^{2}c^{2}}=0\;,\\
\label{c1 eq einst di trans}
-R^{l}_{l} &=\frac{(ab\dot{c})^{.}}{abc} - \lambda^{2} \frac{a^{2}}{2b^{2}c^{2}}=0\;,\\
-R^{0}_{0} &= \frac{\ddot{a}}{a}+\frac{\ddot{b}}{b}+\frac{\ddot{c}}{c}=0\; ,
\end{align}
\eseq
which  differ from the corresponding ones of the homogeneous indices 
(\ref{eeh}-\ref{eeh0}) only  for  the quantity
\begin{equation}
\label{c1 lambda gen}
\lambda(x) = \frac{l\cdot\nabla\wedge l}{l\cdot[m\times n]}\;,
\end{equation}
 no longer being  a constant, but a function of the space coordinates. 
Since (\ref{c1 eq einst di trans}) is a system of ordinary differential 
equations with respect to time where space coordinates 
enter parametrically only, such difference does not affect at all the 
solution of the equations and the following  map. 
Similarly,  the law of alternation of exponents derived for homogeneous 
indices remains valid in the general inhomogeneous case.
Numerical evidences\cite{WeaverIsenbergBerger1998PRL} support the point-like dynamics of the generic cosmological solution at the basis of the BKL conjecture.

\subsubsectionric{Rotation of the Kasner axes\label{sec:rotation}}


Even if the dynamics is quite similar to that of the homogeneous indicess
in vacuum, the  new feature of  the rotation of the 
Kasner axes emerges.
If in the initial epoch the spatial metric is given by (\ref{c1 kasner gen}), 
then in the final one  we  have
\begin{equation}
\label{c1 nuova metrica}
h_{\alpha\beta} = a^{2} l_{\alpha}^{\prime} l_{\beta}^{\prime} 
+ b^{2} m_{\alpha}^{\prime} m_{\beta}^{\prime} 
+ c^{2} n_{\alpha}^{\prime} n_{\beta}^{\prime}\;,
\end{equation}
with   $a,\,b,\,c$ given by a new set of Kasner indexes, 
and some vectors $l^{\prime},\,m^{\prime},\,n^{\prime}$. 
If we project all tensors (including $h_{\alpha\beta}$) in both 
epochs onto the same directions $l,\,m,\,n$,  the turning 
of the Kasner axes can be described as the appearance, in the final 
epoch, of off-diagonal projections $\eta_{lm},\,\eta_{ln},\,\eta_{mn}$, 
which behave in time as linear combinations of the functions 
$a^{2},\,b^{2},\,c^{2}$. 
Such projections do indeed appear \cite{BKL82}, and  induce 
the rotation of the Kasner axes. 

The main effects can be reduced to a rotation of the 
$m-$ and $n-$ axis  by a large  angle, and a rotation of the 
$l-$ axis by a small one, thus neglecting the small changes 
of $l$  the new Kasner axes are related to the old ones as 
\begin{equation}
\label{c1 rot assi di kasner}
l^{\prime} = l\,,\hspace{0.5cm}m^{\prime} = m +\sigma_{m} l\,,\hspace{0.5cm}n^{\prime} = n + \sigma_{n}l\,,
\end{equation}
where the $\sigma_{m}\,,\sigma_{n}$ are of order  unity, 
and are given by 
\bseq
\begin{align}
\nonumber 
\sigma_{m} &= -\frac{2}{p_{2}+3p_{1}}\left\{[l\times m]
	\cdot 	\nabla\frac{p_{1}}{\lambda}+
	\frac{2p_{1}}{\lambda} m\cdot\nabla\wedge l\right\}
	\frac{1}{l\cdot[m\times n]}\;, \\
\sigma_{n} &= \frac{2}{p_{2}+3p_{1}}\left\{[n\times l]
	\cdot \nabla\frac{p_{1}}{\lambda}-
	\frac{2p_{1}}{\lambda} n\cdot\nabla\wedge l\right\}
	\frac{1}{l\cdot[m\times n]}\;,
\end{align}
\eseq
and  can be inferred also from the $0-\alpha$ Einstein equations
which play the role of constraints to the space functions.

The rotation of the Kasner axes (that appears for a matter-filled 
homogeneous space only) is inherent in the inhomogeneous solution already 
in the vacuum case. 
The role played by the matter energy-momentum tensor can be mimicked 
by the terms due to inhomogeneity of the spatial metric in the Einstein equations. 
As in the generalized Kasner solution, in the general inhomogeneous 
approach to the singularity the presence of matter is exhibited in the 
relations between the arbitrary spatial functions which appear in the solution only.

\subsectionric{The fragmentation process\label{sec:frag}}

We will now qualitatively discuss a further mechanism that 
takes place in the inhomogeneous Mixmaster indices in the 
limit towards the singular point: the so-called {\it fragmentation} process \cite{Kochnev1987,Kir93,Montani1995CQG}.

The extension of the BKL mechanism to the general inhomogenous case 
contains the  physical restriction of the  ``local homogeneity'':
in fact, the general derivation  is based on the assumption that 
the spatial variation of all  spatial metric components  possesses 
the same characteristic length, described by a unique parameter $k$, 
which can be regarded as an average wavenumber. 
Nevetheless,  such local homogeneity could cease to be valid, 
 as a  consequence of the asymptotic evolution towards the singularity.\\
The conditions (\ref{c1 vinc kasn gen})
do not require that the functions $p_{a}(x^{\gamma})$ have the same ordering 
in all points of space. 
Indeed, they  can vary their ordering throughout space  an infinite number 
of times without violating  conditions (\ref{c1 vinc kasn gen}), in agreement 
with the oscillatory-like behaviour of their spatial dependence. 
Furthermore, the most important property of the BKL map evolution 
is the strong dependence  on initial values, 
 which produces an exponential divergence of the trajectories 
resulting from its iteration.  

Given a generic initial condition $p_{a}^{0}(x^{\gamma})$, the continuity 
of the three-manifold requires that, at two nearby space points, 
 the Kasner index functions assume correspondingly close values. 
However, for the  mentioned property, the trajectories emerging 
from these two values  exponentially diverge and, since the 
$p_{a}(x^{\gamma})$ vary within the interval $[-1/3;\,1]$ only, 
indeed  the  spatial dependence acquires an increasingly 
oscillatory-like behaviour.\\
In the simplest case, let us assume that, at a fixed instant of time 
$t_{0}$, all the points of the manifold are described by a generalized 
Kasner metric,  the Kasner index functions have the same ordering point 
by point, and   $p_{1}(x^{\gamma})$, $p_{2}(x^{\gamma})$, 
$p_{3}(x^{\gamma})$ are described throughout the whole space, by a narrow 
interval of $u$-values, i.e. $u \in [K,\,K+1]$ for a generic integer $K$. 
We refer to this situation asa  manifold  composed by one ``island''.
We introduce the remainder part of  $u(x^{\gamma})$ as
\begin{equation}
\label{c1 remainder}
X(x^{\gamma}) = u(x^{\gamma})-[u(x^{\gamma})] \, , \qquad X \in [0,\,1)   
\quad \forall x \in \Sigma 	\, ,
\end{equation}
where the square brackets indicate the integer part. 
Thus  the values of the narrow interval  can be written as 
$u^{0}(x^{\gamma}) = K^{0}+ X^{0}(x^{\gamma})$. 
As the evolution goes by, the BKL mechanism induces a transition 
from an epoch to another; the $n$-th epoch is characterized by an
interval $[K-n,\,K-n+1]$, until $K-n=0$; then the era comes to an end
and a new one begins. 
The new  $u^{1}(x^{\gamma})$ starts from $u^{1} = 1/X^{0}$, 
{ i.e.}, takes value in the  interval $[1,\,\infty)$. 
Only very close points  can still be in the same ``island'' of $u$ values; 
distant ones  in space will  be described by very different integer $K^{1}$
and will experience eras of different length. 
As the singular point is reached, more and more  eras take place, 
causing the formation of a greater and greater number of smaller 
and smaller ``islands'', providing the ``fragmentation'' process. 
Our interest is focused to  the value of the parameter $k$, 
which describes the characteristic length and increases as the islands 
get smaller. 
This implies  the progressive increase of the spatial gradients and in principle
could deform  the BKL mechanism. \\
We  argued that this is not the case with  a qualitative explanation: 
the progressive increase of the spatial gradients produces the same 
qualitative effects on all the terms present in the three-dimensional 
Ricci tensor, including the dominant ones. 
In other words,  for each single value of $k$, in every island, a 
condition of the form
\begin{align}
\label{c1 cond k grad}
&\frac{\text{inhomogeneous term}}{\text{dominant term}} 
\sim \frac{k^{2} t^{2K_{i}} f(t)}{k^{2} t^{4p_{1}}} \ll 1\;,\nonumber\\
&K_{i}  = 1 - p_{i}\geq 0,\hspace{0.5cm}f(t)
	= {O}(\ln t,\,\ln^{2} t),\hspace{0.5cm}t\ll 1\nonumber\;,
\end{align}
is still valid, where we call inhomogeneous the terms containing spatial 
gradients of the scale factors, which are evidently absent from the 
dynamics of the homogeneous cosmological indicess.

From this point of view, the fragmentation process does not 
produce any behaviour capable of stopping the iterative scheme of 
the oscillatory regime.

\subsectionric{Hamiltonian formulation and  dry turbulence\label{sec:dryt}}

As mentioned above, a generic cosmological solution is represented 
by a gravitational field with  all its degrees of 
freedom and, therefore, allowing to specify a generic Cauchy problem.\\
In the ADM formalism, the three-metric tensor
corresponding to such a generic indices  reads as 
\begin{equation}
\label{parametrizzazione della metrica}
	h_{\alpha\beta}=e^{q_a}\delta_{ad}O^a_b O^d_c \partial_\alpha y^b \partial_\beta y^c \, , 
\end{equation} 
where $q^a=q^a(x,t)$ and $y^b=y^b(x,t)$ are six scalar functions
and $O^a_b=O^a_b(x)$ is a $SO(3)$ matrix.
Thus the action for the gravitational field is
\bseq
\begin{align}
\label{azione standard}
	S &=\int_{\Sigma \times\nR}dt d^3 x\left(p_a\partial_t q^a
	+\Pi_d\partial_t y^d -NH-N^\alpha H_\alpha\right)\,, \\
	\label{vincoli hamiltoniani} 
	H &=\frac{1}{ \sqrt h}\left[\sum_a (p_a)^2
	-\frac{1}{2}\left(\sum_b p_b\right)^2- h {~}^{(3)}R \right] \\
\label{vincoli hamiltoniani2}	 
	 H_\alpha& =\Pi_a \partial_\alpha y^a +p_a \partial_\alpha q^a 
	+2p_a(O^{-1})^b_a\partial_\alpha O^a_b \,,
\end{align}
\eseq
where $p_a$ and $\Pi_d$ are the conjugate momenta of the variables $q^a$ 
and $y^b$, respectively.

The ten independent components of a generic metric tensor are represented by 
the three scale factors $q^a$, the three degrees of freedom $y^a$, 
the lapse $N$ and the shift-vector $N^\alpha$; by the variation 
of the action (\ref{azione standard}) with respect to  $p_a$, $\Pi_a$, the relations
\begin{align}
	\label {condizione di gauge sullo shift-vector}
	\partial_t y^d &=N^\alpha\partial_\alpha y^d  \\
  \label {condizione di gauge sulla lapse function}
	N& =\frac{\sqrt{h}}{\sum_a p_a}
	\left(N^\alpha\partial_\alpha\sum_b q^b-\partial_t\sum_b q^b\right)
\end{align}
hold.

We remind that a wide class of cosmological models resembling the behavior of the inhomogeneous Mixmaster is the so-called Gowdy cosmology\cite{MCP,IsenbergMoncrief}.

\subsubsectionric{Solution of the super-momentum constraint\label{sec:ssmc}}

By the usual Hamiltonian constraints (\ref{vincoli hamiltoniani}) 
and (\ref{vincoli hamiltoniani2}), the  super-momentum can be 
solved by choosing the function $y^a$ as space coordinates, 
taking  $\eta=t$, and $y^a=y^a(t,x)$, getting
\begin{equation}
	\Pi_b=-p_a\frac{\partial q^a}{\partial y^b}
	-2p_a(O^{-1})^c_a\frac{\partial O^a_c}{ \partial y^b} \,.
\end{equation}
and furthermore 
\bseq
\begin{align}
	\label{cicciaa}
		q^a(t,x) & \to q^a(\eta,y)   \\  
\label{cicciab}		
	p_a(t,x) & \to p'_a(\eta,y)={p_a(\eta,y)/|J|} \\
\label{cicciac}	
	\frac{\partial}{ \partial t} &\rightarrow 
		\frac{\partial y^b}{ \partial t}\frac{\partial}{ \partial y^b}
		+\frac{\partial}{ \partial \eta}   \\ 
\label{cicciad}
		\frac{\partial}{ \partial x^\alpha} & \rightarrow 
		\frac{\partial y^b}{ \partial x^\alpha}\frac{\partial}{ \partial y^b}  
	\,,
\end{align}
\eseq
where $|J|$ denotes the Jacobian of the transformation, 
relation (\ref{cicciaa}) in general  holds for all the scalar quantities, 
while (\ref{cicciab})  for all the scalar densities
and action (\ref{azione standard}) rewrites as
\begin{equation}
	\label{finale non approssimata}
	S=\int_{\Sigma \times\nR}d\eta d^3 y 
	\left(p_a\partial_\eta q^a+2p_a(O^{-1})^c_a\partial_\eta O^a_c-NH\right)\,.
\end{equation}

\subsubsectionric{The Ricci scalar\label{sec:inhricci}}

In this scheme, the potential term appearing in the super-Hamiltonian 
reads  as
\begin{equation}
\label{potU}
	U=\frac{D }{ |J|^2}\ ^{(3)}R=
\sum_a \lambda_a^2 D^{2 Q_a}+
\sum_{b\neq c}D^{Q_b+Q_c}\mathcal{O}\left(\partial q, \left(\partial q\right)^2,y,\eta\right)
\end{equation}
where $D=h |J|$, $h\equiv \exp{\sum_a q^a}$, the $Q_a$ are the anisotropy 
parameters (\ref{qucona}) and $\lambda$ are the functions
\begin{equation}
	\lambda_a^2\equiv \sum_{k,j}\left(O^a_b \vec\nabla O^a_c 
\left(\vec\nabla y^c\wedge	\vec\nabla y^b \right)^2\right)\, .
\end{equation}
Assuming  $y^a(t,x)$ smooth enough (which implies 
smoothness of  the coordinates system as well), then all the gradients 
appearing in the potential $U$ are regular, in the sense 
their behavior is not as strongly divergent as to destroy the billiard 
representation. 
It was shown \reffcite{Kir93} that the spatial gradients logarithmically 
increase in the proper time along the billiard's geodesics and result 
of higher order.
Thus, as $D\to 0$  the spatial curvature ${~}^{(3)}R$ diverges and 
the cosmological singularity appears; in this limit, the first term of 
$U$ dominates all the remaining ones and can be indicesed by the potential
\begin{equation}
\label{UU}
	U=\sum_a{\Theta(Q_a)}\, ,
\end{equation}
resembling the behavior of the Bianchi IX indices.
By (\ref{UU}) the Universe dynamics independently evolves 
in each space point; the point-Universe moves within the  dynamically-closed 
domain $\Pi_Q$ and  near the singularity we have $\partial p_a/ \partial\eta=0$. 
Thus, the term $2p_a(O^{-1})^c_a\partial_\eta O^a_c$ in 
(\ref{finale non approssimata}) behaves as an exact time-derivative and 
can be ruled out of the variational principle. 

The same analysis developed for  the homogeneous Mixmaster model 
in Section \ref{sec:mcl} can be straightforwardly implemented in a 
covariant way (i.e. without any gauge fixing for the lapse function or 
for the shift vector).

The super-Hamiltonian constraint is  solved in the domain $\Pi_Q$ as
\begin{equation}
	\label{hamiltoniana ADM}
	-p_\tau\equiv\mathcal{H}_{ADM} = \sqrt{(\xi^2-1)p_\xi^2+\frac{p_\theta^2}{\xi^2-1}}
\end{equation}
and the reduced action reads as
\begin{equation}
	\label{hamiltoniana ridotta}
	\delta S_{\Pi_Q}=\delta\int d\eta d^3 y
 \left(p_\xi\partial_\eta\xi+p_\theta\partial_\eta\theta
	-\mathcal{H}_{ADM}\partial_\eta\tau\right)=0\,.
\end{equation}
By the asymptotic limit (\ref{UU}) and the Hamilton equations associated 
with (\ref{hamiltoniana ridotta}) we get 
$d\epsilon/ d\eta=\partial\epsilon/\partial\eta=0$  
and therefore 
${\mathcal H}_{ADM} \epsilon(y^a)$
is a constant of motion even in the non-homogeneous case.

\section{The Iwasawa decomposition}

We will now briefly discuss the generalization of the technique due to Chitr\'e and Misner introduced in \cite{cosmbill}. This technique was adopted by the authors to study the dynamics of the multi-dimensional Einstein equations coupled to dilatons and $p$-forms in the neighborhood of a generic singularity (see last section of this chapter). We will limit our discussion to the four dimensional case only.

The Iwasawa decomposition of the spatial metric $h_{\alpha \beta}$ is based on the choice of a unique oriented orthonormal spatial coframe $\{\omega^a\}$, $\omega^a = e^a_\alpha dx^\alpha$ defined by the relations
\begin{equation}
	e^a_\alpha = \sum_\beta D^a_b M^b_\alpha 
\end{equation} 
\begin{equation}\label{Iwasawa}
\Big(\,D^a_b\,\Big) =
\begin{pmatrix}
e^{-b^1} & 0 & 0 \\
0 & e^{-b^2} & 0 \\
0 & 0 & e^{-b^3}
\end{pmatrix} \:,\hspace{5mm}
\Big(\,M^a_\alpha\,\Big) =
\begin{pmatrix}
1 & m_1 & m_2 \\
0 & 1 & m_3 \\
0 & 0 & 1
\end{pmatrix} \:,
\end{equation}
resulting in the following expression for $h_{\alpha \beta}$
\begin{equation}
	h_{\alpha \beta} = \sum_a \exp(-2b^a) M^a_\alpha M^a_\beta
\end{equation}
The existence and the uniqueness of this frame is a consequence of the uniqueness of the QR decomposition in linear algebra; furthermore the matrix $M$ can be viewed as representing the Gram-Schmidt orthogonalization of the spatial coordinate coframe.

In this variables, the Lagrangian of the gravitational field reads
\begin{equation}
	\mathcal{L} = \displaystyle\frac{\sqrt{h} }{N} \left[ \sum_a b^a b^a - \left(\sum_b b^b \right)^2 + \displaystyle\frac{1}{2} \sum_{c<d} \exp[2(b^d-b^c)] \dot{M}^c_\alpha \left(\dot{\bar{M}}^\alpha_d \right)^2 \right] + \sqrt{h} N ^3R\:,
\end{equation}
where $\bar{M}$ denote the inverse matrix of $M$.

As soon as the Legendre transformation is taken and the momenta $\pi_a$ and $P^\alpha_a$ introduced, the gravitational Hamiltonian in vacuum is obtained
\begin{equation}\label{ugglah}
	H = N \left[ \displaystyle\frac{1}{4}\left( \sum_a \pi_a \pi_a - \displaystyle\frac{1}{2} \left(\sum_b \pi_b \right)^2 \right) + \displaystyle\frac{1}{2} \sum_{c<d} \exp[2(b^d-b^c)] \dot{M}^c_\alpha \left(P^\alpha_d \right)^2   - g ^3R\right]\;.
\end{equation}

In \cite{cosmbill} it is argued that, as soon as the existence of a space-like singularity in the past is hypothesized, $b^a$ is expected to be timelike in the vicinity of this singularity (in the sense of the associated Minkowskian metric that endows the $b^a$-space). This allows to introduce a new set of variables, $\lambda$ and the "`orthogonal angular variables"' $\gamma_a$ in which the Hamiltonian \ref{ugglah} reads (the lapse function has been rescaled and set equal to $\exp(2\lambda)$)
\begin{equation}
	H = \displaystyle\frac{1}{4}\left[-\pi_\lambda^2 +\sum_a(\pi_{\gamma_a}) \right] + e^{2\lambda} \sum_A c_A \exp(-2 w_A (\gamma) e^\lambda)\;;
\end{equation}
here $c_A$ denotes some functions of spatial derivatives of the metric, off-diagonal metric variables, and momenta; $w_A(\gamma)$ denote linear forms of the variables $\gamma_a$, i.e. $w_A(\gamma) = \sum_b (w_A)_b \gamma_b$.
It is expected the singularity to appear in the limit $\lambda\to\infty$; in this limit, each term of the summation becomes an infinite wall, and can be described by the usual generalized $\Theta$ function. Like in Chit\`e-Misner picture, the are three dominant terms (in the vacuum case) that describe the billiard table. The coefficients $c_A$ are non negative functions of the variables, but "`generically"' they are positive \cite{cosmbill}. The three dominant terms $c_A e^{2 \lambda}\exp[-2 w_A(\gamma) e^\lambda]$ are assumed to be the only relevant contributions for the asymptotic dynamics, while the other are subdominant and dropped away. The limiting Hamiltonian then reads
\begin{equation}
	H_{asymptotic} = \displaystyle\frac{1}{4}\left[-\pi_\lambda^2 +\sum_a(\pi_{\gamma_a}) \right] + \sum_{A = 1}^3 \Theta[-2w_A(\gamma)]
\end{equation}
As in the previous sections, this Hamiltonian is independent of $M$, $P^\alpha_a$ and $\lambda$, suggesting the existence of asymptotic constants of the motion. The remaining variables describe free motion on the hyperbolic space, constrained by the three walls.

This Hamiltonian formulation of Einstenian theory was adopted in the cosmological framework by Uggla and collaborators in a series of paper \reffcite{Uggla1,Uggla2,Uggla3}. In particular, they split the field equations into decoupled equations for the conformal factor (by means of the conformal Hubble-normalization) and a coupled system of dimensionless equations for quantities associated with the dimensionless conformal metric. This reduced dimensionless system carries the essential information about the problem, while the dimensional one allows to recover the physical metric. One of the main advantages of Hubble-normalization lies in the behaviour of the dynamical variables as the cosmological singularity is reached: the Hubble-normalized variables remain bounded despite the standard field ones do not (for further details see the recent review \reffcite{Uggla1}.\\
When this formalism is applied to the generic field equations in the vicinity of the initial singularity, it can be demonstrated the existence of the so-called \textit{billiard attractor} \reffcite{Uggla2}. The main feature of this attractor is to give a precise meaning to the notion of \textit{piece-wise} BKL approximation. This way the space-like cosmological singularity is \textit{silent}, denoting the decoupling of space points on the causal horizon \reffcite{Uggla3}. Furthermore it was established the existence of a duality of such framework with the cosmological billiard approach of Damour et al. \reffcite{cosmbill}.

%
%
%
%
There are some other interesting features, which we mention here as a reference 
although demanding to the literature for details. 
For the Bianchi type IX there is a very strong limit on global vorticity, 
caused by the difficulty of ``fitting in'' a vortex in the type IX 
compact geometry \reffcite{1969MNRAS.142..129H,1973MNRAS.162..307C,1985MNRAS.213..917B,1986CaJPh..64..152B}.
Moreover, in the inhomogeneous case arises a linearisation 
instability
\reffcite{1979PhR....56..371B}, 
i.e. linearisations around the killing vectors of compact spaces are not stable
and series expansions around them in general do not form the leading order
terms of a series converging to a true solution of the Einstein equations, 
unless staisfying further constraints
%


\subsectionric{Inhomogeneous quantum Mixmaster\label{sec:inhqmix}}

Since the spatial gradients of the configurational variables don't 
play a relevant dynamical role in the asymptotic limit 
$\tau\to\infty$ (indeed the spatial curvature behaves as a potential well), 
then the quantum evolution  takes independently place in each space 
point and the total wave function of the Universe can be represented as
\begin{equation}
\label{funzione d'onda dell'universo}
	\Xi (\tau,\;u,\;v)= \Pi_{x_i} \xi_{x_i} (\tau,\;u,\;v)
\end{equation}
where the product is (heuristically) taken over all the points 
of the spatial hypersurface.\\
However, in the spirit of the long-wavelength approximation, 
the physical meaning of a space point must be recovered on the 
notion of a cosmological horizon; 
in fact, we are dealing with regions where  the inhomogeneity 
effects are negligible and this  corresponds to super-horizon sized 
spatial gradients. 
Even if this request  is statistically  well confirmed to
classical level\cite{Kir93}, in the  quantum sector it acquires a precise 
meaning  if we refer the dynamics to a lattice indicesing the space-time 
in which the spatial gradients of the configurational variables become 
real potential terms. 
In this respect, the geometry of the space-time is expected to acquire 
a discrete structure on the Planck scale  and we believe that a 
regularization of our approach could arrive from a loop quantum gravity 
treatment\cite{Bojinom}.\\
Despite this local homogeneous framework of investigation, the appearance 
near the singularity of high spatial gradients and of a space-time foam 
(like as outlined in classical dynamics\cite{Kir93,Montani1995CQG}) 
can be  recognized in the above quantum picture too.
In fact, the probability that in $n$ different space points (horizons) 
the variables $u$ and  $v$ take values within the same narrow interval
decreases with $n$ as $p^n$, $p$ being the probability in a single point; 
all probabilities are identical to each other and no interference phenomenon 
takes place. 
From a physical point of view, this  consideration indicates that a smooth 
picture of the large scale Universe is forbidden on a probabilistic level 
and different causal regions are expected to be completely disconnected 
from each other during their quantum evolution.
Therefore, if we start with a strongly correlated initial wave function 
$\Xi_0(u,v)\equiv\Xi(\tau_0,u,v)$, its evolution toward the singularity 
induces increasingly irregular distributions, approaching 
(\ref{funzione d'onda dell'universo}) in the asymptotic limit $\tau\to\infty$.

\subsectionric{Multidimensional oscillatory regime\label{sec:multidosc}}

Let us consider a $(d+1)$-dimensional
space-time $(d\ge3)$,
whose associated metric tensor obeys a dynamics described
by a generalized vacuum Einstein equations 
\begin{equation}
^{(d+1)}R_{ik}=0  \;,\qquad  (i,k =0,1,...,d)\,
\, , 
\label{aric}
\end{equation}
where the $(d+1)$-dimensional Ricci tensor
takes its natural form in terms
of the metric components $g_{ik}(x^l)$.\\
In \reffcite{Demaret1986}, it is shown that
the inhomogeneous Mixmaster behaviour finds a direct 
generalization in correspondence to any
value of $d$. 
Moreover, in correspondence to $d>9$, the generalized 
Kasner solution acquires  a generality character, in the sense 
of the number of arbitrary functions, i.e. without a condition 
analogous to ({\c1 lambda gen}).\\ 
In a synchronous reference (described by the usual coordinates
$(t, x^\gamma)$),
the time-evolution of the $d$-dimensional spatial
metric $h_{{\alpha}{\beta }}(t, x^\gamma)$ singles out an iterative 
structure near the  cosmological singularity ($t = 0$). 
Each single stage consists of
intervals of time (Kasner epochs) during which 
$h _{\alpha \beta }$ 
takes the generalized Kasner form
\begin{equation}
h_{\alpha \beta}(t, x^\gamma) =
\sum_{a=1}^{d} t^{2p_{a}}l^{a}_{\alpha}l^{a}_{\beta}
\;,
\label{bric} 
\end{equation}
where the Kasner indeces $p_{a}(x^{\gamma })$  
satisfy 
\begin{equation}
\sum_{a=1}^{d} p_{a}(x^\gamma)\ =
\sum_{a=1}^{d} p_{a}^{2}(x^\gamma)\ = 1 \;,
\label{cric}
\end{equation}
and $ l^{1}(x^\gamma),..., l^{d}(x^\gamma)$
denote $d$ linear independent 1-forms 
$l^{a}={m^{a}_{\alpha} dx^{\alpha}}$, whose components
are arbitrary functions
of the spatial coordinates.\\
In each point of space, the conditions (\ref{cric}) 
define a set of ordered indexes
$\{ p_a\}$ ($p_1\le p_2\le ...\le p_d$) which,
from a geometrical point of
view, fixes one point in $\nR^d$, 
lying on a connected portion
of a $(d-2)$-dimensional sphere.
We note that  conditions (\ref{cric}) 
require $p_1\le 0$ and $  p_{d-1}\ge 0$,
where the equality takes place for the values
$p_1 = \ldots= p_{d-1} = 0 $ and $ p_d =1 $ only.
The following $d$-dimensional BKL map, linking the old Kasner exponents $p_a$ to the new ones $q_a$, holds  \reffcite{1987JSP....48.1269E,Elskens:1987gj}
\begin{equation}
q_1 = \displaystyle\frac{-p_1 - P}{1 + 2p_1 + P}\,,\quad q_{2} = \displaystyle\frac{p_2}{1 + 2p_1 + P}\,,...\,,\quad q_{d-2} = \displaystyle\frac{p_{d-2}}{1 + 2p_1 + P}
\end{equation}
\begin{equation}
q_{d-1} = \displaystyle\frac{p_{d-1} + 2p_1 + P}{1 + 2p_1 + P}\,,\quad q_d = \displaystyle\frac{p_d +2 p_1 + P}{1 + 2p_1 + P}
\end{equation}
where $P = \sum_{a = 2}^{d-2} p_a$.

As shown in \reffcite{Demaret1986su,Hosoya1987} 
(see also \reffcite{Kir93,KirillovMelnikov1995}),
each single step of the 
iterative solution is  stable, in a given 
point of the space, if
\begin{equation}
\label{c4 dem r}
\lim_{t\to 0}t^{2}{~}^{(d)}R^b_a = 0\;.
\end{equation}
The limit (\ref{c4 dem r}) is a sufficient condition 
to disregard the dynamical effects of the spatial gradients 
in the Einstein equations.
An elementary computation shows how  the only  terms capable 
to perturb the Kasner behavior  in $t^{2}\phantom{i}~^{(d)}R$ 
contain the powers $t^{2\alpha_{abc}}$, where $\alpha_{abc}$ 
are related to the Kasner exponents as  
\begin{equation}
\alpha_{abc} = 2p_a + \sum_{d\ne a,b,c} p_d\;,\qquad
(a\ne b, a\ne c, b\ne d) \, , 
\label{eric}
\end{equation}
and for  generic  $l^a$, all possible powers $t^{2\alpha_{abc}}$ appear 
in $t^{2}\phantom{i}^{(d)}R^b_a$. 
This leaves two possibilities for the vanishing of  
$t^{2} \phantom{i}^{(d)}R^b_a$  as $t\to 0$. Either the Kasner exponents can 
be chosen in an open region of the Kasner sphere defined in (\ref{cric}), 
so as to make $\alpha_{abc}$ positive for all triples $a,b,c$, or the conditions 
\begin{equation}
\alpha_{abc}(x^\gamma) > 0 \qquad  \forall (x^1,\ldots,x^d)
\label{dric}
\end{equation} 
are in contradiction with (\ref{cric}), and one must impose 
extra conditions on the functions $l$ and their derivatives.
The second possibility occurs, for instance, for $d=3$, 
since $\alpha_{abc}$ is given by $2 p_{a}$, and one Kasner 
exponent is always negative, i.e. $\alpha_{1,d-1,d}$. 
Thus (\ref{cric}) is a solution of the vacuum Einstein equations 
to leading order if and only if the vector $l^{1}$
obeys the additional condition
\begin{equation}
\label{faklj}
l_{1}\cdot\nabla\wedge l_{1}=0\;,
\end{equation}
and this kills one arbitrary function.\\
It can be shown \cite{Demaret1986su,Hosoya1987}  
that, for $3\le d\le 9$, 
at least the smallest of the quantities 
(\ref{eric}), i.e. $\alpha_{1,d-1,d}$ results to be
always negative
(excluding isolated points $\{p_i\}$ in which it vanishes); 
for $d\ge 10$ an open region exists of the
$(d-2)$-dimensional Kasner sphere where $\alpha_{1,d-1,d}$
takes positive values, 
the so-called Kasner Stability Region (KSR).\\
For $3\le d \le 9$, the
evolution of the
system to the singularity consists of an infinite
number of Kasner epochs, while  for $d\ge 10$,
the existence of the KSR, 
implies a profound modification in the asymptotic dynamics. 
In fact, the indications presented
in \reffcite{Demaret1986,KirillovMelnikov1995} in favor of the ``attractivity''
of the KSR, imply that in each space point
(excluding sets of zero measure)
a final stable Kasner-like regime appears.

In correspondence to any value of $d$, 
the considered iterative scheme contains the right number of
$(d+1)(d-2)$ physically arbitrary functions of the
spatial coordinates,
required to specify generic initial conditions
(on a non-singular
space-like hypersurface). 
In fact,  we have $d^{2}$ functions from the $d$ vectors $l$ and 
$d-2$ Kasner exponents; 
invariance under spatial reparametrizations
allows to eliminate $d$ of  these functions, and other $d$ 
because of the $0a$ Einstein equations.
 This piecewise solution 
describes the asymptotic evolution
of a generic inhomogeneous multidimensional
cosmological indices.
For a cosmological application of the eleven space-times models see \reffcite{Montani2004IJMPDcompactification}. A wide analysis of the dynamics concerning homogeneous and inhomogeneous multidimensional cosmologies see \reffcite{1985JApA....6..137L,1985PhRvD..32.1595B,1986PhLB..167..157L,1987PhLA..123..379K,1987PhLB..195...27S,1987PhRvD..36.2945D,1988PhLB..207...36T,1990ApJ...358...23B,1998PhRvD..58b4001P,2000PhRvD..62l4016C,1986PhLB..167..157L,1985PhRvD..31..929L,1985CQGra...2..829L,1985PhLB..153..134L,1985PhLB..151..105L,1984PhLB..149...79L,1984PhLB..148...43L}.

We summarize some properties about the insertion of $p$-forms 
and dilatons in the gravitational dynamics. 
This is  a  wide topic, and  for review  we refer to \reffcite{cosmbill}.

The inclusion of massless $p$-forms, in a generic multi-dimensional 
indices \cite{DamourHenneauxPRL2000,DamourHenneauxPLB2000,DHJN2001}, can restore chaos when it 
is otherwise suppressed. 
In particular, even though pure gravity is non-chaotic in 11 space-time 
dimensions, the 3-forms of $d+1 = 11$ supergravity\cite{Demaret1985} make  the system 
chaotic   (those $p$-forms are part of 
the low-energy bosonic sector of superstring/M-theory indices\cite{DamourHenneauxPLB2000,DamourHenneauxPRL2000}). \\
The billiard description in the four dimensional case is quite general 
and can be extended to higher space-time dimensions, 
with $p$-forms and dilatons \cite{imk,DamourHenneauxPRL2001,Scalarestringhe}. 
If  there are $n$ dilatons, the billiard is a region of the hyperbolic 
space $\Pi_{d +n-1}$, and in the Hamiltonian each dilaton is equivalent
to the logarithm of a new scale factor.\\ 
The other ingredients that enter 
the billiard definition are the different types of the walls bounding it: 
symmetry walls related to the off-diagonal components of the 
spatial metric, gravitational 
walls related to the spatial curvature, and $p$-form walls 
(electric and magnetic).
All of them  are hyper-planar, and the billiard is a convex polyhedron 
with finitely many vertices, some of which are at infinity.   	

For the cosmological application of the brane framework\cite{Maartens}, see \reffcite{brane1,brane2,brane3}, in which the homogeneous models and the Mixmaster chaos are discussed.

\chapter{Conclusions}

By this work, we resumed many of the
key efforts made over the last four decades  to
characterize, as accurate as possible, the nature of the
asymptotic behavior of the Universe when it approaches
the singularity. In particular, the description of the
oscillatory regime for the Bianchi VIII and IX models
outlines that the singular character of the solution
survives also when the isotropy condition is relaxed,
but the detailed features of the dynamics can be deeply
altered, driving up the smooth behaviour of the FLRW
scale factor to a chaotic evolution of the no longer
equivalent directions.

However, the astonishing result consists of the possibility
to extend the oscillatory regime to the generic inhomogeneous
case, as far as a sub-horizon sized geometry is concerned,
especially in view of the stability this picture acquires
with respect to the presence of a matter fluid.
We think that this very general dynamics does not provide
only the important, but somehow academical, proof about the
existence of the singularity (as a general feature of the
Einstein equations under cosmological hypotheses),
but it also represents the real physical arena
to implement any reliable theory of the Universe birth.
Indeed, both from a classical and from a quantum point of view,
the inhomogeneous Mixmaster offers a scenario of full generality to
investigate the viability of a theoretical conjecture,
without the serious shortcoming of dealing with specific
symmetries (in principle even not appropriate to the
aim of the implementation, i.e. consider the use of the
minisuperspace when the quantization is performed,
disregarding causality).
In this sense, the results we presented here are only the
basic statements for  future developments on the nature
of the generic singularity. An important issue will be
to fix the chaotic features as  expectable
properties of the Universe origin, when a convincing
proposal for the quantization of gravity will acquire
the proper characteristics of a Theory. 
Of course, in a quantum picture,
the chaotic nature of the Universe would be translated into
certain indicators of the state arrangements. The
transition to the classical limit of an expanding
Universe must be addressed as a crucial stage for the early
cosmology investigation.

On the basis of the material included in this work and in view
of the present knowledge of fundamental interactions,
we can formulate the following proposal for the
origin of the actual Universe from the generic singularity.
Far from being a definitive picture of the Universe birth,
it better represents the course grain that would be useful,
in our opinion, as a guideline for future investigations
in this field.

We summarize our point of view into the following
individual, but correlated, steps:

\begin{itemize}
\item
The Universe was born in a quantum configuration, in which
the gravitational field and the kinetic term of the scalar inflaton
field are the only important degrees of freedom. In this scenario,
the scalar field would play the role of a time variable
(in a relational sense) and a Big-Bounce could be inferred.

\item 
The presence of the scalar field, as well as other pure
quantum effects, would imply the absence of real
quantum chaos features. Thus the Universe could reach
configuration of semiclassical nature, as far as
regions of enough large volume are explored. 
However, we
could infer that, without an additional contribution, 
the classical limit could
not arise. Two main terms can be  suggested as good candidates
for this scope, the quadratic potential of small
oscillations and a cosmological constant term; we
propose that these two terms could act as synergic ones to
each other.

\item 
After the classical limit is established, the achievement 
of an isotropic and homogeneous
Universe, up to a large scale, is got by the action of
the classical slow-rolling phase of the scalar field.
In this respect it is crucial that the cosmological
term can act before the three-scalar of curvature becomes
important enough, for instance during a long era of small
oscillations or in the damped oscillation of small
anisotropies.

\end{itemize}

However, we have to stress that the reliability of this
picture is concerned with a strong cosmological term,
to some extents, significantly greater than the one
expected by the current theory of inflation.
Thus the profile of the inflationary Universe
has, in this perspective, to be modified even to include
aspects eventually related to a quantum, or semiclassical
de-Sitter phase.

Finally, we emphasize that the traced ideas rely on the
assumption, made in this review article, to deal only with the
Einstein-Hilbert action. More general proposals for
the quantum interaction of gravity and matter fields
could fix different character for the Universe birth, but
eventually enforce the presented point of view.

We would like to thank all the colleagues who gave valuable feedback 
on our work, and in particular, John Barrow, Dieter Lorenz-Petzold and 
Henk van Elst
\listoffigures
\addtocontents{lof}{The permission to include the figures above was required to the concerned Journals and obtained.}


\begin{thebibliography}{100}
\newcommand{\enquote}[1]{``#1''}
\providecommand{\url}[1]{\texttt{#1}}
\providecommand{\urlprefix}{URL }
\providecommand{\eprint}[2][]{\url{#2}}

\bibitem{abramowitz}
M.~E. D.~T. Abramowitz and I.~A. Stegun.
\newblock \emph{{Handbook of mathematical functions, with formulas, graphs, and
  mathematical tables}} (Dover Publications, 1965).

\bibitem{1983PhRvD..28.1853A}
A.~J. {Accioly}, A.~N. {Vaidya} and M.~M. {Som}.
\newblock \enquote{{Nonminimal coupling and Bianchi type I cosmologies.}}
\newblock \emph{\prd} \textbf{28}, 1853 (1983).

\bibitem{1982ApJ...253....1A}
P.~J. {Adams}, R.~W. {Hellings}, R.~L. {Zimmerman}, H.~{Farhoosh}, D.~I.
  {Levine} and S.~{Zeldich}.
\newblock \enquote{{Inhomogeneous cosmology - Gravitational radiation in
  Bianchi backgrounds}.}
\newblock \emph{\apj} \textbf{253}, 1 (1982).

\bibitem{1997PThPh..98.1225A}
Y.~{Aizawa}, N.~{Koguro} and I.~{Antoniou}.
\newblock \enquote{{Chaos and Singularities in the Mixmaster Universe}.}
\newblock \emph{\ptp} \textbf{98}, 1225 (1997).

\bibitem{Alex}
S.~Alexandrov.
\newblock \enquote{$so(4,c)$-covariant ashtekar-barbero gravity and the immirzi
  parameter.}
\newblock \emph{\cqg} \textbf{17}, 4255 (2000).
\newblock \eprint{arXiv.org/abs/gr-qc/0005085}.

\bibitem{Alexb}
S.~Alexandrov and D.~Vassilevich.
\newblock \enquote{Area spectrum in lorentz covariant loop gravity.}
\newblock \emph{\prd} \textbf{64}, 044023 (2001).
\newblock \eprint{arXiv.org/abs/gr-qc/0103105}.

\bibitem{ACamelia97}
G.~Amelino-Camelia, J.~R. Ellis, N.~E. Mavromatos and D.~V. Nanopoulos.
\newblock \enquote{On the space-time uncertainty relations of liouville strings
  and d branes.}
\newblock \emph{\mpla} \textbf{12}, 2029 (1997).
\newblock \eprint{arXiv.org/abs/hep-th/9701144}.

\bibitem{Uggla3}
L.~Andersson, H.~van Elst, W.~C. Lim and C.~Uggla.
\newblock \enquote{{Asymptotic silence of generic cosmological singularities}.}
\newblock \emph{\prl.} \textbf{94}, 051101 (2005).
\newblock \eprint{arXiv.org/abs/gr-qc/0402051}.

\bibitem{AN67}
D.~V. Anosov.
\newblock \emph{Geodesic Flows on Closed Riemann Manifolds with Negative
  Curvature}, \emph{Proc. Steklov Inst. Math.}, vol.~90 (1967).
\newblock Translation: American Mathematical Society, Providence, 1969.

\bibitem{0264-9381-20-15-307}
P.~S. Apostolopoulos.
\newblock \enquote{Self-similar bianchi models: I. class a models.}
\newblock \emph{Classical and Quantum Gravity} \textbf{20}, 3371 (2003).

\bibitem{Apostolopoulos:2004uq}
P.~S. Apostolopoulos.
\newblock \enquote{{Equilibrium points of the tilted perfect fluid Bianchi
  VI$_h$ state space}.}
\newblock \emph{Gen. Rel. Grav.} \textbf{37}, 937 (2005).
\newblock \eprint{gr-qc/0407040}.

\bibitem{0264-9381-22-21-002}
P.~S. Apostolopoulos.
\newblock \enquote{A geometric description of the intermediate behaviour for
  spatially homogeneous models.}
\newblock \emph{Classical and Quantum Gravity} \textbf{22}, 4425 (2005).

\bibitem{0264-9381-22-2-006}
P.~S. Apostolopoulos.
\newblock \enquote{Self-similar bianchi models: Ii. class b models.}
\newblock \emph{Classical and Quantum Gravity} \textbf{22}, 323 (2005).

\bibitem{2003GReGr..35.2051A}
P.~S. {Apostolopoulos} and M.~{Tsamparlis}.
\newblock \enquote{{Letter: Self-Similar Bianchi Type VIII and IX Models}.}
\newblock \emph{\grg} \textbf{35}, 2051 (2003).
\newblock \eprint{arXiv.org/abs/gr-qc/0305017}.

\bibitem{arfken}
G.~B. Arfken, H.~J. Weber and L.~Ruby.
\newblock \emph{{Mathematical Methods for Physicists}}, vol.~64 (American
  Association of Physics Teachers, 1996).

\bibitem{1998CSF.....9.1813A}
J.~{Argyris}, I.~{Andreadis} and C.~{Ciubotariu}.
\newblock \enquote{{On the influence of noise on the Bianchi IX and VIII
  cosmological models.}}
\newblock \emph{Chaos Solitons and Fractals} \textbf{9}, 1813 (1998).

\bibitem{A89}
V.~I. Arnold.
\newblock \emph{Mathematical Methods of Classical Mechanics} (Springer-Verlag,
  Berlin, 1989).

\bibitem{arlond1968}
V.~I. Arnold and A.~Avez.
\newblock \emph{{Ergodic problems of classical mechanics}} (Benjamin, New York,
  1968).

\bibitem{ADM1962}
R.~Arnowitt, S.~Deser and C.~W. Misner.
\newblock \enquote{The dynamics of general relativity.}  (1962).
\newblock \eprint{arXiv.org/abs/gr-qc/0405109}.

\bibitem{artin1965msq}
E.~Artin.
\newblock \enquote{{Ein Mechanisches System mit quasi-ergodischen Bahnen,
  Collected papers}.} (1965).

\bibitem{ABL}
A.~Ashtekar, M.~Bojowald and J.~Lewandowski.
\newblock \enquote{Mathematical structure of loop quantum cosmology.}
\newblock \emph{Advances in Theoretical and Mathematical Physics} \textbf{7},
  233 (2003).
\newblock \eprint{arXiv.org/abs/gr-qc/0304074}.

\bibitem{ashtekar2004}
A.~Ashtekar and J.~Lewandowski.
\newblock \enquote{{Background independent quantum gravity: A status report}.}
\newblock \emph{\cqg} \textbf{21}, R53 (2004).
\newblock \eprint{arXiv.org/abs/gr-qc/0404018}.

\bibitem{ashtekar2006}
A.~Ashtekar, T.~Pawlowski and P.~Singh.
\newblock \enquote{{Quantum nature of the Big Bang}.}
\newblock \emph{\prl} \textbf{96}, 141301 (2006).
\newblock \eprint{arXiv.org/abs/gr-qc/0602086}.

\bibitem{A06}
A.~Ashtekar, T.~Pawlowski and P.~Singh.
\newblock \enquote{Quantum nature of the big bang: An analytical and numerical
  investigation.}
\newblock \emph{\prd} \textbf{73}, 124038 (2006).
\newblock \eprint{arXiv.org/abs/gr-qc/0604013}.

\bibitem{1991CQGra...8.2191A}
A.~{Ashtekar} and J.~{Samuel}.
\newblock \enquote{{Bianchi cosmologies: the role of spatial topology.}}
\newblock \emph{\cqg} \textbf{8}, 2191 (1991).

\bibitem{Stachel}
A.~{Ashtekar} and J.~{Stachel}, eds.
\newblock \emph{{Conceptual problems of quantum gravity}} (1991).

\bibitem{2000PhRvD..62d3515B}
M.~{Bachmann} and H.-J. {Schmidt}.
\newblock \enquote{{Period-doubling bifurcation in strongly anisotropic Bianchi
  I quantum cosmology}.}
\newblock \emph{\prd} \textbf{62}, 043515 (2000).
\newblock \eprint{arXiv.org/abs/gr-qc/9912068}.

\bibitem{1998Ap&SS.262..145B}
R.~{Bali} and V.~{Chand Jain}.
\newblock \enquote{{Generalized Expanding and Shearing Anisotropic Bianchi Type
  I Magnetofluid Cosmological Model in General Relativity}.}
\newblock \emph{\apss} \textbf{262}, 145 (1998).

\bibitem{Balogh}
C.~B. Balogh.
\newblock \enquote{{Asymptotic expansions of the modified Bessel function of
  the third kind of imaginary order}.}
\newblock \emph{SIAM Journal of Applied Mathematics} \textbf{15}, 1315 (1967).

\bibitem{1990ApJ...358...23B}
A.~{Banerjee}, B.~{Bhui} and S.~{Chatterjee}.
\newblock \enquote{{Bianchi type I cosmological models in higher dimensions}.}
\newblock \emph{\apj} \textbf{358}, 23 (1990).

\bibitem{barrow.cup84}
J.~Barrow.
\newblock \enquote{{Chaotic behaviour and the Einstein equations}.}
\newblock \emph{Classical General Relativity. Edited by WB Bonnor, JN Islam and
  MAH MacCallum, with a foreword by Sir Hermann Bondi. ISBN 0 521 26747 1.
  Library of Congress Catalog Card Number 84-45247: 531'. 14 QC173. 6.
  Published by Cambridge University Press, Cambridge, Great Britain, 1984, p.
  26}  (1984).

\bibitem{BarrowMapPRL1981}
J.~D. Barrow.
\newblock \enquote{Chaos in the einstein equations.}
\newblock \emph{\prl} \textbf{46}, 963 (1981).

\bibitem{Barrow1981PR}
J.~D. Barrow.
\newblock \enquote{Chaotic behavior in general relativity.}
\newblock \emph{\physrep} \textbf{85}, 1 (1982).

\bibitem{1986CaJPh..64..152B}
J.~D. {Barrow}.
\newblock \enquote{{General relativistic cosmological models and the cosmic
  microwave background radiation}.}
\newblock \emph{Canadian Journal of Physics} \textbf{64}, 152 (1986).

\bibitem{Barrow:1995fn}
J.~D. Barrow.
\newblock \enquote{{Why the universe is not anisotropic}.}
\newblock \emph{\prd} \textbf{51}, 3113 (1995).

\bibitem{Barrow:1997sy}
J.~D. Barrow.
\newblock \enquote{{Cosmological limits on slightly skew stresses}.}
\newblock \emph{\prd} \textbf{55}, 7451 (1997).
\newblock \eprint{arXiv.org/abs/gr-qc/9701038}.

\bibitem{Barrow:1997mj}
J.~D. Barrow, P.~G. Ferreira and J.~Silk.
\newblock \enquote{{Constraints on a Primordial Magnetic Field}.}
\newblock \emph{\prl} \textbf{78}, 3610 (1997).
\newblock \eprint{arXiv.org/abs/astro-ph/9701063}.

\bibitem{2000CQGra..17.1435B}
J.~D. {Barrow}, Y.~{Gaspar} and P.~M. {Saffin}.
\newblock \enquote{{Some exact non-vacuum Bianchi $VI_{0}$ and $VII_{0}$
  instantons.}}
\newblock \emph{\cqg} \textbf{17}, 1435 (2000).
\newblock \eprint{arXiv.org/abs/gr-qc/9905100}.

\bibitem{1985MNRAS.213..917B}
J.~D. {Barrow}, R.~{Juszkiewicz} and D.~H. {Sonoda}.
\newblock \enquote{{Universal rotation - How large can it be?}}
\newblock \emph{\mnras} \textbf{213}, 917 (1985).

\bibitem{1980PhRvD..21..336B}
J.~D. {Barrow} and R.~A. {Matzner}.
\newblock \enquote{{Size of a bouncing mixmaster universe}.}
\newblock \emph{\prd} \textbf{21}, 336 (1980).

\bibitem{1986PhR...139....1B}
J.~D. {Barrow} and D.~H. {Sonoda}.
\newblock \enquote{{Asymptotic stability of Bianchi type universes.}}
\newblock \emph{\physrep} \textbf{139}, 1 (1986).

\bibitem{1985PhRvD..32.1595B}
J.~D. {Barrow} and J.~{Stein-Schabes}.
\newblock \enquote{{Kaluza-Klein mixmaster universes}.}
\newblock \emph{\prd} \textbf{32}, 1595 (1985).

\bibitem{1979PhR....56..371B}
J.~D. {Barrow} and F.~J. {Tipler}.
\newblock \enquote{{Analysis of the generic singularity studies by Belinskii,
  Khalatnikov, and Lifschitz.}}
\newblock \emph{\physrep} \textbf{56}, 371 (1979).

\bibitem{BattistiMontani2006PLB}
M.~V. Battisti and G.~Montani.
\newblock \enquote{Evolutionary quantum dynamics of a generic universe.}
\newblock \emph{\plb} \textbf{637}, 203 (2006).
\newblock \eprint{arXiv.org/abs/gr-qc/0604049}.

\bibitem{BM07a}
M.~V. Battisti and G.~Montani.
\newblock \enquote{The big-bang singularity in the framework of a generalized
  uncertainty principle.}
\newblock \emph{\plb} \textbf{656}, 96 (2007).
\newblock \eprint{arXiv.org/abs/gr-qc/0703025}.

\bibitem{BM07b}
M.~V. Battisti and G.~Montani.
\newblock \enquote{Quantum dynamics of the taub universe in a generalized
  uncertainty principle framework.}
\newblock \emph{\prd} \textbf{77}, 023518 (2008).
\newblock \eprint{arXiv.org/abs/0707.2726}.

\bibitem{1993Ap&SS.203..169B}
A.~{Beesham}.
\newblock \enquote{{The isotropization of the vacuum Bianchi type V solution}.}
\newblock \emph{\apss} \textbf{203}, 169 (1993).

\bibitem{1984GReGr..16.1189B}
V.~{Belinskii} and M.~{Francaviglia}.
\newblock \enquote{{Solitonic gravitational waves in Bianchi II cosmologies. II
  - One solitonic perturbations}.}
\newblock \emph{\grg} \textbf{16}, 1189 (1984).

\bibitem{B00}
V.~A. Belinskii.
\newblock In \emph{The Chaotic Universe}, eds. V.~G. Gurzadyan and R.~Ruffini
  (World Scientific, Singapore, 2000), p. 350.

\bibitem{BK69}
V.~A. Belinskii and I.~M. Khalatnikov.
\newblock \enquote{On the nature of the singularities in the general solutions
  of the gravitational equations.}
\newblock \emph{\sovJETP} \textbf{29}, 911 (1969).

\bibitem{BelinskiiKhalatnikov1971gsgSPJ}
V.~A. Belinskii and I.~M. Khalatnikov.
\newblock \enquote{{General solution of the gravitational equations with a
  physical oscillatory singularity}.}
\newblock \emph{\sovJETP} \textbf{32}, 169 (1971).

\bibitem{BK72}
V.~A. Belinskii and I.~M. Khalatnikov.
\newblock \enquote{{Effect of scalar and vector fields on the nature of the
  cosmological singularity}.}
\newblock \emph{\zetf} \textbf{63} (1972).

\bibitem{BK73}
V.~A. Belinskii and I.~M. Khalatnikov.
\newblock \enquote{Effect of scalar and vector fields on the nature of the
  cosmological singularity.}
\newblock \emph{\sovJETP} \textbf{36}, 591 (1973).

\bibitem{BKL1970}
V.~A. Belinskii, I.~M. Khalatnikov and E.~M. Lifshitz.
\newblock \enquote{Oscillatory approach to a singular point in the relativistic
  cosmology.}
\newblock \emph{\advphys} \textbf{19}, 525 (1970).

\bibitem{BKL1971}
V.~A. Belinskii, I.~M. Khalatnikov and E.~M. Lifshitz.
\newblock \enquote{{The Oscillatory Mode of Approach to a Singularity in
  Homogeneous Cosmological Models with Rotating Axes}.}
\newblock \emph{\sovJETP} \textbf{33}, 1061 (1971).

\bibitem{BKL1972}
V.~A. Belinskii, I.~M. Khalatnikov and E.~M. Lifshitz.
\newblock \enquote{{Construction of a general cosmological solution of the
  Einstein equation with a time singularity}.}
\newblock \emph{\sovJETP} \textbf{35}, 838 (1972).

\bibitem{BKL82}
V.~A. Belinskii, I.~M. Khalatnikov and E.~M. Lifshitz.
\newblock \enquote{A general solution of the einstein equations with a time
  singularity.}
\newblock \emph{\advphys} \textbf{31}, 639 (1982).

\bibitem{1982GReGr..14..213B}
V.~A. {Belinskij} and M.~{Francaviglia}.
\newblock \enquote{{Solitonic gravitational waves in Bianchi II cosmologies. 1.
  The general framework.}}
\newblock \emph{\grg} \textbf{14}, 213 (1982).

\bibitem{BeniniMontani2005CQG}
R.~Benini, A.~A. Kirillov and G.~Montani.
\newblock \enquote{Oscillatory regime in the multidimensional homogeneous
  cosmological models induced by a vector field.}
\newblock \emph{\cqg} \textbf{22}, 1483 (2005).
\newblock \eprint{arXiv.org/abs/gr-qc/0505027}.

\bibitem{BeniniMontani2004PRD}
R.~Benini and G.~Montani.
\newblock \enquote{Frame-independence of the inhomogeneous mixmaster chaos via
  misner-chitre'-like variables.}
\newblock \emph{\prd} \textbf{70}, 103527 (2004).
\newblock \eprint{arXiv.org/abs/gr-qc/0411044}.

\bibitem{BeniniMontani2007CQG}
R.~Benini and G.~Montani.
\newblock \enquote{Inhomogeneous quantum mixmaster: From classical toward
  quantum mechanics.}
\newblock \emph{\cqg} \textbf{24}, 387 (2007).
\newblock \eprint{arXiv.org/abs/gr-qc/0612095}.

\bibitem{1997PhRvD..55.1896B}
R.~{Bergamini}, P.~{Sedici} and P.~{Verrocchio}.
\newblock \enquote{{Inflation for Bianchi type IX models}.}
\newblock \emph{\prd} \textbf{55}, 1896 (1997).
\newblock \eprint{arXiv.org/abs/gr-qc/9612001}.

\bibitem{Berger1989PRD}
B.~K. Berger.
\newblock \enquote{Quantum chaos in the mixmaster universe.}
\newblock \emph{\prd} \textbf{39}, 2426 (1989).

\bibitem{1990CQGra...7..203B}
B.~K. {Berger}.
\newblock \enquote{{Numerical study of initially expanding mixmaster
  universes.}}
\newblock \emph{\grg} \textbf{7}, 203 (1990).

\bibitem{1991GReGr..23.1385B}
B.~K. {Berger}.
\newblock \enquote{{Comments on the computation of Lyapunov exponents for the
  Mixmaster universe}.}
\newblock \emph{\grg} \textbf{23}, 1385 (1991).

\bibitem{1993PhRvD..47.3222B}
B.~K. {Berger}.
\newblock \enquote{{Note on time reversible chaos in mixmaster dynamics}.}
\newblock \emph{\prd} \textbf{47}, 3222 (1993).

\bibitem{1994PhRvD..49.1120B}
B.~K. {Berger}.
\newblock \enquote{{How to determine approximate mixmaster parameters from
  numerical evolution of Einstein's equations}.}
\newblock \emph{\prd} \textbf{49}, 1120 (1994).
\newblock \eprint{arXiv.org/abs/gr-qc/9308016}.

\bibitem{1996CQGra..13.1273B}
B.~K. {Berger}.
\newblock \enquote{{Comment on the ''chaotic'' singularity in some magnetic
  Bianchi $VI_{0}$ cosmologies.}}
\newblock \emph{\cqg} \textbf{13}, 1273 (1996).
\newblock \eprint{arXiv.org/abs/gr-qc/9512005}.

\bibitem{Berger1999PRD}
B.~K. Berger.
\newblock \enquote{Influence of scalar fields on the approach to a cosmological
  singularity.}
\newblock \emph{\prd} \textbf{61}, 023508 (1999).
\newblock \eprint{arXiv.org/abs/gr-qc/9907083}.

\bibitem{Berger01wd}
B.~K. Berger.
\newblock \enquote{Approach to the singularity in spatially inhomogeneous
  cosmologies.}  (2001).
\newblock \eprint{arXiv.org/abs/gr-qc/0106009}.

\bibitem{berger2002nas}
B.~K. Berger.
\newblock \enquote{{Numerical Approaches to Spacetime Singularities}.}
\newblock \emph{Living Reviews in Relativity} \textbf{5} (2002).
\newblock \eprint{arXiv.org/abs/gr-qc/0201056}.

\bibitem{2004CQGra..21S..81B}
B.~K. {Berger}.
\newblock \enquote{{Hunting local Mixmaster dynamics in spatially inhomogeneous
  cosmologies}.}
\newblock \emph{\cqg} \textbf{21}, 81 (2004).
\newblock \eprint{arXiv.org/abs/gr-qc/0312095}.

\bibitem{Berger98MPLA}
B.~K. Berger, D.~Garfinkle, J.~Isenberg, V.~Moncrief and M.~Weaver.
\newblock \enquote{The singularity in generic gravitational collapse is
  spacelike, local and oscillatory.}
\newblock \emph{\mpla} \textbf{13}, 1565 (1998).
\newblock \eprint{arXiv.org/abs/gr-qc/9805063}.

\bibitem{BGS}
B.~K. Berger, D.~Garfinkle and E.~Strasser.
\newblock \enquote{{New algorithm for Mixmaster dynamics}.}
\newblock \emph{\cqg} \textbf{14}, L29 (1997).
\newblock \eprint{arXiv.org/abs/gr-qc/9609072}.

\bibitem{berger1993nic}
B.~K. Berger and V.~Moncrief.
\newblock \enquote{{Numerical investigation of cosmological singularities}.}
\newblock \emph{\prd} \textbf{48}, 4676 (1993).
\newblock \eprint{arXiv.org/abs/gr-qc/9307032}.

\bibitem{2000PhRvD..62b3509B}
B.~K. {Berger} and V.~{Moncrief}.
\newblock \enquote{{Exact U(1) symmetric cosmologies with local mixmaster
  dynamics}.}
\newblock \emph{\prd} \textbf{62}, 023509 (2000).
\newblock \eprint{arXiv.org/abs/gr-qc/0001083}.

\bibitem{2000PhRvD..62l3501B}
B.~K. {Berger} and V.~{Moncrief}.
\newblock \enquote{{Signature for local mixmaster dynamics in U(1) symmetric
  cosmologies}.}
\newblock \emph{\prd} \textbf{62}, 123501 (2000).
\newblock \eprint{arXiv.org/abs/gr-qc/0006071}.

\bibitem{1992PhRvD..46.1551B}
A.~L. {Berkin}.
\newblock \enquote{{Coleman-Weinberg symmetry breaking in a Bianchi type-I
  universe}.}
\newblock \emph{\prd} \textbf{46}, 1551 (1992).

\bibitem{1986PhRvD..33.3581B}
D.~{Bernard}.
\newblock \enquote{{Hadamard singularity and quantum states in Bianchi type-I
  space-time}.}
\newblock \emph{\prd} \textbf{33}, 3581 (1986).

\bibitem{BIA97}
L.~Bianchi.
\newblock \emph{Sugli spazi a tre dimensioni che ammettono un gruppo continuo
  di movimenti} (Roma, 1897), \emph{Opere}, vol.~9.
\newblock Edizione cremonese edn.

\bibitem{Bianchi}
L.~Bianchi.
\newblock \enquote{Sugli spazi a tre dimensioni che ammettono un gruppo
  continuo di movimenti.}
\newblock \emph{Mem. Mat. Fis. Soc. Ital. Sci.} \textbf{11}, 267 (1898).

\bibitem{BI95}
M.~Biesiada.
\newblock \enquote{Searching for invariant description of chaos in general
  relativity.}
\newblock \emph{\cqg} \textbf{12}, 715 (1995).

\bibitem{1991PhLA..160..123B}
M.~{Biesiada} and M.~{Szyd{\l}owski}.
\newblock \enquote{{Mixmaster cosmological models as disturbed Toda lattices.}}
\newblock \emph{\pla} \textbf{160}, 123 (1991).

\bibitem{BGOB88}
S.~Bleher, C.~Grebogi, E.~Ott and R.~Brown.
\newblock \enquote{Fractal boundaries for exit in hamiltonian dynamics.}
\newblock \emph{\pra} \textbf{38}, 930 (1988).

\bibitem{BOG89}
S.~Bleher, E.~Ott and C.~Grebogi.
\newblock \enquote{Routes to chaotic scattering.}
\newblock \emph{\prl} \textbf{63}, 919 (1989).

\bibitem{BlyIsh}
W.~F. Blyth and C.~J. Isham.
\newblock \enquote{Quantization of a friedmann universe filled with a scalar
  field.}
\newblock \emph{\prd} \textbf{11}, 768 (1975).

\bibitem{BN73}
O.~Bogoyavlenskii and S.~Novikov.
\newblock \enquote{Properties of cosmological models of bianchi ix type from
  the point of view of the qualitative theory of differential equations.}
\newblock \emph{\sovJETP} \textbf{37}, 747 (1973).

\bibitem{Bo01}
M.~Bojowald.
\newblock \enquote{Absence of singularity in loop quantum cosmology.}
\newblock \emph{\prl} \textbf{86}, 5227 (2001).
\newblock \eprint{arXiv.org/abs/gr-qc/0102069}.

\bibitem{Bo03}
M.~Bojowald.
\newblock \enquote{Inflation from quantum geometry.}
\newblock \emph{\prl} \textbf{89}, 261301 (2002).
\newblock \eprint{arXiv.org/abs/gr-qc/0206054}.

\bibitem{Bojoiso}
M.~Bojowald.
\newblock \enquote{Isotropic loop quantum cosmology.}
\newblock \emph{\cqg} \textbf{19}, 2717 (2002).
\newblock \eprint{arXiv.org/abs/gr-qc/0202077}.

\bibitem{Bojinom}
M.~Bojowald.
\newblock \enquote{Loop quantum cosmology and inhomogeneities.}
\newblock \emph{\grg} \textbf{38}, 1771 (2006).
\newblock \eprint{arXiv.org/abs/gr-qc/0609034}.

\bibitem{Boj07}
M.~Bojowald.
\newblock \enquote{Singularities and quantum gravity.}
\newblock \emph{AIP Conference Proceedings} \textbf{910}, 294 (2007).
\newblock \eprint{arXiv.org/abs/gr-qc/0702144}.

\bibitem{Bo04}
M.~Bojowald and G.~Date.
\newblock \enquote{Quantum suppression of the generic chaotic behavior close to
  cosmological singularities.}
\newblock \emph{\prl} \textbf{92}, 071302 (2004).
\newblock \eprint{arXiv.org/abs/gr-qc/0311003}.

\bibitem{2004CQGra..21.3541B}
M.~{Bojowald}, G.~{Date} and G.~{Mortuza Hossain}.
\newblock \enquote{{The Bianchi IX model in loop quantum cosmology}.}
\newblock \emph{\cqg} \textbf{21}, 3541 (2004).
\newblock \eprint{arXiv.org/abs/gr-qc/0404039}.

\bibitem{Bo04a}
M.~Bojowald, G.~Date and K.~Vandersloot.
\newblock \enquote{Homogeneous loop quantum cosmology: the role of the spin
  connection.}
\newblock \emph{\cqg} \textbf{21}, 1253 (2004).
\newblock \eprint{arXiv.org/abs/gr-qc/0311004}.

\bibitem{2004CQGra..21.4087B}
B.~{Bolen}, L.~{Bombelli} and A.~{Corichi}.
\newblock \enquote{{Semiclassical states in quantum cosmology: Bianchi I
  coherent states}.}
\newblock \emph{\cqg} \textbf{21}, 4087 (2004).
\newblock \eprint{arXiv.org/abs/gr-qc/0404004}.

\bibitem{1984GReGr..16.1119B}
J.~M. {Bradley} and E.~{Sviestins}.
\newblock \enquote{{Some rotating, time-dependent Bianchi type VIII cosmologies
  with heat flow}.}
\newblock \emph{\grg} \textbf{16}, 1119 (1984).

\bibitem{2007MNRAS.377.1473B}
M.~{Bridges}, J.~D. {McEwen}, A.~N. {Lasenby} and M.~P. {Hobson}.
\newblock \enquote{{Markov chain Monte Carlo analysis of Bianchi $VII_{h}$
  models}.}
\newblock \emph{\mnras} \textbf{377}, 1473 (2007).
\newblock \eprint{arXiv.org/abs/astro-ph/0605325}.

\bibitem{BK94}
J.~D. Brown and K.~V. Kucha{\v r}.
\newblock \enquote{Dust as a standard of space and time in canonical quantum
  gravity.}
\newblock \emph{\prd} \textbf{51}, 5600 (1995).
\newblock \eprint{arXiv.org/abs/gr-qc/9409001}.

\bibitem{BruThi05}
J.~Brunnemann and T.~Thiemann.
\newblock \enquote{On (cosmological) singularity avoidance in loop quantum
  gravity.}
\newblock \emph{\cqg} \textbf{23}, 1395 (2006).
\newblock \eprint{arXiv.org/abs/gr-qc/0505032}.

\bibitem{1986GReGr..18.1263B}
M.~H. {Bugalho}, A.~{Rica da Silva} and J.~{Sousa Ramos}.
\newblock \enquote{{The order of chaos on a Bianchi-IX cosmological model}.}
\newblock \emph{\grg} \textbf{18}, 1263 (1986).

\bibitem{BurdTavakol1993PRD}
A.~Burd and R.~Tavakol.
\newblock \enquote{Invariant lyapunov exponents and chaos in cosmology.}
\newblock \emph{\prd} \textbf{47}, 5336 (1993).

\bibitem{1989grg..conf..324B}
A.~B. {Burd}, N.~{Buric} and G.~F.~R. {Ellis}.
\newblock \enquote{{Numerical Analysis of Chaos in Bianchi IX Models}.}
\newblock In \emph{\grg} (1989), p. 324.

\bibitem{1990GReGr..22..349B}
A.~B. {Burd}, N.~{Buric} and G.~F.~R. {Ellis}.
\newblock \enquote{{A numerical analysis of chaotic behaviour in Bianchi IX
  models}.}
\newblock \emph{\grg} \textbf{22}, 349 (1990).

\bibitem{BBT}
A.~B. Burd, N.~Buric and R.~K. Tavakol.
\newblock \enquote{Chaos, entropy and cosmology.}
\newblock \emph{\cqg} \textbf{8}, 123 (1991).

\bibitem{1978PhLA...67...19C}
N.~{Caderni} and R.~{Fabbri}.
\newblock \enquote{{Neutrino viscosity in Bianchi type IX universes}.}
\newblock \emph{\pla} \textbf{67}, 19 (1978).

\bibitem{1997IJMPD...6..491C}
S.~{Capozziello}, G.~{Marmo}, C.~{Rubano} and P.~{Scudellaro}.
\newblock \enquote{{N{\"o}ther symmetries in Bianchi universes.}}
\newblock \emph{\ijmpd} \textbf{6}, 491 (1997).
\newblock \eprint{arXiv.org/abs/gr-qc/9606050}.

\bibitem{1980PhLA...75..333C}
M.~{Carmeli} and C.~{Charach}.
\newblock \enquote{{Inhomogeneous generalization of some Bianchi models.}}
\newblock \emph{\pla} \textbf{75}, 333 (1980).

\bibitem{1983AnPhy.150..392C}
M.~{Carmeli}, C.~{Charach} and A.~{Feinstein}.
\newblock \enquote{{Inhomogeneous mixmaster universes: some exact solutions.}}
\newblock \emph{Annals of Physics} \textbf{150}, 392 (1983).

\bibitem{2005PhRvL..94k1302C}
D.~{Cartin} and G.~{Khanna}.
\newblock \enquote{{Absence of Preclassical Solutions in Bianchi I Loop Quantum
  Cosmology}.}
\newblock \emph{\prl} \textbf{94}, 111302 (2005).
\newblock \eprint{arXiv.org/abs/gr-qc/0501016}.

\bibitem{2005PhRvD..72h4008C}
D.~{Cartin} and G.~{Khanna}.
\newblock \enquote{{Separable sequences in Bianchi I loop quantum cosmology}.}
\newblock \emph{\prd} \textbf{72}, 084008 (2005).
\newblock \eprint{arXiv.org/abs/gr-qc/0506024}.

\bibitem{1985PhRvD..31..742C}
M.~{Castagnino} and F.~D. {Mazzitelli}.
\newblock \enquote{{Weak and strong quantum vacua in Bianchi type I
  universes.}}
\newblock \emph{\prd} \textbf{31}, 742 (1985).

\bibitem{2006MNRAS.369..598C}
L.~{Cay{\'o}n}, A.~J. {Banday}, T.~{Jaffe}, H.~K. {Eriksen}, F.~K. {Hansen},
  K.~M. {Gorski} and J.~{Jin}.
\newblock \enquote{{No Higher Criticism of the Bianchi-corrected Wilkinson
  Microwave Anisotropy Probe data}.}
\newblock \emph{\mnras} \textbf{369}, 598 (2006).
\newblock \eprint{arXiv.org/abs/astro-ph/0602023}.

\bibitem{1999PhRvD..59d3501C}
J.~L. {Cervantes-Cota} and P.~A. {Chauvet}.
\newblock \enquote{{Can induced gravity isotropize Bianchi type I, V, or IX
  universes?}}
\newblock \emph{\prd} \textbf{59}, 043501 (1999).
\newblock \eprint{arXiv.org/abs/gr-qc/9810064}.

\bibitem{2001GReGr..33..767C}
J.~L. {Cervantes-Cota} and M.~{Nahmad}.
\newblock \enquote{{Isotropization of Bianchi Models and a New FRW Solution in
  Brans-Dicke Theory}.}
\newblock \emph{\grg} \textbf{33}, 767 (2001).
\newblock \eprint{arXiv.org/abs/gr-qc/0005032}.

\bibitem{1996IJMPD...5...65C}
S.~{Chakraborty} and G.~C. {Nandy}.
\newblock \enquote{{Homogeneous perfect fluid in Brans-Dicke-Bianchi type V
  cosmological model.}}
\newblock \emph{\ijmpd} \textbf{5}, 65 (1996).

\bibitem{2001PhRvD..64l7502C}
S.~{Chakraborty} and B.~C. {Paul}.
\newblock \enquote{{Inflation in Bianchi models and the cosmic no-hair
  theorem}.}
\newblock \emph{\prd} \textbf{64}, 127502 (2001).

\bibitem{1997Ap&SS.253..205C}
S.~{Chakraborty} and A.~{Roy}.
\newblock \enquote{{A Study of Variable G and {$\Lambda$} in Bianchi Type II
  Cosmological Models}.}
\newblock \emph{\apss} \textbf{253}, 205 (1997).

\bibitem{2007Ap&SS.tmp..329C}
N.~{Chakravarty} and L.~{Biswas}.
\newblock \enquote{{Bianchi type I cosmological model with variable
  {$\Lambda$}-term}.}
\newblock \emph{\apss} p. 329 (2007).

\bibitem{1996PhRvL..76..857C}
S.~{Chakravarty} and M.~J. {Ablowitz}.
\newblock \enquote{{Integrability, Monodromy Evolving Deformations, and
  Self-Dual Bianchi IX Systems}.}
\newblock \emph{\prl} \textbf{76}, 857 (1996).

\bibitem{1982PhRvD..26.3367C}
C.~{Charach}.
\newblock \enquote{{Feynman propagators and particle creation in linearly
  expanding Bianchi type-I Universes.}}
\newblock \emph{\prd} \textbf{26}, 3367 (1982).

\bibitem{1992CQGra...9.1923C}
P.~{Chauvet}, J.~{Cervantes-Cota} and H.~N. {N{\'u}{\~n}ez-Y{\'e}pez}.
\newblock \enquote{{Vacuum-analytic solutions in Bianchi-type universes.}}
\newblock \emph{\cqg} \textbf{9}, 1923 (1992).

\bibitem{2000PhRvD..62l4016C}
C.-M. {Chen}, T.~{Harko} and M.~K. {Mak}.
\newblock \enquote{{Bianchi type I cosmologies in arbitrary dimensional dilaton
  gravities}.}
\newblock \emph{\prd} \textbf{62}, 124016 (2000).
\newblock \eprint{arXiv.org/abs/hep-th/0004096}.

\bibitem{ChernoffBarrow1983PRL}
D.~F. Chernoff and J.~D. Barrow.
\newblock \enquote{Chaos in the mixmaster universe.}
\newblock \emph{\prl} \textbf{50}, 134 (1983).

\bibitem{2005CQGra..22.1763C}
C.~{Cherubini}, D.~{Bini}, M.~{Bruni} and Z.~{Perjes}.
\newblock \enquote{{The speciality index as invariant indicator in the BKL
  mixmaster dynamics}.}
\newblock \emph{\cqg} \textbf{22}, 1763 (2005).
\newblock \eprint{arXiv.org/abs/gr-qc/0408040}.

\bibitem{1997PhLB..408...47C}
T.~{Chiba}, S.~{Mukohyama} and T.~{Nakamura}.
\newblock \enquote{{Anisotropy of the cosmic background radiation implies the
  violation of the strong energy condition in Bianchi type I universe.}}
\newblock \emph{\plb} \textbf{408}, 47 (1997).
\newblock \eprint{arXiv.org/abs/gr-qc/9707043}.

\bibitem{1987PhLA..121....7C}
L.~P. {Chimento} and M.~S. {Mollerach}.
\newblock \enquote{{Dirac equation in Bianchi I metrics.}}
\newblock \emph{Physics Letters A} \textbf{121}, 7 (1987).

\bibitem{2007PhRvD..75b4029C}
D.-W. {Chiou}.
\newblock \enquote{{Loop quantum cosmology in Bianchi typeI models: Analytical
  investigation}.}
\newblock \emph{\prd} \textbf{75}, 024029 (2007).
\newblock \eprint{arXiv.org/abs/gr-qc/0609029}.

\bibitem{2007PhRvD..76h4015C}
D.-W. {Chiou} and K.~{Vandersloot}.
\newblock \enquote{{Behavior of nonlinear anisotropies in bouncing Bianchi I
  models of loop quantum cosmology}.}
\newblock \emph{\prd} \textbf{76}, 084015 (2007).
\newblock \eprint{arXiv.org/abs/0707.2548}.

\bibitem{1972PhRvD...6.3390C}
D.~M. {Chitre}.
\newblock \enquote{{High-frequency sound waves to eliminate a horizon in the
  mixmaster universe.}}
\newblock \emph{\prd} \textbf{6}, 3390 (1972).

\bibitem{C72}
D.~M. Chitre.
\newblock \emph{Investigation of vanishing of a horizon for Bianchi type IX
  (The mixmaster universe) ~~}.
\newblock Ph.D. thesis, University of Maryland (1972).
\newblock Technical Report No. 72-125.

\bibitem{1988PhRvD..38.2392C}
P.~{Chmielowski} and D.~N. {Page}.
\newblock \enquote{{Probability of Bianchi type-I inflation}.}
\newblock \emph{\prd} \textbf{38}, 2392 (1988).

\bibitem{1995PhRvD..52.5445C}
H.~T. {Cho} and A.~D. {Speliotopoulos}.
\newblock \enquote{{Gravitational waves in Bianchi type-I universes: The
  classical theory}.}
\newblock \emph{\prd} \textbf{52}, 5445 (1995).
\newblock \eprint{arXiv.org/abs/gr-qc/9504046}.

\bibitem{1995JPhA...28..657C}
F.~{Christiansen}, H.~H. {Rugh} and S.~E. {Rugh}.
\newblock \enquote{{Non-integrability of the mixmaster universe.}}
\newblock \emph{Journal of Physics A Mathematical General} \textbf{28}, 657
  (1995).

\bibitem{2002CQGra..19.1013C}
T.~{Christodoulakis}, T.~{Gakis} and G.~O. {Papadopoulos}.
\newblock \enquote{{Conditional symmetries and the quantization of Bianchi Type
  I vacuum cosmologies with and without cosmological constant }.}
\newblock \emph{\cqg} \textbf{19}, 1013 (2002).
\newblock \eprint{arXiv.org/abs/gr-qc/0106065}.

\bibitem{2006JMP....47d2505C}
T.~{Christodoulakis}, T.~{Grammenos}, C.~{Helias}, P.~G. {Kevrekidis} and
  A.~{Spanou}.
\newblock \enquote{{Decoupling of the general scalar field mode and the
  solution space for Bianchi type I and V cosmologies coupled to perfect fluid
  sources}.}
\newblock \emph{Journal of Mathematical Physics} \textbf{47}, 2505 (2006).
\newblock \eprint{arXiv.org/abs/gr-qc/0506132}.

\bibitem{2002CMaPh.226..377C}
T.~{Christodoulakis}, E.~{Korfiatis} and G.~O. {Papadopoulos}.
\newblock \enquote{{Automorphism Inducing Diffeomorphisms,Invariant
  Characterization of Homogeneous 3-Spaces and Hamiltonian Dynamics of Bianchi
  Cosmologies}.}
\newblock \emph{Communications in Mathematical Physics} \textbf{226}, 377
  (2002).
\newblock \eprint{arXiv.org/abs/gr-qc/0107050}.

\bibitem{2001PhLB..501..264C}
T.~{Christodoulakis} and G.~O. {Papadopoulos}.
\newblock \enquote{{Conditional symmetries, the true degree of freedom and GCT
  invariant wave functions for the general Bianchi type II vacuum cosmology}.}
\newblock \emph{\plb} \textbf{501}, 264 (2001).
\newblock \eprint{arXiv.org/abs/gr-qc/0009074}.

\bibitem{Coleman}
S.~Coleman.
\newblock \enquote{{Why there is nothing rather than something: A theory of the
  cosmological constant}.}
\newblock \emph{\ncb} \textbf{310}, 643 (1988).

\bibitem{CW73}
S.~Coleman and E.~Weinberg.
\newblock \enquote{Radiative corrections as the origin of spontaneous symmetry
  breaking.}
\newblock \emph{\prd} \textbf{7}, 1888 (1973).

\bibitem{brane3}
A.~Coley.
\newblock \enquote{No chaos in brane-world cosmology.}
\newblock \emph{\cqg} \textbf{19}, L45 (2002).
\newblock \eprint{arXiv.org/abs/hep-th/0110117}.

\bibitem{2004CQGra..21.4193C}
A.~A. {Coley} and S.~{Hervik}.
\newblock \enquote{{Bianchi cosmologies: a tale of two tilted fluids}.}
\newblock \emph{\cqg} \textbf{21}, 4193 (2004).
\newblock \eprint{arXiv.org/abs/gr-qc/0406120}.

\bibitem{1992CQGra...9..651C}
A.~A. {Coley} and J.~{Wainwright}.
\newblock \enquote{{Qualitative analysis of two-fluid Bianchi cosmologies.}}
\newblock \emph{\cqg} \textbf{9}, 651 (1992).

\bibitem{1979PhR....56...65C}
C.~B. {Collins} and G.~F.~R. {Ellis}.
\newblock \enquote{{Singularities in Bianchi cosmologies.}}
\newblock \emph{\physrep} \textbf{56}, 65 (1979).

\bibitem{1973MNRAS.162..307C}
C.~B. {Collins} and S.~W. {Hawking}.
\newblock \enquote{{The rotation and distortion of the Universe}.}
\newblock \emph{\mnras} \textbf{162}, 307 (1973).

\bibitem{CH73}
C.~B. Collins and S.~W. Hawking.
\newblock \enquote{Why is the universe isotropic?}
\newblock \emph{\apj} \textbf{180}, 317 (1973).

\bibitem{1993JPhA...26.5795C}
G.~{Contopoulos}, B.~{Grammaticos} and A.~{Ramani}.
\newblock \enquote{{Painlev{\'e} analysis for the mixmaster universe model.}}
\newblock \emph{Journal of Physics A Mathematical General} \textbf{26}, 5795
  (1993).

\bibitem{CGR94}
G.~Contopoulos, B.~Grammaticos and A.~Ramani.
\newblock In \emph{Deterministic Chaos in General Relativity}, eds. D.~Hobill,
  D.~Bernstein, D.~Simpkins and M.~Welge (1994), p. 423.

\bibitem{1994JPhA...27.5357C}
G.~{Contopoulos}, B.~{Grammaticos} and A.~{Ramani}.
\newblock \enquote{{The mixmaster universe model, revisited.}}
\newblock \emph{Journal of Physics A Mathematical General} \textbf{27}, 5357
  (1994).

\bibitem{sinai}
I.~P. Cornfeld, S.~V. Fomin and I.~A.~G. Sinai.
\newblock \emph{{Ergodic theory}} (Springer, Berlin, 1982).

\bibitem{CornishLevin1997PRD}
N.~J. Cornish and J.~J. Levin.
\newblock \enquote{The mixmaster universe: A chaotic farey tale.}
\newblock \emph{\prd} \textbf{55}, 7489 (1997).
\newblock \eprint{arXiv.org/abs/gr-qc/9612066}.

\bibitem{CornishLevin1997PRL}
N.~J. Cornish and J.~J. Levin.
\newblock \enquote{The mixmaster universe is chaotic.}
\newblock \emph{\prl} \textbf{78}, 998 (1997).
\newblock \eprint{arXiv.org/abs/gr-qc/9605029}.

\bibitem{1999magr.meet..616C}
N.~J. {Cornish} and J.~J. {Levin}.
\newblock \enquote{{The Mixmaster Universe is Unambiguously Chaotic}.}
\newblock In \emph{Recent Developments in Theoretical and Experimental General
  Relativity, Gravitation, and Relativistic Field Theories}, eds. T.~{Piran}
  and R.~{Ruffini} (1999), p. 616.

\bibitem{Corvino}
G.~Corvino and G.~Montani.
\newblock \enquote{Dark matter prediction from canonical quantum gravity with
  frame fixing.}
\newblock \emph{\mpla} \textbf{19}, 2777 (2004).
\newblock \eprint{arXiv.org/abs/gr-qc/0410026}.

\bibitem{1994JPhA...27.1625C}
S.~{Cotsakis} and P.~G.~L. {Leach}.
\newblock \enquote{{Painlev{\'e} analysis of the mixmaster universe.}}
\newblock \emph{Journal of Physics A Mathematical General} \textbf{27}, 1625
  (1994).

\bibitem{Cot97cm}
S.~Cotsakis, R.~L. Lemmer and P.~G.~L. Leach.
\newblock \enquote{Adiabatic invariants and mixmaster catastrophes.}
\newblock \emph{\prd} \textbf{57}, 4691 (1998).
\newblock \eprint{arXiv.org/abs/gr-qc/9712027}.

\bibitem{1993CQGra..10.1607C}
O.~{Coussaert} and M.~{Henneaux}.
\newblock \enquote{{Bianchi cosmological models and gauge symmetries.}}
\newblock \emph{\cqg} \textbf{10}, 1607 (1993).
\newblock \eprint{arXiv.org/abs/gr-qc/9301001}.

\bibitem{Graham1991PRA1}
A.~Csord\'as, R.~Graham and P.~Sz\'epfalusy.
\newblock \enquote{Level statistics of a noncompact cosmological billiard.}
\newblock \emph{\pra} \textbf{44}, 1491 (1991).

\bibitem{1995RpMP...36...75C}
R.~{Cushman} and J.~{Sniatycki}.
\newblock \enquote{{Local integrability of the mixmaster model.}}
\newblock \emph{Reports on Mathematical Physics} \textbf{36}, 75 (1995).

\bibitem{DamourHenneauxPRL2000}
T.~Damour and M.~Henneaux.
\newblock \enquote{Chaos in superstring cosmology.}
\newblock \emph{\prl} \textbf{85}, 920 (2000).
\newblock \eprint{arXiv.org/abs/hep-th/0003139}.

\bibitem{DamourHenneauxPLB2000}
T.~Damour and M.~Henneaux.
\newblock \enquote{Oscillatory behavior in homogeneous string cosmology
  models.}
\newblock \emph{\plb} \textbf{488}, 108 (2000).
\newblock \eprint{arXiv.org/abs/hep-th/0006171}.

\bibitem{DamourHenneauxPRL2001}
T.~Damour and M.~Henneaux.
\newblock \enquote{$e_{10}$, $be_{10}$ and arithmetical chaos in superstring
  cosmology.}
\newblock \emph{\prl} \textbf{86}, 4749 (2001).
\newblock \eprint{arXiv.org/abs/hep-th/0012172}.

\bibitem{DHJN2001}
T.~Damour, M.~Henneaux, B.~Julia and H.~Nicolai.
\newblock \enquote{Hyperbolic kac-moody algebras and chaos in kaluza-klein
  models.}
\newblock \emph{\plb} \textbf{509}, 323 (2001).
\newblock \eprint{arXiv.org/abs/hep-th/0103094}.

\bibitem{cosmbill}
T.~Damour, M.~Henneaux and H.~Nicolai.
\newblock \enquote{{Cosmological billiards}.}
\newblock \emph{\cqg} \textbf{20}, 020 (2003).
\newblock \eprint{arXiv.org/abs/hep-th/0212256}.

\bibitem{2005PhRvD..72f7301D}
G.~{Date}.
\newblock \enquote{{Preclassical solutions of the vacuum Bianchi I loop quantum
  cosmology}.}
\newblock \emph{\prd} \textbf{72}, 067301 (2005).
\newblock \eprint{arXiv.org/abs/gr-qc/0505030}.

\bibitem{deBernardis2000Na}
P.~de~Bernardis.
\newblock \enquote{A flat universe from high-resolution maps of the cosmic
  microwave background radiation.}
\newblock \emph{Nature} \textbf{404}, 955 (2000).
\newblock \eprint{arXiv.org/abs/astro-ph/0004404}.

\bibitem{MG01}
A.~de~Moura and C.~Grebogi.
\newblock \enquote{Output functions and fractal dimensions in dynamical
  systems.}
\newblock \emph{\prl} \textbf{86}, 2778 (2001).

\bibitem{MG02}
A.~de~Moura and C.~Grebogi.
\newblock \enquote{Countable and uncountable boundaries in chaotic scattering.}
\newblock \emph{\pre} \textbf{66}, 046214 (2002).

\bibitem{OAST02}
H.~P. de~Oliveira, A.~M.~O. de~Almeida, I.~D. Soares and E.~V. Tonini.
\newblock \enquote{Homoclinic chaos in the dynamics of a general bianchi
  type-ix model.}
\newblock \emph{\prd} \textbf{D65}, 083511 (2002).
\newblock \eprint{arXiv.org/abs/gr-qc/0202047}.

\bibitem{1993PhLB..300...44D}
P.~D. {D'Eath}, S.~W. {Hawking} and O.~{Oberg{\'o}n}.
\newblock \enquote{{Supersymmetric Bianchi models and the square root of the
  Wheeler-DeWitt equation.}}
\newblock \emph{\plb} \textbf{300}, 44 (1993).

\bibitem{1993PhLB..299..223D}
J.~{Demaret} and Y.~{de Rop}.
\newblock \enquote{{The fractal nature of the power spectrum as an indicator of
  chaos in the Bianchi IX cosmological model.}}
\newblock \emph{\plb} \textbf{299}, 223 (1993).

\bibitem{Demaret1985}
J.~Demaret, J.~L. Hanquin, M.~Henneaux and P.~Spindel.
\newblock \enquote{Cosmological models in eleven-dimensional supergravity.}
\newblock \emph{\ncb} \textbf{252}, 538 (1985).

\bibitem{Demaret1986}
J.~Demaret, J.~L. Hanquin, M.~Henneaux, P.~Spindel and A.~Taormina.
\newblock \enquote{{The fate of the mixmaster behaviour in vacuum inhomogeneous
  Kaluza-Klein cosmological models}.}
\newblock \emph{\plb} \textbf{175}, 129 (1986).

\bibitem{Demaret1986su}
J.~Demaret, M.~Henneaux and P.~Spindel.
\newblock \enquote{Non-oscillatory behaviour in vacuum kaluza-klein
  cosmologies.}
\newblock \emph{\plb} \textbf{164}, 27 (1985).

\bibitem{1996JPhA...29...59D}
J.~{Demaret} and C.~{Scheen}.
\newblock \enquote{{Painlev{\'e} singularity analysis of the perfect fluid
  Bianchi type-IX relativistic cosmological model.}}
\newblock \emph{Journal of Physics A Mathematical General} \textbf{29}, 59
  (1996).

\bibitem{1987PhRvD..36.2945D}
M.~{Demianski}, M.~{Heller} and M.~{Szydlowski}.
\newblock \enquote{{Dynamics of multidimensional generalization of Bianchi
  type-IX cosmological models}.}
\newblock \emph{\prd} \textbf{36}, 2945 (1987).

\bibitem{DeWitt1967I}
B.~S. DeWitt.
\newblock \enquote{Quantum theory of gravity i: The canonical theory.}
\newblock \emph{Physical Review} \textbf{160}, 1113 (1967).

\bibitem{DeWitt1967II}
B.~S. DeWitt.
\newblock \enquote{Quantum theory of gravity ii: The manifestly covariant
  theory.}
\newblock \emph{Physical Review} \textbf{162}, 1195 (1967).

\bibitem{DeWitt1967III}
B.~S. DeWitt.
\newblock \enquote{Quantum theory of gravity iii: Applications of the covariant
  theory.}
\newblock \emph{Physical Review} \textbf{162}, 1239 (1967).

\bibitem{2004GReGr..36.2635D}
L.~{di Menza} and T.~{Lehner}.
\newblock \enquote{{The Chaotic Mixmaster and the Suppression of Chaos in
  Scalar-Tensor Cosmologies}.}
\newblock \emph{\grg} \textbf{36}, 2635 (2004).

\bibitem{Dirac64}
P.~Dirac.
\newblock Lecture on Quantum Mechanics, Belfer Graduate of Science (New York -
  Yeshiva University, 1964).

\bibitem{1971ZhETF..60.1201D}
A.~G. {Doroshkevich}, V.~N. {Lukash} and I.~D. {Novikov}.
\newblock \enquote{{Impossibility of mixing in a cosmological model of the
  Bianchi IX type.}}
\newblock \emph{Zhurnal Eksperimental noi i Teoreticheskoi Fiziki} \textbf{60},
  1201 (1971).

\bibitem{1971SvA....14..763D}
A.~G. {Doroshkevich} and I.~D. {Novikov}.
\newblock \enquote{{Mixmaster Universes and the Cosmological Problem.}}
\newblock \emph{Soviet Astronomy} \textbf{14}, 763 (1971).

\bibitem{2006GReGr..38.1003E}
G.~F.~R. {Ellis}.
\newblock \enquote{{The Bianchi models: Then and now}.}
\newblock \emph{\grg} \textbf{38}, 1003 (2006).

\bibitem{El-Ma}
G.~F.~R. Ellis and M.~A.~H. MacCallum.
\newblock \enquote{A class of homogeneous cosmological models.}
\newblock \emph{Communications in Mathematical Physics} \textbf{12}, 108
  (1969).

\bibitem{1983PhRvD..28.....E}
Y.~{Elskens}.
\newblock \enquote{{Alternative descriptions of the discrete mixmaster
  universe.}}
\newblock \emph{\prd} \textbf{28} (1983).

\bibitem{1987JSP....48.1269E}
Y.~{Elskens}.
\newblock \enquote{{Ergodic theory of the mixmaster universe in higher
  space-time dimensions. II.}}
\newblock \emph{Journal of Statistical Physics} \textbf{48}, 1269 (1987).

\bibitem{Elskens:1987gj}
Y.~Elskens and M.~Henneaux.
\newblock \enquote{Ergodic theory of the mixmaster model in higher space-time
  dimensions.}
\newblock \emph{\ncb} \textbf{290}, 111 (1987).

\bibitem{2004PhRvD..69f3514E}
J.~K. {Erickson}, D.~H. {Wesley}, P.~J. {Steinhardt} and N.~{Turok}.
\newblock \enquote{{Kasner and mixmaster behavior in universes with equation of
  state $w{\geq}1$}.}
\newblock \emph{\prd} \textbf{69}, 063514 (2004).
\newblock \eprint{arXiv.org/abs/hep-th/0312009}.

\bibitem{Espo94}
G.~Esposito.
\newblock \emph{Quantum Gravity, Quantum Cosmology and Lorentzian Geometries},
  \emph{Lecture Notes in Physics}, vol. M 12 (Springer-Verlag, Berlin, 1994).

\bibitem{1987A&A...179...11F}
R.~{Fabbri} and M.~{Tamburrano}.
\newblock \enquote{{Polarization of the cosmic background radiation in magnetic
  Bianchi type-II cosmologies}.}
\newblock \emph{\aap} \textbf{179}, 11 (1987).

\bibitem{2004CQGra..21.1609F}
S.~{Fay}.
\newblock \enquote{{Isotropization of Bianchi class A models with a minimally
  coupled scalar field and a perfect fluid}.}
\newblock \emph{\grg} \textbf{21}, 1609 (2004).
\newblock \eprint{arXiv.org/abs/gr-qc/0402104}.

\bibitem{2005GReGr..37.1233F}
S.~{Fay}.
\newblock \enquote{{Isotropisation of flat homogeneous Bianchi type I model
  with a non minimally coupled and massive scalar field}.}
\newblock \emph{General Relativity and Gravitation} \textbf{37}, 1233 (2005).
\newblock \eprint{arXiv.org/abs/gr-qc/0509070}.

\bibitem{2005GReGr..37.1097F}
S.~{Fay} and T.~{Lehner}.
\newblock \enquote{{Bianchi type IX asymptotical behaviours with a massive
  scalar field: chaos strikes back}.}
\newblock \emph{\grg} \textbf{37}, 1097 (2005).
\newblock \eprint{arXiv.org/abs/gr-qc/0509071}.

\bibitem{Fee}
G.~J. Fee.
\newblock \emph{Homogeneous Spacetimes}.
\newblock Master's thesis (1979).

\bibitem{FF92}
K.~Ferraz and G.~Francisco.
\newblock \enquote{Mixmaster numerical behavior and generalizations.}
\newblock \emph{\prd} \textbf{45}, 1158 (1992).

\bibitem{1990STIN...9215945F}
K.~{Ferraz}, G.~{Francisco} and G.~E.~A. {Matsas}.
\newblock \enquote{{Mixmaster: A weakly chaotic model?}}
\newblock \emph{NASA STI/Recon Technical Report N} \textbf{92}, 15945 (1990).

\bibitem{FFM91}
K.~Ferraz, G.~Francisco and G.~E.~A. Matsas.
\newblock \enquote{Chaotic and nonchaotic behavior in the mixmaster dynamics.}
\newblock \emph{\pla} \textbf{156}, 407  (1991).

\bibitem{Fis-Mar}
A.~E. Fischer and J.~E. Marsden.
\newblock \enquote{The einstein equations of evolution - a geometric approach.}
\newblock \emph{Journal of Mathematical Physics} \textbf{13}, 546 (1972).

\bibitem{1992GReGr..24..679F}
K.~{Fi{\v s}er}, K.~{Rosquist} and C.~{Uggla}.
\newblock \enquote{{Bianchi type V perfect fluid cosmologies}.}
\newblock \emph{\grg} \textbf{24}, 679 (1992).

\bibitem{2000GReGr..32.1255F}
V.~N. {Folomeev} and V.~T. {Gurovich}.
\newblock \enquote{{Classical and Quantum Evolution of the Bianchi Type I
  Model}.}
\newblock \emph{\grg} \textbf{32}, 1255 (2000).
\newblock \eprint{arXiv.org/abs/gr-qc/0001065}.

\bibitem{2000GrCos...6...19F}
V.~N. {Folomeev} and V.~T. {Gurovich}.
\newblock \enquote{{Quantum evolution of Bianchi type I model.}}
\newblock \emph{Gravitation and Cosmology} \textbf{6}, 19 (2000).

\bibitem{FranciscoMatsas1988}
G.~Francisco and G.~E.~A. Matsas.
\newblock \enquote{{Qualitative and numerical study of Bianchi IX models}.}
\newblock \emph{\grg} \textbf{20}, 1047 (1988).

\bibitem{1993CQGra..10.1147F}
Y.~{Furihata} and K.~{Sato}.
\newblock \enquote{{Regge calculus model of Bianchi IX inflation.}}
\newblock \emph{\cqg} \textbf{10}, 1147 (1993).

\bibitem{1986PThPh..75...59F}
T.~{Furusawa}.
\newblock \enquote{{Quantum Chaos of Mixmaster Universe}.}
\newblock \emph{\ptp} \textbf{75}, 59 (1986).

\bibitem{Furusawa1985PTP}
T.~Furusawa.
\newblock \enquote{Quantum chaos of mixmaster universe. 2.}
\newblock \emph{\ptp} \textbf{76}, 67 (1986).

\bibitem{2003CQGra..20.5425G}
R.~{Garc{\'{\i}}a-Salcedo} and N.~{Bret{\'o}n}.
\newblock \enquote{{Nonlinear electrodynamics in Bianchi spacetimes}.}
\newblock \emph{\cqg} \textbf{20}, 5425 (2003).
\newblock \eprint{arXiv.org/abs/hep-th/0212130}.

\bibitem{GerochPenrose}
R.~Geroch, E.~H. Kronheimer and R.~Penrose.
\newblock \enquote{Ideal points in space-time.}
\newblock \emph{Royal Society of London Proceedings Series A} \textbf{327}, 545
  (1972).

\bibitem{GerochWald}
R.~Geroch, C.~B. Liang and R.~M. Wald.
\newblock \enquote{Singular boundaries of space-times.}
\newblock \emph{Journal of Mathematical Physics} \textbf{23}, 432 (1982).

\bibitem{1985PhLB..152..171G}
E.~{Giannetto} and P.~{Salucci}.
\newblock \enquote{{Quark-hadron phase transition in Bianchi type I
  cosmology.}}
\newblock \emph{\plb} \textbf{152}, 171 (1985).

\bibitem{ThiAGQ1}
K.~Giesel and T.~Thiemann.
\newblock \enquote{Algebraic quantum gravity (aqg). i: Conceptual setup.}
\newblock \emph{\cqg} \textbf{24}, 2465 (2007).
\newblock \eprint{arXiv.org/abs/gr-qc/0607099}.

\bibitem{ThiAGQ2}
K.~Giesel and T.~Thiemann.
\newblock \enquote{Algebraic quantum gravity (aqg). ii: Semiclassical
  analysis.}
\newblock \emph{\cqg} \textbf{24}, 2499 (2007).
\newblock \eprint{arXiv.org/abs/gr-qc/0607100}.

\bibitem{ThiAGQ3}
K.~Giesel and T.~Thiemann.
\newblock \enquote{Algebraic quantum gravity (aqg). iii: Semiclassical
  perturbation theory.}
\newblock \emph{\cqg} \textbf{24}, 2565 (2007).
\newblock \eprint{arXiv.org/abs/gr-qc/0607101}.

\bibitem{marolf}
D.~Giulini and D.~Marolf.
\newblock \enquote{On the generality of refined algebraic quantization.}
\newblock \emph{\cqg} \textbf{16}, 2479 (1999).
\newblock \eprint{arXiv.org/abs/gr-qc/9812024}.

\bibitem{1997GrCos...3...85G}
S.~A. {Gogilidze}, A.~M. {Khvedelidze}, D.~M. {Mladenov} and V.~N. {Perbuhin}.
\newblock \enquote{{Hamiltonian analysis of Bianchi IX cosmology.}}
\newblock \emph{Gravitation and Cosmology} \textbf{3}, 85 (1997).

\bibitem{1995CQGra..12.1287G}
G.~{Gonz{\'a}lez} and R.~S. {Tate}.
\newblock \enquote{{Classical analysis of Bianchi types I and II in Ashtekar
  variables }.}
\newblock \emph{\cqg} \textbf{12}, 1287 (1995).
\newblock \eprint{arXiv.org/abs/gr-qc/9412015}.

\bibitem{Got-Dem}
M.~J. Gotay and J.~Demaret.
\newblock \enquote{Quantum cosmological singularities.}
\newblock \emph{\prd} \textbf{28}, 2402 (1983).

\bibitem{grahamsuper}
R.~Graham.
\newblock \enquote{Supersymmetric bianchi type ix cosmology.}
\newblock \emph{\prl} \textbf{67}, 1381 (1991).

\bibitem{Graham1994Chaos}
R.~Graham.
\newblock \enquote{Chaos and quantum chaos in cosmological models.}  (1994).
\newblock \eprint{arXiv.org/abs/gr-qc/9403030}.

\bibitem{Graham1991PRA2}
R.~Graham, R.~H\"ubner, P.~Sz\'epfalusy and G.~Vattay.
\newblock \enquote{Level statistics of a noncompact integrable billiard.}
\newblock \emph{\pra} \textbf{44}, 7002 (1991).

\bibitem{grahamhartlehawk}
R.~Graham and H.~Luckock.
\newblock \enquote{Hartle-hawking state for the bianchi type ix model in
  supergravity.}
\newblock \emph{\prd} \textbf{49}, R4981 (1994).

\bibitem{1996PhRvD..54.2589G}
R.~{Graham} and R.~{Paternoga}.
\newblock \enquote{{Physical states of Bianchi type IX quantum cosmologies
  described by the Chern-Simons functional}.}
\newblock \emph{\prd} \textbf{54}, 2589 (1996).
\newblock \eprint{arXiv.org/abs/gr-qc/9603027}.

\bibitem{grahamquantumbirth}
R.~Graham and P.~Sz\'epfalusy.
\newblock \enquote{Quantum creation of a generic universe.}
\newblock \emph{\prd} \textbf{42}, 2483 (1990).

\bibitem{GOPJ84}
C.~Grebogi, E.~Ott, S.~Pelikan and J.~Yorke.
\newblock \enquote{Chaos, strange attractors, and fractal basin boundaries in
  nonlinear dynamics.}
\newblock \emph{Physica D} \textbf{13}, 261 (1984).

\bibitem{Grishchuk1975ZETF}
L.~P. Grishchuk, A.~G. Doroshkevich and V.~M. Yudin.
\newblock \enquote{Long gravitational waves in a closed universe. (in
  russian).}
\newblock \emph{\zetf} \textbf{69}, 1857 (1975).

\bibitem{1986PhRvD..33.1204G}
{\O}.~{Gr{\o}n}.
\newblock \enquote{{Transition of rotating Bianchi type-IX cosmological model
  into an inflationary era}.}
\newblock \emph{\prd} \textbf{33}, 1204 (1986).

\bibitem{gross1988}
D.~J. Gross and P.~F. Mende.
\newblock \enquote{{String theory beyond the Planck scale}.}
\newblock \emph{\ncb} \textbf{303}, 407 (1988).

\bibitem{MCP}
B.~Grubi$\check{s}$i\'{c} and V.~Moncrief.
\newblock \enquote{Asymptotic behavior of the
  $t^{3}\ifmmode\times\else\texttimes\fi{}r$ gowdy space-times.}
\newblock \emph{\prd} \textbf{47}, 2371 (1993).
\newblock \eprint{arXiv.org/abs/gr-qc/9209006}.

\bibitem{1994PhRvD..49.2792G}
B.~{Grubi{\v s}i{\'C}} and V.~{Moncrief}.
\newblock \enquote{{Mixmaster spacetime, Geroch's transformation, and constants
  of motion}.}
\newblock \emph{\prd} \textbf{49}, 2792 (1994).
\newblock \eprint{arXiv.org/abs/gr-qc/9309007}.

\bibitem{G81}
A.~H. Guth.
\newblock \enquote{Inflationary universe: A possible solution to the horizon
  and flatness problems.}
\newblock \emph{\prd} \textbf{23}, 347 (1981).

\bibitem{GP82}
A.~H. Guth and S.~Pi.
\newblock \enquote{Fluctuations in the new inflationary universe.}
\newblock \emph{\prl} \textbf{49}, 1110 (1982).

\bibitem{1993Ap&SS.199..289G}
E.~{Guzman}.
\newblock \enquote{{Isotropization in Bianchi type-I vacuum cosmology}.}
\newblock \emph{\apss} \textbf{199}, 289 (1993).

\bibitem{1996IJTP...35..419G}
E.~{Guzman}.
\newblock \enquote{{Isotropization in Bianchi type IX vacuum cosmology.}}
\newblock \emph{\ijtp} \textbf{35}, 419 (1996).

\bibitem{Haag}
R.~Haag.
\newblock \emph{Local quantum physics} (Springer, New York, 1992).

\bibitem{H91c}
J.~J. Halliwell.
\newblock \enquote{Global spacetime symmetries in the functional schr\"odinger
  picture.}
\newblock \emph{\prd} \textbf{43}, 2590 (1991).

\bibitem{1987GReGr..19...73H}
P.~{Halpern}.
\newblock \enquote{{Chaos in the long-term behavior of some Bianchi-type VIII
  models.}}
\newblock \emph{\grg} \textbf{19}, 73 (1987).

\bibitem{Halpern2002PRD}
P.~Halpern.
\newblock \enquote{Exact solutions of five dimensional anisotropic
  cosmologies.}
\newblock \emph{\prd} \textbf{66}, 027503 (2002).
\newblock \eprint{arXiv.org/abs/gr-qc/0203055}.

\bibitem{Halpern2003GRG}
P.~Halpern.
\newblock \enquote{{The Mixmaster universe in five dimensions}.}
\newblock \emph{\grg} \textbf{35}, 251 (2003).
\newblock \eprint{arXiv.org/abs/gr-qc/0207056}.

\bibitem{DASI1}
N.~W. Halverson.
\newblock \enquote{Dasi first results: a measurement of the cosmic microwave
  background angular power spectrum.}
\newblock Tech. rep. (2001).
\newblock \eprint{arXiv.org/abs/astro-ph/0104489}.

\bibitem{Han}
M.~Han and Y.~Ma.
\newblock \enquote{Master constraint operator in loop quantum gravity.}
\newblock \emph{\plb} \textbf{635}, 225 (2006).
\newblock \eprint{arXiv.org/abs/gr-qc/0510014}.

\bibitem{1984CQGra...1..291H}
J.~L. {Hanquin} and J.~{Demaret}.
\newblock \enquote{{Exact solutions for inhomogeneous generalisations of some
  vacuum Bianchi models}.}
\newblock \emph{\cqg} \textbf{1}, 291 (1984).

\bibitem{Ruff}
C.~O. Hans and R.~Ruffini.
\newblock \emph{Gravitation and Spacetime} (WW Norton and Company, New York,
  1994).

\bibitem{brane1}
T.~Harko and M.~K. Mak.
\newblock \enquote{Anisotropy in bianchi-type brane cosmologies.}
\newblock \emph{\cqg} \textbf{21}, 1489 (2004).
\newblock \eprint{arXiv.org/abs/gr-qc/0401069}.

\bibitem{har88}
J.~B. Hartle.
\newblock In \emph{Highlights in Gravitation and Cosmology}, eds. A.~K.
  B.~Iyer, J.~V. Narlikar and C.~Vishveshuara (Cambridge Univ. Press,
  Cambridge, 1988).

\bibitem{HartleHawking1983PRD}
J.~B. Hartle and S.~W. Hawking.
\newblock \enquote{Wave function of the universe.}
\newblock \emph{\prd} \textbf{28}, 2960 (1983).

\bibitem{1969MNRAS.142..129H}
S.~{Hawking}.
\newblock \enquote{{On the rotation of the Universe}.}
\newblock \emph{\mnras} \textbf{142}, 129 (1969).

\bibitem{HE}
S.~Hawking and G.~F.~R. Ellis.
\newblock \emph{{The large scale structure of space-time}} (Cambridge
  University Press, 1973).

\bibitem{HH94}
J.~F. Heagy and S.~M. Hammel.
\newblock \emph{Physica D} \textbf{70}, 140 (1994).

\bibitem{2006CQGra..23.3463H}
J.~M. {Heinzle} and C.~{Uggla}.
\newblock \enquote{{Dynamics of the spatially homogeneous Bianchi type I
  Einstein Vlasov equations}.}
\newblock \emph{Classical and Quantum Gravity} \textbf{23}, 3463 (2006).
\newblock \eprint{arXiv.org/abs/gr-qc/0512031}.

\bibitem{Uggla1}
J.~M. Heinzle, C.~Uggla and N.~Rohr.
\newblock \enquote{{The cosmological billiard attractor}.}  (2007).
\newblock \eprint{arXiv.org/abs/gr-qc/0702141}.

\bibitem{HT}
M.~Henneaux and C.~Teitelboim.
\newblock \emph{Quantization of Gauge Systems} (Princeton Univ. Press, 1992).

\bibitem{1991PhLB..256..359H}
A.~B. {Henriques}, J.~M. {Mour{\~a}o} and P.~M. {S{\'a}}.
\newblock \enquote{{Inflation in a Bianchi-IX cosmological model. The roles of
  primordial shear and gauge fields.}}
\newblock \emph{\plb} \textbf{256}, 359 (1991).

\bibitem{2000CQGra..17.2765H}
S.~{Hervik}.
\newblock \enquote{{The Bianchi type I minisuperspace model }.}
\newblock \emph{\cqg} \textbf{17}, 2765 (2000).
\newblock \eprint{arXiv.org/abs/gr-qc/0003084}.

\bibitem{2001GReGr..33...65H}
C.~G. {Hewitt}, R.~{Bridson} and J.~{Wainwright}.
\newblock \enquote{{The Asymptotic Regimes of Tilted Bianchi II Cosmologies}.}
\newblock \emph{\grg} \textbf{33}, 65 (2001).
\newblock \eprint{arXiv.org/abs/gr-qc/0008037}.

\bibitem{1993CQGra..10...99H}
C.~G. {Hewitt} and J.~{Wainwright}.
\newblock \enquote{{A dynamical systems approach to Bianchi cosmologies:
  orthogonal models of class B.}}
\newblock \emph{\cqg} \textbf{10}, 99 (1993).

\bibitem{1991NYASA.631...15H}
D.~{Hobill}.
\newblock \enquote{{Sources of chaos in Mixmaster cosmologies}.}
\newblock \emph{New York Academy Sciences Annals} \textbf{631}, 15 (1991).

\bibitem{HobillNum1991}
D.~Hobill, D.~Bernstein, M.~Welge and D.~Simkins.
\newblock \enquote{The mixmaster cosmology as a dynamical system.}
\newblock \emph{\cqg} \textbf{8}, 1155 (1991).

\bibitem{Hobill1994}
D.~Hobill, A.~Burd and A.~Coley.
\newblock \emph{Deterministic Chaos in General Relativity} (Plenum Press, New
  York, 1994).

\bibitem{Horava}
P.~Horava.
\newblock \enquote{On a covariant hamilton-jacobi framework for the
  einstein-maxwell theory.}
\newblock \emph{\cqg} \textbf{8}, 2069 (1991).

\bibitem{2004GReGr..36..799H}
J.~T. {Horwood} and J.~{Wainwright}.
\newblock \enquote{{Asymptotic Regimes of Magnetic Bianchi Cosmologies}.}
\newblock \emph{\grg} \textbf{36}, 799 (2004).
\newblock \eprint{arXiv.org/abs/gr-qc/0309083}.

\bibitem{Hosoya1987}
A.~Hosoya, L.~G. Jensen and J.~A. Stein-Schabes.
\newblock \enquote{{The critical dimension for chaotic cosmology}.}
\newblock \emph{\ncb} \textbf{283}, 657 (1987).

\bibitem{1973PhRvD...8.1048H}
B.~L. {Hu}.
\newblock \enquote{{Scalar waves in the mixmaster universe. I. The Helmholtz
  equation in a fixed background.}}
\newblock \emph{\prd} \textbf{8}, 1048 (1973).

\bibitem{1974PhRvD...9.3263H}
B.~L. {Hu}.
\newblock \enquote{{Scalar waves in the mixmaster universe. II. Particle
  creation.}}
\newblock \emph{\prd} \textbf{9}, 3263 (1974).

\bibitem{1975PhRvD..12.1551H}
B.~L. {Hu}.
\newblock \enquote{{Numerical examples from perturbation analysis of the
  mixmaster universe}.}
\newblock \emph{\prd} \textbf{12}, 1551 (1975).

\bibitem{1978PhRvD..18..969H}
B.~L. {Hu}.
\newblock \enquote{{Gravitational waves in a Bianchi type-I universe}.}
\newblock \emph{\prd} \textbf{18}, 969 (1978).

\bibitem{1986PhRvD..34.2535H}
B.~L. {Hu} and D.~J. {O'Connor}.
\newblock \enquote{{Mixmaster inflation.}}
\newblock \emph{\prd} \textbf{34}, 2535 (1986).

\bibitem{1972PhRvL..29.1616H}
B.~L. {Hu} and T.~{Regge}.
\newblock \enquote{{Perturbations on the mixmaster universe.}}
\newblock \emph{\prl} \textbf{29}, 1616 (1972).

\bibitem{Huang}
K.~Huang.
\newblock \emph{Statistical mechanics} (Wiley, New York, 1987).

\bibitem{1998PhRvD..58h4007H}
W.~Huang.
\newblock \enquote{Semiclassical gravitation and quantization for the bianchi
  type-i universe with large anisotropy.}
\newblock \emph{\prd} \textbf{58}, 084007 (1998).
\newblock \eprint{arXiv.org/abs/hep-th/0209092}.

\bibitem{1988PhLA..129..429H}
W.-H. {Huang}.
\newblock \enquote{{Bianchi type I cosmological model with bulk viscosity.}}
\newblock \emph{\pla} \textbf{129}, 429 (1988).

\bibitem{1995PhRvD..51..928I}
J.~{Ib{\'a}{\~n}ez}, R.~J. {van den Hoogen} and A.~A. {Coley}.
\newblock \enquote{{Isotropization of scalar field Bianchi models with an
  exponential potential}.}
\newblock \emph{\prd} \textbf{51}, 928 (1995).

\bibitem{ImponenteMontani2001IJMPD}
G.~P. Imponente and G.~Montani.
\newblock \enquote{Covariant formulation of the invariant measure for the
  mixmaster dynamics.}
\newblock \emph{\ijmpd} \textbf{11}, 1321 (2001).
\newblock \eprint{arXiv.org/abs/gr-qc/0106028}.

\bibitem{ImponenteMontani2001PRD}
G.~P. Imponente and G.~Montani.
\newblock \enquote{On the covariance of the mixmaster chaoticity.}
\newblock \emph{\prd} \textbf{63}, 103501 (2001).
\newblock \eprint{arXiv.org/abs/astro-ph/0102067}.

\bibitem{ImponenteMontani2001NPPS}
G.~P. Imponente and G.~Montani.
\newblock \enquote{On the quantum origin of the mixmaster chaos covariance.}
\newblock \emph{Nuclear Physics Proceeding Supplement} \textbf{104}, 193
  (2002).
\newblock \eprint{arXiv.org/abs/gr-qc/0111051}.

\bibitem{ImponenteMontani2003IJMPD}
G.~P. Imponente and G.~Montani.
\newblock \enquote{Mixmaster chaoticity as semiclassical limit of the canonical
  quantum dynamics.}
\newblock \emph{\ijmpd} \textbf{12}, 977 (2003).
\newblock \eprint{arXiv.org/abs/gr-qc/0404106}.

\bibitem{ImponenteMontani2002JKPS}
G.~P. Imponente and G.~Montani.
\newblock \enquote{The mixmaster model as a cosmological framework and aspects
  of its quantum dynamics.}
\newblock \emph{Journal of Korean Physical Society} \textbf{42}, S54 (2003).
\newblock \eprint{arXiv.org/abs/gr-qc/0201102}.

\bibitem{IM03}
G.~P. Imponente and G.~Montani.
\newblock \enquote{Quasi isotropic inflationary solution.}
\newblock \emph{\ijmpd} \textbf{12}, 1845 (2003).
\newblock \eprint{arXiv.org/abs/gr-qc/0307048}.

\bibitem{ImponenteMontani2004PA}
G.~P. Imponente and G.~Montani.
\newblock \enquote{Bianchi ix chaoticity: Bkl map and continuous flow.}
\newblock \emph{Physica A} \textbf{338}, 282 (2004).
\newblock \eprint{arXiv.org/abs/gr-qc/0401086}.

\bibitem{IsenbergMoncrief}
J.~Isenberg and V.~Moncrief.
\newblock \enquote{{Asymptotic behavior of the gravitational field and the
  nature of singularities in gowdy spacetimes}.}
\newblock \emph{Annals of Physics} \textbf{199}, 84 (1990).
\newblock \eprint{arXiv.org/abs/gr-qc/9209006}.

\bibitem{ishamtime}
C.~J. Isham.
\newblock \enquote{{Canonical Quantum Gravity and the Problem of Time}.}
  (1992).
\newblock \eprint{arXiv.org/abs/gr-qc/9210011}.

\bibitem{imk}
V.~D. Ivashchuk, V.~N. Mel'Nikov and A.~A. Kirillov.
\newblock \enquote{{Stochastic properties of multidimensional cosmological
  models near a singular point}.}
\newblock \emph{\sjetp Letters} \textbf{60}, 235 (1994).

\bibitem{2006A&A...460..393J}
T.~R. {Jaffe}, A.~J. {Banday}, H.~K. {Eriksen}, K.~M. {G{\'o}rski} and F.~K.
  {Hansen}.
\newblock \enquote{{Bianchi type VIIh models and the WMAP 3-year data}.}
\newblock \emph{\aap} \textbf{460}, 393 (2006).
\newblock \eprint{arXiv.org/abs/astro-ph/0606046}.

\bibitem{2006ApJ...644..701J}
T.~R. {Jaffe}, S.~{Hervik}, A.~J. {Banday} and K.~M. {G{\'o}rski}.
\newblock \enquote{{On the Viability of Bianchi Type $VII_{h}$ Models with Dark
  Energy}.}
\newblock \emph{\apj} \textbf{644}, 701 (2006).
\newblock \eprint{arXiv.org/abs/astro-ph/0512433}.

\bibitem{2005PhRvD..71f4007J}
Y.~{Jin} and K.-I. {Maeda}.
\newblock \enquote{{Chaos of Yang-Mills field in class A Bianchi spacetimes}.}
\newblock \emph{\prd} \textbf{71}, 064007 (2005).
\newblock \eprint{arXiv.org/abs/gr-qc/0412060}.

\bibitem{Johnson}
R.~A. Johnson.
\newblock \enquote{The bundle boundary in some special cases.}
\newblock \emph{Journal of Mathematical Physics} \textbf{18}, 898 (1977).

\bibitem{2000PhRvD..62j4017J}
M.~{Joy} and V.~C. {Kuriakose}.
\newblock \enquote{{First order phase transitions in a Bianchi type-I
  universe}.}
\newblock \emph{\prd} \textbf{62}, 104017 (2000).
\newblock \eprint{arXiv.org/abs/hep-th/0008234}.

\bibitem{algebra}
T.~W. Judson.
\newblock \emph{{Abstract Algebra: Theory and Applications}} (PWS Pub. Co.,
  1994).

\bibitem{2001PhRvD..64j7301K}
W.~Kao.
\newblock \enquote{Bianchi type-i space and the stability of the inflationary
  friedmann-robertson-walker solution.}
\newblock \emph{\prd} \textbf{64}, 107301 (2001).
\newblock \eprint{arXiv.org/abs/hep-th/0104166}.

\bibitem{kasner}
E.~Kasner.
\newblock \enquote{Geometrical theorems on einstein's cosmological equations.}
\newblock \emph{\advphys} \textbf{43}, 217 (1921).

\bibitem{Kempf97}
A.~Kempf and G.~Mangano.
\newblock \enquote{Minimal length uncertainty relation and ultraviolet
  regularisation.}
\newblock \emph{\prd} \textbf{55}, 7909 (1997).
\newblock \eprint{arXiv.org/abs/hep-th/9612084}.

\bibitem{Kempf95}
A.~Kempf, G.~Mangano and R.~B. Mann.
\newblock \enquote{Hilbert space representation of the minimal length
  uncertainty relation.}
\newblock \emph{\prd} \textbf{52}, 1108 (1995).
\newblock \eprint{arXiv.org/abs/hep-th/9412167}.

\bibitem{KHA03}
I.~M. Khalatnikov, A.~Y. Kamenshchik, M.~Martellini and A.~A. Starobinsky.
\newblock \enquote{Quasi-isotropic solution of the einstein equations near a
  cosmological singularity for a two-fluid cosmological model.}
\newblock \emph{Journal on Cosmology and Astroparticle Physics} \textbf{03003},
  001 (2003).
\newblock \eprint{arXiv.org/abs/gr-qc/0301119}.

\bibitem{KHA02}
I.~M. Khalatnikov, A.~Y. Kamenshchik and A.~A. Starobinsky.
\newblock \enquote{Comment about quasi-isotropic solution of einstein equations
  near cosmological singularity.}
\newblock \emph{\cqg} \textbf{19}, 3845 (2002).
\newblock \eprint{arXiv.org/abs/gr-qc/0204045}.

\bibitem{LK1963}
I.~M. Khalatnikov and E.~M. Lifshitz.
\newblock \enquote{{Investigations in relativistic cosmology}.}
\newblock \emph{\advphys} \textbf{12}, 185 (1963).

\bibitem{K8385}
I.~M. Khalatnikov and oth.
\newblock \enquote{On the stochastic proprieties of relativistic cosmological
  models near singularity.}
\newblock \emph{\pzetf} \textbf{38}, 79 (1983).
\newblock [JETP Lett., {\bf 38}, 91 (1983)]; {\it J. Stat. Phys.} {\bf 38}, 97
  (1985).

\bibitem{Kheyfets2006CQG}
A.~Kheyfets, W.~A. Miller and R.~Vaulin.
\newblock \enquote{Quantum geometrodynamics of the bianchi ix cosmological
  model.}
\newblock \emph{\cqg} \textbf{23}, 4333 (2006).
\newblock \eprint{arXiv.org/abs/gr-qc/0512040}.

\bibitem{Kijo}
J.~Kijowski and G.~Magli.
\newblock \enquote{Unconstrained hamiltonian formulation of general relativity
  with thermo-elastic sources.}
\newblock \emph{\cqg} \textbf{15}, 3891 (1998).
\newblock \eprint{arXiv.org/abs/gr-qc/9709021}.

\bibitem{Kirillov1992JETPL}
A.~A. Kirillov.
\newblock \enquote{Quantum birth of a universe near a cosmological
  singularity.}
\newblock \emph{\JETPl} \textbf{55}, 561 (1992).

\bibitem{Kir93}
A.~A. Kirillov.
\newblock \enquote{On the question of the characteristics of the spatial
  distribution of metric inhomogeneities in a general solution to einstein
  equations in the vicinity of a cosmological singularity.}
\newblock \emph{\sovJETP} \textbf{76}, 355 (1993).

\bibitem{1998GrCos...4...23K}
A.~A. {Kirillov}.
\newblock \enquote{{Quantum behaviour of inhomogeneous Mixmaster model near the
  singularity.}}
\newblock \emph{Gravitation and Cosmology, Supplement issue} \textbf{4}, 23
  (1998).

\bibitem{Kochnev1987}
A.~A. Kirillov and A.~A. Kochnev.
\newblock \enquote{Cellular structure of space in the vicinity of a time
  singularity in the einstein equations.}
\newblock \emph{\pzetf} \textbf{46}, 345 (1987).

\bibitem{KirillovMelnikov1995}
A.~A. Kirillov and V.~N. Melnikov.
\newblock \enquote{Dynamics of inhomogeneities of the metric in the vicinity of
  a singularity in multidimensional cosmology.}
\newblock \emph{\prd} \textbf{52}, 723 (1995).
\newblock \eprint{arXiv.org/abs/gr-qc/9408004}.

\bibitem{KirillovMontani1997PRD}
A.~A. Kirillov and G.~Montani.
\newblock \enquote{Description of statistical properties of the mixmaster
  universe.}
\newblock \emph{\prd} \textbf{56}, 6225 (1997).

\bibitem{KirillovMontani1997JL}
A.~A. Kirillov and G.~Montani.
\newblock \enquote{Origin of a classical space in quantum inhomogeneous
  models.}
\newblock \emph{\JETPl} \textbf{66}, 475 (1997).

\bibitem{KirillovMontani2002PRD}
A.~A. Kirillov and G.~Montani.
\newblock \enquote{Quasi-isotropization of the inhomogeneous mixmaster universe
  induced by an inflationary process.}
\newblock \emph{\prd} \textbf{66}, 064010 (2002).
\newblock \eprint{arXiv.org/abs/gr-qc/0209054}.

\bibitem{1993CQGra..10..703K}
Y.~{Kitada} and K.~{Maeda}.
\newblock \enquote{{Cosmic no-hair theorem in homogeneous spacetimes. 1.
  Bianchi models.}}
\newblock \emph{\cqg} \textbf{10}, 703 (1993).

\bibitem{1988PThPh..80.1024K}
H.~{Kodama}.
\newblock \enquote{{Specialization of Ashtekar's Formalism to Bianchi
  Cosmology}.}
\newblock \emph{\ptp} \textbf{80}, 1024 (1988).

\bibitem{1989grg..conf..613K}
H.~{Kodama}.
\newblock \enquote{{Bianchi Cosmology in Terms of the Ashtekar Variables}.}
\newblock In \emph{\grg} (1989), p. 613.

\bibitem{2002PThPh.107..305K}
H.~{Kodama}.
\newblock \enquote{{Phase Space of Compact Bianchi Models with Fluid}.}
\newblock \emph{\ptp} \textbf{107}, 305 (2002).
\newblock \eprint{arXiv.org/abs/gr-qc/0109064}.

\bibitem{KT90}
E.~W. Kolb and M.~S. Turner.
\newblock \emph{The Early Universe} (Adison-Wesley, Reading, New York, 1990).

\bibitem{2003GReGr..35..475K}
A.~{Krasi{\'n}ski}, C.~G. {Behr}, E.~{Sch{\"u}cking}, F.~B. {Estabrook}, H.~D.
  {Wahlquist}, G.~F.~R. {Ellis}, R.~{Jantzen} and W.~{Kundt}.
\newblock \enquote{{``Golden Oldie'': The Bianchi Classification in the
  Sch{\"u}cking-Behr Approach}.}
\newblock \emph{\grg} \textbf{35}, 475 (2003).

\bibitem{1987PhLA..123..379K}
K.~D. {Krori} and M.~{Barua}.
\newblock \enquote{{Higher-dimensional Bianchi type I cosmologies.}}
\newblock \emph{\pla} \textbf{123}, 379 (1987).

\bibitem{1999GReGr..31.1423K}
K.~D. {Krori}, K.~{Pathak} and S.~{Dutta}.
\newblock \enquote{{Particle-like Behaviour of Some Bianchi Cosmologies: A
  Fundamental Unity in Nature?}}
\newblock \emph{\grg} \textbf{31}, 1423 (1999).

\bibitem{kuchar1972}
K.~Kucha{\v r}.
\newblock \enquote{{Bubble-time canonical formalism for geometrodynamics.}}
\newblock \emph{Journal of Mathematical Physics} \textbf{13}, 768 (1972).

\bibitem{Kuchar1980}
K.~Kucha{\v r}.
\newblock \enquote{{Canonical Methods of Quantization}.}
\newblock In \emph{Quantum Gravity II}, eds. C.~J. Isham, R.~Penrose and D.~W.
  Sciama (1981), p. 329.

\bibitem{KT91}
K.~Kucha{\v r} and C.~Torre.
\newblock \enquote{Gaussian reference fluid and interpretation of quantum
  geometrodynamics.}
\newblock \emph{\prd} \textbf{43}, 419 (1991).

\bibitem{1989PhRvD..40.3982K}
K.~V. {Kucha{\v r}} and M.~P. {Ryan}.
\newblock \enquote{{Is minisuperspace quantization valid? Taub in mixmaster.}}
\newblock \emph{\prd} \textbf{40}, 3982 (1989).

\bibitem{LL4}
L.~D. Landau and E.~M. Lifshitz.
\newblock \emph{Relativistic Quantum Mechanics} (Pergamon Press, Elmsford, New
  York, 1973).

\bibitem{LF}
L.~D. Landau and E.~M. Lifshitz.
\newblock \emph{Classical Theory of Fields} (Addison-Wesley, New York, 1975),
  fourth edn.

\bibitem{1995PhyD...87...70L}
A.~{Latifi}, M.~{Musette} and R.~{Conte}.
\newblock \enquote{{Nonintegrability of the Bianchi IX model.}}
\newblock \emph{Physica D} \textbf{87}, 70 (1995).

\bibitem{1997CQGra..14.2281L}
V.~G. {Leblanc}.
\newblock \enquote{{Asymptotic states of magnetic Bianchi I cosmologies.}}
\newblock \emph{Classical and Quantum Gravity} \textbf{14}, 2281 (1997).

\bibitem{1998CQGra..15.1607L}
V.~G. {Leblanc}.
\newblock \enquote{{Bianchi II magnetic cosmologies.}}
\newblock \emph{\cqg} \textbf{15}, 1607 (1998).

\bibitem{MAXIMA}
A.~T. Lee \emph{et~al.}
\newblock \enquote{A high spatial resolution analysis of the maxima-1 cosmic
  microwave background anisotropy data.}
\newblock \emph{\apj} pp. L1--L6 (2001).
\newblock \eprint{astro-ph/0104459}.

\bibitem{LOST}
J.~Lewandowski, A.~Okolow, H.~Sahlmann and T.~Thiemann.
\newblock \enquote{Uniqueness of diffeomorphism invariant states on
  holonomy-flux algebras.}
\newblock \emph{Communications in Mathematical Physics} \textbf{267}, 703
  (2006).
\newblock \eprint{arXiv.org/abs/gr-qc/0504147}.

\bibitem{LK1960}
E.~M. Lifshitz and I.~M. Khalatnikov.
\newblock \emph{\zetf} \textbf{39}, 149 (1960).

\bibitem{LLK73}
E.~M. Lifshitz, I.~M. Lifshitz and I.~M. Khalatnikov.
\newblock \emph{\sovJETP} \textbf{32}, 173 (1971).

\bibitem{Linde1}
A.~D. Linde.
\newblock \enquote{Chaotic inflation.}
\newblock \emph{\plb} \textbf{129}, 177 (1983).

\bibitem{livine}
E.~R. Livine.
\newblock \enquote{Towards a covariant loop quantum gravity.}  (2006).
\newblock \eprint{arXiv.org/abs/gr-qc/0608135}.

\bibitem{Lorenz:1980fk}
D.~Lorenz.
\newblock \enquote{{An exact Bianchi type II cosmological model with matter and
  an electromagnetic field}.}
\newblock \emph{\pla} \textbf{79}, 19 (1980).

\bibitem{1984PhLB..148...43L}
D.~{Lorenz-Petzold}.
\newblock \enquote{{Higher-dimensional cosmologies.}}
\newblock \emph{\plb} \textbf{148}, 43 (1984).

\bibitem{1984PhLB..149...79L}
D.~{Lorenz-Petzold}.
\newblock \enquote{{Kaluza-Klein-Bianchi-Kantowski-Sachs cosmologies.}}
\newblock \emph{\plb} \textbf{149}, 79 (1984).

\bibitem{1985PhLB..151..105L}
D.~{Lorenz-Petzold}.
\newblock \enquote{{Anisotropic supergravity cosmologies.}}
\newblock \emph{\plb} \textbf{151}, 105 (1985).

\bibitem{1985CQGra...2..829L}
D.~{Lorenz-Petzold}.
\newblock \enquote{{Bianchi-Kantowski-Sachs solutions of N = 1 supergravity in
  d = 11 dimensions for a torus space.}}
\newblock \emph{\cqg} \textbf{2}, 829 (1985).

\bibitem{1985PhRvD..31..929L}
D.~{Lorenz-Petzold}.
\newblock \enquote{{Higher-dimensional extensions of Bianchi type-I
  cosmologies}.}
\newblock \emph{\prd} \textbf{31}, 929 (1985).

\bibitem{1985PhLB..153..134L}
D.~{Lorenz-Petzold}.
\newblock \enquote{{Higher-dimensional perfect fluid cosmologies.}}
\newblock \emph{\plb} \textbf{153}, 134 (1985).

\bibitem{1985JApA....6..137L}
D.~{Lorenz-Petzold}.
\newblock \enquote{{Higher-dimensional vacuum Bianchi-Mixmaster cosmologies}.}
\newblock \emph{Journal of Astrophysics and Astronomy} \textbf{6}, 137 (1985).

\bibitem{1986PhLB..167..157L}
D.~{Lorenz-Petzold}.
\newblock \enquote{{Higher-dimensional Bianchi cosmologies.}}
\newblock \emph{\plb} \textbf{167}, 157 (1986).

\bibitem{1987JMP....28.1382L}
D.~{Lorenz-Petzold}.
\newblock \enquote{{Comment on the two ``new'' classes of Bianchi type II
  solutions}.}
\newblock \emph{Journal of Mathematical Physics} \textbf{28}, 1382 (1987).

\bibitem{1987PhRvD..35.3760L}
J.~{Louko}.
\newblock \enquote{{Fate of singularities in Bianchi type-III quantum
  cosmology}.}
\newblock \emph{\prd} \textbf{35}, 3760 (1987).

\bibitem{Scalarestringhe}
H.~L{\"u} and C.~N. Pope.
\newblock \enquote{p-brane solitons in maximal supergravities.}
\newblock \emph{\ncb} \textbf{465}, 127 (1996).
\newblock \eprint{arXiv.org/abs/hep-th/9512012}.

\bibitem{L74}
V.~N. Lukash.
\newblock \enquote{Gravitational waves that conserve the homogeneity of space.}
\newblock \emph{\sovJETP} \textbf{40}, 792 (1974).

\bibitem{LU03}
J.~P. Luminet, J.~R. Weeks, A.~Riazuelo, R.~Lehoucq and J.~P. Uzan.
\newblock \enquote{Dodecahedral space topology as an explanation for weak
  wide-angle temperature correlations in the cosmic microwave background.}
\newblock \emph{Nature} \textbf{425}, 593 (2003).
\newblock \eprint{arXiv.org/abs/astro-ph/0310253}.

\bibitem{LYAP07}
A.~M. Lyapunov.
\newblock \emph{Annals of Mathematical Studies} \textbf{17} (1907).

\bibitem{Maartens}
R.~Maartens.
\newblock \enquote{{Brane-World Gravity}.}
\newblock \emph{Living Reviews in Relativity} \textbf{7} (2004).
\newblock \eprint{arXiv.org/abs/gr-qc/0312059}.

\bibitem{1990GReGr..22..595M}
R.~{Maartens} and S.~D. {Maharaj}.
\newblock \enquote{{Collision-free gases in Bianchi space-times}.}
\newblock \emph{\grg} \textbf{22}, 595 (1990).

\bibitem{1992CQGra...9.1525M}
R.~{Maartens} and M.~F. {Wolfaardt}.
\newblock \enquote{{Some exact Bianchi solutions.}}
\newblock \emph{\cqg} \textbf{9}, 1525 (1992).

\bibitem{1971Natur.230..112M}
M.~A.~H. {MacCallum}.
\newblock \enquote{{Cosmology-Problems of the Mixmaster universe}.}
\newblock \emph{\nat} \textbf{230}, 112 (1971).

\bibitem{1972PhLA...40..385M}
M.~A.~H. {MacCallum}.
\newblock \enquote{{On ''diagonal'' Bianchi cosmologies.}}
\newblock \emph{\pla} \textbf{40}, 385 (1972).

\bibitem{MAC79}
M.~A.~H. MacCallum.
\newblock \emph{General relativity: an Einstein centenary survey} (Cambridge
  Univ. Press, Cambridge, 1979), s. w. hawking and w. israel edn.

\bibitem{1998JPhA...31.2031M}
A.~J. {Maciejewski} and M.~{Szydlowski}.
\newblock \enquote{{On the integrability of Bianchi cosmological models.}}
\newblock \emph{Journal of Physics A Mathematical General} \textbf{31}, 2031
  (1998).
\newblock \eprint{arXiv.org/abs/gr-qc/9702045}.

\bibitem{1996gr.qc.....7020M}
S.~{Major} and L.~{Smolin}.
\newblock \enquote{{Mixmaster quantum cosmology in terms of physical
  dynamics}.}  (1996).
\newblock \eprint{arXiv.org/abs/gr-qc/9607020}.

\bibitem{2002IJMPD..11..447M}
M.~K. {Mak}, T.~{Harko} and L.~Z. {Fang}.
\newblock \enquote{{Bianchi Type I Universes with Causal Bulk Viscous
  Cosmological Fluid}.}
\newblock \emph{International Journal of Modern Physics D} \textbf{11}, 447
  (2002).
\newblock \eprint{arXiv.org/abs/gr-qc/0110069}.

\bibitem{2004PhRvD..70h4040M}
J.~{Malecki}.
\newblock \enquote{{Inflationary quantum cosmology: General framework and exact
  Bianchi typeI solution}.}
\newblock \emph{\prd} \textbf{70}, 084040 (2004).
\newblock \eprint{arXiv.org/abs/gr-qc/0407114}.

\bibitem{1993PhRvD..48.3704M}
N.~{Manojlovi{\'c}} and G.~A. {Mena Marug{\'a}n}.
\newblock \enquote{{Nonperturbative canonical quantization of minisuperspace
  models: Bianchi types I and II}.}
\newblock \emph{\prd} \textbf{48}, 3704 (1993).
\newblock \eprint{arXiv.org/abs/gr-qc/9304041}.

\bibitem{1993CQGra..10..559M}
N.~{Manojlovi{\'c}} and A.~{Mikovi{\'c}}.
\newblock \enquote{{Canonical analysis of the Bianchi models in the Ashtekar
  formulation.}}
\newblock \emph{\cqg} \textbf{10}, 559 (1993).

\bibitem{2000JMP....41.4777M}
N.~{Manojlovi{\'c}} and A.~{Mikovi{\'c}}.
\newblock \enquote{{Belinskii-Zakharov formulation for Bianchi models and
  Painlev{\'e} III equation}.}
\newblock \emph{Journal of Mathematical Physics} \textbf{41}, 4777 (2000).
\newblock \eprint{arXiv.org/abs/math-ph/0002037}.

\bibitem{1995CQGra..12.1441M}
D.~{Marolf}.
\newblock \enquote{{Observables and a Hilbert space for Bianchi IX.}}
\newblock \emph{\cqg} \textbf{12}, 1441 (1995).
\newblock \eprint{arXiv.org/abs/gr-qc/9409049}.

\bibitem{1994GReGr..26..307M}
A.~{Mazumder}.
\newblock \enquote{{Solutions of LRS Bianchi I space-time filled with a perfect
  fluid}.}
\newblock \emph{General Relativity and Gravitation} \textbf{26}, 307 (1994).

\bibitem{2006MNRAS.369.1858M}
J.~D. {McEwen}, M.~P. {Hobson}, A.~N. {Lasenby} and D.~J. {Mortlock}.
\newblock \enquote{{Non-Gaussianity detections in the Bianchi $VII_{h}$
  corrected WMAP one-year data made with directional spherical wavelets}.}
\newblock \emph{\mnras} \textbf{369}, 1858 (2006).
\newblock \eprint{arXiv.org/abs/astro-ph/0510349}.

\bibitem{1991CQGra...8.1173M}
C.~B.~G. {McIntosh} and J.~D. {Steele}.
\newblock \enquote{{All vacuum Bianchi I metrics with a homothety.}}
\newblock \emph{Classical and Quantum Gravity} \textbf{8}, 1173 (1991).

\bibitem{Mercuri2}
S.~Mercuri and G.~Montani.
\newblock \enquote{Dualism between physical frames and time in quantum
  gravity.}
\newblock \emph{\mpla} \textbf{19}, 1519 (2004).
\newblock \eprint{arXiv.org/abs/gr-qc/0312077}.

\bibitem{Mercuri}
S.~Mercuri and G.~Montani.
\newblock \enquote{Revised canonical quantum gravity via the frame fixing.}
\newblock \emph{\ijmpd} \textbf{13}, 165 (2004).
\newblock \eprint{arXiv.org/abs/gr-qc/0310077}.

\bibitem{1994PhRvD..50.2431M}
P.~G. {Miedema}.
\newblock \enquote{{Density perturbations in Bianchi type-I universes}.}
\newblock \emph{\prd} \textbf{50}, 2431 (1994).

\bibitem{1992CQGra...9S.183M}
P.~G. {Miedema} and W.~A. {van Leeuwen}.
\newblock \enquote{{Perturbations in Bianchi I universes.}}
\newblock \emph{Classical and Quantum Gravity} \textbf{9}, 183 (1992).

\bibitem{1993PhRvD..47.3151M}
P.~G. {Miedema} and W.~A. {van Leeuwen}.
\newblock \enquote{{Cosmological perturbations in Bianchi type-I universes}.}
\newblock \emph{\prd} \textbf{47}, 3151 (1993).

\bibitem{1985A&A...151....7M}
E.~{Milaneschi} and R.~{Fabbri}.
\newblock \enquote{{Polarization of the microwave background radiation in
  Bianchi type-I cosmological models with a homogeneous magnetic field}.}
\newblock \emph{\aap} \textbf{151}, 7 (1985).

\bibitem{Misner1969PRL}
C.~W. Misner.
\newblock \enquote{Mixmaster universe.}
\newblock \emph{\prl} \textbf{22}, 1071 (1969).

\bibitem{Misner1969PR}
C.~W. Misner.
\newblock \enquote{Quantum cosmology i.}
\newblock \emph{Physical Review} \textbf{186}, 1319 (1969).

\bibitem{1994gr.qc.....5068M}
C.~W. {Misner}.
\newblock \enquote{{The Mixmaster cosmological metrics}.}  (1994).
\newblock \eprint{arXiv.org/abs/gr-qc/9405068}.

\bibitem{MTW}
C.~W. Misner, K.~S. Thorne and J.~A. Wheeler.
\newblock \emph{Gravitation} (New York, 1973).

\bibitem{2003Ap&SS.283...67M}
G.~{Mohanty} and B.~{Mishra}.
\newblock \enquote{{Scale invariant theory for Bianchi Type VIII and IX
  space-times with perfect fluid}.}
\newblock \emph{\apss} \textbf{283}, 67 (2003).

\bibitem{2003Ap&SS.288..523M}
G.~{Mohanty}, S.~K. {Sahu} and P.~K. {Sahoo}.
\newblock \enquote{{Massive scalar field in the Bianchi Type I space time}.}
\newblock \emph{\apss} \textbf{288}, 523 (2003).

\bibitem{1991PhRvD..44.2375M}
V.~{Moncrief} and M.~P. {Ryan}.
\newblock \enquote{{Amplitude-real-phase exact solutions for quantum mixmaster
  universes.}}
\newblock \emph{\prd} \textbf{44}, 2375 (1991).

\bibitem{Montani1995CQG}
G.~Montani.
\newblock \enquote{On the general behavior of the universe near the
  cosmological singularity.}
\newblock \emph{\cqg} \textbf{12}, 2505 (1995).

\bibitem{Montani1999CQG}
G.~Montani.
\newblock \enquote{On the quasi-isotropic solution in the presence of
  ultrarelativistic matter and a scalar field.}
\newblock \emph{\cqg} \textbf{16}, 723 (1999).

\bibitem{M00a}
G.~Montani.
\newblock In \emph{The Chaotic Universe}, \emph{Advanced Series in Astrophysics
  and Cosmology}, vol.~10, eds. V.~G. Gurzadyan and R.~Ruffini (Singapore,
  2000). World scientific edn.

\bibitem{Montani2000CQGultrarelativistic}
G.~Montani.
\newblock \enquote{On the generic cosmological solution in the presence of
  ultrarelativistic matter and a scalar field.}
\newblock \emph{\cqg} \textbf{17}, 2205 (2000).

\bibitem{Montani2000CQGquasiisotropic}
G.~Montani.
\newblock \enquote{Quasi-isotropic solution containing an electromagnetic and a
  real massless scalar field.}
\newblock \emph{\cqg} \textbf{17}, 2197 (2000).

\bibitem{Montani2001NCB}
G.~Montani.
\newblock \enquote{{Nonstationary corrections to the Mixmaster model invariant
  measure}.}
\newblock \emph{Nuovo Cimento B} \textbf{116} (2001).

\bibitem{Montani2002NPB}
G.~Montani.
\newblock \enquote{Canonical quantization of gravity without 'frozen
  formalism'.}
\newblock \emph{\ncb} \textbf{634}, 370 (2002).
\newblock \eprint{arXiv.org/abs/gr-qc/0205032}.

\bibitem{Montani2003IJMPD}
G.~Montani.
\newblock \enquote{Cosmological issues for revised canonical quantum gravity.}
\newblock \emph{\ijmpd} \textbf{12}, 1445 (2003).
\newblock \eprint{arXiv.org/abs/gr-qc/0307069}.

\bibitem{Montani2004IJMPDminisuperspace}
G.~Montani.
\newblock \enquote{Minisuperspace model for revised canonical quantum gravity.}
\newblock \emph{\ijmpd} \textbf{13}, 1703 (2004).
\newblock \eprint{arXiv.org/abs/gr-qc/0404105}.

\bibitem{Montani2004IJMPDcompactification}
G.~Montani.
\newblock \enquote{A scenario for the dimensional compactification in
  eleven-dimensional space-time.}
\newblock \emph{\ijmpd} \textbf{D13}, 1029 (2004).
\newblock \eprint{arXiv.org/abs/gr-qc/0404030}.

\bibitem{MPVb}
J.~Moser and E.~P. ans S.~Varadhan.
\newblock \emph{Dynamical System, Theory and Applications}, vol.~38
  (Springer-Verlag, Berlin, 1975).

\bibitem{cit13barrow}
J.~Moser, E.~Phillips and S.~Varadhan.
\newblock \emph{{Ergodic theory}} (Courant Institute of Mathematical Sciences,
  New York University, New York, 1975).

\bibitem{Motter2003PRL}
A.~E. Motter.
\newblock \enquote{Relativistic chaos is coordinate invariant.}
\newblock \emph{\prl} \textbf{91}, 231101 (2003).
\newblock \eprint{arXiv.org/abs/gr-qc/0305020}.

\bibitem{MotterLetelier2001PLA}
A.~E. Motter and P.~S. Letelier.
\newblock \enquote{Mixmaster chaos.}
\newblock \emph{\pla} \textbf{285}, 127 (2001).
\newblock \eprint{arXiv.org/abs/gr-qc/0011001}.

\bibitem{1983JApA....4..295M}
G.~{Mukherjee}.
\newblock \enquote{{A Bianchi type I tilted universe}.}
\newblock \emph{Journal of Astrophysics and Astronomy} \textbf{4}, 295 (1983).

\bibitem{1977PThPh..58..842N}
H.~{Nariai}.
\newblock \enquote{{On a Quantized Scalar Field in Some Bianchi-Type I
  Universe. II ---DeWitt's Two Vacuum States Connected Causally---}.}
\newblock \emph{\ptp} \textbf{58}, 842 (1977).

\bibitem{1977PThPh..57...67N}
H.~{Nariai}.
\newblock \enquote{{Propagators for a Scalar Field in Some Bianchi-Type I
  Universe}.}
\newblock \emph{Progress of Theoretical Physics} \textbf{57}, 67 (1977).

\bibitem{1989GReGr..21..211N}
B.~K. {Nayak} and B.~K. {Sahoo}.
\newblock \enquote{{Bianchi Type V models with a matter distribution admitting
  anisotropic pressure and heat flow}.}
\newblock \emph{\grg} \textbf{21}, 211 (1989).

\bibitem{BOOM}
C.~B. Netterfield \emph{et~al.}
\newblock \enquote{A measurement by boomerang of multiple peaks in the angular
  power spectrum of the cosmic microwave background.}
\newblock \emph{\apj} \textbf{571}, 604 (2002).
\newblock \eprint{arXiv.org/abs/astro-ph/0104460}.

\bibitem{Nic}
H.~Nicolai, K.~Peeters and M.~Zamaklar.
\newblock \enquote{Loop quantum gravity: An outside view.}
\newblock \emph{\cqg} \textbf{22}, R193 (2005).
\newblock \eprint{arXiv.org/abs/hep-th/0501114}.

\bibitem{1997JMP....38.2611N}
U.~S. {Nilsson} and C.~{Uggla}.
\newblock \enquote{{Stationary Bianchi type II perfect fluid models}.}
\newblock \emph{Journal of Mathematical Physics} \textbf{38}, 2611 (1997).
\newblock \eprint{arXiv.org/abs/gr-qc/9702039}.

\bibitem{2000GReGr..32.1319N}
U.~S. {Nilsson}, C.~{Uggla} and J.~{Wainwright}.
\newblock \enquote{{A Dynamical Systems Approach to Geodesics in Bianchi
  Cosmologies}.}
\newblock \emph{\grg} \textbf{32}, 1319 (2000).
\newblock \eprint{arXiv.org/abs/gr-qc/9908062}.

\bibitem{2000GReGr..32.1981N}
U.~S. {Nilsson}, C.~{Uggla} and J.~{Wainwright}.
\newblock \enquote{{A Dynamical Systems Approach to Geodesics in Bianchi
  Cosmologies}.}
\newblock \emph{\grg} \textbf{32}, 1981 (2000).

\bibitem{1993PhRvD..48.5642O}
O.~{Obreg{\'o}n}, J.~{Pullin} and M.~P. {Ryan}.
\newblock \enquote{{Bianchi cosmologies: New variables and a hidden
  supersymmetry}.}
\newblock \emph{\prd} \textbf{48}, 5642 (1993).
\newblock \eprint{arXiv.org/abs/gr-qc/9308001}.

\bibitem{1981JMaPh..22..623O}
O.~{Obregon} and M.~P. {Ryan}, Jr.
\newblock \enquote{{Bianchi type IX cosmological models with homogeneous spinor
  fields.}}
\newblock \emph{Journal of Mathematics and Physics} \textbf{22}, 623 (1981).

\bibitem{1996IJTP...35.1381O}
O.~{Obregon} and J.~{Socorro}.
\newblock \enquote{{ $ \Psi= W e^{ \Phi}$ quantum cosmological solutions for
  class A Bianchi models.}}
\newblock \emph{\ijtp} \textbf{35}, 1381 (1996).
\newblock \eprint{arXiv.org/abs/gr-qc/9506021}.

\bibitem{O01}
D.~Oriti.
\newblock \enquote{Spacetime geometry from algebra: spin foam models for
  non-perturbative quantum gravity.}
\newblock \emph{Reports of Progress in Physics} \textbf{64}, 1489 (2001).
\newblock \eprint{arXiv.org/abs/gr-qc/0106091}.

\bibitem{2007gr.qc.....3101O}
C.~{Ortiz}, E.~{Mena-Barboza}, M.~{Sabido} and J.~{Socorro}.
\newblock \enquote{{(Non)commutative isotropization in Bianchi I with
  Barotropic perfect fluid and \$$\backslash$Lambda\$ Cosmological}.}  (2007).
\newblock \eprint{gr-qc/0703101}.

\bibitem{Ott}
E.~Ott.
\newblock \emph{Chaos in dynamical systems} (Cambridge Univ. Press, Cambridge,
  1993).

\bibitem{1982GReGr..14..549P}
T.~{Padmanabhan}.
\newblock \enquote{{Quantum stationary states in the Bianchi universes.}}
\newblock \emph{\grg} \textbf{14}, 549 (1982).

\bibitem{Palm}
E.~Palm.
\newblock \enquote{{On the zeros of Bessel functions of pure imaginary order}.}
\newblock \emph{Quarterly Journal of Mechanics and Applied Mathematics}
  \textbf{10}, 500 (1957).

\bibitem{0264-9381-7-4-008}
S.~L. Parnovsky.
\newblock \enquote{A general solution of gravitational equations near their
  singularities.}
\newblock \emph{Classical and Quantum Gravity} \textbf{7}, 571 (1990).

\bibitem{1989IJTP...28..415P}
A.~C. {Patra}, M.~N. {Sinha Roy} and D.~{Ray}.
\newblock \enquote{{Quantum stationary state of class A Bianchi universe.}}
\newblock \emph{\ijtp} \textbf{28}, 415 (1989).

\bibitem{Penrose1964}
R.~Penrose.
\newblock \enquote{Gravitational collapse and space-time singularities.}
\newblock \emph{\prl} \textbf{14}, 57 (1965).

\bibitem{Per}
A.~Perez.
\newblock \enquote{Spin foam models for quantum gravity.}
\newblock \emph{\cqg} \textbf{20}, R43 (2003).
\newblock \eprint{arXiv.org/abs/gr-qc/0301113}.

\bibitem{PE77}
Y.~B. Pesin.
\newblock \enquote{Lyapunov characteristic numbers and smooth ergodic theory.}
\newblock \emph{UMN (Russian Mathematical Surveys)} \textbf{32}, 55 (1977).

\bibitem{PY79}
G.~Pianigiani and J.~Yorke.
\newblock \enquote{Expanding maps on sets which are almost invariant: decay and
  chaos.}
\newblock \emph{Transactions of the American Mathematical Society}
  \textbf{252}, 351 (1979).

\bibitem{Podolsky:2007vu}
D.~Podolsky.
\newblock \enquote{{General asymptotic solutions of the Einstein equations and
  phase transitions in quantum gravity}.}  (2007).
\newblock \eprint{arXiv:0704.0354 [hep-th]}.

\bibitem{1998PhRvD..58b4001P}
J.~M. {Pons} and L.~C. {Shepley}.
\newblock \enquote{{Dimensional reduction and gauge group reduction in
  Bianchi-type cosmology}.}
\newblock \emph{\prd} \textbf{58}, 024001 (1998).
\newblock \eprint{arXiv.org/abs/gr-qc/9805030}.

\bibitem{2007MNRAS.380.1387P}
A.~{Pontzen} and A.~{Challinor}.
\newblock \enquote{{Bianchi model CMB polarization and its implications for CMB
  anomalies}.}
\newblock \emph{\mnras} \textbf{380}, 1387 (2007).
\newblock \eprint{arXiv.org/abs/0706.2075}.

\bibitem{2006Ap&SS.301..127P}
A.~{Pradhan} and P.~{Pandey}.
\newblock \enquote{{Some Bianchi Type I Viscous Fluid Cosmological Models with
  a Variable Cosmological Constant}.}
\newblock \emph{\apss} \textbf{301}, 127 (2006).

\bibitem{2005Ap&SS.298..419P}
A.~{Pradhan}, S.~K. {Srivastav} and M.~K. {Yadav}.
\newblock \enquote{{Some Homogeneous Bianchi type IX Viscous Fluid Cosmological
  Models with a Varying {$\Lambda$}}.}
\newblock \emph{\apss} \textbf{298}, 419 (2005).
\newblock \eprint{arXiv.org/abs/gr-qc/0307039}.

\bibitem{1985PhLA..107...33P}
S.~A. {Pritomanov}.
\newblock \enquote{{Quantum effects in 'mixmaster universe'}.}
\newblock \emph{\pla} \textbf{107}, 33 (1985).

\bibitem{DASI2}
C.~Pryke and oth.
\newblock \enquote{Cosmological parameter extraction from the first season of
  observations with dasi.}
\newblock \emph{\apj} \textbf{568}, 46  (2002).
\newblock \eprint{arXiv.org/abs/astro-ph/0104490}.

\bibitem{PU91}
J.~Pullin.
\newblock \enquote{Relativity and gravitation: Classical and quantum.}
\newblock In \emph{Proceedings of the VIIth SILARG Symposium}, ed. D.~et~al.
  (1991).

\bibitem{Puzio1994CQG}
R.~Puzio.
\newblock \enquote{On the square root of the laplace-beltrami operator as a
  hamiltonian.}
\newblock \emph{\cqg} \textbf{11}, 609 (1994).

\bibitem{1988JMaPh..29..449R}
S.~{Ram}.
\newblock \enquote{{Bianchi type $VI_{0}$ space-times with perfect fluid
  source.}}
\newblock \emph{Journal of Mathematics and Physics} \textbf{29}, 449 (1988).

\bibitem{1989JMaPh..30..757R}
S.~{Ram}.
\newblock \enquote{{Spatially homogeneous cosmological models of Bianchi type
  III.}}
\newblock \emph{Journal of Mathematics and Physics} \textbf{30}, 757 (1989).

\bibitem{1987Ap&SS.136...17R}
D.~R.~K. {Reddy} and R.~{Venkateswarlu}.
\newblock \enquote{{Bianchi type-I universe in the presence of zero-mass scalar
  fields}.}
\newblock \emph{\apss} \textbf{136}, 17 (1987).

\bibitem{1996JMP....37..438R}
A.~D. {Rendall}.
\newblock \enquote{{The initial singularity in solutions of the Einstein-Vlasov
  system of Bianchi type I}.}
\newblock \emph{Journal of Mathematical Physics} \textbf{37}, 438 (1996).
\newblock \eprint{arXiv.org/abs/gr-qc/9505017}.

\bibitem{1997CQGra..14.2341R}
A.~D. {Rendall}.
\newblock \enquote{{Global dynamics of the mixmaster model.}}
\newblock \emph{\cqg} \textbf{14}, 2341 (1997).
\newblock \eprint{arXiv.org/abs/gr-qc/9703036}.

\bibitem{2000gr.qc.....6035R}
H.~{Ringstrom}.
\newblock \enquote{{The Bianchi IX attractor}.}
\newblock \emph{Annales Henri Poincare} \textbf{2}, 405 (2001).
\newblock \eprint{arXiv.org/abs/gr-qc/0006035}.

\bibitem{1993PhRvD..47.1396R}
V.~{Romano} and D.~{Pav{\'o}n}.
\newblock \enquote{{Causal dissipative Bianchi cosmology}.}
\newblock \emph{\prd} \textbf{47}, 1396 (1993).

\bibitem{1984CQGra...1...81R}
K.~{Rosquist}.
\newblock \enquote{{Regularised field equations for Bianchi type VI spatially
  homogeneous cosmology}.}
\newblock \emph{\cqg} \textbf{1}, 81 (1984).

\bibitem{1990CQGra...7..625R}
K.~{Rosquist}, C.~{Uggla} and R.~T. {Jantzen}.
\newblock \enquote{{Extended dynamics and symmetries in perfect fluid Bianchi
  cosmologies.}}
\newblock \emph{\cqg} \textbf{7}, 625 (1990).

\bibitem{1990CQGra...7..611R}
K.~{Rosquist}, C.~{Uggla} and R.~T. {Jantzen}.
\newblock \enquote{{Extended dynamics and symmetries in vacuum Bianchi
  cosmologies.}}
\newblock \emph{\cqg} \textbf{7}, 611 (1990).

\bibitem{1989grg..conf..351R}
K.~{Rosquist}, C.~{Uggla}, H.~{von Zur-M{\"u}hlen} and R.~T. {Jantzen}.
\newblock \enquote{{Geometrized Field Equations for Bianchi Cosmology}.}
\newblock In \emph{\grg} (1989), p. 351.

\bibitem{1985PhLB..159..256R}
T.~{Rothman} and M.~S. {Madsen}.
\newblock \enquote{{Bianchi I inflation: assumptions and inconsistencies.}}
\newblock \emph{\plb} \textbf{159}, 256 (1985).

\bibitem{R91b}
C.~Rovelli.
\newblock \enquote{Time in quantum gravity: An hypothesis.}
\newblock \emph{\prd} \textbf{43}, 442 (1991).

\bibitem{R91a}
C.~Rovelli.
\newblock \enquote{What is observable in classical and quantum gravity?}
\newblock \emph{\cqg} \textbf{8}, 297 (1991).

\bibitem{Ro97}
C.~Rovelli.
\newblock \enquote{Loop quantum gravity.}
\newblock \emph{Living Reviews in Relativity} \textbf{1}, 1 (1998).
\newblock \eprint{arXiv.org/abs/gr-qc/9710008}.

\bibitem{rovelli2004qg}
C.~Rovelli.
\newblock \emph{Quantum Gravity} (Cambridge University Press, 2004).

\bibitem{RovThi}
C.~Rovelli and T.~Thiemann.
\newblock \enquote{The immirzi parameter in quantum general relativity.}
\newblock \emph{\prd} \textbf{57}, 1009 (1998).
\newblock \eprint{arXiv.org/abs/gr-qc/9705059}.

\bibitem{1997CQGra..14.2845R}
S.~R. {Roy} and S.~K. {Banerjee}.
\newblock \enquote{{A Bianchi type II cosmological model of Petrov type D
  representing an imperfect fluid with a source-free magnetic field.}}
\newblock \emph{\cqg} \textbf{14}, 2845 (1997).

\bibitem{1985Ap&SS.108..195R}
S.~R. {Roy} and S.~{Narain}.
\newblock \enquote{{Inhomogeneous generalisation of Bianchi type-VI(0)
  cosmological model of perfect fluid distribution in general relativity}.}
\newblock \emph{\apss} \textbf{108}, 195 (1985).

\bibitem{1986Ap&SS.128..447R}
S.~R. {Roy}, S.~{Narain} and J.~P. {Singh}.
\newblock \enquote{{Comments on homogeneous and inhomogeneous Bianchi
  type-$VI_{0}$ solutions}.}
\newblock \emph{\apss} \textbf{128}, 447 (1986).

\bibitem{1991Ap&SS.181...61R}
S.~R. {Roy} and A.~{Prasad}.
\newblock \enquote{{Inhomogeneous generalizations of Bianchi type VIh models
  with perfect fluid}.}
\newblock \emph{\apss} \textbf{181}, 61 (1991).

\bibitem{1995IJPAM..26..169R}
S.~R. {Roy} and A.~{Prasad}.
\newblock \enquote{{Some inhomogeneous cosmological models conformal to Bianchi
  type V universes in general relativity.}}
\newblock \emph{Indian Journal of Pure Applied Mathematics} \textbf{26}, 169
  (1995).

\bibitem{1982AZh....59.1044R}
V.~A. {Ruban}.
\newblock \enquote{{The dynamical role of a primordial electromagnetic field in
  spatially homogeneous, diagonal Bianchi type I-IX cosmologies}.}
\newblock \emph{\azh} \textbf{59}, 1044 (1982).

\bibitem{RU94}
S.~E. Rugh.
\newblock In \emph{Deterministic Chaos in General Relativity}, eds. D.~Hobill,
  A.~Burd and A.~Coley (Singapore, 1994). World scientific edn.

\bibitem{1990PhLA..147..353R}
S.~E. {Rugh} and B.~J.~T. {Jones}.
\newblock \enquote{{Chaotic behaviour and oscillating three-volumes in Bianchi
  IX universes.}}
\newblock \emph{\pla} \textbf{147}, 353 (1990).

\bibitem{1997gr.qc.....9012R}
M.~P. {Ryan}, {Jr.} and S.~M. {Waller}.
\newblock \enquote{{On the Hamiltonian Formulation of Class B Bianchi
  Cosmological Models}.}  (1997).
\newblock \eprint{arXiv.org/abs/gr-qc/9709012}.

\bibitem{Ry-Sh}
M.~P. Ryan and L.~C. Shepley.
\newblock \emph{Homogeneous Relativistic Cosmologies} (Princeton University
  Press, 1975).

\bibitem{1971AnPhy..65..506R}
M.~P. {Ryan}, Jr.
\newblock \enquote{{Qualitative cosmology: Diagrammatic solutions for Bianchi
  type IX universes with expansion, rotation, and shear. I. The symmetric
  case.}}
\newblock \emph{Annals of Physics} \textbf{65}, 506 (1971).

\bibitem{1971AnPhy..65..541R}
M.~P. {Ryan}, Jr.
\newblock \enquote{{Qualitative cosmology: Diagrammatic solutions for Bianchi
  type IX universes with expansion, rotation, and shear. II. The general
  case.}}
\newblock \emph{Annals of Physics} \textbf{65}, 541 (1971).

\bibitem{1992PhLB..287...61S}
P.~M. {S{\'a}} and A.~B. {Henriques}.
\newblock \enquote{{The dynamics of an inflationary Bianchi-IX model in the
  presence of a rotating fluid.}}
\newblock \emph{\plb} \textbf{287}, 61 (1992).

\bibitem{1999PhLB..465..101S}
P.~M. {Saffin}.
\newblock \enquote{{Tunnelling with Bianchi IX instantons}.}
\newblock \emph{\plb} \textbf{465}, 101 (1999).
\newblock \eprint{arXiv.org/abs/gr-qc/9906001}.

\bibitem{1981PhRvD..24..305S}
A.~{Sagnotti} and B.~{Zwiebach}.
\newblock \enquote{{Electromagnetic waves in a Bianchi type-I Universe.}}
\newblock \emph{\prd} \textbf{24}, 305 (1981).

\bibitem{2005MPLA...20.2127S}
B.~{Saha}.
\newblock \enquote{{Bianchi Type i Universe with Viscous Fluid}.}
\newblock \emph{\mpla} \textbf{20}, 2127 (2005).
\newblock \eprint{arXiv.org/abs/gr-qc/0409104}.

\bibitem{2006PhyD..219..168S}
B.~{Saha} and V.~{Rikhvitsky}.
\newblock \enquote{{Bianchi type I universe with viscous fluid and a
  {$\Lambda$} term: A qualitative analysis}.}
\newblock \emph{Physica D} \textbf{219}, 168 (2006).
\newblock \eprint{arXiv.org/abs/gr-qc/0410056}.

\bibitem{1983NCimB..77...62S}
P.~{Salucci} and R.~{Fabbri}.
\newblock \enquote{{The cosmological evolution of general Bianchi models in the
  adiabatic regime}.}
\newblock \emph{Nuovo Cimento B Serie} \textbf{77}, 62 (1983).

\bibitem{samuel}
J.~Samuel.
\newblock \enquote{Is barbero's hamiltonian formulation a gauge theory of
  lorentzian gravity?}
\newblock \emph{\cqg} \textbf{17}, L141 (2000).
\newblock \eprint{arXiv.org/abs/gr-qc/0005095}.

\bibitem{1999IJMPA..14.4473S}
V.~A. {Savchenko}, T.~P. {Shestakova} and G.~M. {Vereshkov}.
\newblock \enquote{{Quantum Geometrodynamics of the Bianchi-Ix Model in
  Extended Phase Space}.}
\newblock \emph{\ijmpa} \textbf{14}, 4473 (1999).
\newblock \eprint{arXiv.org/abs/gr-qc/9909047}.

\bibitem{1997PhLB..399..207S}
C.~{Scheen} and J.~{Demaret}.
\newblock \enquote{{Analytic structure and chaos in Bianchi type-IX
  relativistic dynamics.}}
\newblock \emph{\plb} \textbf{399}, 207 (1997).

\bibitem{1995CQGra..12.1099S}
J.~{Schirmer}.
\newblock \enquote{{Hamiltonian reduction of Bianchi cosmologies}.}
\newblock \emph{\cqg} \textbf{12}, 1099 (1995).
\newblock \eprint{arXiv.org/abs/gr-qc/9408008}.

\bibitem{1982AN....303..227S}
H.-J. {Schmidt}.
\newblock \enquote{{Inhomogeneous cosmological models containing homogeneous
  inner hypersurface geometry - Changes of the Bianchi type}.}
\newblock \emph{Astronomische Nachrichten} \textbf{303}, 227 (1982).

\bibitem{1980A&A....87..236S}
J.~H.~M.~M. {Schmitt}.
\newblock \enquote{{The anisotropic microwave background in Bianchi V models}.}
\newblock \emph{\aap} \textbf{87}, 236 (1980).

\bibitem{STN02}
J.~Schneider and Z.~Neufeld.
\newblock \enquote{Dynamics of "leaking" hamiltonian systems.}
\newblock \emph{\pre} \textbf{66}, 066218 (2002).

\bibitem{ST87}
J.~Silk and M.~S. Turner.
\newblock \enquote{Double inflation.}
\newblock \emph{\prd} \textbf{35}, 419 (1987).

\bibitem{1995Ap&SS.225...57S}
J.~K. {Singh} and S.~{Ram}.
\newblock \enquote{{A class of new Bianchi 1 perfect fluid space-times}.}
\newblock \emph{\apss} \textbf{225}, 57 (1995).

\bibitem{1982GReGr..14..751S}
H.~{Sirousse-Zia}.
\newblock \enquote{{Fluctuations produced by the cosmological constant in the
  empty Bianchi type-IX model}.}
\newblock \emph{\grg} \textbf{14}, 751 (1982).

\bibitem{1983ApJ...268..513S}
R.~J. {Slagter}.
\newblock \enquote{{Numerical solutions of high-frequency perturbations in
  Bianchi type IX models}.}
\newblock \emph{\apj} \textbf{268}, 513 (1983).

\bibitem{1984ApJ...286..379S}
R.~J. {Slagter}.
\newblock \enquote{{Behavior of higher modes of gravitational waves and
  gauge-invariant density perturbations in Bianchi IX cosmological models}.}
\newblock \emph{\apj} \textbf{286}, 379 (1984).

\bibitem{Smo}
L.~Smolin.
\newblock \enquote{An invitation to loop quantum gravity.}  (2004).
\newblock \eprint{arXiv.org/abs/gr-qc/0408048}.

\bibitem{2006PhRvD..73f9901S}
I.~D. {Soares} and T.~J. {Stuchi}.
\newblock \enquote{{Erratum: Homoclinic chaos in axisymmetric Bianchi IX
  cosmologies reexamined [Phys. Rev. D 72, 083516 (2005)]}.}
\newblock \emph{\prd} \textbf{73}, 069901 (2006).

\bibitem{Spergel:2003cb}
D.~N. Spergel \emph{et~al.}
\newblock \enquote{First year wilkinson microwave anisotropy probe (wmap)
  observations: Determination of cosmological parameters.}
\newblock \emph{\apj Supplement} \textbf{148}, 175 (2003).
\newblock \eprint{arXiv.org/abs/astro-ph/0302209}.

\bibitem{Stanco}
F.~Stancu.
\newblock \emph{Group Theory in Subnuclear Physics} (Calendon Press, Oxford,
  1996).

\bibitem{STA79}
A.~A. Starobinsky.
\newblock \enquote{{Relict gravitation radiation spectrum and initial state of
  the universe}.}
\newblock \emph{\JETPl} \textbf{30}, 682 (1979).

\bibitem{Star83}
A.~A. Starobinsky.
\newblock \enquote{Isotropization of arbitrary cosmological expansion given an
  effective cosmological constant.}
\newblock \emph{\pzetf} \textbf{37}, 55 (1983).

\bibitem{MacCallexsol}
H.~Stephani, D.~Kramer, M.~A.~H. MacCallum, C.~Hoenselaers and E.~Herlt.
\newblock \emph{{Exact solutions of Einstein's field equations}} (Cambridge,
  UK, 2003).

\bibitem{SZ93}
M.~Szydlowski.
\newblock \enquote{Toward an invariant measure of chaotic behavior in general
  relativity.}
\newblock \emph{\pla} \textbf{176}, 22 (1993).

\bibitem{1996GrCos...2...92S}
M.~{Szydlowski}.
\newblock \enquote{{Hidden chaos in the Bianchi IX cosmology.}}
\newblock \emph{Gravitation and Cosmology} \textbf{2}, 92 (1996).

\bibitem{1997GReGr..29..185S}
M.~{Szydlowski}.
\newblock \enquote{{Chaos Hidden Behind Time Parametrization in the Mixmaster
  Cosmology}.}
\newblock \emph{\grg} \textbf{29}, 185 (1997).

\bibitem{1991PhRvD..44.2369S}
M.~{Szydlowski} and M.~{Biesiada}.
\newblock \enquote{{Chaos in mixmaster models.}}
\newblock \emph{\prd} \textbf{44}, 2369 (1991).

\bibitem{1987PhLB..195...27S}
M.~{Szydlowski}, M.~{Biesiada} and J.~{Szczesny}.
\newblock \enquote{{Multidimensional mixmaster models.}}
\newblock \emph{\plb} \textbf{195}, 27 (1987).

\bibitem{1994AcC....20...17S}
M.~{Szydlowski} and M.~{Heller}.
\newblock \enquote{{Lyapunov Characteristic Timescale in the Mixmaster World
  Models}.}
\newblock \emph{Acta Cosmologica} \textbf{20}, 17 (1994).

\bibitem{SK93}
M.~Szydlowski and A.~Krawiec.
\newblock \enquote{Average rate of separation of trajectories near the
  singularity in mixmaster models.}
\newblock \emph{\prd} \textbf{47}, 5323 (1993).

\bibitem{SL90}
M.~Szydlowski and A.~Lapeta.
\newblock \enquote{Pseudo-riemannian manifold of mixmaster dynamical systems.}
\newblock \emph{\pla} \textbf{148}, 239 (1990).

\bibitem{SS94}
M.~Szydlowski and J.~Szczesny.
\newblock \enquote{Invariant chaos in mixmaster cosmology.}
\newblock \emph{\prd} \textbf{50} (1994).

\bibitem{1979IJTP...18..371T}
G.~E. {Tauber}.
\newblock \enquote{{Classification of Bianchi cosmologies in conformal flat
  spacetimes.}}
\newblock \emph{\ijtp} \textbf{18}, 371 (1979).

\bibitem{Terras}
A.~Terras.
\newblock \emph{Harmonic analysis on symmetric spaces and applications I}
  (Springer-Verlag, New York, 1985).

\bibitem{Thi}
T.~Thiemann.
\newblock \emph{Modern Canonical General Relativity} (Cambridge University
  Press, 2005).

\bibitem{Thirev}
T.~Thiemann.
\newblock \enquote{Loop quantum gravity: An inside view.}  (2006).
\newblock \eprint{arXiv.org/abs/hep-th/0608210}.

\bibitem{Thimascon}
T.~Thiemann.
\newblock \enquote{Quantum spin dynamics. viii: The master constraint.}
\newblock \emph{\cqg} \textbf{23}, 2249 (2006).
\newblock \eprint{arXiv.org/abs/gr-qc/0510011}.

\bibitem{2007arXiv0710.5723T}
P.~{Tod} and C.~{L{\"u}bbe}.
\newblock \enquote{{A global conformal extension theorem for perfect fluid
  Bianchi space-times}.}  (2007).
\newblock \eprint{arXiv.org/abs/0710.5723}.

\bibitem{DT94}
K.~Tomita and N.~Deruelle.
\newblock \enquote{Nonlinear behaviors of cosmological inhomogeneities with a
  standard fluid and inflationary matter.}
\newblock \emph{\prd} \textbf{50}, 7216 (1994).

\bibitem{2000CQGra..17.2215T}
C.~G. {Tsagas} and R.~{Maartens}.
\newblock \enquote{{Cosmological perturbations on a magnetized Bianchi I
  background.}}
\newblock \emph{\cqg} \textbf{17}, 2215 (2000).
\newblock \eprint{arXiv.org/abs/gr-qc/9912044}.

\bibitem{1988PhLB..207...36T}
P.~{Turkowski}.
\newblock \enquote{{A multidimensional extension of the Bianchi type-IX
  cosmology.}}
\newblock \emph{\plb} \textbf{207}, 36 (1988).

\bibitem{1990PhRvD..42..404U}
C.~{Uggla}, K.~{Rosquist} and R.~T. {Jantzen}.
\newblock \enquote{{Geometrizing the dynamics of Bianchi cosmology.}}
\newblock \emph{\prd} \textbf{42}, 404 (1990).

\bibitem{Uggla2}
C.~Uggla, H.~van Elst, J.~Wainwright and G.~F.~R. Ellis.
\newblock \enquote{{The past attractor in inhomogeneous cosmology}.}
\newblock \emph{\prd} \textbf{68}, 103502 (2003).
\newblock \eprint{arXiv.org/abs/gr-qc/0304002}.

\bibitem{1990CQGra...7.1365U}
C.~{Uggla} and H.~{von Zur-M{\"u}hlen}.
\newblock \enquote{{Compactified and reduced dynamics for locally rotationally
  symmetric Bianchi type IX perfect fluid models.}}
\newblock \emph{\cqg} \textbf{7}, 1365 (1990).

\bibitem{Un-Wa89}
W.~G. Unruh and R.~M. Wald.
\newblock \enquote{Time and the interpretation of canonical quantum gravity.}
\newblock \emph{\prd} \textbf{40}, 2598 (1989).

\bibitem{2007CQGra..24..931V}
B.~{Vakili}, N.~{Khosravi} and H.~R. {Sepangi}.
\newblock \enquote{{Bianchi spacetimes in noncommutative phase space}.}
\newblock \emph{\cqg} \textbf{24}, 931 (2007).
\newblock \eprint{arXiv.org/abs/gr-qc/0701075}.

\bibitem{2007PhLB..651...79V}
B.~{Vakili} and H.~R. {Sepangi}.
\newblock \enquote{{Generalized uncertainty principle in Bianchi type I quantum
  cosmology}.}
\newblock \emph{\plb} \textbf{651}, 79 (2007).
\newblock \eprint{arXiv.org/abs/0706.0273}.

\bibitem{brane2}
R.~J. van~den Hoogen, A.~A. Coley and Y.~He.
\newblock \enquote{Bianchi ix brane-world cosmologies.}
\newblock \emph{\prd} \textbf{68}, 023502 (2003).
\newblock \eprint{arXiv.org/abs/gr-qc/0212094}.

\bibitem{1999PhRvD..59j7302V}
R.~J. {van den Hoogen} and I.~{Olasagasti}.
\newblock \enquote{{Isotropization of scalar field Bianchi type-IX models with
  an exponential potential}.}
\newblock \emph{\prd} \textbf{59}, 107302 (1999).

\bibitem{1989GReGr..21..413V}
W.~A. {van Leeuwen}, P.~G. {Miedema} and S.~H. {Wiersma}.
\newblock \enquote{{Viscous Bianchi universes with a magnetic field. I. Effect
  of magnetoviscosity on the metric}.}
\newblock \emph{\grg} \textbf{21}, 413 (1989).

\bibitem{Vilkenin88}
A.~Vilenkin.
\newblock \enquote{Quantum cosmology and the initial state of the universe.}
\newblock \emph{\prd} \textbf{37}, 888 (1988).

\bibitem{Vilkenin89}
A.~Vilenkin.
\newblock \enquote{Interpretation of the wave function of the universe.}
\newblock \emph{\prd} \textbf{39}, 1116 (1989).

\bibitem{2006PhRvD..74f3508V}
S.~D.~P. {Vitenti} and D.~{M{\"u}ller}.
\newblock \enquote{{Numerical Bianchi typeI solutions in semiclassical
  gravitation}.}
\newblock \emph{\prd} \textbf{74}, 063508 (2006).
\newblock \eprint{arXiv.org/abs/gr-qc/0604127}.

\bibitem{1998CQGra..15..331W}
J.~{Wainwright}, A.~A. {Coley}, G.~F.~R. {Ellis} and M.~{Hancock}.
\newblock \enquote{{On the isotropy of the universe: do Bianchi $VII_{h}$
  cosmologies isotropize?}}
\newblock \emph{\cqg} \textbf{15}, 331 (1998).

\bibitem{wald}
R.~M. Wald.
\newblock \emph{General Relativity} (University of Chicago Press, Chicago,
  1984).

\bibitem{WeaverIsenbergBerger1998PRL}
M.~Weaver, J.~Isenberg and B.~K. Berger.
\newblock \enquote{Mixmaster behavior in inhomogeneous cosmological
  spacetimes.}
\newblock \emph{\prl} \textbf{80}, 2984 (1998).
\newblock \eprint{arXiv.org/abs/gr-qc/9712055}.

\bibitem{1987JMaPh..28.1658W}
E.~{Weber}.
\newblock \enquote{{Global qualitative study of Bianchi universes in the
  presence of a cosmological constant.}}
\newblock \emph{Journal of Mathematics and Physics} \textbf{28}, 1658 (1987).

\bibitem{witten1997dsa}
E.~Witten.
\newblock \enquote{{Duality, spacetime and quantum mechanics}.}
\newblock \emph{Physics Today} \textbf{50}, 28 (1997).

\bibitem{Woj}
M.~Wojtkowski.
\newblock \enquote{{Invariant families of cones and Lyapunov exponents}.}
\newblock \emph{Ergodic Theory and Dynamical Systems} \textbf{5}, 145 (1985).

\bibitem{1990PhRvD..41.2444X}
L.~{Xuefeng} and R.~M. {Wald}.
\newblock \enquote{{Proof of the closed-universe recollapse conjecture for
  general Bianchi type IX cosmologies.}}
\newblock \emph{\prd} \textbf{41}, 2444 (1990).

\bibitem{2001PThPh.106..323Y}
H.~{Yamazaki} and T.~{Hara}.
\newblock \enquote{{Dirac Decomposition of Wheeler-DeWitt Equation in the
  Bianchi Class A Models}.}
\newblock \emph{\ptp} \textbf{106}, 323 (2001).

\bibitem{1990PhRvD..41.1047Y}
J.~{Yokoyama} and K.-I. {Maeda}.
\newblock \enquote{{Quantum cosmological approach to the cosmic no-hair
  conjecture in the Bianchi type IX spacetime.}}
\newblock \emph{\prd} \textbf{41}, 1047 (1990).

\bibitem{Y72}
J.~W.~J. York.
\newblock \enquote{Role of conformal three-geometry in the dynamics of
  gravitation.}
\newblock \emph{\prl} \textbf{28}, 1082 (1972).

\bibitem{Z83}
A.~Zardecki.
\newblock \enquote{Modeling in chaotic relativity.}
\newblock \emph{\prd} \textbf{28}, 1235 (1983).

\bibitem{1985PhRvD..31..718Z}
A.~{Zardecki}.
\newblock \enquote{{Inflation and reheating in Bianchi type-IX cosmology}.}
\newblock \emph{\prd} \textbf{31}, 718 (1985).

\bibitem{Zel-Nov}
I.~B. Zeldovich and I.~D. Novikov.
\newblock \emph{{Relativistic astrophysics. Volume 2. The structure and
  evolution of the universe (Revised and enlarged edition)}} (University of
  Chicago Press, Chicago, IL, 1983).

\bibitem{1970JETPL..11..197Z}
Y.~B. {Zel'Dovich}.
\newblock \enquote{{''Mixmaster universe'' and the cold variant.}}
\newblock \emph{Soviet Journal of Experimental and Theoretical Physics Letters}
  \textbf{11}, 197 (1970).

\end{thebibliography}
\end{document}